\pdfoutput=1
\documentclass[cernpreprint,texlive=2011,txfonts,UKenglish]{latex/atlasdoc}

\usepackage{latex/atlaspackage}

\usepackage{latex/atlasphysics}

\graphicspath{{logos/}{figures/}}

\usepackage{Coupling2015-defs}

\usepackage{latex/atlascover}
\usepackage{multirow}
\usepackage{rotating}
\usepackage{amssymb}
\usepackage{tabularx}
\usepackage{tabularew}
\usepackage{adjustbox}
\usepackage{cite}
\hypersetup{pdftitle={ATLAS+CMS draft},pdfauthor={The ATLAS and CMS Collaborations}}

\AtlasTitle{Measurements of the Higgs boson production and decay rates and constraints on its couplings from a combined ATLAS and CMS analysis of the LHC $pp$ collision data at $\sqrt{s}=$ 7 and 8~TeV}

\author{The ATLAS and CMS Collaborations}

\date{\today}

\AtlasRefCode{ATLAS-HIGG-2015-07 }
\AtlasJournalRef{JHEP08(2016)045}
\AtlasDOI{10.1007/JHEP08(2016)045}

 \PreprintIdNumber{CERN-EP-2016-100}

\AtlasAbstract{Combined ATLAS and CMS measurements of the Higgs boson production and decay rates, as well as constraints on its couplings to vector bosons and fermions, are presented. The combination is based on the analysis of five production processes, namely gluon fusion, vector boson fusion, and associated production with a $W$ or a $Z$ boson or a pair of top quarks, and of the six decay modes $H \to ZZ, WW$, $\gamma\gamma, \tau\tau, \bb$, and $\mu\mu$. All results are reported assuming a value of~$125.09$~GeV for the Higgs boson mass, the result of the combined measurement by the ATLAS and CMS experiments. The analysis uses the CERN LHC proton--proton collision data recorded by the ATLAS and CMS experiments in 2011 and 2012, corresponding to integrated luminosities per experiment of approximately~5~\ifb\ at $\sqrt{s}=7$~TeV and~20~\ifb\ at $\sqrt{s} = 8$~TeV. The Higgs boson production and decay rates measured by the two experiments are combined within the context of three generic parameterisations: two based on cross sections and branching fractions, and one on ratios of coupling modifiers. Several interpretations of the measurements with more model-dependent parameterisations are also given. The combined signal yield relative to the Standard Model prediction is measured to be $1.09 \pm 0.11$. The combined measurements lead to observed significances for the vector boson fusion production process and for the $\Htt$~decay of~$5.4$ and $5.5$~standard deviations, respectively. The data are consistent with the Standard Model predictions for all parameterisations considered.
}

 \AtlasCoverSupportingNote{Supporting note ATLAS-CMS Higgs coupling combination}{https://cds.cern.ch/record/2042100}
\AtlasCoverCommentsDeadline{May 10, 2016}

\AtlasCoverAnalysisTeam{
T. Adye,  A. Armbruster, O. Arnaez, S. Banerjee, P. Bortignon, M. Chen, A. David, M. De Gruttola, M. Duehrssen, E. Feng, D. Froidevaux, S. Gadatsch, A. Gilbert, D. Gillberg, G. Gomez Ceballos, E. Gross, R. Harrington, J. Heikkil\"a, A. Holzner, X. Janssen, H. Li, C. Mariotti, A. Massironi, N. Morange, R. Mudd, P. Musella, G. Petrucciani, M. Pieri, D. Puigh, T. Vazquez Schroeder, R. Tanaka, N. Wardle, H. Yang, F. Zhang
}
\AtlasCoverEdBoardMember{
N. Berger, A. Nisati~(chair), G. Redlinger, M. Schumacher }

\AtlasCoverEgroupEditors{atlas-higg-2015-07-editors@cern.ch}

\AtlasCoverEgroupEdBoard{atlas-HIGG-2015-07-editorial-board@cern.ch}

\begin{document}
\author{The ATLAS and CMS Collaborations}

\maketitle

\newpage
\section{Introduction}
\label{sec:Introduction}

The elucidation of the mechanism of electroweak (EW) symmetry breaking has been one of the main goals driving the design of the ATLAS~\cite{ATLASdetector} and CMS~\cite{CMSdetector} experiments at the CERN LHC.
In the Standard Model (SM) of particle physics~\cite{Glashow1961579,Weinberg19671264,Salam1968367,tHooft:1972fi}, the breaking of the symmetry is achieved through the introduction of a complex doublet scalar field, leading
to the prediction of the existence of one physical neutral scalar particle, commonly known as the Higgs boson~\cite{Englert:1964et,Higgs:1964ia,Higgs:1964pj,Guralnik:1964eu,Higgs:1966ev,Kibble:1967sv}. 
Through Yukawa interactions, the Higgs scalar field can also account for fermion masses~\cite{Weinberg19671264,Nambu:1961fr}. 
While the SM does not predict the value of the Higgs boson mass, $m_H$, the production cross sections and decay branching fractions~(\BR) of the Higgs boson can be precisely calculated once its mass is known.

In 2012, the ATLAS and CMS Collaborations reported the observation of a new particle with a mass of approximately 125~GeV and Higgs-boson-like properties~\cite{ATLAS:2012obs,CMS:2012obs,CMSLong2013}. 
Subsequent results from both experiments, summarised in~Refs.~\cite{ATLAS_combination,CMS_combination,Chatrchyan:2012jja,atlas_spin_paper,Khachatryan:2014kca}, established that all measurements of the properties of the new particle, including its spin, CP~properties, and coupling strengths to SM~particles, are consistent within the uncertainties with those expected for the SM~Higgs boson. ATLAS and CMS have published a combined measurement of the Higgs boson mass~\cite{ATLASCMSHmass}, using  LHC~\runone\ data for the \Hyy\ and \HZZ\ channels, where \runone\ indicates the LHC proton--proton~($pp$) data taking period in 2011 and 2012 at centre-of-mass energies $\sqrt{s} = 7$ and 8~TeV.
The combined mass measurement is
\begin{linenomath}\begin{equation}
  \label{final_mHresult}
    \mH=125.09\pm0.21 (\mathrm{stat.}) \pm 0.11 (\mathrm{syst.}) \mathrm{\gev},
\end{equation}\end{linenomath}
where the total uncertainty is dominated by the statistical component.
The Higgs boson mass is assumed to be $\mH = 125.09$~GeV for all analyses presented in this paper.

This paper reports the first ATLAS and CMS combined measurements of the Higgs boson production and decay rates as well as constraints on its couplings to SM~particles. These measurements yield the most precise and comprehensive experimental results on these quantities to date.
The main production processes studied are gluon fusion (\aggF), vector boson fusion~(\aVBF), and associated production with vector bosons (\aWH~and~\aZH, denoted together as~\aVH) or a pair of top quarks (\attH). 
The decay channels considered are those to bosons, $\HZZ$, $\HWW$, and $\Hyy$; and to fermions, $\Htt$, $\Hbb$, and $\Hmm$. Throughout this paper, $Z$ and $W$ indicate both real and virtual vector bosons, and no distinction is made between particles and antiparticles.

All analyses used in the combination are based on the complete \runone\ collision data collected by the ATLAS and CMS experiments. These data correspond to integrated luminosities per experiment of approximately~5~\ifb\ at~$\sqrt{s} =~7$~TeV (recorded in~2011) and~20~\ifb\ at~$\sqrt{s} = 8$~TeV (recorded in~2012). The results of the ATLAS and CMS individual combinations based on the \runone\ data are reported in Refs.~\cite{ATLAS_combination,CMS_combination}. 

Unless otherwise stated, in this paper it is assumed, as in~Refs.~\cite{ATLAS_combination,CMS_combination}, that the particle under study is a single SM-like Higgs boson state, i.e.~a CP-even scalar particle with the tensor coupling structure of the SM for its interactions.
The Higgs boson width, predicted to be approximately 4 MeV in the SM, is assumed to be small enough that the narrow-width approximation is valid and that the Higgs boson production and decay mechanisms can be factorised. These assumptions are corroborated by tests of the spin and CP properties of the Higgs boson~\cite{atlas_spin_paper,Khachatryan:2014kca} and by studies of its width~\cite{CMS_combination,Aad:2015xua,CMSoffshell,Khachatryan:2015mma}. The Higgs boson signal modelling is based on the hypothesis of a SM Higgs boson in terms of its production and decay kinematics. Measurements of differential production cross sections~\cite{Aad:2014lwa,Aad:2014tca,Khachatryan:2015rxa,Khachatryan:2015yvw} support these assumptions within the current statistical uncertainties. The inherent model dependence related to these hypotheses applies to all results presented here; the reliance on this model has a negligible impact for small deviations from the~SM, but  could be important for significant deviations from the SM~predictions.

The results presented here for each experiment separately are slightly different from those reported in~Refs.~\cite{ATLAS_combination,CMS_combination}. 
Some small variations with respect to the earlier results are related to a different choice for the value of the Higgs boson mass.
Other differences 
arise from minor modifications to the signal parameterisation and to the treatment of systematic uncertainties. 
These modifications are introduced in the present analysis to allow a fully consistent and correlated treatment of the dominant theoretical uncertainties in the signal modelling between the two experiments.

This paper is organised as described below. Section~\ref{sec:TheoryFramework} briefly reviews the theoretical calculations of Higgs boson production and decay, and  the modelling of the Higgs boson signal in Monte Carlo (MC) simulation; it also introduces the formalisms of signal strengths and coupling modifiers used for the interpretation of the data. Section~\ref{sec:CombinationProcedure}  gives an overview of the analyses included in this combination, describes the statistical procedure used, together with the treatment of systematic uncertainties, and summarises modifications to the individual analyses for the combination. Section~\ref{sec:GenericParameterisation} describes the parameterisation of the measured signal yields in generic terms and reports the results using three distinct parameterisations. Section~\ref{sec:SignalStrength} compares the measured Higgs boson yields to the SM~predictions for different production processes and decay modes, and reports the results of a test for the possible presence of multiple mass-degenerate states. Section~\ref{sec:CouplingFits} studies the couplings of the Higgs boson to probe for possible deviations from the SM~predictions, using various assumptions motivated in many cases by beyond the~SM~(BSM)~physics scenarios. Finally, Section~\ref{sec:Conclusion} presents a summary of the results.

\section{Higgs boson phenomenology and interpretation framework}
\label{sec:TheoryFramework}

This section briefly reviews Higgs boson phenomenology and introduces the most important aspects of the interpretation framework used to combine the measurements and to assess their compatibility with the SM~predictions. The dominant production processes and major decay modes of the SM Higgs boson, along with the theoretical predictions for the cross sections and branching fractions, are presented. 
The main features of the MC generators used to simulate Higgs boson production and decay in each experiment are described. 
Finally, the formalisms of two widely used frameworks, based on signal strengths and coupling modifiers, for the interpretation of the Higgs boson measurements at the LHC, are introduced.

\subsection{Higgs boson production and decay}
\label{sec:Theory}

In the SM, Higgs boson production at the LHC mainly occurs through the following processes, listed in order of decreasing cross section at the \runone\ centre-of-mass energies:
\begin{itemize}
\item gluon fusion production $gg\to H$ (Fig.~\ref{fig:feyn_ggFVBF}a);
\item vector boson fusion production $qq \to qqH$ (Fig.~\ref{fig:feyn_ggFVBF}b);
\item associated production with a $W$ boson, $qq \to WH$ (Fig.~\ref{fig:feyn_prod}a), or with a $Z$ boson, $pp \to ZH$, including a small~($\sim 8\%$) but less precisely known contribution from $gg \to ZH$~(\aggZH) (Figs.~\ref{fig:feyn_prod}a, \ref{fig:feyn_prod}b, and~\ref{fig:feyn_prod}c);
\item associated production with a pair of top quarks, $qq,gg \to ttH$ (Fig.~\ref{fig:feyn_ttH}).
\end{itemize}

\begin{figure}[hbt]
\centering
\begin{tabular}{cc}
    \includegraphics[width=0.25\textwidth]{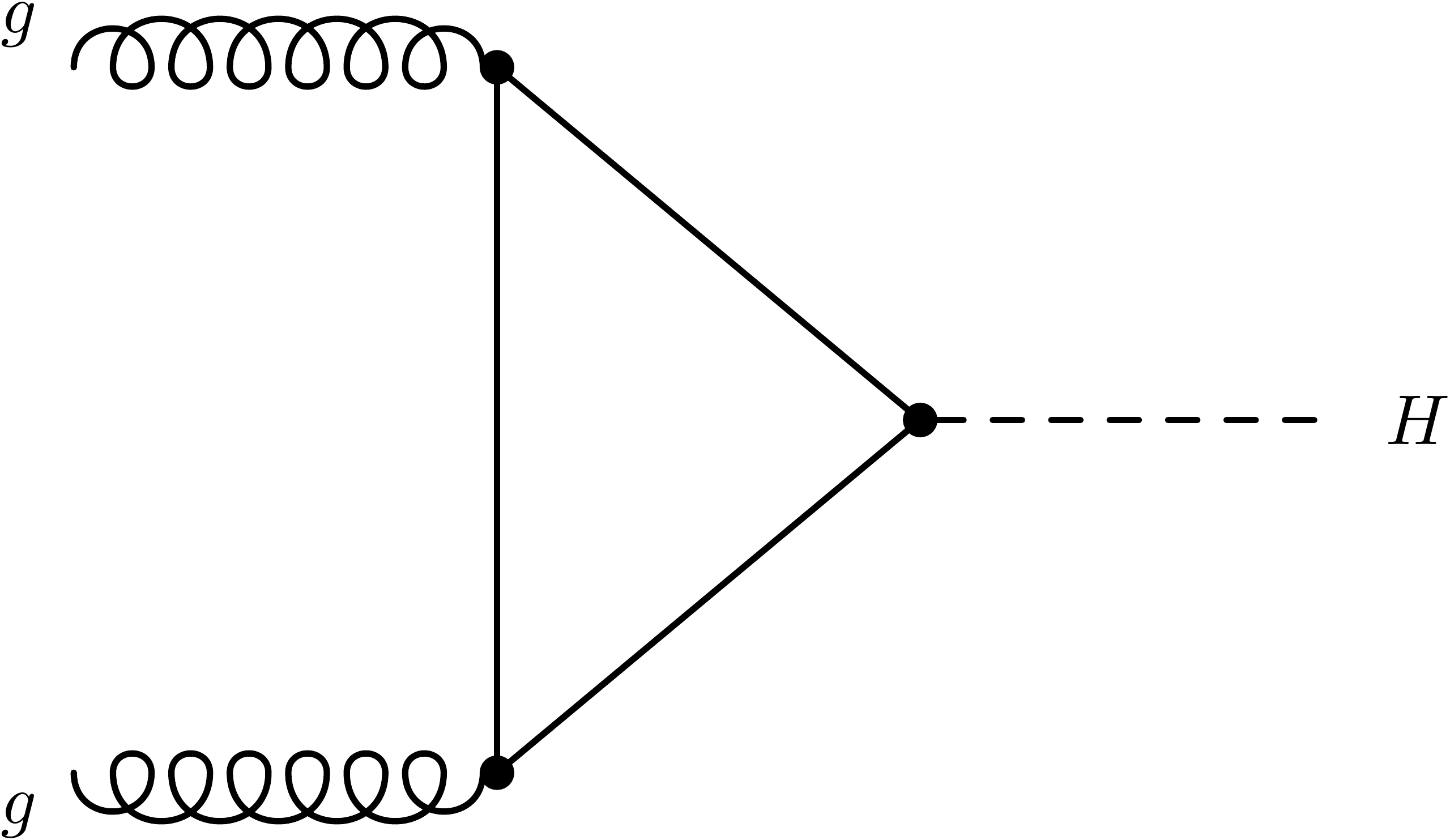}&
    \includegraphics[width=0.25\textwidth]{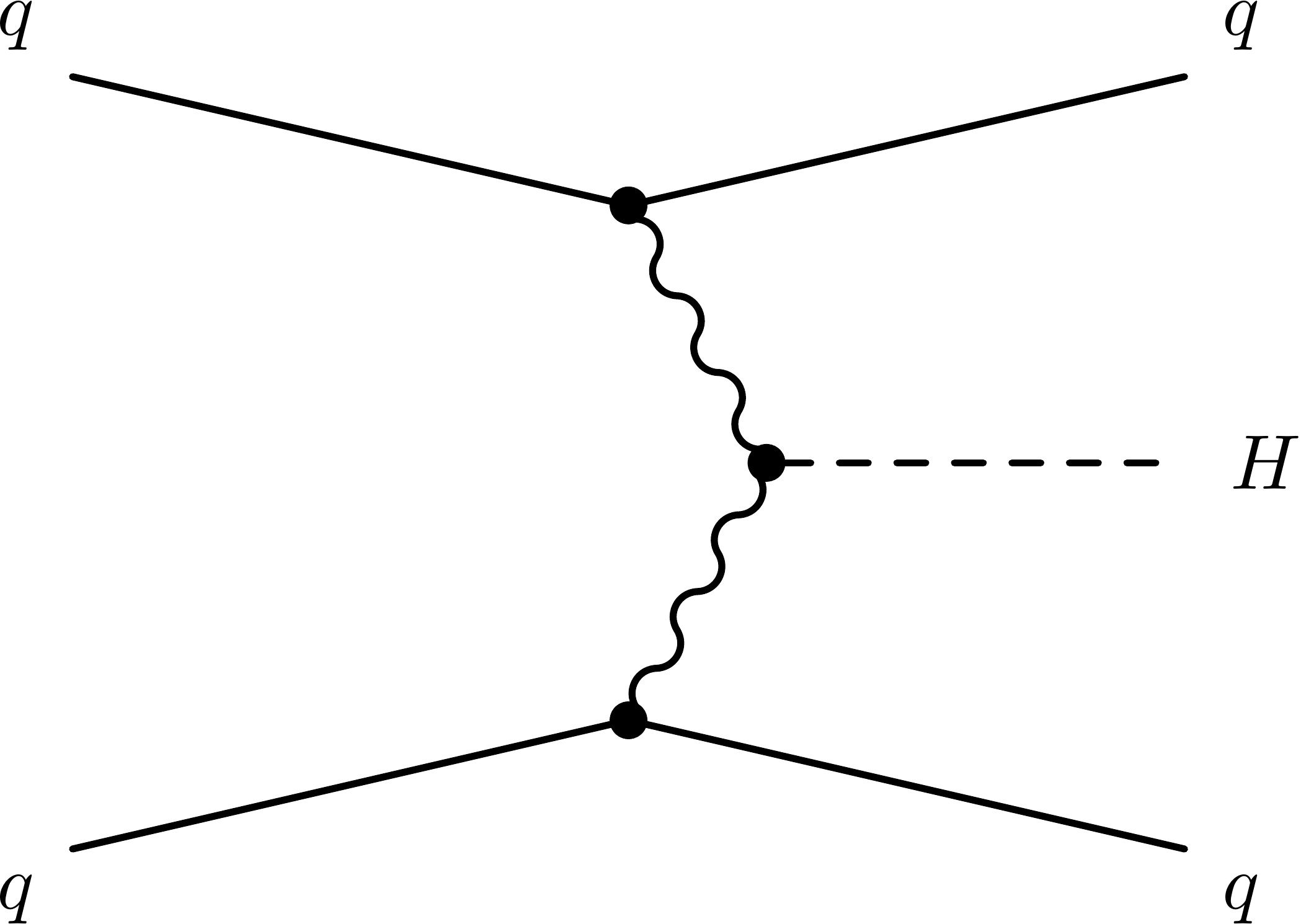}\\
    (a)&(b)\\
\end{tabular}
\caption{Examples of leading-order Feynman diagrams for Higgs boson production via the (a) \aggF\ and (b) \aVBF\ production processes.}
\label{fig:feyn_ggFVBF}
\end{figure}

\begin{figure}[hbt]
\centering
\begin{tabular}{ccc}
    \includegraphics[width=0.25\textwidth]{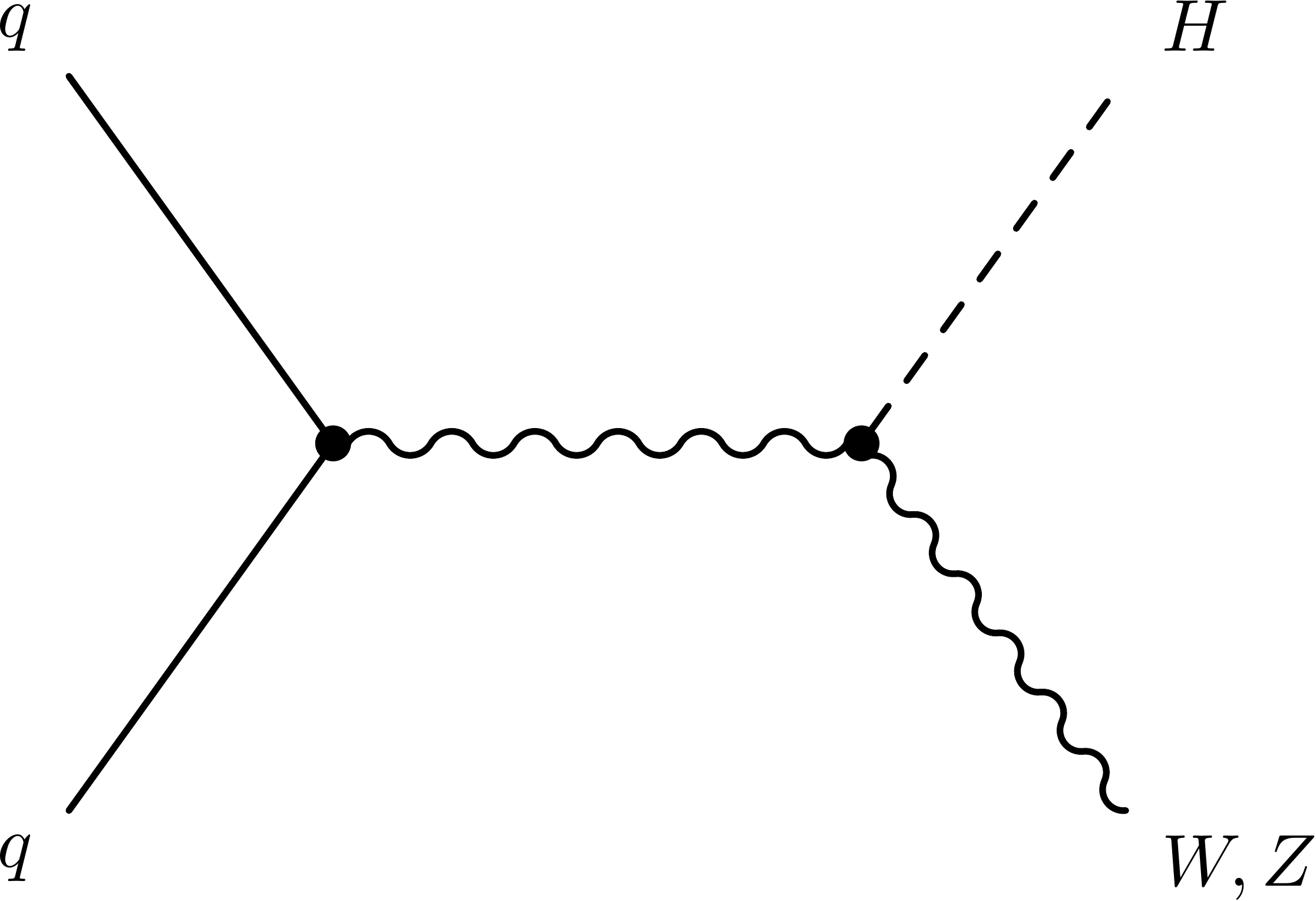}&
    \includegraphics[width=0.25\textwidth]{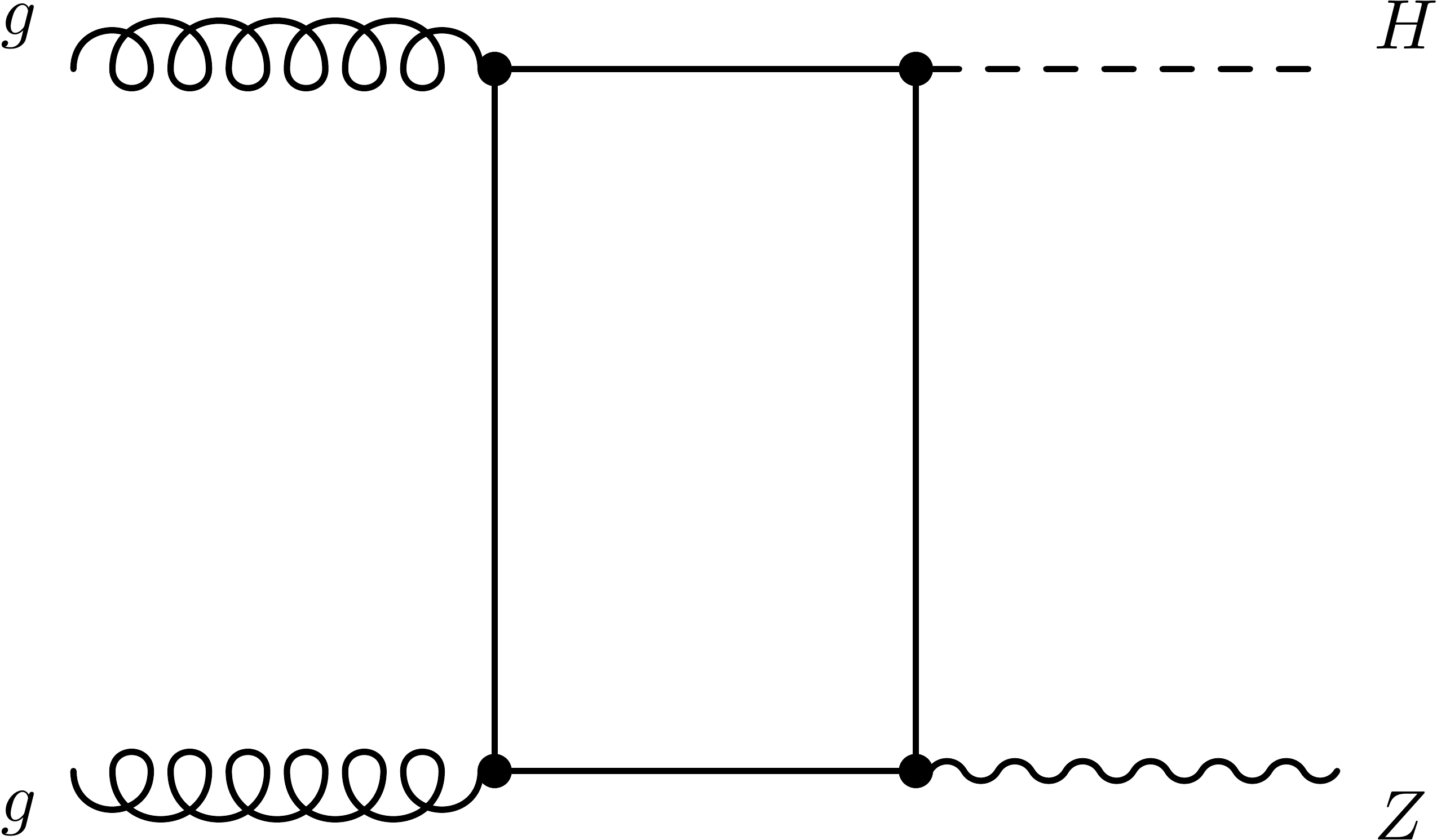}&
    \includegraphics[width=0.25\textwidth]{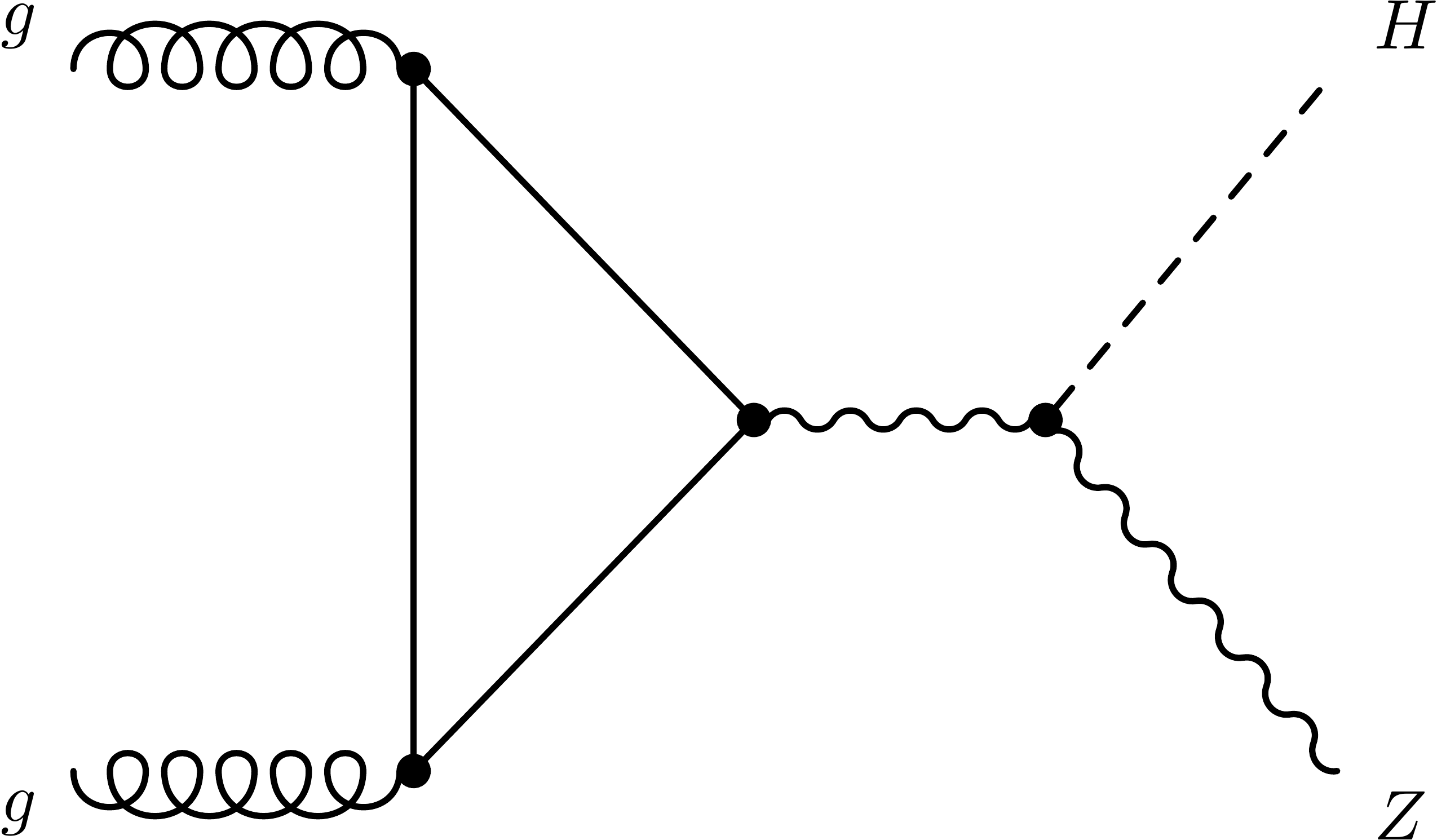}\\
    (a)&(b)&(c)\\
\end{tabular}
  \caption{Examples of leading-order Feynman diagrams for Higgs boson production via the (a) $qq \to VH$ and (b,~c)~\ggZH\ production processes.}
\label{fig:feyn_prod}
\end{figure}

\begin{figure}[hbt]
\centering
\begin{tabular}{ccc}
    \includegraphics[width=0.25\textwidth]{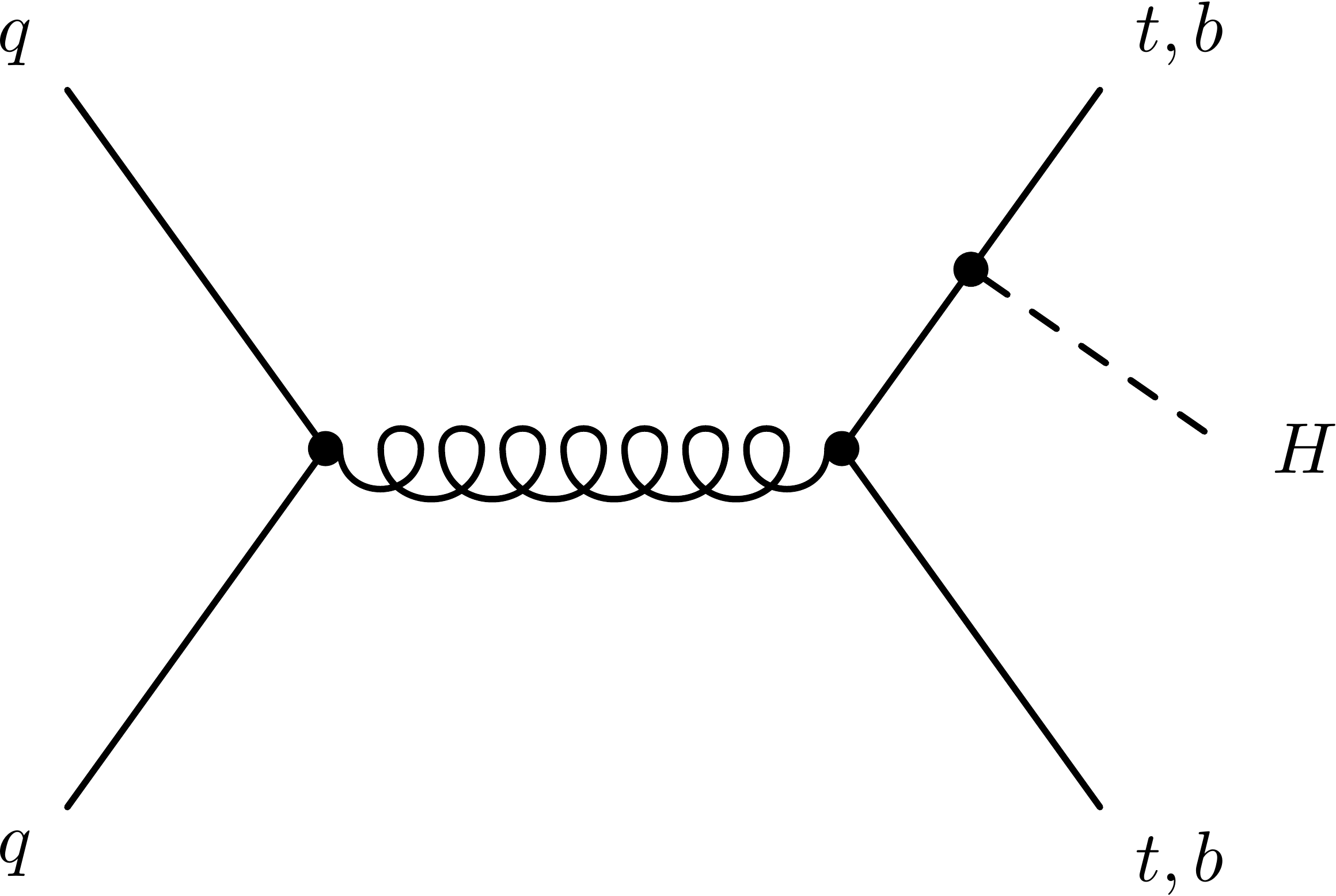}&
    \includegraphics[width=0.25\textwidth]{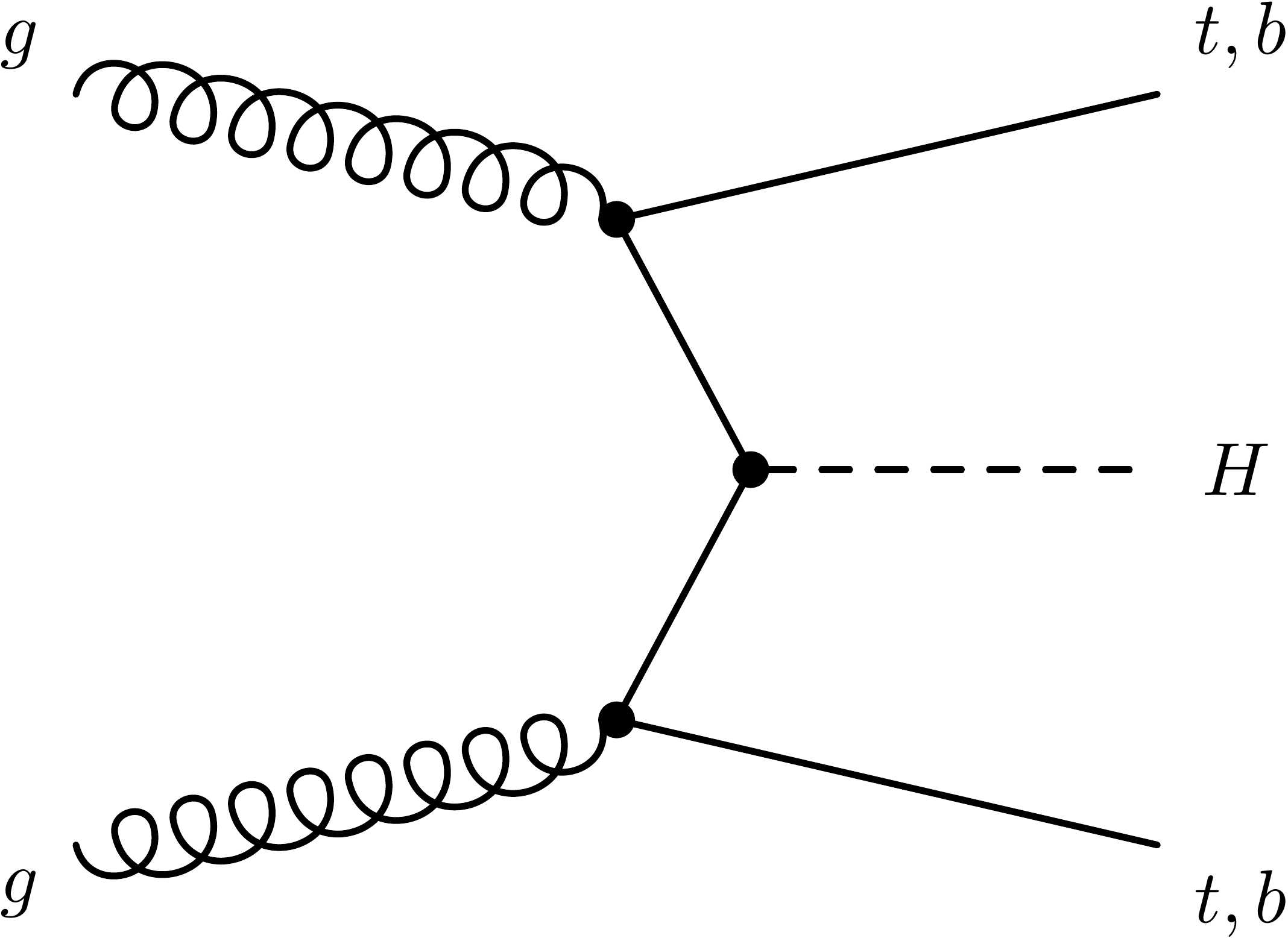}&
    \includegraphics[width=0.25\textwidth]{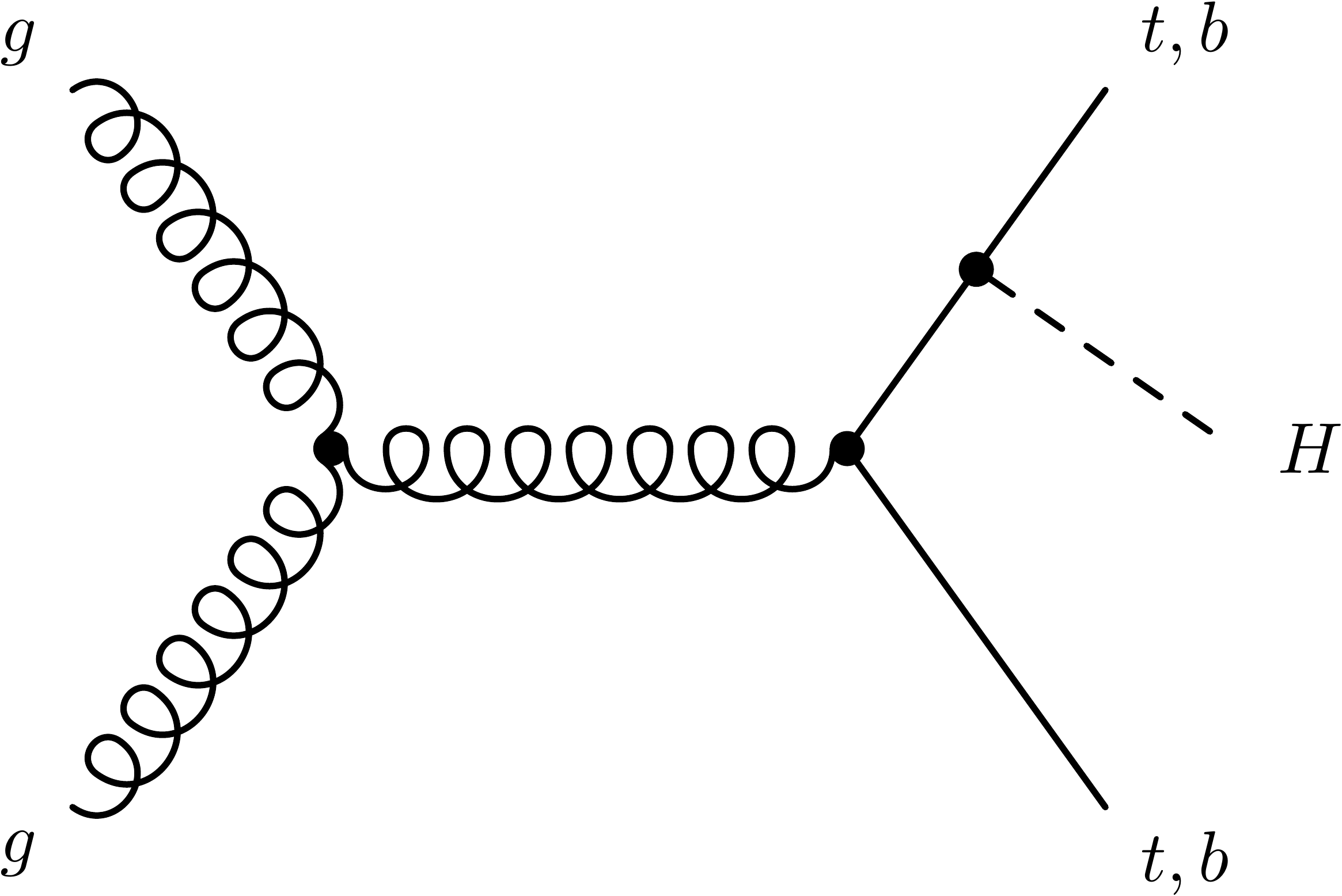}\\
    (a)&(b)&(c)\\
\end{tabular}
\caption{Examples of leading-order Feynman diagrams for Higgs boson production via the \ttHnobar\ and \bbHnobar\ processes.}
\label{fig:feyn_ttH}
\end{figure}

Other less important production processes in the SM, which are not the target of a direct search but are included in the combination, are $qq,gg \to bbH$~(\abbH), also shown in Fig.~\ref{fig:feyn_ttH}, and production in association with a single top quark (\atH), shown in Fig.~\ref{fig:feyn_tH}. 
The latter process proceeds through either $qq/qb \to tHb/tHq^{\prime}$ (\atHq) (Figs.~\ref{fig:feyn_tH}a and~\ref{fig:feyn_tH}b) or $gb \to tHW$ (\atHW) (Figs.~\ref{fig:feyn_tH}c and~\ref{fig:feyn_tH}d) production. 

Examples of leading-order~(LO) Feynman diagrams for the Higgs boson decays considered in the combination are shown in Figs.~\ref{fig:feyn_hVVff} and~\ref{fig:feyn_hgg}. 
The decays to $W$ and $Z$ bosons (Fig.~\ref{fig:feyn_hVVff}a) and to fermions (Fig.~\ref{fig:feyn_hVVff}b) proceed through tree-level processes whereas the $\Hyy$ decay is mediated by $W$ boson or heavy quark loops (Fig.~\ref{fig:feyn_hgg}).

\begin{figure}[hbt]
\centering
\begin{tabular}{cccc}
    \includegraphics[width=0.225\textwidth]{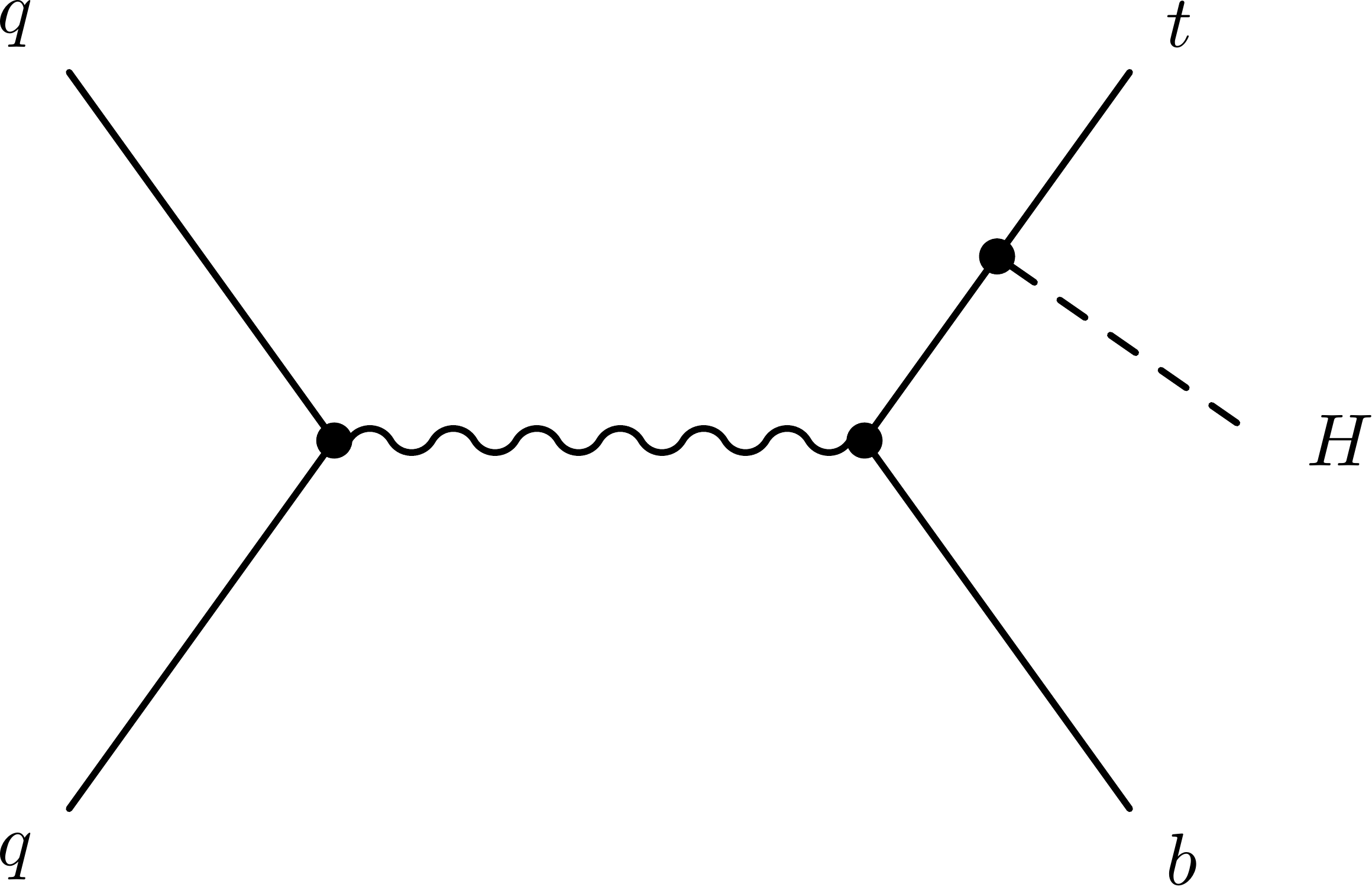}&
    \includegraphics[width=0.225\textwidth]{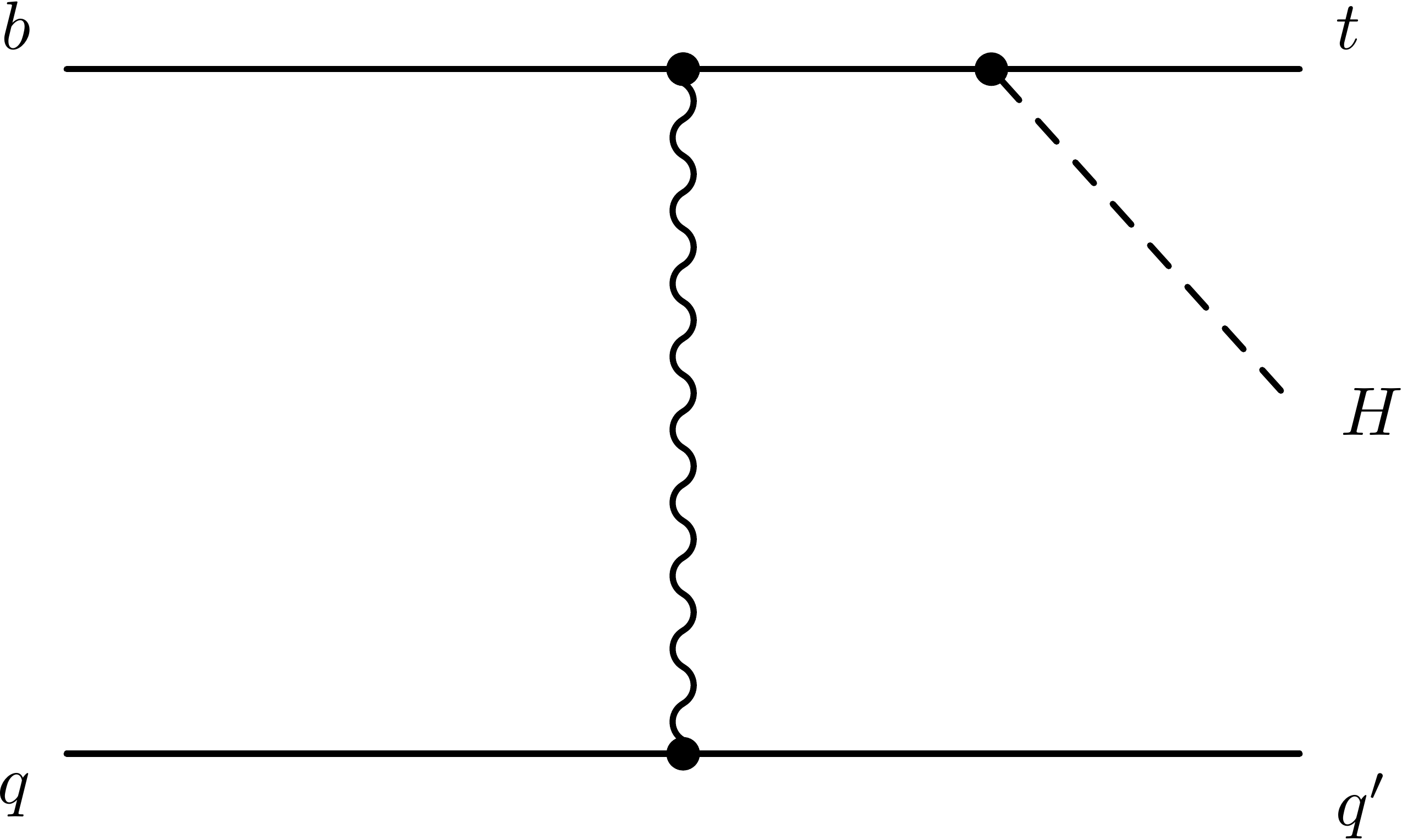}&
    \includegraphics[width=0.225\textwidth]{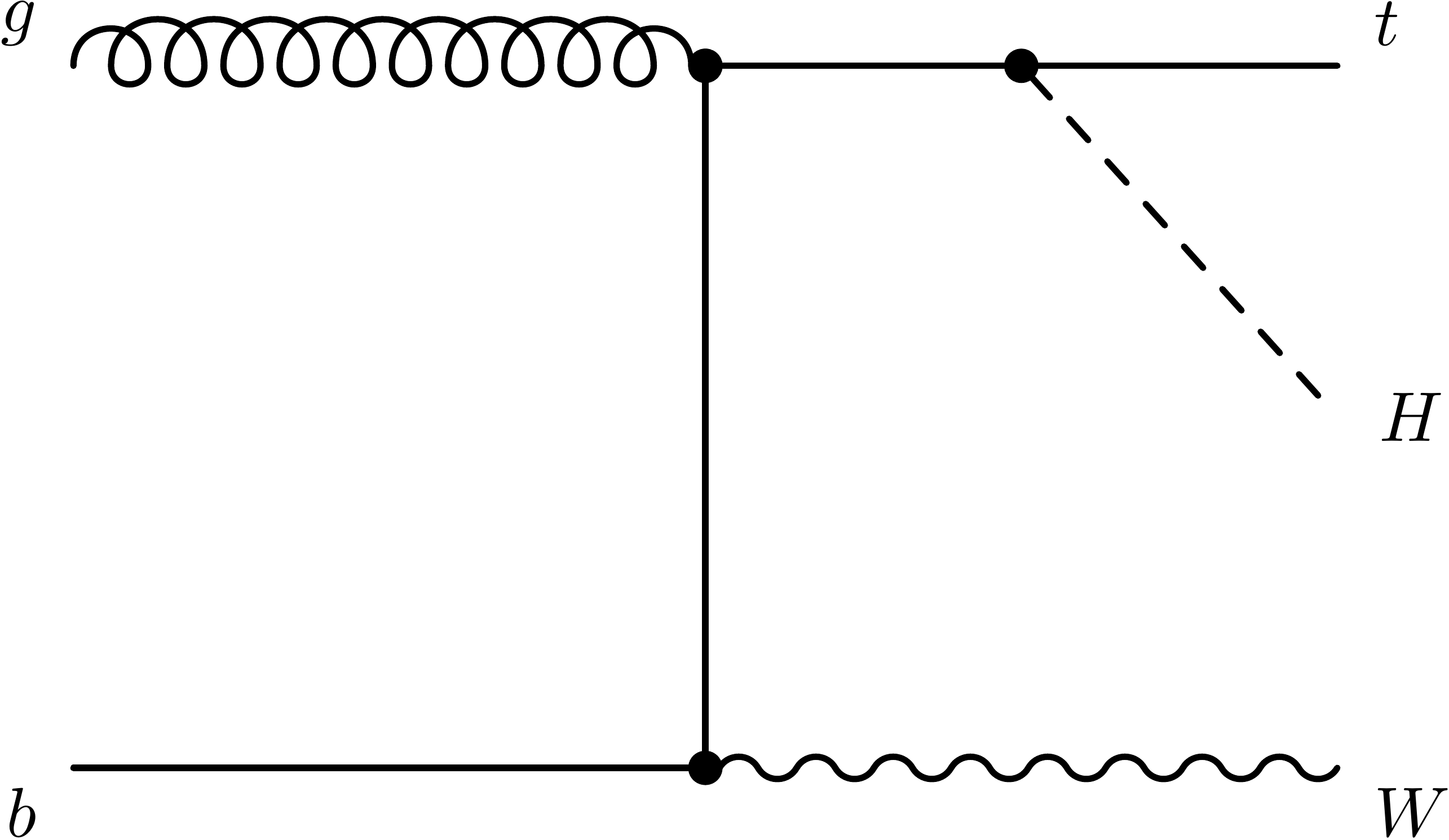}&
    \includegraphics[width=0.225\textwidth]{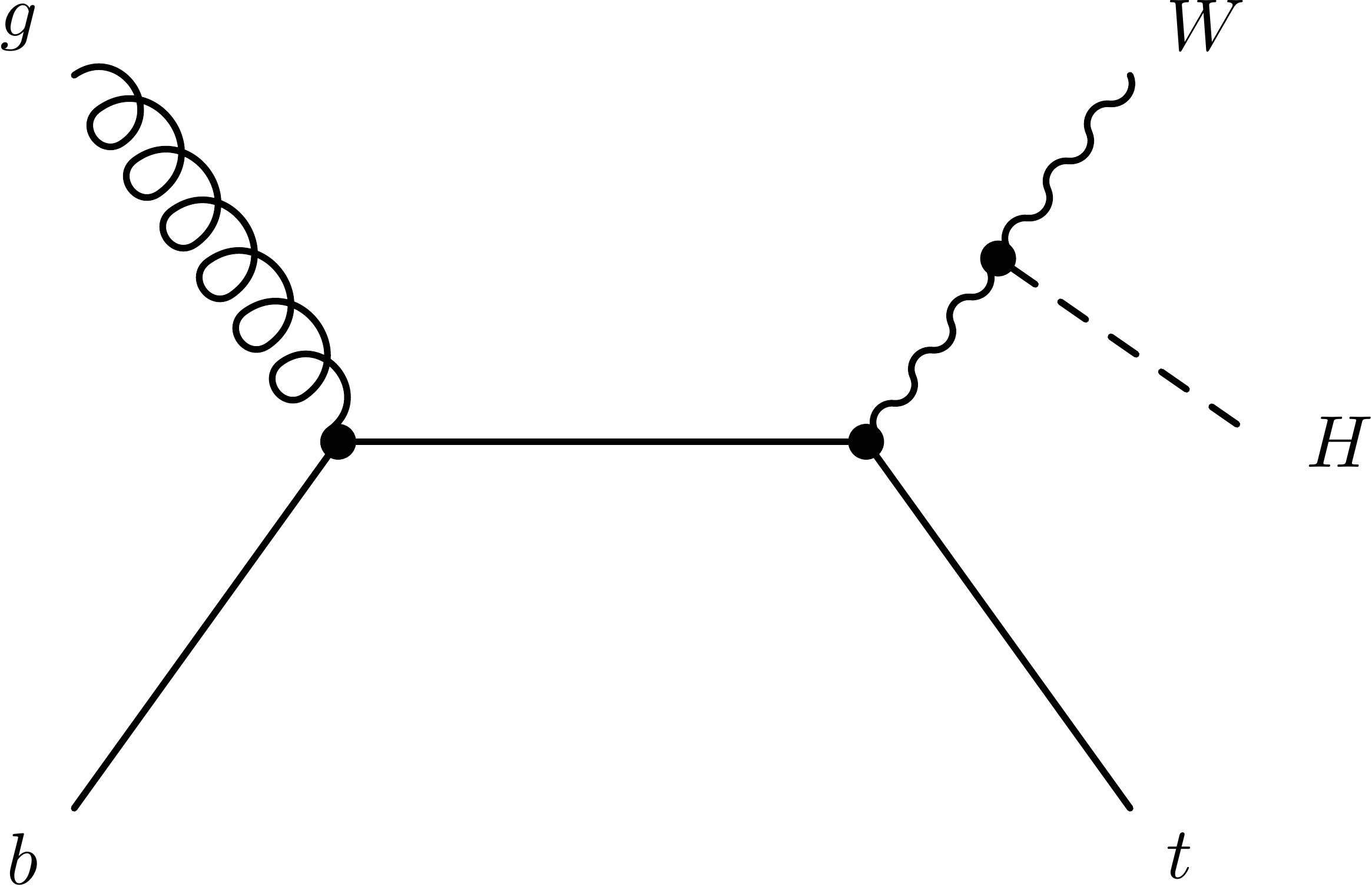}\\
    (a)&(b)&(c)&(d)\\
\end{tabular}
\caption{Examples of leading-order Feynman diagrams for Higgs boson production in association with a single top quark via the (a,~b)~\atHq\ and (c,~d)~\atHW\ production processes.}
\label{fig:feyn_tH}
\end{figure}

\begin{figure}[hbt]
\centering
\begin{tabular}{cc}
    \includegraphics[width=0.25\textwidth]{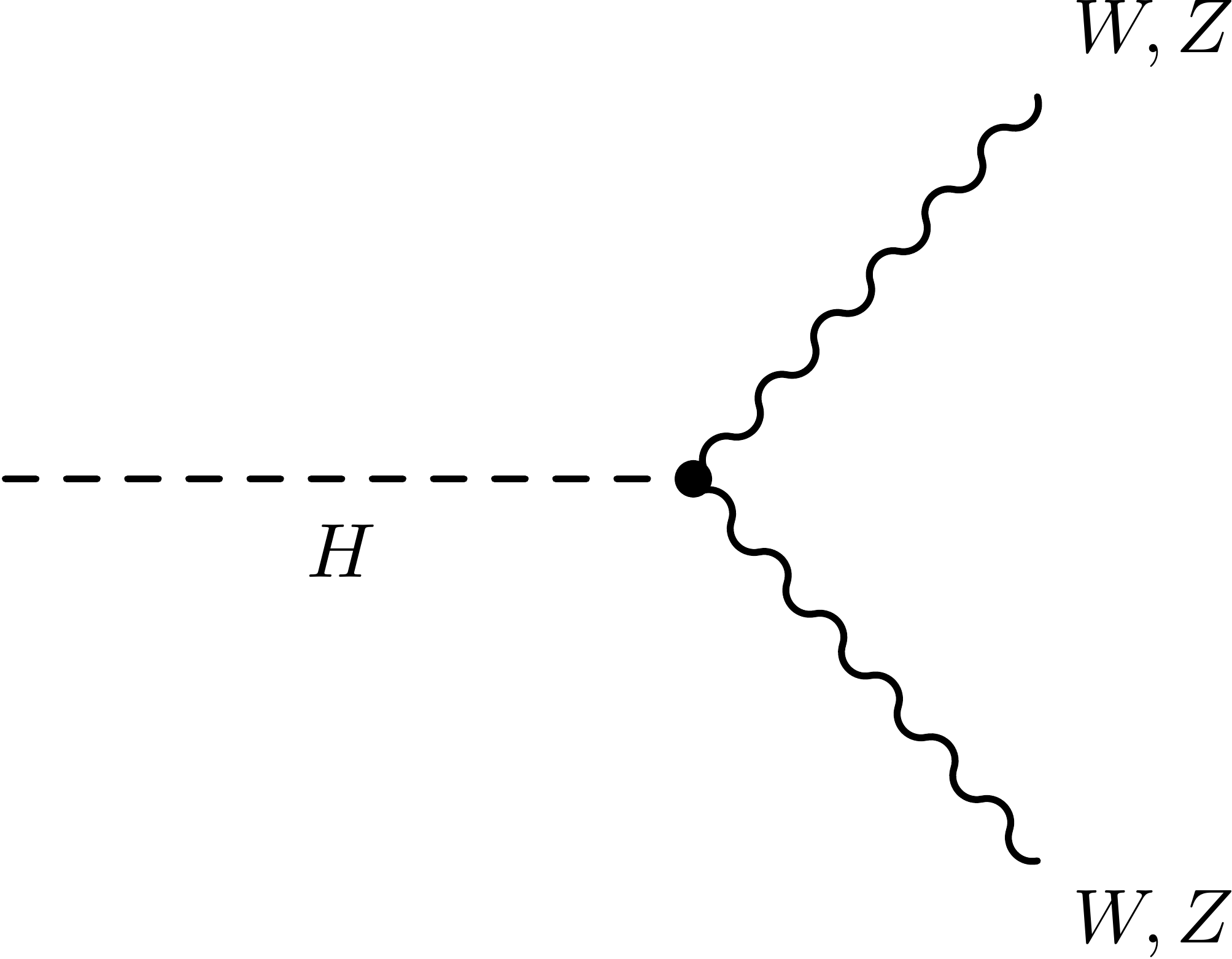}&
    \includegraphics[width=0.25\textwidth]{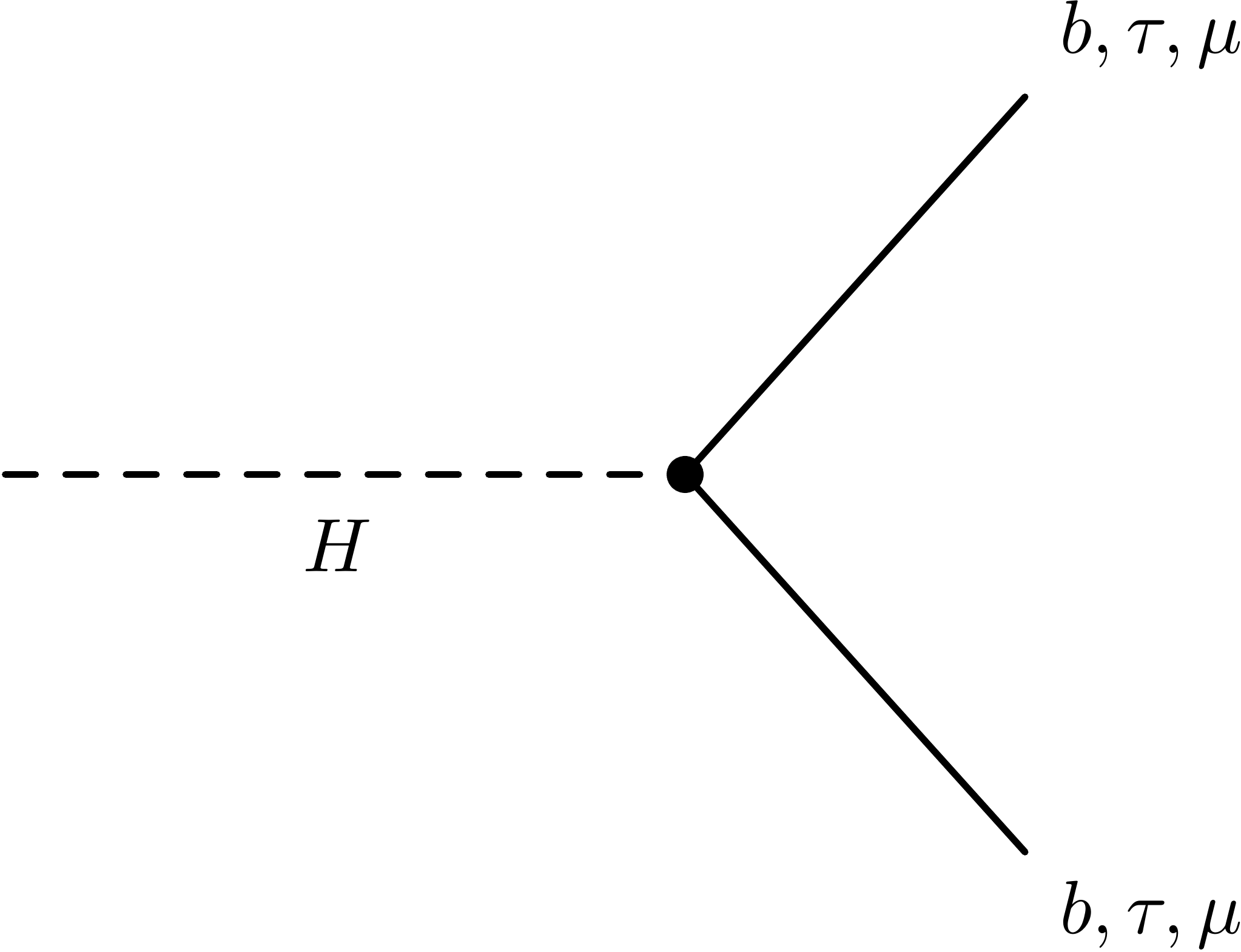}\\
    (a)&(b)\\
\end{tabular}
\caption{Examples of leading-order Feynman diagrams for Higgs boson decays (a) to $W$ and $Z$ bosons and (b) to fermions.}
\label{fig:feyn_hVVff}
\end{figure}

\begin{figure}[hbt!]
\centering
\begin{tabular}{ccc}
    \includegraphics[width=0.25\textwidth]{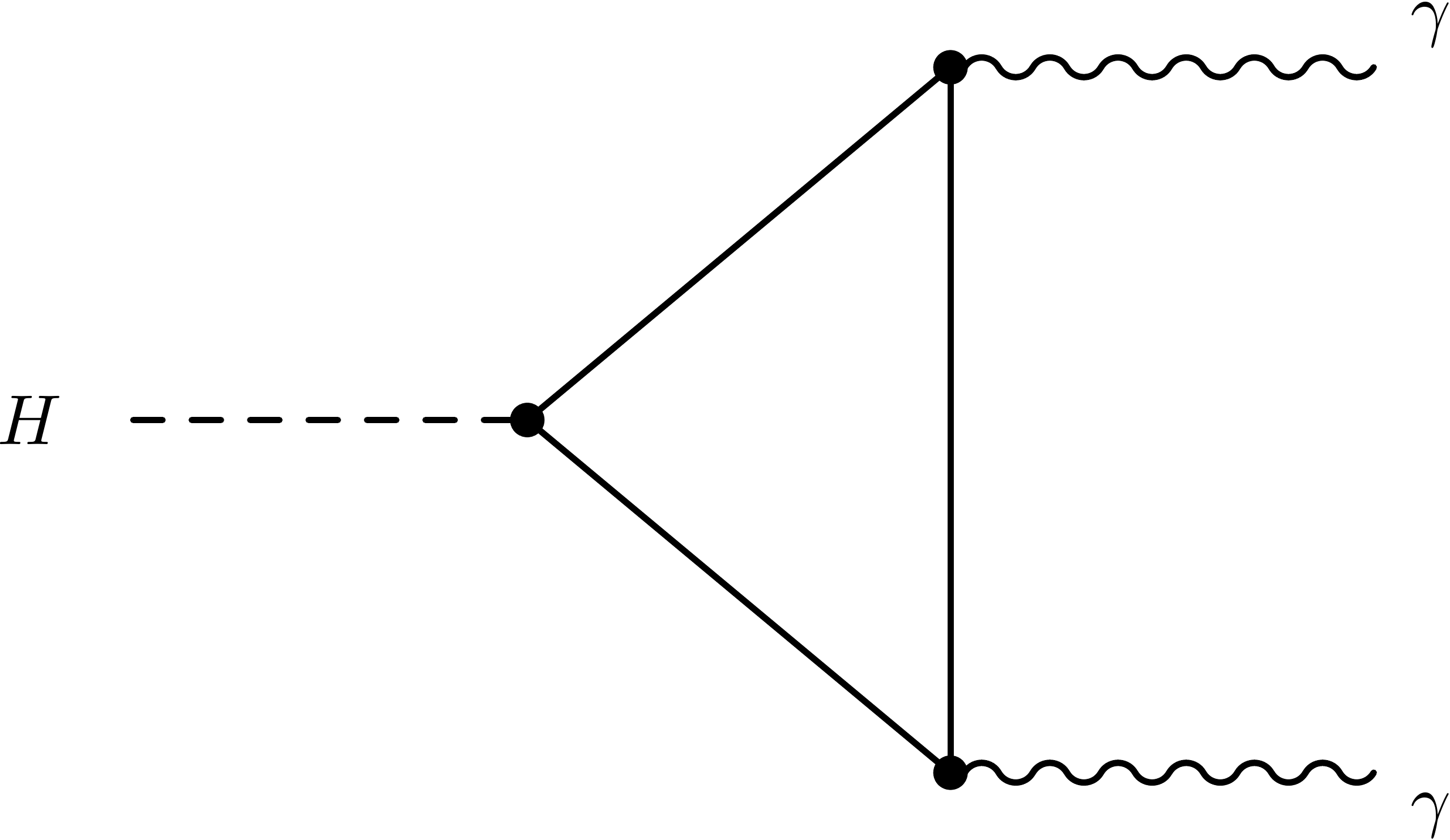}&
    \includegraphics[width=0.25\textwidth]{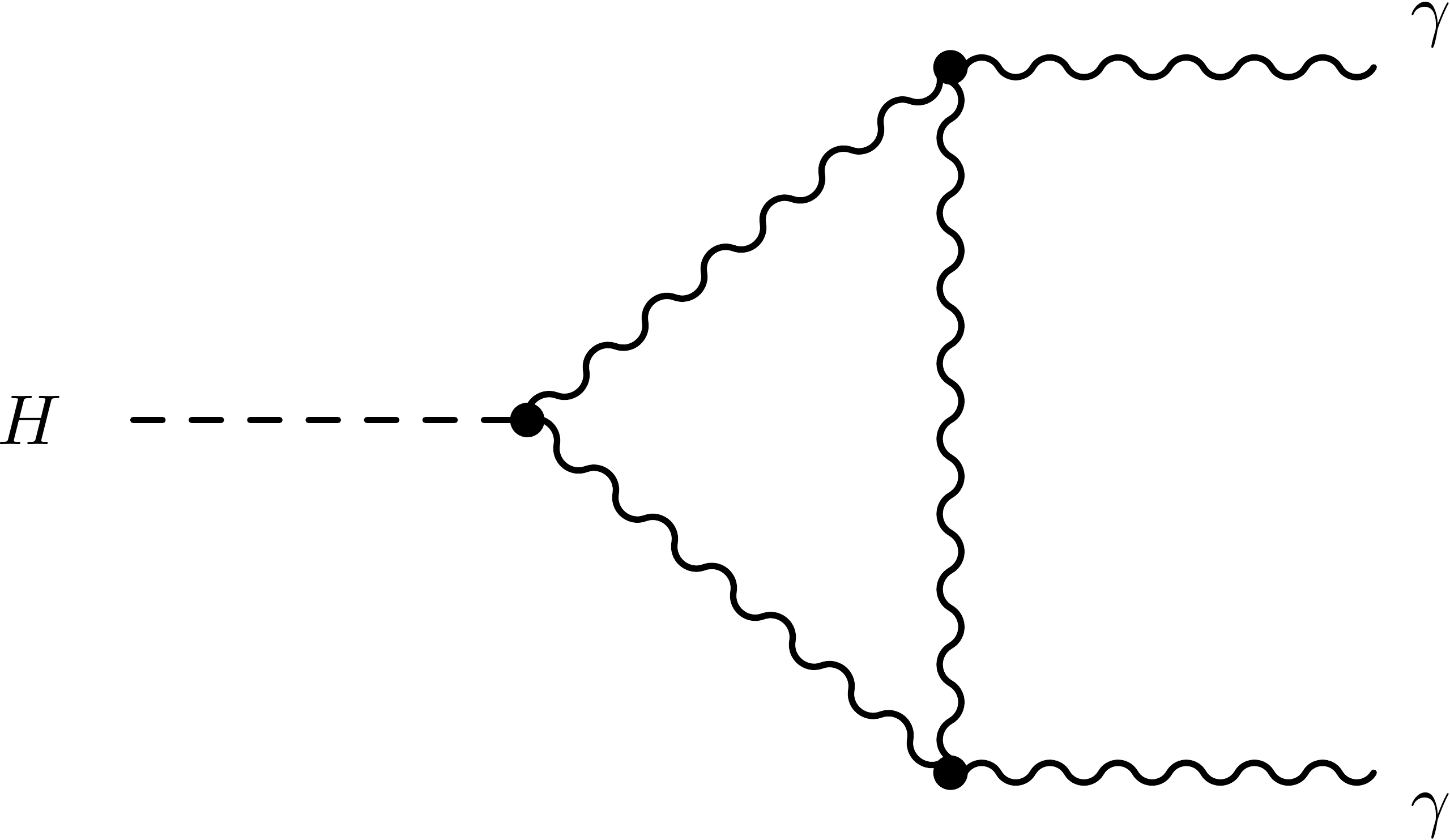}&
    \includegraphics[width=0.25\textwidth]{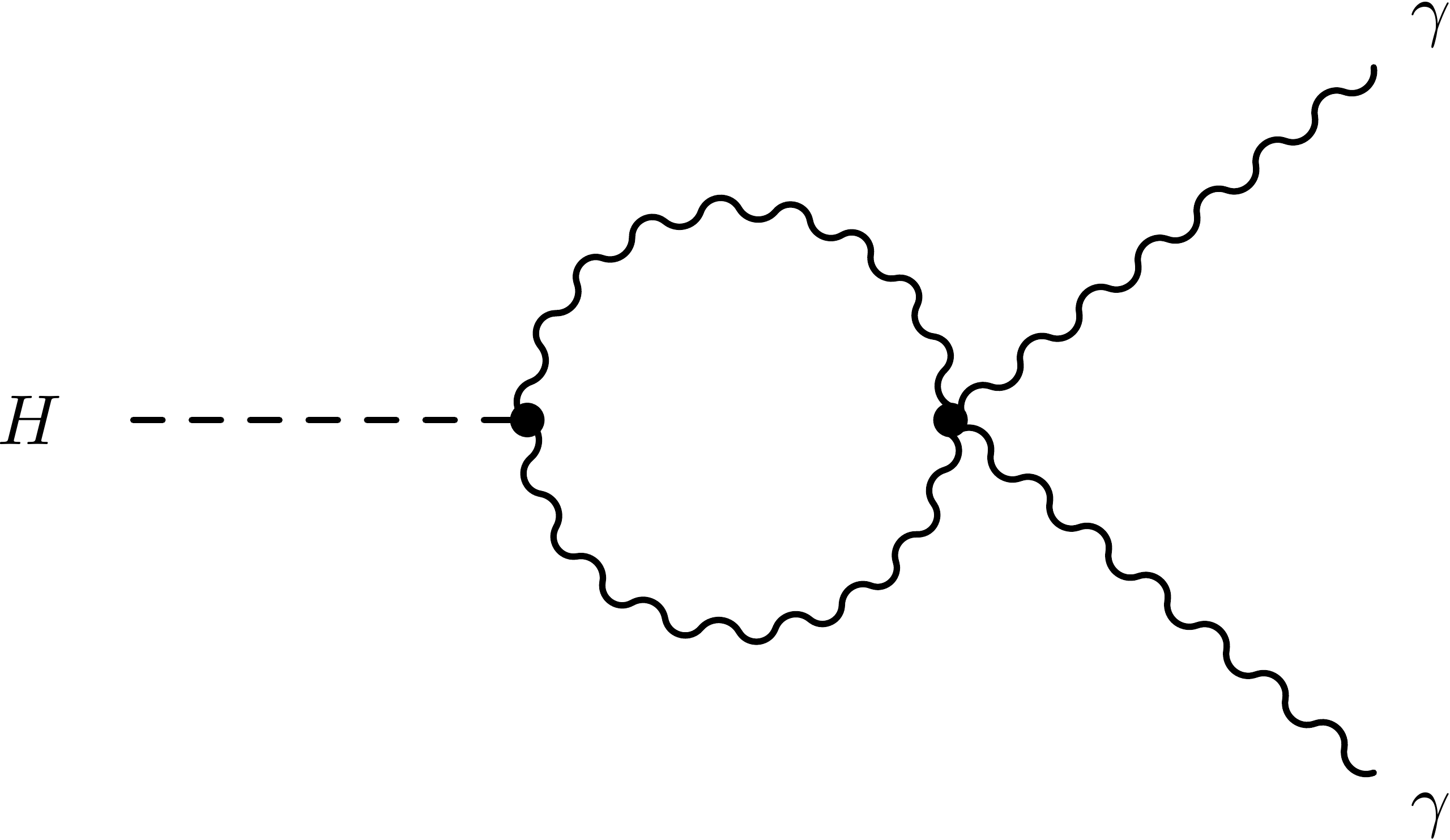}\\
     & & \\
    (a)&(b)&(c)\\
\end{tabular}
  \caption{Examples of leading-order Feynman diagrams for Higgs boson decays to a pair of photons.}
  \label{fig:feyn_hgg}
\end{figure}

The SM Higgs boson production cross sections and decay branching fractions are taken from Refs.~\cite{Dittmaier:2011ti,Dittmaier:2012vm,Heinemeyer:2013tqa} and are based on the extensive theoretical work documented 
in Refs.~\cite{Georgi:1977gs,Djouadi:1991tka,Dawson:1990zj,Spira:1995rr,Harlander:2002wh,Anastasiou:2002yz,Ravindran:2003um,Catani:2003zt,Actis:2008ug,Anastasiou:2008tj,deFlorian:2009hc,deFlorian:2012yg,Bozzi:2005wk,deFlorian:2011xf,Grazzini:2013mca,Stewart:2011cf,Banfi:2012jm,EllisGaillard:1976jg,Djouadi:1997yw,Bredenstein:2006rh,Bredenstein:2006ha,Denner:2011mq,Cahn:1983ip,Ciccolini:2007jr,Ciccolini:2007ec,Bolzoni:2010xr,Botje:2011sn,Lai:2010vv,Martin:2009iq,Ball:2010de,Glashow:1978ab,Brein:2003wg,Denner:2011id,Englert:2013vua,Ferrera:2014lca,Raitio:1978pt,Beenakker:2002nc,Dawson:2003zu,Harlander:2003ai,Dittmaier:2003ej,Dawson:2003kb,Harlander:2011aa,Farina:2012xp,Zeppenfeld:2000td,Duhrssen:2004cv}. 
The inclusive cross sections and branching fractions for the most important production and decay modes are summarised with their overall uncertainties in Tables~\ref{tab:SMCrossSections} and~\ref{tab:SMBranchingFractions} for a Higgs boson mass $\mH=125.09$~\gev.
The SM predictions of the branching fractions for $H \to gg, \cc$, and~$Z\gamma$ are included for completeness. 
Although not an explicit part of the searches, they impact the combination through their contributions to the Higgs boson width and, at a small level, through their expected yields in some of the individual analyses.

\begin{table}[htbp]
\caption{Standard Model predictions for the Higgs boson production cross sections together with their theoretical uncertainties. 
The value of the Higgs boson mass is assumed to be  $m_H=125.09$~GeV and the predictions are obtained by linear interpolation between those at 125.0 and 125.1~GeV from Ref.~\cite{Heinemeyer:2013tqa} except for the \atH~cross section, which is taken from Ref.~\cite{Aad:2014lma}. 
The $pp \to ZH$~cross section, calculated at NNLO in QCD, includes  both the quark-initiated, i.e.~$qq \to ZH$ or~$qg \to ZH$, and the $gg\to ZH$ contributions. 
The contribution from the $gg \to~ZH$~production process, calculated only at NLO in QCD and indicated separately in brackets, is given with a theoretical uncertainty assumed to be~30\%. 
The uncertainties in the cross sections are evaluated as the sum in quadrature of the uncertainties resulting from variations of the QCD scales, parton distribution functions, and $\alpha_{\text s}$. 
The uncertainty in the \atH\ cross section is calculated following the procedure of Ref.~\cite{Alwall:2014hca}. 
The order of the theoretical calculations for the different production processes is also indicated. In the case of \abbH~production, the values are given for the mixture of five-flavour~(5FS) and four-flavour~(4FS) schemes recommended in~Ref.~\cite{Harlander:2011aa}.
}\vspace*{-0.2cm}
\setlength\extrarowheight{2pt}
\begin{center}
\begin{tabular}{clr@{\hskip 0.4ex}lr@{\hskip 0.4ex}lc} \\ \hline\hline
 &  \multicolumn{1}{c}{Production }  & \multicolumn{4}{c}{Cross section [pb]} & \hsd Order of \hsd \\ \cline{3-6}
 &  \multicolumn{1}{c}{process}     &   \multicolumn{2}{c}{$\sqrt{s}=7$ TeV}   &  \multicolumn{2}{c}{$\sqrt{s}=8$ TeV} & calculation  \\ \hline
 &  \aggF       &   $15.0$&${} \pm 1.6$      &  $19.2$&${}\pm 2.0$    & {\small\sc NNLO(QCD) + NLO(EW) } \\
 &  \aVBF       &   $1.22$&${}\pm 0.03$     &  $1.58$&${}\pm 0.04$   & {\small\sc NLO(QCD+EW) + approx.~NNLO(QCD) } \\
 &  \aWH        &   $0.577$&${}\pm 0.016$   &  $0.703$&${}\pm 0.018$ & {\small\sc NNLO(QCD) + NLO(EW) }  \\
 &  \aZH        &   $0.334$&${}\pm 0.013$   &  $0.414$&${}\pm 0.016$ & {\small\sc NNLO(QCD) + NLO(EW) }  \\ 
 & [\aggZH]       &   $0.023$&${}\pm 0.007$    &   $0.032$&${}\pm 0.010$      & {\small\sc NLO(QCD) } \\
 &  \attH       &   $0.086$&${}\pm 0.009$   &  $0.129$&${}\pm 0.014$ & {\small\sc NLO(QCD) } \\ 
 &  \atH        &   $0.012$&${}\pm 0.001$   &  $0.018$&${}\pm 0.001$ & {\small\sc NLO(QCD) }   \\ 
 &  \abbH       &   $0.156$&${}\pm 0.021$   &  $0.203$&${}\pm 0.028$ & {\small\sc 5FS NNLO(QCD) + 4FS NLO(QCD) }  \\ \hline
 &  Total       &   $17.4 $&${}\pm 1.6$     &  $22.3$&${}\pm 2.0$    &
 \\ \hline\hline
\end{tabular}
\end{center}
\label{tab:SMCrossSections}
\end{table}

\begin{table}[htbp]
\caption{Standard Model predictions for the decay branching fractions of a Higgs boson with a mass of 125.09~GeV, together with their uncertainties~\cite{Heinemeyer:2013tqa}. Included are decay modes that are either directly studied or important for the combination because of their contributions to the Higgs boson width.
}
\setlength\extrarowheight{2pt}
\begin{center}
\begin{tabular}{clcr@{\hskip 0.4ex}l} \\ \hline\hline
 &    Decay mode &\multicolumn{3}{c}{Branching fraction [\%]}  \\  \hline
 &     $\Hbb$       & &   $57.5$&${}\pm 1.9$    \\
 &     $\HWW$       & &   $21.6$&${}\pm 0.9$   \\
 &     $\Hgg$       & &   $8.56$&${}\pm 0.86$  \\
 &     $\Htt$       & &   $6.30$&${}\pm 0.36$  \\
 &     $\Hcc$       & &   $2.90$&${}\pm 0.35$  \\
 &     $\HZZ$       & &   $2.67$&${}\pm 0.11$  \\
 &     $\Hyy$       & &   $0.228$&${}\pm 0.011$  \\
 &     $\Hzg$       & &   $0.155$&${}\pm 0.014$  \\
 &     $\Hmm$       & &   ~~$0.022$&${}\pm 0.001$  \\ 
\hline\hline
\end{tabular} 
\end{center}
\label{tab:SMBranchingFractions}
\end{table}

\subsection{Signal Monte Carlo simulation}
\label{sec:Generators}

All analyses use MC samples to model the Higgs boson production and decay kinematics, to estimate the acceptance and selection efficiency, and to describe the distributions of variables used to discriminate between signal and background events.
The main features of the signal simulation are summarised here; for more details, the reader is referred to the individual publications:
\begin{itemize} 

\item for \aggF\ and \aVBF\ production, both experiments use {\sc Powheg}~\cite{Nason:2004rx,Frixione:2007vw,Alioli:2008tz,Alioli:2010xd,Bagnaschi:2011tu} for the event generation, interfaced either to {\sc Pythia8}~\cite{Sjostrand:2007gs} (ATLAS) or {\sc Pythia6.4}~\cite{Sjostrand:2006za} (CMS) for the simulation of the parton shower, the hadronisation, and the underlying event, collectively referred to in the following as~UEPS.

\item for \aWH\ and \aZH\ production, both experiments use LO event generators for all quark-initiated processes, namely {\sc Pythia8} in ATLAS and {\sc Pythia6.4} in~CMS. 
A prominent exception is the $\Hbb$ decay channel, for which ATLAS uses {\sc Powheg} interfaced to~{\sc Pythia8}, while CMS uses {\sc Powheg} interfaced to~{\sc Herwig++}~\cite{Bahr:2008pv}. 
The \aggZH\ production process is also considered, even though it contributes only approximately~8\% of the total \aZH~production cross section in the SM, because it is expected to yield a relatively hard Higgs boson transverse momentum (\pT) spectrum, enhancing the contribution to the most sensitive categories in the $\Hbb$ decay channel. 
Both experiments therefore include \aggZH\ production as a separate process in the \aVH\ analysis for the $\Hbb$ channel. 
ATLAS uses {\sc Powheg} interfaced to {\sc Pythia8}
while CMS uses a reweighted $qq \to ZH$~sample to model the \aggZH~contribution, including next-to-leading order~(NLO) effects~\cite{Englert:2013vua,Ferrera:2014lca}. 
For the other channels, the contribution from this process is only accounted for as a correction to the overall signal cross section. 

\item for \attH\ production, ATLAS uses the NLO calculation of the {\sc HELAC-Oneloop} package~\cite{Bevilacqua:2011xh} interfaced to {\sc Powheg}, often referred to as~{\sc Powhel}~\cite{Powhel}, while CMS simulates this process with the LO {\sc Pythia6.4} program.

\item within the SM, the contribution from \atH\ production to analyses searching for \attH\ production \break is small, but in certain BSM~scenarios it may become large through interference effects \break (see~Section~\ref{sec:kappas}). 
The \atH\ production processes are simulated in both experiments using \break {\sc MadGraph5\_aMC@NLO}~\cite{Alwall:2014hca} interfaced to {\sc Herwig++} 
in the case of~\atHW~production, while the \atHq~production process is simulated using {\sc MadGraph}~\cite{Maltoni:2002qb} interfaced to {\sc Pythia8} in ATLAS and {\sc MadGraph5\_aMC@NLO} interfaced to~{\sc Pythia6.4} in~CMS.

\item finally, \abbH\ production contributes approximately 1\% to the total Higgs boson cross section in the~SM. 
It is studied using {\sc Pythia8} in ATLAS and {\sc Pythia6.4} and {\sc MadGraph5\_aMC@NLO} in CMS, for the categories most sensitive to this production process in the various channels. 
Given that the selection efficiencies of \abbH\ production are similar to those of the \aggF\ process, the latter process is used to model the \abbH\ signal for all decay channels, with an approximate correction to account for the difference in overall efficiency.
\end{itemize} 

Table~\ref{tab:generator} summarises the event generators used by ATLAS and CMS for the $\sqrt{s}=8$~TeV data analyses.
For each production process and decay mode, the cross section and branching fraction used correspond to the higher-order state-of-the-art theoretical calculations, namely the values given in~Tables~\ref{tab:SMCrossSections} and~\ref{tab:SMBranchingFractions}.

Furthermore, the \pT\ distribution of the Higgs boson in the \aggF\ process, which in many cases affects categorisation and selection efficiencies, is reweighted to match the {\sc HRes2.1} prediction~\cite{Bozzi:2005wk,deFlorian:2011xf,Grazzini:2013mca}, which accounts for next-to-next-to-leading-order~(NNLO) 
and next-to-next-to-leading-logarithmic (NNLL) QCD corrections. 
In addition, the Higgs boson \pT\ spectrum in \ggF\ events with two or more jets is reweighted to match the prediction of the {\sc Powheg MiNLO} H+2-jet generator~\cite{Campbell:2006xx}.
This consistent treatment by the two experiments of the most prominent theoretical aspects of Higgs boson production and decay is quite important since all theoretical uncertainties in the various signal processes described in~Table~\ref{tab:generator} are treated as correlated for the combination (see~Section~\ref{sec:CombinationProcedure}). 
The impact of using different generators for the less sensitive channels is negligible compared to their dominant sources of uncertainty.

\begin{table}[htb]
\caption{Summary of the event generators used by ATLAS and CMS to model the Higgs boson production processes and decay channels at $\sqrt{s}=8$~TeV.}
\setlength\extrarowheight{2pt}
\begin{center}
\begin{tabular}{clccc}\hline\hline
\multicolumn{2}{c}{Production\hsd}  &\hsd& \multicolumn{2}{c}{Event generator}           \\ \cline{4-5} 
\multicolumn{2}{c}{process\hsd}    & & \hsd  ATLAS \hsd   & \hsd   CMS \hsd                     \\ \hline  
& \aggF                   && {\sc Powheg}~\cite{Nason:2004rx,Frixione:2007vw,Alioli:2008tz,Alioli:2010xd,Bagnaschi:2011tu}    &  {\sc Powheg}                     \\  
& \aVBF                  &&  {\sc Powheg}        &  {\sc Powheg}                \\  
& \aWH                   && {\sc Pythia8}~\cite{Sjostrand:2007gs}           &   {\sc Pythia6.4}~\cite{Sjostrand:2006za}                         \\
& \aZH\ ($qq \to ZH$ or $qg \to ZH$) && {\sc Pythia8}           &    {\sc Pythia6.4}                    \\
& \aggZH\ ($gg \to ZH$)     && {\sc Powheg}            &   See text                                             \\
& \attH                && {\sc Powhel}~\cite{Bevilacqua:2011xh}            &   {\sc Pythia6.4}                     \\
& \atHq\ ($qb\to tHq$) && {\sc MadGraph}~\cite{Maltoni:2002qb}          & {\small\sc aMC@NLO}~\cite{Alwall:2014hca}                                      \\
& \atHW\  ($gb\to tHW$)  &&{\small\sc aMC@NLO} &   {\small\sc aMC@NLO}                    \\
& \abbH                &&  {\sc Pythia8}           &    {\sc Pythia6.4}, {\small\sc aMC@NLO}                \\
\hline\hline
\end{tabular}
\end{center}
\label{tab:generator}
\end{table}

\subsection{Signal strengths}
\label{sec:Mu}

The signal strength $\mu$, defined as the ratio of the measured Higgs boson rate to its SM prediction, is used to characterise the Higgs boson yields.  
For a specific production process and decay mode~$i\to H\to f$, the signal strengths for the production, $\mu_i$, and for the decay, $\mu^f$, are defined as
\begin{equation}
\mu_i = \frac{\sigma_i}{(\sigma_i)_\SM}\hspace*{0.5cm} {\rm and}\hspace*{0.5cm} \mu^f = \frac{\BR^f}{(\BR^f)_\SM}.
\label{eq:mui_f}
\end{equation}
Here $\sigma_i\; (i=\aggF,\aVBF,\aWH,\aZH,\attH)$ and $\BR^f\; (f = ZZ, WW, \gamma\gamma, \tau\tau, \bb, \mu\mu)$ are respectively the production cross section for $i\to H$ and the decay branching fraction for $H\to f$. 
The subscript~``SM'' refers to their respective SM~predictions, so by definition, $\mu_i=1$ and $\mu^f=1$ in the~SM. 
Since $\sigma_i$ and $\BR^f$ cannot be separated without additional assumptions, only the product of $\mu_i$ and $\mu^f$ can be measured experimentally, leading to a signal strength~$\mu_i^f$ for the combined production and decay:
\begin{equation}
  \mu_i^f =  \frac{\sigma_i\cdot \BR^f}{(\sigma_i)_\SM \cdot (\BR^f)_\SM} = \mu_i\cdot\mu^f.
\label{eq:muif}
\end{equation}

The ATLAS and CMS data are combined and analysed using this signal strength formalism and the results are presented in~Section~\ref{sec:SignalStrength}. 
For all these signal strength fits, as well as for the generic parameterisations presented in~Section~\ref{sec:sigBR}, the parameterisations of the expected yields in each analysis category are performed with a set of assumptions, which are needed because some production processes or decay modes, which are not specifically searched for, contribute to other channels. These assumptions are the following: for the production processes, the \abbH~signal strength is assumed to be the same as for~\aggF, the \atH\ signal strength is assumed to be the same as for~\attH, and the \aggZH\  signal strength is assumed to be the same as for quark-initiated \aZH~production; for the Higgs boson decays, the~$\Hgg$ and~$\Hcc$ signal strengths are assumed to be the same as for $\Hbb$~decays, and the $\Hzg$ signal strength is assumed to be the same as for $\Hyy$~decays.

\subsection{Coupling modifiers}
\label{sec:kappas}

Based on a LO-motivated framework~\cite{Heinemeyer:2013tqa} ($\Cc$-framework), coupling modifiers have been proposed to interpret the LHC data by introducing specific modifications of the Higgs boson couplings related to BSM physics. 
Within the assumptions already mentioned in~Section~\ref{sec:Introduction}, the production and decay of the Higgs boson can be factorised, such that the cross section times branching fraction of an individual channel $\sigma(\mathit{i}\to H\to\mathit{f})$ contributing to a measured signal yield can be parameterised as:
\begin{equation}
\sigma_i\cdot {\BR}^f = \frac{\sigma_{\mathit{i}}(\vec\Cc)\cdot\Gamma^{\mathit{f}}(\vec\Cc)}{\Gamma_{H}},
\end{equation}
where $\Gamma_H$~is the total width of the Higgs boson and $\Gamma^{\mathit{f}}$ is the partial width for Higgs boson decay to the final state~$f$. 
A set of coupling modifiers, $\vec\Cc$, is introduced to parameterise possible deviations from the SM~predictions of the Higgs boson couplings to SM~bosons and fermions. 
For a given production process or decay mode, denoted~``$j$'', a coupling modifier $\kappa_j$ is defined such that:
\begin{equation}
\label{eq:kappa}
  \Cc_j^2=\sigma_j/\sigma_j^\SM\ \ \ {\rm or}\ \ \  \Cc_j^2=\Gamma^j/\Gamma^j_\SM,
\end{equation}
where all $\kappa_j$ values equal unity in the~SM; here, by construction, the SM cross sections and branching fractions include the best available higher-order QCD and EW corrections. 
This higher-order accuracy is not necessarily preserved for $\Cc_j$~values different from unity, but the dominant higher-order QCD corrections factorise to a large extent from any rescaling of the coupling strengths and are therefore assumed to remain valid over the entire range of $\Cc_j$~values considered in this paper.
Different production processes and decay modes probe different coupling modifiers, as can be visualised from the Feynman diagrams shown in Figs.~\ref{fig:feyn_ggFVBF}--\ref{fig:feyn_hgg}. Individual coupling modifiers, corresponding to tree-level Higgs boson couplings to the different particles, are introduced, as well as two effective coupling modifiers, $\kappa_g$ and  $\kappa_\gamma$, which describe the loop processes for~\aggF\ production and~$\Hyy$ decay. This is possible because BSM~particles that might be present in these loops are not expected to appreciably change the kinematics of the corresponding process. The~$gg \to H$ and~$\Hyy$ loop processes can thus be studied, either through these effective coupling modifiers, thereby providing sensitivity to potential BSM~particles in the loops, or through the coupling modifiers corresponding to the SM~particles.
In contrast, the \ggZH~process, which occurs at~LO through box and triangular loop diagrams (Figs.~\ref{fig:feyn_prod}b and~\ref{fig:feyn_prod}c), is always taken into account, within the limitations of the framework, by resolving the loop in terms of the corresponding coupling modifiers,~$\Cc_{Z}$ and~$\Cc_{t}$.
 
Contributions from interference effects between the different diagrams provide some sensitivity to the relative signs of the Higgs boson couplings to different particles. As discussed in~Section~\ref{sec:ModelK3}, such effects are potentially largest for the $\Hyy$~decays, but may also be significant in the case of~\aggZH\ and~\atH~production. The \aggF~production process, when resolved in terms of its SM~structure, provides sensitivity, although limited, to the relative signs of~$\Cc_t$ and~$\Cc_b$ through the $t$--$b$~interference. The relative signs of the coupling modifiers $\Cc_\tau$ and $\Cc_\mu$ with respect to other coupling modifiers are not considered in this paper, since the current sensitivity to possible interference terms is negligible.

\begin{table}[hbt]
\caption{Higgs boson production cross sections $\sigma_{i}$, partial decay widths $\Gamma^{f}$, and total decay width (in the absence of BSM~decays) parameterised as a function of the $\Cc$ coupling modifiers as discussed in the text, including higher-order QCD and EW corrections to the inclusive cross sections and decay partial widths. 
The coefficients in the expression for~$\Gamma_{H}$ do not sum exactly to unity because some contributions that are negligible or not relevant to the analyses presented in this paper are not shown. 
}  
\begin{center}
\setlength\extrarowheight{3pt}%
\begin{tabular}{lcccl}\hline\hline
 & & & Effective & \multicolumn{1}{c}{Resolved} \\
Production & Loops & Interference & scaling factor & \multicolumn{1}{c}{scaling factor}\\
\hline
$\sigma(\aggF)$ & $\checkmark$ & $t$--$b$ & $\Cc_{g}^2$ &  $ 1.06 \cdot\Cc_{t}^2 + 0.01 \cdot \Cc_{b}^2 - 0.07\cdot\Cc_{t}\Cc_{b}$  \\
$\sigma(\aVBF)$     & --            &  --    &    &   $ 0.74 \cdot \Cc_{W}^2 + 0.26 \cdot \Cc_{Z}^2$ \\
$\sigma(\aWH)$             & --            &  --    &  &   $\Cc_{W}^2$\\
$\sigma(qq/qg \to ZH)$ & --            &  --     &  &   $\Cc_{Z}^2$\\
$\sigma(gg \to ZH)$       & $\checkmark$ & $t$--$Z$  &  &   $2.27\cdot\Cc_{Z}^2 + 0.37 \cdot\Cc_{t}^2 - 1.64 \cdot \Cc_{Z}\Cc_{t} $\\
$\sigma(\attH)$            & --            &  --    & &    $\Cc_{t}^2$ \\
$\sigma(gb \to tHW)$     & --            & $t$--$W$   &   & $ 1.84 \cdot \Cc_{t}^2 + 1.57 \cdot \Cc_{W}^2 - 2.41 \cdot \Cc_{t}\Cc_{W}$ \\
$\sigma(qq/qb \to tHq)$ & --          & $t$--$W$  &   &    $ 3.40 \cdot \Cc_{t}^2 + 3.56 \cdot \Cc_{W}^2 - 5.96 \cdot \Cc_{t}\Cc_{W}$ \\
$\sigma(\abbH)$            & --            &  --     &   & $\Cc_{b}^2$ \\ [2pt] \hline
Partial decay width~~ \\
\hline
$\Gamma^{ZZ}$            & --             &  --     &   & $\Cc_{Z}^2$ \\
$\Gamma^{WW}$            & --             &  --     &   & $\Cc_{W}^2$ \\
$\Gamma^{\gamma\gamma}$  & $\checkmark$  & $t$--$W$ & $\Cc_{\gamma}^2$&  $ 1.59 \cdot \Cc_{W}^2 + 0.07 \cdot \Cc_{t}^2 -0.66 \cdot \Cc_{W} \Cc_{t}$  \\
$\Gamma^{\tau\tau}$      & --             &  --     &   & $\Cc_{\tau}^2$ \\
$\Gamma^{bb}$      & --             &  --     &   & $\Cc_{b}^2$ \\
$\Gamma^{\mu\mu}$        & --             &  --    &   & $\Cc_{\mu}^2$ \\ [2pt] \hline
\multicolumn{2}{l}{Total width ($\BRbsm=0$)} \\ \hline
& & &  &   $0.57 \cdot \Cc_{b}^2 + 0.22 \cdot \Cc_{W}^2 + 0.09 \cdot \Cc_{g}^2 +$\\
$\Gamma_{H}$  & $\checkmark$  & --  &  $\Cc_{H}^2$ & $0.06 \cdot \Cc_{\tau}^2 + 0.03 \cdot \Cc_{Z}^2 + 0.03 \cdot \Cc_{c}^2 + $  \\
& & &  &   $0.0023 \cdot \Cc_{\gamma}^2 +~0.0016 \cdot \Cc_{(Z\gamma)}^2 +$\\
& & &  &   $0.0001 \cdot \Cc_{s}^2 + 0.00022 \cdot \Cc_{\mu}^2$\\ [2pt] \hline
\hline
\end{tabular}
\end{center}
\label{tab:kexpr}
\end{table}

As an example of the possible size of such interference effects, the \atH\ cross section is small in the~SM, approximately~14\% of the \attH~cross section, because of destructive interference between diagrams involving the couplings to the $W$~boson and the top quark, as shown in Table~\ref{tab:kexpr}. However, the interference becomes constructive for negative values of the product $\Cc_W \cdot \Cc_t$. 
In the specific case where $\Cc_W \cdot \Cc_t = -1$, the~\atHW\ and \atHq\ cross sections increase by factors of 6 and 13, respectively, so that the \atH~process displays some sensitivity to the relative sign between the $W$ boson and top quark couplings, despite its small SM~cross section. 

The relations among the coupling modifiers, the production cross sections $\sigma_i$, and partial decay widths~$\Gamma^f$ are derived within this context, as shown in~Table~\ref{tab:kexpr}, and are used as a parameterisation to extract the coupling modifiers from the measurements. 
The coefficients are derived from Higgs production cross sections and decay rates evaluated including the best available higher-order QCD and EW corrections (up to NNLO QCD and NLO EW precision), as indicated in Tables~\ref{tab:SMCrossSections} and~\ref{tab:SMBranchingFractions}. 
The numerical values are obtained from~Ref.~\cite{Heinemeyer:2013tqa} and are given for $\rts = 8\TeV$ and $m_H = 125.09\GeV$ (they are similar for $\rts = 7\TeV$). 
The current LHC data are insensitive to the coupling modifiers $\Cc_c$ and $\Cc_s$, and have limited sensitivity to $\Cc_\mu$. 
Thus, in the following, it is assumed that $\Cc_c$~varies as~$\Cc_t$, $\Cc_s$~as~$\Cc_b$, and $\Cc_\mu$~as~$\Cc_\tau$. 
Other coupling modifiers ($\Cc_u$, $\Cc_d$, and $\Cc_e$) are irrelevant for the combination provided they are of order unity. 
When probing the total width, the partial decay width $\Gamma^{gg}$ is assumed to vary as $\Cc_g^2$. 
These assumptions are not the same as those described for the signal strength framework in~Section~\ref{sec:Mu}, so the two parameterisations are only approximately equivalent. The two sets of assumptions have a negligible impact on the measurements reported here provided that the unmeasured parameters do not deviate strongly from unity. 

Changes in the values of the couplings will result in a variation of the Higgs boson width. 
A new modifier, $\Cc_H$, defined as $\Cc_H^2=\sum_j \BR^j_\SM \Cc_j^2$ and assumed to be positive without loss of generality, is introduced to characterise this variation. 
In the case where the SM~decays of the Higgs boson are the only ones allowed, the relation $\Cc_H^2=\Gamma_H/\Gamma_H^\SM$ holds. 
If instead deviations from the~SM are introduced in the decays, the width $\Gamma_{H}$ can be expressed as:
\begin{equation}
  \Gamma_{H} = \frac{\Cc_H^2\cdot\Gamma_H^\SM}{1-\BRbsm},
\end{equation} 
where $\BRbsm$ indicates the total branching fraction into BSM~decays. 
Such BSM decays can be of three types: decays into BSM~particles that are invisible to the detector because they do not appreciably interact with ordinary matter, decays into BSM particles that are not detected because they produce event topologies that are not searched for, or modifications of the decay branching fractions into SM~particles in the case of channels that are not directly measured, such as~$H \to cc$. 
Although direct and indirect experimental constraints on the Higgs boson width exist, they are either model dependent or are not stringent enough to constrain the present fits, and are therefore not included in the combinations. 
Since $\Gamma_{H}$ is not experimentally constrained in a model-independent manner with sufficient precision, only ratios of coupling strengths can be measured in the most generic parameterisation considered in the $\Cc$-framework.

\section{Combination procedure and experimental inputs}
\label{sec:CombinationProcedure}

The individual ATLAS and CMS analyses of the Higgs boson production and decay rates are combined using the profile likelihood method described in Section~\ref{sec:CombinationStatistics}. The combination is based on simultaneous fits to the data from both experiments taking into account the correlations between systematic uncertainties within each experiment and between the two experiments. 
The analyses included in the combination, the statistical procedure used, the treatment of systematic uncertainties, and the changes made to the analyses for the combination are summarised in this section.

\subsection{Overview of input analyses}
\label{sec:Analyses}

The individual analyses included in the combination were published separately by each experiment. 
Most of these analyses examine a specific Higgs boson decay mode, with categories related to the various production processes. 
They are $\Hyy$~\cite{Aad:2014eha,CMS:Hgamgam}, $\Hzz$~\cite{Aad:2014eva,CMS:H4l}, $\Hww$~\cite{ATLAS:2014aga,ATLAS:VHWW,CMS:HWW}, $\Htt$~\cite{Aad:2015vsa,CMS:Htautau}, $\Hbb$~\cite{Aad:2014xzb,CMS:VHbb}, and $\Hmm$~\cite{Aad:2014xva,CMS:Hmumu}. 
The \attH\ production process was also studied separately~\cite{ATLAS:ttHhad,ATLAS:ttHlep,Aad:2014lma,Chatrchyan:2013yea,CMS:ttH} and the results are included in the combination. 
The $\Hmm$ analysis is included in the combination fit only for the measurement of the corresponding decay signal strength reported in Section~\ref{sec:ProductionDecay} and for the specific parameterisation of the coupling analysis described in Section~\ref{sec:ModelK1}. 
It provides constraints on the coupling of the Higgs boson to second-generation fermions, but offers no relevant constraints for other parameterisations.  
The ATLAS~\cite{ATLAS_combination} and CMS~\cite{CMS_combination} individual combined publications take into account other results, such as upper limits on the $\Hzg$ decay~\cite{Aad:2014fia,Chatrchyan:2013vaa}, results on  \aVBF~production in the $\Hbb$~decay channel~\cite{Khachatryan:2015bnx}, constraints on off-shell Higgs boson production~\cite{Aad:2015xua, CMSoffshell}, and upper limits on invisible Higgs boson decays~\cite{Aad:2014iia,Aad:2015uga,Chatrchyan:2014tja}. These results are not considered further here since they were not included in both combined publications of the individual experiments. In the case of the $\Hbb$~decay mode, the \aggF~production process is not considered by either experiment because of the overwhelming QCD multijet background.

Almost all input analyses are based on the concept of event categorisation. For each decay mode, events are classified in different categories, based on their kinematic characteristics and their detailed properties. 
This categorisation increases the sensitivity of the analysis and also allows separation of the different production processes on the basis of exclusive selections that identify the decay products of the particles produced in association with the Higgs boson: $W$ or $Z$ boson decays, \aVBF\ jets, and so on.
A total of approximately~600 exclusive categories addressing the five production processes explicitly considered are defined for the five main decay channels. The exception is $\Hbb$, for which only the \aVH\ and \attH\ production processes are used in the combination for the reasons stated above.

The signal yield in a category~$k$,~$n_{\rm signal}(k)$, can be expressed as a sum over all possible Higgs boson production processes~$i$, with cross section~$\sigma_i$, and decay modes~$f$, with branching fraction~$\BR^f$: 
\begin{equation}
\begin{split}
n_{\rm signal}(k) & =  \mathcal{L}(k) \cdot \sum_i \sum_f \left\{\sigma_i \cdot A_i^{f,{\text SM}}(k) \cdot \varepsilon_i^f(k) \cdot \BR^f\right\} \\
                  & = \mathcal{L}(k) \cdot \sum_i \sum_f \mu_i\mu^f \left\{\sigma_i^\SM \cdot A_i^{f,{\text SM}}(k) \cdot \varepsilon_i^f(k) \cdot \BR^f_\SM\right\},   
\end{split}
\label{eq:nsig}
\end{equation}
where $ \mathcal{L}(k)$ represents the integrated luminosity, $A_i^{f,{\text SM}}(k)$ the detector acceptance assuming SM Higgs boson production and decay, and $\varepsilon_i^f(k)$ the overall selection efficiency for the signal category~$k$. 
The symbols $\mu_i$ and $\mu^f$ are the production and decay signal strengths, respectively, defined in Section~\ref{sec:Mu}. 
As Eq.~(\ref{eq:nsig}) shows, the measurements considered in this paper are only sensitive to the products of the cross sections and branching fractions, $\sigma_i\cdot\BR^f$. 

In the ideal case, each category would only contain signal events from a given production process and decay mode.
Most decay modes approach this ideal case, but, in the case of the production processes, the categories are much less pure and there is significant cross-contamination in most channels.

\subsection{Statistical treatment}
\label{sec:CombinationStatistics} 

The overall statistical methodology used in the combination to extract the parameters of interest in various parameterisations is the same as that used for the individual ATLAS and CMS combinations, as published in~Refs.~\cite{ATLAS_combination,CMS_combination}. 
It was developed by the ATLAS and CMS Collaborations and is described in~Ref.~\cite{LHC-HCG}. 
Some details of this procedure are important for this combination and are briefly reviewed here.

The statistical treatment of the data is based on the standard LHC data modelling and handling toolkits: {\sc RooFit}~\cite{Verkerke:2003ir}, {\sc RooStats}~\cite{Moneta:2010pm}, and  {\sc HistFactory}~\cite{Cranmer:2012sba}. 
The parameters of interest, $\vec\alpha$, e.g.\ signal strengths ($\mu$), coupling modifiers ($\kappa$), production cross sections, branching fractions, or ratios of the above quantities, are estimated, together with their corresponding confidence intervals, via the profile likelihood ratio test statistic $\Lambda(\vec\alpha)$~\cite{Cowan:2010st}. 
The latter depends on one or more parameters of interest, as well as on the nuisance parameters, {$\vec\theta$}, which reflect various experimental or theoretical uncertainties:  
\begin{equation}
  \Lambda(\vec\alpha) = \frac{L\big(\vec\alpha\,,\,\hat{\hat{\vec\theta}}(\vec\alpha)\big)}
                              {L(\hat{\vec\alpha},\hat{\vec\theta})\label{eq:LH}}.
\end{equation}

The likelihood functions in the numerator and denominator of this equation are constructed using products of signal and background probability density functions (pdfs) of the discriminating variables. 
The pdfs are obtained from simulation for the signal and from both data and simulation for the background, as described in Refs.~\cite{ATLAS_combination, CMS_combination}. 
The vectors $\hat{\vec\alpha}$ and $\hat{\vec\theta}$ represent the unconditional maximum likelihood estimates of the parameter values, while $\hat{\hat{\vec\theta}}$ denotes the conditional maximum likelihood estimate for given values of the parameters of interest $\vec\alpha$.  Systematic uncertainties and their correlations are a subset of the nuisance parameters $\vec\theta$, described by likelihood functions associated with the estimate of the corresponding parameter.

As an example of a specific choice of parameters of interest, the parameterisation considered in Section~\ref{sec:ModelK3} assumes that all fermion couplings are scaled by $\kappa_F$ and all weak vector boson couplings by $\kappa_V$. The likelihood ratio is therefore a function of the two parameters of interest, $\kappa_F$ and $\kappa_V$, and the profile likelihood ratio is expressed as:
\begin{equation}\label{eq:testkFkV}
  \Lambda(\kappa_F,\kappa_V) = \frac{L\big(\kappa_F,\kappa_V,\hat{\hat{\vec\theta}}(\kappa_F,\kappa_V)\big)} {L(\hat\kappa_F,\hat\kappa_V,\hat{\vec\theta})} \;.
\end{equation}

Likelihood fits are performed to determine the parameters of interest and their uncertainties, using the data to obtain the observed values and Asimov data sets to determine the predicted values in the~SM. 
An Asimov data set~\cite{Cowan:2010st} is a pseudo-data distribution that is equal to the signal plus background prediction for given values of the parameters of interest and of all nuisance parameters, and does not include statistical fluctuations. 
It is a representative data set of a given parameterisation that yields a result corresponding to the median of an ensemble of pseudo-experiments generated from the same parameterisation.
A pre-fit Asimov data set is meant to represent the predictions of the theory, and all parameters are fixed to their estimates prior to the fit to the data. 

These fits are rather challenging, involving many parameters of interest and a very large number of nuisance parameters. 
All the fit results were independently cross-checked to a very high level of precision by ATLAS and CMS, both for the combination and for the individual results. In particular, fine likelihood scans of all the parameters of interest were inspected to verify the convergence and stability of the fits.

For all results presented in this paper, unless otherwise stated, the negative log-likelihood estimator $q(\vec\alpha) = -2\ln\Lambda(\vec\alpha)$ is assumed to follow a $\chi^2$ distribution (asymptotic approximation). 
The 1$\sigma$ and 2$\sigma$ confidence level (CL) intervals for one-dimensional measurements are defined by requiring $q(\alpha_i)=1$ and $q(\alpha_i)=4$, respectively. In the case of disjoint intervals, the uncertainties corresponding only to the interval around the best fit value with $q(\alpha_i) < 1$ are also given for some parameterisations.
The 68\%~(95\%) confidence level regions for two-dimensional scans are defined at $q(\alpha_i) = 2.30~(5.99)$. 
For the derivation of the upper limit on $\BRbsm$ in Section~\ref{sec:ModelK2}, the test statistic $\tilde t(\alpha)$ of Ref.~\cite{Cowan:2010st} is used to account for the constraint $\alpha=\BRbsm\geq 0$. 
This is equivalent to the confidence interval estimation method of Ref.~\cite{FeldmanCousins}. 
The upper limit at~95\%~CL corresponds to $\tilde t(\alpha)=3.84$.
The $p$-values, characterising the compatibility of a fit result with a given hypothesis, are likewise computed in the asymptotic approximation.

\subsection{Treatment of systematic uncertainties}
\label{sec:SystematicUncertainties}

The treatment of the systematic uncertainties and of their correlations is a crucial aspect of the combination of Higgs boson coupling measurements. 
The details of the chosen methodology for treating systematic uncertainties, characterised by nuisance parameters, are given in Ref.~\cite{LHC-HCG}. 
The combined analysis presented here incorporates approximately 4200 nuisance parameters. 
A large fraction of these are statistical in nature, i.e. related to the finite size of the simulated samples used to model the expected signals and backgrounds, but are classified as part of the systematic uncertainties, as described below.

Nuisance parameters can be associated with a single analysis category or can be correlated between categories, channels, and/or experiments.
A very important and delicate part of this combination is the estimation of the correlations between the various sources of systematic uncertainty, both between the various channels and between the two experiments. 
The correlations within each experiment are modelled following the procedure adopted for their individual combinations. 
The systematic uncertainties that are correlated between the two experiments are theoretical systematic uncertainties affecting the signal yield, certain theoretical systematic uncertainties in the background predictions, and a part of the experimental uncertainty related to the measurement of the integrated luminosity.

The main sources of theoretical uncertainties affecting the signal yield are the following: missing higher-order QCD corrections (estimated through variation of the QCD scales, i.e.\ renormalisation and factorisation scale) and uncertainties in parton distribution functions~(PDF), in the treatment of~UEPS, and in Higgs boson branching fractions. These uncertainties apply both to the inclusive cross sections and to the acceptances and selection efficiencies in the various categories. 
The PDF uncertainties in the inclusive rates are correlated between the two experiments for a given production process, but are treated as uncorrelated between different processes, except for the \aWH,~\aZH, and \aVBF~production processes, where they are assumed to be fully correlated. A cross-check with the full PDF~correlation matrix, as given in Ref.~\cite{Heinemeyer:2013tqa}, yields differences no larger than~1\% for the generic parameterisations discussed in~Section~\ref{sec:GenericParameterisation}. 
Similarly, QCD scale and UEPS uncertainties are assumed to be correlated between the two experiments in the same production processes and to be uncorrelated between different processes. 
The effects of correlations between Higgs boson branching fractions were determined to be negligible in general, and are ignored in the fits, except for the uncertainties in the branching fractions to $WW$ and $ZZ$, which are assumed to be fully correlated. 
When measuring ratios, however, there are cases, e.g.\ the measurements of ratios of coupling modifiers described in~Section~\ref{sec:kappaParam}, where such uncertainties become the dominant theoretical uncertainties, and in these cases the full branching fraction correlation model specified in~Ref.~\cite{Heinemeyer:2013tqa} was applied. 
Other theoretical uncertainties in the signal acceptance and selection efficiencies are also usually small. 
They are estimated and treated in very different manners by the two experiments and therefore are assumed to be uncorrelated between ATLAS and CMS. It was verified that treating them as correlated would have a negligible impact on the results.

Whereas the signal selection criteria are quite inclusive in most channels, this is not the case for the backgrounds, which are often restricted to very limited regions of phase space and which are often treated differently by the two experiments. 
For these reasons, the ATLAS and CMS background modelling uncertainties cannot be easily correlated, even though such correlations should be considered for channels where they represent significant contributions to the overall systematic uncertainty. 
Obvious examples are those where the background estimates are obtained from simulation, as is the case for the $ZZ$~continuum background in the $\HZZ$~channel, and for the $ttW$ and $ttZ$ backgrounds in the $ttH$ multi-lepton channel. 
For these two cases, the background cross section uncertainties are treated as fully correlated between the two experiments. 
Other more complex examples are the $WW$~continuum background in the $\HWW$ channel, the $ttbb$~background in the $ttH, H \to \bb$~channel, and the $Wbb$~background in the $WH, H\to \bb$ channel. 
In these cases, it was verified that the choice of not implementing correlations in the background modelling uncertainties between the two experiments has only a small impact on the measurements. 
The most significant impact was found for the $ttbb$~background in the $ttH, H \to \bb$~channel, for which the choice of different correlation models between the two experiments yields an impact below~10\% of the total uncertainty in the signal strength measurement in this specific channel. 

Finally, all experimental systematic uncertainties are treated independently by the two experiments, reflecting independent assessments of these uncertainties, except for the integrated luminosity uncertainties, which are treated as partially correlated through the contribution arising from the imperfect knowledge of the beam currents in the LHC~accelerator. 

The various sources of uncertainties 
can be broadly classified in four groups:
\begin{enumerate}
\item uncertainties (labelled as "stat" in the following) that are statistical in nature. 
In addition to the data, these include 
the statistical uncertainties in certain background control regions and certain fit parameters used to describe backgrounds measured from data, but they exclude the finite size of MC~simulation samples;
\item theoretical uncertainties affecting the Higgs boson signal (labelled as "thsig" in the following);
\item theoretical uncertainties affecting background processes only (these are not correlated with any of the signal theoretical uncertainties and are labelled as "thbgd" in the following);
\item all other uncertainties (labelled as "expt" in the following), which include the experimental uncertainties and those related to the finite size of the MC simulation samples.
\end{enumerate}

Some of the results are provided with a full breakdown of the uncertainties into these four categories, but, in most cases, the uncertainties are divided only into their statistical and  systematic~(syst) components. 
In some cases, as in Section~\ref{sec:GenericParameterisation}, when considering ratios of cross sections or coupling strengths,  the theoretical systematic uncertainties are very small, because the signal normalisation uncertainties, which are in general dominant, do not affect the measurements. The precision with which the uncertainties and their components are quoted is typically 
of order 1\% relative to the SM~prediction.

As mentioned above, the Higgs boson mass is fixed, for all results reported in this paper, at the measured value of~$125.09\GeV$. 
The impact of the Higgs boson mass uncertainty ($\pm 0.24$ GeV) on the measurements has two main sources. One is the dependence of the $\sigma\cdot\BR$ product on the mass. This dependence has an impact only on the measurements of the signal strengths and of the coupling modifiers, in which the SM~signal yield predictions enter directly. The associated uncertainties are up to~4\% for the signal strengths and~2\% for the coupling modifiers. The other source of uncertainty is the dependence of the measured yields on the mass, arising from the fit to the mass spectra in the high-resolution~$\Hyy$ and~$\HZZ$ decay channels. In principle, this uncertainty affects all the measurements, including those related to the generic parameterisations, and is expected to be of the same order as the first one, namely~1\% to~2\%. In practice, since the measured masses in the $\Hyy$ and $\HZZ$~decay channels, resulting from the combination of ATLAS and CMS data, agree within~100~MeV, this uncertainty is less than~1\% for all combined ATLAS and CMS measurements reported in this paper. 
Additional uncertainties of approximately~1\% in the measurements of the Higgs boson signal strengths and coupling modifiers arise from the uncertainty in the LHC~beam energy, which is estimated to be~$~0.66\%$ at 8~TeV~\cite{LHCbeamenergy}.
The uncertainties in the Higgs boson mass and the LHC beam energy are much smaller than the statistical uncertainties in the measurements and are neglected in the following.

\subsection{Analysis modifications for the combination}
\label{sec:Changes}

There are some differences in the treatment of signal and background in the combined analysis compared to the published analyses from each experiment. 
The differences are larger for~CMS than for~ATLAS, mainly because the CMS~analyses were published earlier, before some refinements for the SM Higgs boson predictions were made available.
The main differences are the following:
\begin{itemize}
\item ATLAS now uses the Stewart-Tackmann prescription~\cite{Stewart:2011cf} for the jet bin uncertainties in the \mbox{$\HWW$} channel instead of the jet-veto-efficiency procedure~\cite{Banfi:2012jm};
\item CMS now includes the \abbH, \atH, and \aggZH\ production processes in the signal model for all the channels in which they are relevant;
\item CMS now uses the signal cross section calculations from~Ref.~\cite{Heinemeyer:2013tqa} for all channels;
\item CMS now adopts a unified prescription for the treatment of the Higgs boson~\pT\ in the \aggF~production process, as described in~Section~\ref{sec:Generators};
\item The cross sections for the dominant backgrounds were adjusted to the most recent theoretical calculations in the cases where they are estimated from simulation ($ZZ$~background in the \mbox{$\HZZ$} channel and~$ttZ$ and $ttW$ backgrounds in the $\attH$~channels); 
\item Both experiments have adopted the same correlation scheme for some of the signal theoretical uncertainties: for example, the treatment of the  PDF~uncertainties in the signal production cross sections now follows a common scheme for all decay channels, as described in~Section~\ref{sec:SystematicUncertainties}.
\end{itemize}

The total effect of these modifications is small, both for the expected and observed results. All measurements differ from the individual combined results by less than approximately 10\% of the total uncertainty for CMS and by even less for ATLAS.

Table~\ref{tab:inputs} gives an overview of the Higgs boson decay and production processes that are combined in the following. 
To provide a snapshot of the relative importance of the various channels, the results from the analysis presented in this paper (Tables~\ref{tab:muProduction} and~\ref{tab:muDecay} in~Section~\ref{sec:ProductionDecay}) are shown separately for each experiment, as measurements of the overall signal strengths~$\mu$, for each of the six decay channels and for the $\attH$~production process. 
The total observed and expected statistical significances  for~$\mH=125.09\GeV$ are also shown, except for the $\Hmm$~channel, which has very low sensitivity. 
These results are quite close to those published for the individual analyses by each experiment, which are cited in~Table~\ref{tab:inputs}. 
For several decay channels, these refer only to the most sensitive analyses, e.g.\ the \aVH\ analysis for the $\Hbb$~decay channel. 
Even though they are less sensitive, the \attH\ analyses have a contribution from all the decay channels, and this is one of the reasons for quoting this production process specifically in this table. 
As stated above, the differences between the analysis in this paper and the published ones are also in part due to the different values assumed 
for the Higgs boson mass, and to adjustments in the various analyses for the purposes of this combination, mostly in terms of the signal modelling and of the treatment of the correlations of the signal theoretical uncertainties between different channels.

\begin{table}[htbp]
\vspace{-2mm}
\caption{Overview of the decay channels analysed in this paper. The \attH\ production process, which has contributions from all decay channels, is also shown.
To show the relative importance of the various channels, the results from the combined analysis presented in this paper for~$\mH=125.09\GeV$ (Tables~\ref{tab:muProduction} and~\ref{tab:muDecay} in~Section~\ref{sec:ProductionDecay}) are reported as observed signal strengths~$\mu$ with their measured uncertainties. The expected uncertainties are shown in parentheses. 
Also shown are the observed statistical significances, together with the expected significances in parentheses, except for the $\Hmm$~channel, which has very low sensitivity. 
For most decay channels, only the most sensitive analyses are quoted as references, e.g.\ the \aggF\ and \aVBF\ analyses for the $\Hww$~decay channel or the \aVH\ analysis for the $\Hbb$~decay channel. 
Although not exactly the same, the results are close to those from the individual publications, in which slightly different values for the Higgs boson mass were assumed and in which the signal modelling and signal uncertainties were slightly different, as discussed in the text.
}
\label{tab:inputs}
\setlength\extrarowheight{3pt}
\begin{center}
\small
\vspace{-3mm}
\begin{tabular}{lcccr@{\hskip 0.5ex}lr@{\hskip 0.5ex}lccc} \hline\hline
Channel          & \multicolumn{2}{c}{References for}                                &~~&  \multicolumn{4}{c}{Signal strength [$\mu$]} & \multicolumn{2}{c}{Signal significance [$\sigma$]}  &~\\[-3pt]
                 & \multicolumn{2}{c}{individual publications}                       &~&  \multicolumn{6}{c}{from results in this paper (Section~\ref{sec:ProductionDecay})} &~\\
\cline{2-3}\cline{5-10}
                 & ATLAS                                        & CMS                &~& \multicolumn{2}{c}{ATLAS}        & \multicolumn{2}{c}{CMS}          & ATLAS &  CMS   &~\\ \hline
$\Hyy$           & \cite{Aad:2014eha}                           & \cite{CMS:Hgamgam} &~& $1.14$ & $^{+0.27}_{-0.25}     $ & $1.11$ & $^{+0.25}_{-0.23}     $ &  5.0  &  5.6  &~ \\
                 &                                              &                    &~& & $\Bgl({}^{+0.26}_{-0.24}\Big)$ & & $\Bgl({}^{+0.23}_{-0.21}\Big)$ & (4.6) & (5.1)  &~\\[2pt] \hline
$\Hzz$           & \cite{Aad:2014eva}                           & \cite{CMS:H4l}     &~& $1.52$ & $^{+0.40}_{-0.34}     $ & $1.04$ & $^{+0.32}_{-0.26}     $ &  7.6  &  7.0   &~\\
                 &                                              &                    &~& & $\Bgl({}^{+0.32}_{-0.27}\Big)$ & & $\Bgl({}^{+0.30}_{-0.25}\Big)$ & (5.6) & (6.8)  &~\\[2pt] \hline
$\HWW$           & \cite{ATLAS:2014aga,ATLAS:VHWW}              & \cite{CMS:HWW}     &~& $1.22$ & $^{+0.23}_{-0.21}     $ & $0.90$ & $^{+0.23}_{-0.21}     $ &  6.8  &  4.8   &~\\
                 &                                              &                    &~& & $\Bgl({}^{+0.21}_{-0.20}\Big)$ & & $\Bgl({}^{+0.23}_{-0.20}\Big)$ & (5.8) & (5.6)  &~\\[2pt] \hline
$\Htt$           & \cite{Aad:2015vsa}                           & \cite{CMS:Htautau} &~& $1.41$ & $^{+0.40}_{-0.36}     $ & $0.88$ & $^{+0.30}_{-0.28}     $ &  4.4  &  3.4   &~\\
                 &                                              &                   &~ & & $\Bgl({}^{+0.37}_{-0.33}\Big)$ & & $\Bgl({}^{+0.31}_{-0.29}\Big)$ & (3.3) & (3.7)  &~\\[2pt] \hline
$\Hbb$           & \cite{Aad:2014xzb}                           & \cite{CMS:VHbb}    &~& $0.62$ & $^{+0.37}_{-0.37}     $ & $0.81$ & $^{+0.45}_{-0.43}     $ &  1.7  &  2.0   &~\\
                 &                                              &                    &~& & $\Bgl({}^{+0.39}_{-0.37}\Big)$ & & $\Bgl({}^{+0.45}_{-0.43}\Big)$ & (2.7) & (2.5)  &~\\[2pt] \hline
$\Hmm$           & \cite{Aad:2014xva}                           & \cite{CMS:Hmumu}   &~& $-0.6$ & $^{+3.6 }_{-3.6 }     $ & $0.9 $ & $^{+3.6 }_{-3.5 }     $ &       &        &~\\
                 &                                              &                    &~& & $\Bgl({}^{+3.6 }_{-3.6 }\Big)$ & & $\Bgl({}^{+3.3 }_{-3.2 }\Big)$ &       &        &~\\[2pt] \hline\hline
$ttH$ production & \cite{ATLAS:ttHhad,ATLAS:ttHlep,Aad:2014lma} & \cite{CMS:ttH}     &~& $1.9 $ & $^{+0.8 }_{-0.7 }     $ & $2.9 $ & $^{+1.0 }_{-0.9 }     $ &  2.7  &  3.6   &~\\
                 &                                              &                    &~& & $\Bgl({}^{+0.7 }_{-0.7 }\Big)$ & & $\Bgl({}^{+0.9 }_{-0.8 }\Big)$ & (1.6) & (1.3)  &~\\[2pt] \hline\hline
\end{tabular}
\end{center}
\end{table}

\section{Generic parameterisations of experimental results}
\label{sec:GenericParameterisation}

This section describes three generic parameterisations and presents their results. 
The first two are based on cross sections and branching fractions, either expressed as independent products $\sigma_i\cdot \BR^f$ for each channel~$i\to H\to f$, or as ratios of cross sections and branching fractions plus one reference $\sigma_i\cdot \BR^f$~product.  
In these parameterisations, the theoretical uncertainties in the signal inclusive cross sections for the various production processes do not affect the measured observables, in contrast to measurements of signal strengths, such as those described in~Section~\ref{sec:Mu}. 
These analyses lead to the most model-independent results presented in this paper and test, with minimal assumptions, the compatibility of the measurements with the~SM.
The third generic parameterisation is derived from the one described in Section~\ref{sec:kappas} and is based on ratios of coupling modifiers. 
None of these parameterisations incorporate any assumption about the Higgs boson total width other than the narrow-width approximation. Some theoretical and experimental systematic uncertainties largely cancel in the parameterisations involving ratios but at the current level of sensitivity the impact is small. 

\begin{table}[thb]
\caption{
Parameters of interest in the two generic parameterisations described in Sections~\ref{sec:sigBR9} and~\ref{sec:kappaParam}. 
For both parameterisations, the $gg \to H \to ZZ$ channel is chosen as a reference, expressed through the first row in the table. 
All other measurements are expressed as ratios of cross sections or branching fractions in the first column and of coupling modifiers in the second column. 
There are fewer parameters of interest in the case of the coupling parameterisation, in which the ratios of cross sections for the \aWH, \aZH, and \aVBF\ processes can all be expressed as functions of the two parameters, $\Rr_{Zg}$ and~$\Rr_{WZ}$. The slightly different additional assumptions in each parameterisation are discussed in the text.
}
\begin{center}
\setlength\extrarowheight{3pt}%
\begin{tabular}{cc} \hline\hline
\hsc $\sigma$ and \BR\ ratio parameterisation \hsc  & \hsc Coupling modifier ratio parameterisation \hsc   \\ \hline
 $\sigma(gg\to H\to ZZ)$         &      $\Cc_{gZ}= \Cc_{g}\cdot\Cc_{Z} / \Cc_{H} $         \\
 $\sigma_{\aVBF}/\sigma_{\aggF}$   &               \\
 $\sigma_{\aWH}/\sigma_{\aggF}$        &                 \\
 $\sigma_{\aZH}/\sigma_{\aggF}$        &  $\Rr_{Zg} = \Cc_{Z} / \Cc_{g}$                 \\
 $\sigma_{\attH}/\sigma_{\aggF}$       &    $\Rr_{tg} = \Cc_{t} / \Cc_{g}$                 \\  
 $\BR^{WW}/\BR^{ZZ}$         &    $\Rr_{WZ} = \Cc_{W} / \Cc_{Z}$                   \\
 $\BR^{\gamma\gamma}/\BR^{ZZ}$ &  $\Rr_{\gamma Z} = \Cc_{\gamma} / \Cc_{Z}$         \\
 $\BR^{\tau\tau}/\BR^{ZZ}$     &   $\Rr_{\tau Z} = \Cc_{\tau} / \Cc_{Z}$                   \\
 $\BR^{bb}/\BR^{ZZ}$           &     $\Rr_{bZ} = \Cc_{b} / \Cc_{Z}$                    \\ [2pt] \hline
\hline
\end{tabular}
\end{center}
\label{tab:genericmodels}
\end{table}

Table~\ref{tab:genericmodels} gives an overview of the parameters of interest for the two generic parameterisations involving ratios which are described in more detail in~Sections~\ref{sec:sigBR9}~and~\ref{sec:kappaParam}. The first row makes explicit that the $gg\to H\to ZZ$~channel is chosen as a reference. The $\Rr_{Zg} = \Cc_{Z} / \Cc_{g}$ term in the fourth row is related to the ratio of the \aZH\ and \aggF\ production cross sections. Once $\Rr_{WZ} = \Cc_{W} / \Cc_{Z}$~is also specified, the \aVBF, \aWH, and \aZH~production cross sections are fully defined. This explains the smaller number of independent parameters of interest in the coupling modifier ratio parameterisation compared to the parameterisation based mostly on ratios of cross sections and branching fractions. In addition, these two parameterisations rely on slightly different assumptions and approximations, which are summarised in Sections~\ref{sec:Mu} and~\ref{sec:kappas}. These  approximations are due to the fact that one cannot experimentally constrain all possible Higgs boson production processes and decay modes, in particular those that are expected to be small in the~SM, but might be enhanced, should specific BSM~physics scenarios be realised in nature.

\subsection{Parameterisations using cross sections and branching fractions}
\label{sec:sigBR}

\subsubsection{Parameterisation using independent products of cross sections and branching fractions}
\label{sec:sigBR5x5}

\begin{table*}[htb]
\centering
\caption{The signal parameterisation used to express the $\sigma_i \cdot \BR^f$ values for each specific channel~$i \to H\to f$. The values labelled with a~"$-$" are not measured and are therefore fixed to the SM~predictions.  
}
\label{tab:generic5x5}
\renewcommand{\arraystretch}{1.1}
\vspace{4mm}
\begin{tabular}{c|ccccc}
\hline\hline
Production process & \multicolumn{5}{|c}{Decay channel}\\
\hline
      &  $\Hyy$  &  $\Hzz$  &  $\Hww$  &  $\Htt$  &  $\Hbb$ \\
\hline
\aggF  & $(\sigma \cdot \BR)_{\aggF}^{\gamma\gamma}$ & $(\sigma \cdot \BR)_{\aggF}^{ZZ}$ & $(\sigma \cdot \BR)_{\aggF}^{WW}$ & $(\sigma \cdot \BR)_{\aggF}^{\tau\tau}$ & $-$ \\
\aVBF  & $(\sigma \cdot \BR)_{\aVBF}^{\gamma\gamma}$ & $(\sigma \cdot \BR)_{\aVBF}^{ZZ}$ & $(\sigma \cdot \BR)_{\aVBF}^{WW}$ & $(\sigma \cdot \BR)_{\aVBF}^{\tau\tau}$ & $-$ \\
\aWH   & $(\sigma \cdot \BR)_{\aWH}^{\gamma\gamma}$ & $(\sigma \cdot \BR)_{\aWH}^{ZZ}$ & $(\sigma \cdot \BR)_{\aWH}^{WW}$ & $(\sigma \cdot \BR)_{\aWH}^{\tau\tau}$ & $(\sigma \cdot \BR)_{\aWH}^{bb}$ \\
\aZH   & $(\sigma \cdot \BR)_{\aZH}^{\gamma\gamma}$ & $(\sigma \cdot \BR)_{\aZH}^{ZZ}$ & $(\sigma \cdot \BR)_{\aZH}^{WW}$ & $(\sigma \cdot \BR)_{\aZH}^{\tau\tau}$ & $(\sigma \cdot \BR)_{\aZH}^{bb}$ \\
\attH  & $(\sigma \cdot \BR)_{\attH}^{\gamma\gamma}$ & $(\sigma \cdot \BR)_{\attH}^{ZZ}$ & $(\sigma \cdot \BR)_{\attH}^{WW}$ & $(\sigma \cdot \BR)_{\attH}^{\tau\tau}$ & $(\sigma \cdot \BR)_{\attH}^{bb}$ \\
\hline\hline
\end{tabular}
\end{table*}

In a very generic approach, one can extract for each specific channel~$i \to H\to f$ a measurement of the product~$\sigma_i \cdot \BR^f$  and then compare it to the theoretical prediction. Based on all the categories considered in the various analyses and on the five production processes (\aggF, \aVBF, \aWH, \aZH, and \attH) and five main decay channels ($\HZZ$, $\HWW$, $\Hyy$, $\Htt$, and~$\Hbb$) considered in this paper, there are in principle 25~such independent products to be measured. In practice, as already mentioned, the~\aggF\ and \aVBF~production processes are not probed in the case of the $\Hbb$~decay mode and are assumed to have the values predicted by the~SM, so the fit is performed with 23~parameters of interest, which are specified in~Table~\ref{tab:generic5x5}. The individual experiments cannot provide constraints on all the parameters of interest because of the low overall expected and observed yields in the current data. Even when combining the ATLAS and CMS data, the \aZH,~\aWH, and \attH~production processes cannot be measured with meaningful precision in the $\Hzz$~decay channel. The fit results are therefore quoted only for the remaining 20~parameters and for the combined ATLAS and CMS data.

\begin{table*}[htb]
  \scriptsize
\centering
\caption{Best fit values of $\sigma_i \cdot \BR^f$ for each specific channel~$i \to H\to f$, as obtained from the generic parameterisation with 23~parameters for the combination of the ATLAS and CMS measurements, using the $\sqrt{s}=7$ and 8~TeV data. The cross sections are given for $\sqrt{s}=8$~TeV, assuming the SM values for~$\sigma_i(7\TeV)/\sigma_i(8\TeV)$. 
The results are shown together with their total uncertainties and their breakdown into statistical and systematic components. 
The expected uncertainties in the measurements are displayed in parentheses. The SM~predictions~\protect\cite{Heinemeyer:2013tqa} and the ratios of the results to these SM~predictions are also shown.
The values labelled with a~"$-$" are either not measured with a meaningful precision and therefore not quoted, in the case of the $\Hzz$ decay channel for the \aWH,~\aZH, and \attH~production processes, or not measured at all and therefore fixed to their corresponding SM~predictions, in the case of the $\Hbb$~decay mode for the \aggF\ and \aVBF~production processes.
}
\label{tab:generic5x5results}
\begin{adjustbox}{width=\textwidth}%
\setlength\extrarowheight{3pt}%
\setlength\tabcolsep{1pt}%
\renewcommand{\arraystretch}{1.1}%
\begin{tabular}{l@{\hskip 2pt}|@{\hskip 2pt}l@{\hskip 2pt}|rlcc|rlcc|rlcc|rlcc|rlcc}
\multicolumn{22}{c}{~~~}\\ \hline\hline
\multicolumn{2}{l|}{\multirow{2}{*}{\begin{tabular}{@{}l}Production\\[-3pt] process\\\mbox{}\end{tabular}}} &  \multicolumn{20}{c}{Decay mode}\\
\cline{3-22}
\multicolumn{2}{l|}{} & \multicolumn{4}{c|}{$\Hyy$ [fb]}   &  \multicolumn{4}{c|}{$\Hzz$ [fb]}  &  \multicolumn{4}{c|}{$\Hww$ [pb]}  &  \multicolumn{4}{c|}{$\Htt$ [fb]} &  \multicolumn{4}{c}{$\Hbb$ [pb]}  \\ \cline{3-22}
\multicolumn{2}{l|}{} & \multicolumn{2}{c}{Best fit}  & \multicolumn{2}{c|}{Uncertainty}  &  \multicolumn{2}{c}{Best fit} & \multicolumn{2}{c|}{Uncertainty}  &  \multicolumn{2}{c}{Best fit}  & \multicolumn{2}{c|}{Uncertainty}  &  \multicolumn{2}{c}{Best fit}  & \multicolumn{2}{c|}{Uncertainty}  &  \multicolumn{2}{c}{Best fit}  & \multicolumn{2}{c}{Uncertainty}  \\[-3pt]
\multicolumn{2}{l|}{} & \multicolumn{2}{c}{value} &  Stat  &  \multicolumn{1}{c|}{Syst}  &  \multicolumn{2}{c}{value} &  Stat  &  \multicolumn{1}{c|}{Syst}  &  \multicolumn{2}{c}{value} &  Stat  &  \multicolumn{1}{c|}{Syst}  &  \multicolumn{2}{c}{value} &  Stat  &  \multicolumn{1}{c|}{Syst}  &  \multicolumn{2}{c}{value} &  Stat  &  Syst  \\
\hline\hline
\renewcommand{\arraystretch}{1.8}
\aggF
& Measured
& $48.0$ & $       ^{+10.0}_{-9.7}      $ & $       ^{+9.4} _{-9.4}      $ & $       ^{+3.2} _{-2.3}      $ 
& $580 $ & $       ^{+170} _{-160}      $ & $       ^{+170} _{-160}      $ & $       ^{+40}  _{-40}       $ 
& $3.5 $ & $       ^{+0.7} _{-0.7}      $ & $       ^{+0.5} _{-0.5}      $ & $       ^{+0.5} _{-0.5}      $ 
& $1300$ & $       ^{+700} _{-700}      $ & $       ^{+400} _{-400}      $ & $       ^{+500} _{-500}      $ 
& \multicolumn{4}{c}{$-$}                                                                                   
\\
&
&        & $\Bgl({}^{+9.7} _{-9.5} \Big)$ & $\Big({}^{+9.4} _{-9.4} \Big)$ & $\Big({}^{+2.5} _{-1.6} \Big)$ 
&        & $\Bgl({}^{+150} _{-130} \Big)$ & $\Big({}^{+140} _{-130} \Big)$ & $\Big({}^{+30}  _{-20}  \Big)$ 
&        & $\Bgl({}^{+0.7} _{-0.7} \Big)$ & $\Big({}^{+0.5} _{-0.5} \Big)$ & $\Big({}^{+0.5} _{-0.5} \Big)$ 
&        & $\Bgl({}^{+700} _{-700} \Big)$ & $\Big({}^{+400} _{-400} \Big)$ & $\Big({}^{+500} _{-500} \Big)$ 
& \multicolumn{4}{c}{$-$}                                                                                   
\\
& Predicted
&$44    $& $\pm 5     $                   &                                &                                
&$510   $& $\pm 60    $                   &                                &                                
&$4.1   $& $\pm 0.5   $                   &                                &                                
&$1210  $& $\pm 140   $                   &                                &                                
&$11.0  $& $\pm 1.2   $                   &                                &                                
\\
& Ratio
& $1.10$ & $       ^{+0.23}_{-0.22}     $ & $       ^{+0.22}_{-0.21}     $ & $       ^{+0.07}_{-0.05}     $ 
& $1.13$ & $       ^{+0.34}_{-0.31}     $ & $       ^{+0.33}_{-0.30}     $ & $       ^{+0.09}_{-0.07}     $ 
& $0.84$ & $       ^{+0.17}_{-0.17}     $ & $       ^{+0.12}_{-0.12}     $ & $       ^{+0.12}_{-0.11}     $ 
& $1.0 $ & $       ^{+0.6} _{-0.6}      $ & $       ^{+0.4} _{-0.4}      $ & $       ^{+0.4} _{-0.4}      $ 
& \multicolumn{4}{c}{$-$}                                                                                   
\\[2pt] \hline
\aVBF
& Measured
& $4.6 $ & $       ^{+1.9} _{-1.8}      $ & $       ^{+1.8} _{-1.7}      $ & $       ^{+0.6} _{-0.5}      $ 
& $3   $ & $       ^{+46}  _{-26}       $ & $       ^{+46}  _{-25}       $ & $       ^{+7}   _{-7}        $ 
& $0.39$ & $       ^{+0.14}_{-0.13}     $ & $       ^{+0.13}_{-0.12}     $ & $       ^{+0.07}_{-0.05}     $ 
& $125 $ & $       ^{+39}  _{-37}       $ & $       ^{+34}  _{-32}       $ & $       ^{+19}  _{-18}       $ 
& \multicolumn{4}{c}{$-$}                                                                                   
\\
&
&        & $\Bgl({}^{+1.8} _{-1.6} \Big)$ & $\Big({}^{+1.7} _{-1.6} \Big)$ & $\Big({}^{+0.5} _{-0.4} \Big)$ 
&        & $\Bgl({}^{+60}  _{-39}  \Big)$ & $\Big({}^{+60}  _{-39}  \Big)$ & $\Big({}^{+8}   _{-5}   \Big)$ 
&        & $\Bgl({}^{+0.15}_{-0.13}\Big)$ & $\Big({}^{+0.13}_{-0.12}\Big)$ & $\Big({}^{+0.07}_{-0.06}\Big)$ 
&        & $\Bgl({}^{+39}  _{-37}  \Big)$ & $\Big({}^{+34}  _{-32}  \Big)$ & $\Big({}^{+19}  _{-18}  \Big)$ 
& \multicolumn{4}{c}{$-$}                                                                                   
\\
& Predicted
&$3.60  $& $\pm 0.20  $                   &                                &                                
&$42.2  $& $\pm 2.0   $                   &                                &                                
&$0.341 $& $\pm 0.017 $                   &                                &                                
&$100   $& $\pm 6     $                   &                                &                                
&$0.91  $& $\pm 0.04  $                   &                                &                                
\\
& Ratio
& $1.3 $ & $       ^{+0.5} _{-0.5}      $ & $       ^{+0.5} _{-0.5}      $ & $       ^{+0.2} _{-0.1}      $ 
& $0.1 $ & $       ^{+1.1} _{-0.6}      $ & $       ^{+1.1} _{-0.6}      $ & $       ^{+0.2} _{-0.2}      $ 
& $1.2 $ & $       ^{+0.4} _{-0.4}      $ & $       ^{+0.4} _{-0.3}      $ & $       ^{+0.2} _{-0.2}      $ 
& $1.3 $ & $       ^{+0.4} _{-0.4}      $ & $       ^{+0.3} _{-0.3}      $ & $       ^{+0.2} _{-0.2}      $ 
& \multicolumn{4}{c}{$-$}                                                                                   
\\[2pt] \hline
\aWH
& Measured
& $0.7 $ & $       ^{+2.1} _{-1.9}      $ & $       ^{+2.1} _{-1.8}      $ & $       ^{+0.3} _{-0.3}      $ 
& \multicolumn{4}{c|}{$-$}                                                                                  
& $0.24$ & $       ^{+0.18}_{-0.16}     $ & $       ^{+0.15}_{-0.14}     $ & $       ^{+0.10}_{-0.08}     $ 
& $-64 $ & $       ^{+64}  _{-61}       $ & $       ^{+55}  _{-50}       $ & $       ^{+32}  _{-34}       $ 
& $0.42$ & $       ^{+0.21}_{-0.20}     $ & $       ^{+0.17}_{-0.16}     $ & $       ^{+0.12}_{-0.11}     $ 
\\
&
&        & $\Bgl({}^{+1.9} _{-1.8} \Big)$ & $\Big({}^{+1.9} _{-1.8} \Big)$ & $\Big({}^{+0.1} _{-0.1} \Big)$ 
& \multicolumn{4}{c|}{$-$}                                                                                  
&        & $\Bgl({}^{+0.16}_{-0.14}\Big)$ & $\Big({}^{+0.14}_{-0.13}\Big)$ & $\Big({}^{+0.08}_{-0.07}\Big)$ 
&        & $\Bgl({}^{+67}  _{-64}  \Big)$ & $\Big({}^{+60}  _{-54}  \Big)$ & $\Big({}^{+30}  _{-32}  \Big)$ 
&        & $\Bgl({}^{+0.22}_{-0.21}\Big)$ & $\Big({}^{+0.18}_{-0.17}\Big)$ & $\Big({}^{+0.12}_{-0.11}\Big)$ 
\\
& Predicted
&$1.60  $& $\pm 0.09  $                   &                                &                                
&$18.8  $& $\pm 0.9   $                   &                                &                                
&$0.152 $& $\pm 0.007 $                   &                                &                                
&$44.3  $& $\pm 2.8   $                   &                                &                                
&$0.404 $& $\pm 0.017 $                   &                                &                                
\\
& Ratio
& $0.5 $ & $       ^{+1.3} _{-1.2}      $ & $       ^{+1.3} _{-1.1}      $ & $       ^{+0.2} _{-0.2}      $ 
& \multicolumn{4}{c|}{$-$}                                                                                  
& $1.6 $ & $       ^{+1.2} _{-1.0}      $ & $       ^{+1.0} _{-0.9}      $ & $       ^{+0.6} _{-0.5}      $ 
& $-1.4$ & $       ^{+1.4} _{-1.4}      $ & $       ^{+1.2} _{-1.1}      $ & $       ^{+0.7} _{-0.8}      $ 
& $1.0 $ & $       ^{+0.5} _{-0.5}      $ & $       ^{+0.4} _{-0.4}      $ & $       ^{+0.3} _{-0.3}      $ 
\\[2pt] \hline
\aZH
& Measured
& $0.5 $ & $       ^{+2.9} _{-2.4}      $ & $       ^{+2.8} _{-2.3}      $ & $       ^{+0.5} _{-0.2}      $ 
& \multicolumn{4}{c|}{$-$}                                                                                  
& $0.53$ & $       ^{+0.23}_{-0.20}     $ & $       ^{+0.21}_{-0.19}     $ & $       ^{+0.10}_{-0.07}     $ 
& $58  $ & $       ^{+56}  _{-47}       $ & $       ^{+52}  _{-44}       $ & $       ^{+20}  _{-16}       $ 
& $0.08$ & $       ^{+0.09}_{-0.09}     $ & $       ^{+0.08}_{-0.08}     $ & $       ^{+0.04}_{-0.04}     $ 
\\
&
&        & $\Bgl({}^{+2.3} _{-1.9} \Big)$ & $\Big({}^{+2.3} _{-1.9} \Big)$ & $\Big({}^{+0.1} _{-0.1} \Big)$ 
& \multicolumn{4}{c|}{$-$}                                                                                  
&        & $\Bgl({}^{+0.17}_{-0.14}\Big)$ & $\Big({}^{+0.16}_{-0.14}\Big)$ & $\Big({}^{+0.05}_{-0.04}\Big)$ 
&        & $\Bgl({}^{+49}  _{-40}  \Big)$ & $\Big({}^{+46}  _{-38}  \Big)$ & $\Big({}^{+16}  _{-12}  \Big)$ 
&        & $\Bgl({}^{+0.10}_{-0.09}\Big)$ & $\Big({}^{+0.09}_{-0.08}\Big)$ & $\Big({}^{+0.05}_{-0.04}\Big)$ 
\\
& Predicted
&$0.94  $& $\pm 0.06  $                   &                                &                                
&$11.1  $& $\pm 0.6   $                   &                                &                                
&$0.089 $& $\pm 0.005 $                   &                                &                                
&$26.1  $& $\pm 1.8   $                   &                                &                                
&$0.238 $& $\pm 0.012 $                   &                                &                                
\\
& Ratio
& $0.5 $ & $       ^{+3.0} _{-2.5}      $ & $       ^{+3.0} _{-2.5}      $ & $       ^{+0.5} _{-0.2}      $ 
& \multicolumn{4}{c|}{$-$}                                                                                  
& $5.9 $ & $       ^{+2.6} _{-2.2}      $ & $       ^{+2.3} _{-2.1}      $ & $       ^{+1.1} _{-0.8}      $ 
& $2.2 $ & $       ^{+2.2} _{-1.8}      $ & $       ^{+2.0} _{-1.7}      $ & $       ^{+0.8} _{-0.6}      $ 
& $0.4 $ & $       ^{+0.4} _{-0.4}      $ & $       ^{+0.3} _{-0.3}      $ & $       ^{+0.2} _{-0.2}      $ 
\\[2pt] \hline
\attH
& Measured
& $0.64$ & $       ^{+0.48}_{-0.38}     $ & $       ^{+0.48}_{-0.38}     $ & $       ^{+0.07}_{-0.04}     $ 
& \multicolumn{4}{c|}{$-$}                                                                                  
& $0.14$ & $       ^{+0.05}_{-0.05}     $ & $       ^{+0.04}_{-0.04}     $ & $       ^{+0.03}_{-0.03}     $ 
& $-15 $ & $       ^{+30}  _{-26}       $ & $       ^{+26}  _{-22}       $ & $       ^{+15}  _{-15}       $ 
& $0.08$ & $       ^{+0.07}_{-0.07}     $ & $       ^{+0.04}_{-0.04}     $ & $       ^{+0.06}_{-0.06}     $ 
\\
&
&        & $\Bgl({}^{+0.45}_{-0.34}\Big)$ & $\Big({}^{+0.44}_{-0.33}\Big)$ & $\Big({}^{+0.10}_{-0.05}\Big)$ 
& \multicolumn{4}{c|}{$-$}                                                                                  
&        & $\Bgl({}^{+0.04}_{-0.04}\Big)$ & $\Big({}^{+0.04}_{-0.04}\Big)$ & $\Big({}^{+0.02}_{-0.02}\Big)$ 
&        & $\Bgl({}^{+31}  _{-26}  \Big)$ & $\Big({}^{+26}  _{-22}  \Big)$ & $\Big({}^{+16}  _{-13}  \Big)$ 
&        & $\Bgl({}^{+0.07}_{-0.06}\Big)$ & $\Big({}^{+0.04}_{-0.04}\Big)$ & $\Big({}^{+0.06}_{-0.05}\Big)$ 
\\
& Predicted
&$0.294 $& $\pm 0.035 $                   &                                &                                
&$3.4   $& $\pm 0.4   $                   &                                &                                
&$0.0279$& $\pm 0.0032$                   &                                &                                
&$8.1   $& $\pm 1.0   $                   &                                &                                
&$0.074 $& $\pm 0.008 $                   &                                &                                
\\
& Ratio
& $2.2 $ & $       ^{+1.6} _{-1.3}      $ & $       ^{+1.6} _{-1.3}      $ & $       ^{+0.2} _{-0.1}      $ 
& \multicolumn{4}{c|}{$-$}                                                                                  
& $5.0 $ & $       ^{+1.8} _{-1.7}      $ & $       ^{+1.5} _{-1.5}      $ & $       ^{+1.0} _{-0.9}      $ 
& $-1.9$ & $       ^{+3.7} _{-3.3}      $ & $       ^{+3.2} _{-2.7}      $ & $       ^{+1.9} _{-1.8}      $ 
& $1.1 $ & $       ^{+1.0} _{-1.0}      $ & $       ^{+0.5} _{-0.5}      $ & $       ^{+0.8} _{-0.8}      $ 
\\[2pt]

\hline\hline
\end{tabular}
\end{adjustbox}
\end{table*}

Table~\ref{tab:generic5x5results} presents, for the combination of ATLAS and CMS, the fit results for each $\sigma_i \cdot \BR^f$~product along with its statistical and systematic uncertainties. The corresponding SM~predictions are also given. The ratios of the fit results to SM~predictions are included in~Table~\ref{tab:generic5x5results} and displayed in~Fig.~\ref{fig:plot_A15PDvertical}. Figure~\ref{fig:plot_A15PDvertical} additionally shows the theoretical uncertainties in the SM~predictions for the fitted parameters.  
In almost all cases, the dominant uncertainty is statistical. The results presented in~Table~\ref{tab:generic5x5results} and Fig.~\ref{fig:plot_A15PDvertical} clearly exhibit which decay modes are probed best for each production process, and conversely which production processes are probed best for each decay mode. With the current sensitivity of the combination, six of the $\sigma_i \cdot \BR^f$~products can be measured with a precision better than~40\%, namely the $\Hyy$, $\HZZ$, and $\HWW$ decay modes for the~\aggF~production process, and the $\Hyy$, $\HWW$, and $\Htt$ decay modes for the~\aVBF~production process. Because of the sizeable cross-contamination between the~\aggF\ and \aVBF~categories, the corresponding results are significantly anticorrelated, as illustrated by the measured correlation matrix in Fig.~\ref{fig:correlation5x5} of~Appendix~\ref{sec:app_correlations}.

\subsubsection{Parameterisation using ratios of cross sections and branching fractions}
\label{sec:sigBR9}
 
If there is only one Higgs boson, each row or column in~Table~\ref{tab:generic5x5} can be derived from the others by identical ratios of cross sections for the rows and of branching fractions for the columns. Therefore, in a second generic approach, ratios of cross sections and of branching fractions can be extracted from a combined fit to the data by normalising the yield of any specific channel $i \to H\to f$ to a reference process. In this paper, the $gg \to H \to ZZ$~channel is chosen as the reference 
because it has very little background and is one of the channels with the smallest overall and systematic uncertainties. The $gg \to H \to WW$~channel, which has the smallest overall uncertainty but larger systematic uncertainties, is used as an alternate reference for comparison, and the corresponding results are reported in~Appendix~\ref{sec:app_B1WW}.

The product of the cross section and the branching fraction of $i\to H\to f$ can then be expressed using the ratios as:
\begin{equation}
\sigma_i\cdot \BR^f = \sigma(gg\to H\to ZZ) \cdot \left(\frac{\sigma_i}{\sigma_{\aggF}}\right)\cdot \left(\frac{\BR^f}{\BR^{ZZ}}\right),
\label{eq:sigratio}
\end{equation}
where $\sigma(gg\to H\to ZZ) = \sigma_{\aggF}\cdot \BR^{ZZ}$ in the narrow-width approximation. 
With~$\sigma(gg\to H\to ZZ)$ constraining the overall normalisation, the ratios in~Eq.~(\ref{eq:sigratio}) can be determined separately, based on the five production processes (\aggF, \aVBF, \aWH, \aZH, and \attH) and five decay modes ($\HZZ$, $\HWW$, $\Hyy$, $\Htt$, and~$\Hbb$). 
The combined fit results can be presented as a function of nine parameters of interest: one reference cross section times branching fraction,~$\sigma(gg\to H\to ZZ)$, four ratios of production cross sections,~$\sigma_i/\sigma_{\aggF}$, and four ratios of branching fractions,~$\BR^f/\BR^{ZZ}$, as reported in the left column of Table~\ref{tab:genericmodels}. 

\begin{figure}[htb]
\centering
\vspace{10mm}
\includegraphics[width=1.0\textwidth]{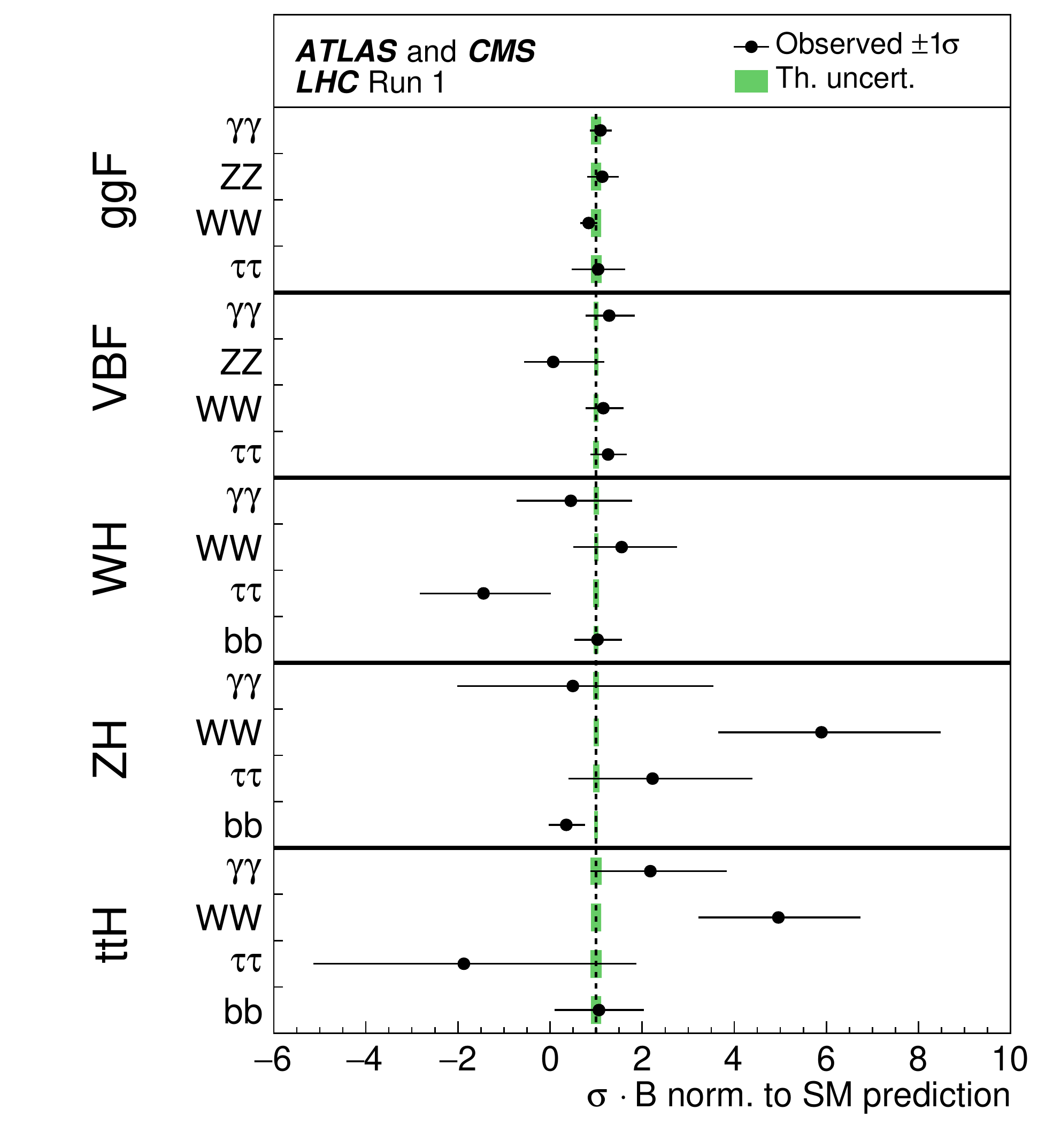}
\caption{Best fit values of $\sigma_i \cdot \BR^f$ for each specific channel~$i \to H\to f$, as obtained from the generic parameterisation with 23~parameters for the combination of the ATLAS and CMS measurements. The error bars indicate the $1\sigma
$~intervals. The fit results are normalised to the SM predictions for the various parameters and the shaded bands indicate the theoretical uncertainties in these predictions.
Only 20~parameters are shown because some are either not measured with a meaningful precision, in the case of the $\Hzz$ decay channel for the \aWH,~\aZH, and \attH~production processes, or not measured at all and therefore fixed to their corresponding SM~predictions, in the case of the $\Hbb$~decay mode for the \aggF\ and \aVBF~production processes. 
}
\label{fig:plot_A15PDvertical}
\end{figure}

\clearpage

Expressing the measurements through ratios of cross sections and branching fractions has the advantage that the ratios are independent of the theoretical predictions for the inclusive production cross sections and decay branching fractions of the Higgs boson. 
In particular, they  are not subject to the dominant signal theoretical uncertainties in the inclusive cross sections for the various production processes. 
These measurements will therefore remain valid when, for example, improved theoretical calculations of Higgs boson production cross sections become available. 
The remaining theoretical uncertainties are 
those 
related to the acceptances and selection efficiencies in the various categories, for which SM~Higgs boson production and decay kinematics are assumed in the simulation, based on the MC~generators discussed in~Section~\ref{sec:Generators}. 

Table~\ref{tab:B1ZZ_results} shows the results of the fit to the data with a breakdown of the uncertainties into their statistical and systematic components. The full breakdown of the uncertainties into the four components is shown in Table~\ref{tab:B1ZZ_results_full} of Appendix~\ref{sec:app_B1WW}, while the measured correlation matrix can be found in Fig.~\ref{fig:correlationB1ZZ} of~Appendix~\ref{sec:app_correlations}. The results are shown for the combination of ATLAS and CMS and also separately for each experiment. 
They are illustrated in~Fig.~\ref{fig:plot_B1ZZ}, where the fit result for each parameter is normalised to the corresponding SM~prediction. 
Also shown in~Fig.~\ref{fig:plot_B1ZZ} are the theoretical uncertainties in the SM~predictions for the fitted parameters.
For the ratios of branching fractions, the theoretical uncertainties in the predictions are barely visible since they are below~5\%.
Compared to the results of the fit of Table~\ref{tab:generic5x5results}, where the \aggF~parameters are independent for each decay mode, the uncertainties in~$\sigma(gg\to H\to ZZ)$ are reduced by almost a factor of~two in Table~\ref{tab:B1ZZ_results}, owing to the contributions from the other decay channels (mainly $\Hyy$ and~$\HWW$). The total uncertainty in~$\sigma(gg\to H\to ZZ)$ is approximately~18\%, with its main contribution coming from the statistical uncertainty. The total relative systematic uncertainty is only~$\sim$4\%. 
Appendix~\ref{sec:app_B1WW} shows the results obtained when choosing the $\HWW$~decay mode as an alternative reference process. 
This yields a smaller total uncertainty of approximately~15\% in~$\sigma(gg\to H\to WW)$, but with a much larger contribution of~$\sim$11\% from the systematic uncertainties. 
The ratios $\sigma_{\aVBF}/\sigma_{\aggF}$, $\BR^{WW}/\BR^{ZZ}$,~$\BR^{\gamma\gamma}/\BR^{ZZ}$, and~$\BR^{\tau\tau}/\BR^{ZZ}$ are measured with relative uncertainties of~$30$--$40$\%.

\begin{table}[htbp]
  \scriptsize
  \centering
  \caption{Best fit values of $\sigma(gg\to H\to ZZ)$, $\sigma_i/\sigma_{\aggF}$, and $\BR^f/\BR^{ZZ}$, as obtained from the generic parameterisation with nine parameters for the combined analysis of the $\sqrt{s}=7$ and 8~TeV data. The  values involving cross sections are given for $\sqrt{s}=8$~TeV, assuming the SM values for~$\sigma_i(7\TeV)/\sigma_i(8\TeV)$. 
The results are reported for the combination of ATLAS and CMS and also separately for each experiment, together with their total uncertainties and their breakdown into statistical and systematic components. 
The expected uncertainties in the measurements are displayed in parentheses. 
The SM predictions~\protect\cite{Heinemeyer:2013tqa} are also shown with their total uncertainties. 
}
  \label{tab:B1ZZ_results}
  \begin{adjustbox}{width=\textwidth}%
    \setlength\extrarowheight{3pt}%
    \setlength\tabcolsep{3pt}%
    \begin{tabular}{l|r@{\hskip 0.5ex}l|r@{\hskip 0.5ex}lcc|r@{\hskip 0.5ex}lcc|r@{\hskip 0.5ex}lcc}\hline\hline
       Parameter  & \multicolumn{2}{c|}{SM prediction}   & \multicolumn{2}{c}{Best fit} &  \multicolumn{2}{c|}{Uncertainty}  & \multicolumn{2}{c}{Best fit} &  \multicolumn{2}{c|}{Uncertainty} & \multicolumn{2}{c}{Best fit} &  \multicolumn{2}{c}{Uncertainty}     \\[-3pt]
                                        &  &       & \multicolumn{2}{c}{value}  & Stat                  & Syst & \multicolumn{2}{c}{value} & Stat                  & Syst &  \multicolumn{2}{c}{value} & Stat           & Syst \\
      \hline & & & \multicolumn{4}{c|}{ATLAS+CMS} & \multicolumn{4}{c|}{ATLAS} &  \multicolumn{4}{c}{CMS}  \\
\multirow{2}{*}{\begin{tabular}{@{}l}$\sigma(gg\to$\\$H\to ZZ)$ [pb]\end{tabular}} & $0.51$  & $\pm 0.06$                 
& 0.59  & $       ^{+0.11} _{-0.10}      $ & $       ^{+0.11} _{-0.10}      $ & $       ^{+0.02} _{-0.02}      $          
& 0.77  & $       ^{+0.19} _{-0.17}      $ & $       ^{+0.19} _{-0.16}      $ & $       ^{+0.05} _{-0.03}      $          
& 0.44  & $       ^{+0.14} _{-0.12}      $ & $       ^{+0.13} _{-0.11}      $ & $       ^{+0.05} _{-0.03}      $ \\       
& &&    & $\Bgl({}^{+0.11} _{-0.10} \Big)$ & $\Big({}^{+0.11} _{-0.09} \Big)$ & $\Big({}^{+0.03} _{-0.02} \Big)$          
&       & $\Bgl({}^{+0.16} _{-0.14} \Big)$ & $\Big({}^{+0.16} _{-0.13} \Big)$ & $\Big({}^{+0.03} _{-0.02} \Big)$          
&       & $\Bgl({}^{+0.15} _{-0.13} \Big)$ & $\Big({}^{+0.15} _{-0.13} \Big)$ & $\Big({}^{+0.04} _{-0.03} \Big)$ \\[1mm]  
$\sigma_{\aVBF}/\sigma_{\aggF}$                                    & $0.082$ & $\pm 0.009$                                
& 0.109 & $       ^{+0.034}_{-0.027}     $ & $       ^{+0.029}_{-0.024}     $ & $       ^{+0.018}_{-0.013}     $          
& 0.079 & $       ^{+0.035}_{-0.026}     $ & $       ^{+0.030}_{-0.023}     $ & $       ^{+0.019}_{-0.012}     $          
& 0.138 & $       ^{+0.073}_{-0.051}     $ & $       ^{+0.061}_{-0.046}     $ & $       ^{+0.039}_{-0.023}     $ \\       
& &&    & $\Bgl({}^{+0.029}_{-0.024}\Big)$ & $\Big({}^{+0.024}_{-0.020}\Big)$ & $\Big({}^{+0.016}_{-0.012}\Big)$          
&       & $\Bgl({}^{+0.042}_{-0.031}\Big)$ & $\Big({}^{+0.036}_{-0.028}\Big)$ & $\Big({}^{+0.022}_{-0.014}\Big)$          
&       & $\Bgl({}^{+0.043}_{-0.033}\Big)$ & $\Big({}^{+0.037}_{-0.029}\Big)$ & $\Big({}^{+0.023}_{-0.015}\Big)$ \\[1mm]  
$\sigma_{\aWH}/\sigma_{\aggF}$                                     & $0.037$ & $\pm 0.004$                                
& 0.031 & $       ^{+0.028}_{-0.026}     $ & $       ^{+0.024}_{-0.022}     $ & $       ^{+0.015}_{-0.014}     $          
& 0.054 & $       ^{+0.036}_{-0.026}     $ & $       ^{+0.031}_{-0.023}     $ & $       ^{+0.020}_{-0.013}     $          
& 0.005 & $       ^{+0.044}_{-0.037}     $ & $       ^{+0.037}_{-0.028}     $ & $       ^{+0.023}_{-0.024}     $ \\       
& &&    & $\Bgl({}^{+0.021}_{-0.017}\Big)$ & $\Big({}^{+0.019}_{-0.015}\Big)$ & $\Big({}^{+0.011}_{-0.007}\Big)$          
&       & $\Bgl({}^{+0.033}_{-0.022}\Big)$ & $\Big({}^{+0.029}_{-0.020}\Big)$ & $\Big({}^{+0.015}_{-0.009}\Big)$          
&       & $\Bgl({}^{+0.032}_{-0.022}\Big)$ & $\Big({}^{+0.027}_{-0.020}\Big)$ & $\Big({}^{+0.017}_{-0.010}\Big)$ \\[1mm]  
$\sigma_{\aZH}/\sigma_{\aggF}$                                     & $0.0216$ & $\pm 0.0024$                              
& 0.066 & $       ^{+0.039}_{-0.031}     $ & $       ^{+0.032}_{-0.025}     $ & $       ^{+0.023}_{-0.018}     $          
& 0.013 & $       ^{+0.028}_{-0.014}     $ & $       ^{+0.021}_{-0.012}     $ & $       ^{+0.018}_{-0.007}     $          
& 0.123 & $       ^{+0.076}_{-0.053}     $ & $       ^{+0.063}_{-0.046}     $ & $       ^{+0.044}_{-0.026}     $ \\       
& &&    & $\Bgl({}^{+0.016}_{-0.011}\Big)$ & $\Big({}^{+0.014}_{-0.010}\Big)$ & $\Big({}^{+0.009}_{-0.004}\Big)$          
&       & $\Bgl({}^{+0.027}_{-0.014}\Big)$ & $\Big({}^{+0.023}_{-0.013}\Big)$ & $\Big({}^{+0.014}_{-0.005}\Big)$          
&       & $\Bgl({}^{+0.024}_{-0.013}\Big)$ & $\Big({}^{+0.020}_{-0.012}\Big)$ & $\Big({}^{+0.014}_{-0.006}\Big)$ \\[1mm]  
$\sigma_{\attH}/\sigma_{\aggF}$                                    & $0.0067$ & $\pm 0.0010$                              
& 0.022 & $       ^{+0.007}_{-0.006}     $ & $       ^{+0.005}_{-0.005}     $ & $       ^{+0.004}_{-0.003}     $          
& 0.013 & $       ^{+0.007}_{-0.005}     $ & $       ^{+0.005}_{-0.004}     $ & $       ^{+0.004}_{-0.003}     $          
& 0.034 & $       ^{+0.016}_{-0.012}     $ & $       ^{+0.012}_{-0.010}     $ & $       ^{+0.010}_{-0.006}     $ \\       
& &&    & $\Bgl({}^{+0.004}_{-0.004}\Big)$ & $\Big({}^{+0.003}_{-0.003}\Big)$ & $\Big({}^{+0.003}_{-0.002}\Big)$          
&       & $\Bgl({}^{+0.006}_{-0.004}\Big)$ & $\Big({}^{+0.005}_{-0.004}\Big)$ & $\Big({}^{+0.004}_{-0.003}\Big)$          
&       & $\Bgl({}^{+0.007}_{-0.005}\Big)$ & $\Big({}^{+0.005}_{-0.004}\Big)$ & $\Big({}^{+0.004}_{-0.004}\Big)$ \\[1mm]  
$\BR^{WW}/\BR^{ZZ}$                                                & $8.09$  & $\pm <0.01$                                
& 6.7   & $       ^{+1.6}  _{-1.3}       $ & $       ^{+1.5}  _{-1.2}       $ & $       ^{+0.6}  _{-0.5}       $          
& 6.5   & $       ^{+2.1}  _{-1.6}       $ & $       ^{+2.0}  _{-1.4}       $ & $       ^{+0.8}  _{-0.6}       $          
& 7.1   & $       ^{+2.9}  _{-2.1}       $ & $       ^{+2.6}  _{-1.8}       $ & $       ^{+1.3}  _{-0.9}       $ \\       
& &&    & $\Bgl({}^{+2.2}  _{-1.7}  \Big)$ & $\Big({}^{+2.0}  _{-1.6}  \Big)$ & $\Big({}^{+0.9}  _{-0.7}  \Big)$          
&       & $\Bgl({}^{+3.5}  _{-2.4}  \Big)$ & $\Big({}^{+3.3}  _{-2.2}  \Big)$ & $\Big({}^{+1.2}  _{-0.9}  \Big)$          
&       & $\Bgl({}^{+3.2}  _{-2.2}  \Big)$ & $\Big({}^{+2.9}  _{-2.0}  \Big)$ & $\Big({}^{+1.4}  _{-1.0}  \Big)$ \\[1mm]  
$\BR^{\gamma\gamma}/\BR^{ZZ}$                                      & $0.0854$ & $\pm 0.0010$                              
& 0.069 & $       ^{+0.018}_{-0.014}     $ & $       ^{+0.018}_{-0.014}     $ & $       ^{+0.004}_{-0.003}     $          
& 0.062 & $       ^{+0.024}_{-0.018}     $ & $       ^{+0.023}_{-0.017}     $ & $       ^{+0.007}_{-0.005}     $          
& 0.079 & $       ^{+0.034}_{-0.023}     $ & $       ^{+0.032}_{-0.023}     $ & $       ^{+0.010}_{-0.006}     $ \\       
& &&    & $\Bgl({}^{+0.025}_{-0.019}\Big)$ & $\Big({}^{+0.024}_{-0.019}\Big)$ & $\Big({}^{+0.006}_{-0.004}\Big)$          
&       & $\Bgl({}^{+0.040}_{-0.027}\Big)$ & $\Big({}^{+0.039}_{-0.027}\Big)$ & $\Big({}^{+0.010}_{-0.006}\Big)$          
&       & $\Bgl({}^{+0.035}_{-0.025}\Big)$ & $\Big({}^{+0.034}_{-0.024}\Big)$ & $\Big({}^{+0.008}_{-0.005}\Big)$ \\[1mm]  
$\BR^{\tau\tau}/\BR^{ZZ}$                                          & $2.36$  & $\pm 0.05$                                 
& 1.8   & $       ^{+0.6}  _{-0.5}       $ & $       ^{+0.5}  _{-0.4}       $ & $       ^{+0.3}  _{-0.2}       $          
& 2.2   & $       ^{+1.1}  _{-0.7}       $ & $       ^{+0.9}  _{-0.6}       $ & $       ^{+0.6}  _{-0.4}       $          
& 1.6   & $       ^{+0.9}  _{-0.6}       $ & $       ^{+0.8}  _{-0.5}       $ & $       ^{+0.5}  _{-0.3}       $ \\       
& &&    & $\Bgl({}^{+0.9}  _{-0.7}  \Big)$ & $\Big({}^{+0.8}  _{-0.6}  \Big)$ & $\Big({}^{+0.5}  _{-0.3}  \Big)$          
&       & $\Bgl({}^{+1.5}  _{-1.0}  \Big)$ & $\Big({}^{+1.3}  _{-0.9}  \Big)$ & $\Big({}^{+0.8}  _{-0.5}  \Big)$          
&       & $\Bgl({}^{+1.2}  _{-0.9}  \Big)$ & $\Big({}^{+1.0}  _{-0.7}  \Big)$ & $\Big({}^{+0.7}  _{-0.4}  \Big)$ \\[1mm]  
$\BR^{bb}/\BR^{ZZ}$                                                & $21.5$  & $\pm 1.0$                                  
& 4.2   & $       ^{+4.4}  _{-2.6}       $ & $       ^{+2.8}  _{-2.0}       $ & $       ^{+3.4}  _{-1.6}       $          
& 9.6   & $       ^{+10.1} _{-5.7}       $ & $       ^{+7.4}  _{-4.4}       $ & $       ^{+6.9}  _{-3.6}       $          
& 3.7   & $       ^{+4.1}  _{-2.4}       $ & $       ^{+3.1}  _{-2.0}       $ & $       ^{+2.7}  _{-1.4}       $ \\       
& &&    & $\Bgl({}^{+16.8} _{-9.0}  \Big)$ & $\Big({}^{+13.9} _{-7.9}  \Big)$ & $\Big({}^{+9.5}  _{-4.4}  \Big)$          
&       & $\Bgl({}^{+29.3} _{-11.8} \Big)$ & $\Big({}^{+24.2} _{-10.5} \Big)$ & $\Big({}^{+16.6} _{-5.3}  \Big)$          
&       & $\Bgl({}^{+29.4} _{-11.9} \Big)$ & $\Big({}^{+23.4} _{-10.4} \Big)$ & $\Big({}^{+17.8} _{-5.9}  \Big)$ \\[1mm]  

      \hline\hline
    \end{tabular}
  \end{adjustbox}
\end{table}

\begin{figure}[ht!]
\centering
\vspace{8mm}
\includegraphics[height=0.80\textheight,width=1.0\textwidth]{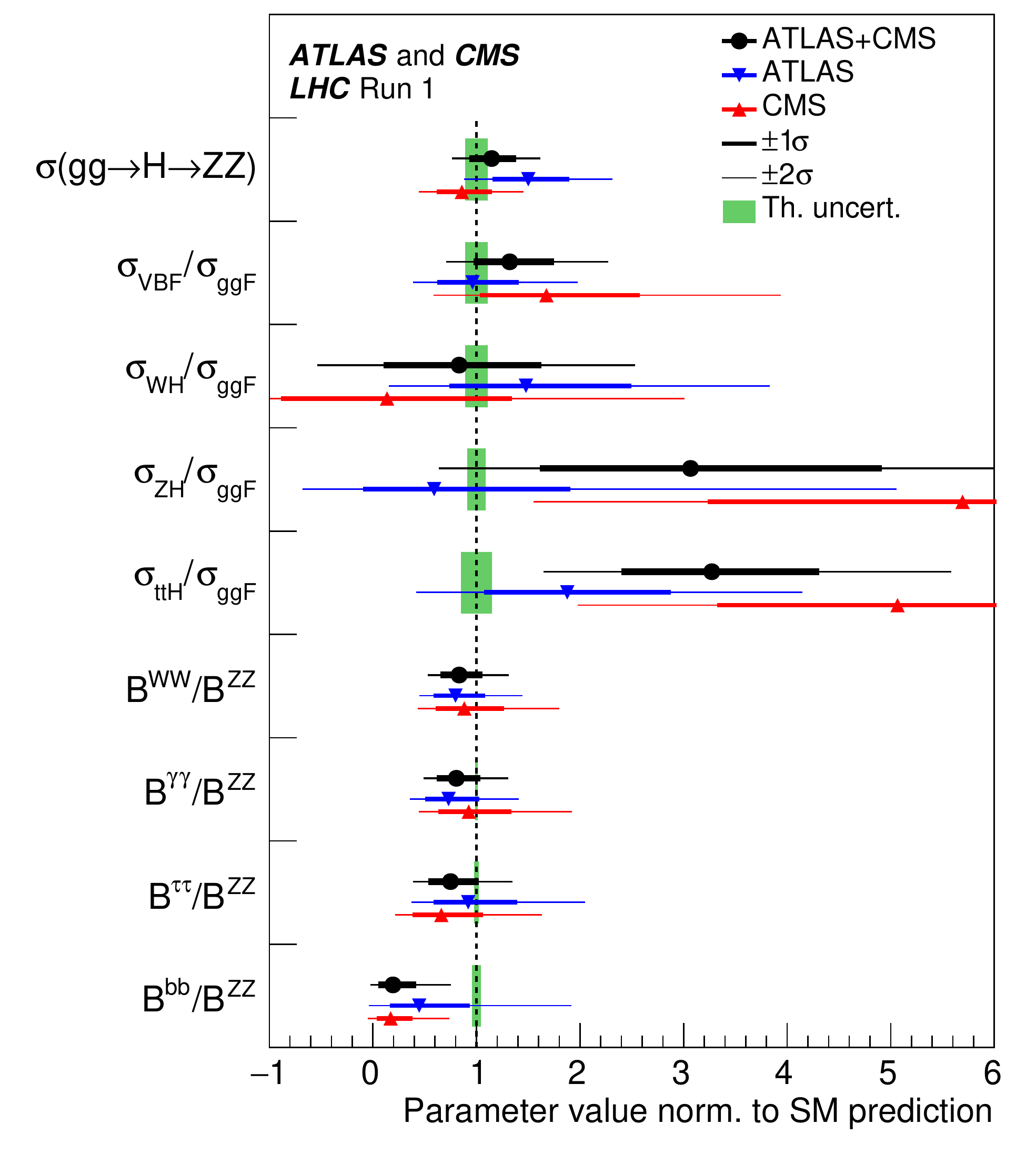}
\caption{Best fit values of the $\sigma(gg\to H\to ZZ)$ cross section and of ratios of cross sections and branching fractions, as obtained from the generic parameterisation with nine parameters and tabulated in~Table~\ref{tab:B1ZZ_results} for the combination of the ATLAS and CMS measurements. 
Also shown are the results from each experiment. 
The values involving cross sections are given for $\sqrt{s}=8$~TeV, assuming the SM values for $\sigma_i(7~\TeV)/\sigma_i(8~\TeV)$.
The error bars indicate the $1\sigma$~(thick lines) and $2\sigma$~(thin lines) intervals. 
The fit results are normalised to the SM predictions for the various parameters and the shaded bands indicate the theoretical uncertainties in these predictions.
}
\label{fig:plot_B1ZZ}
\end{figure}


\begin{figure}[ht!]
\centering
\vspace{-4mm}
\includegraphics[width=1.0\textwidth]{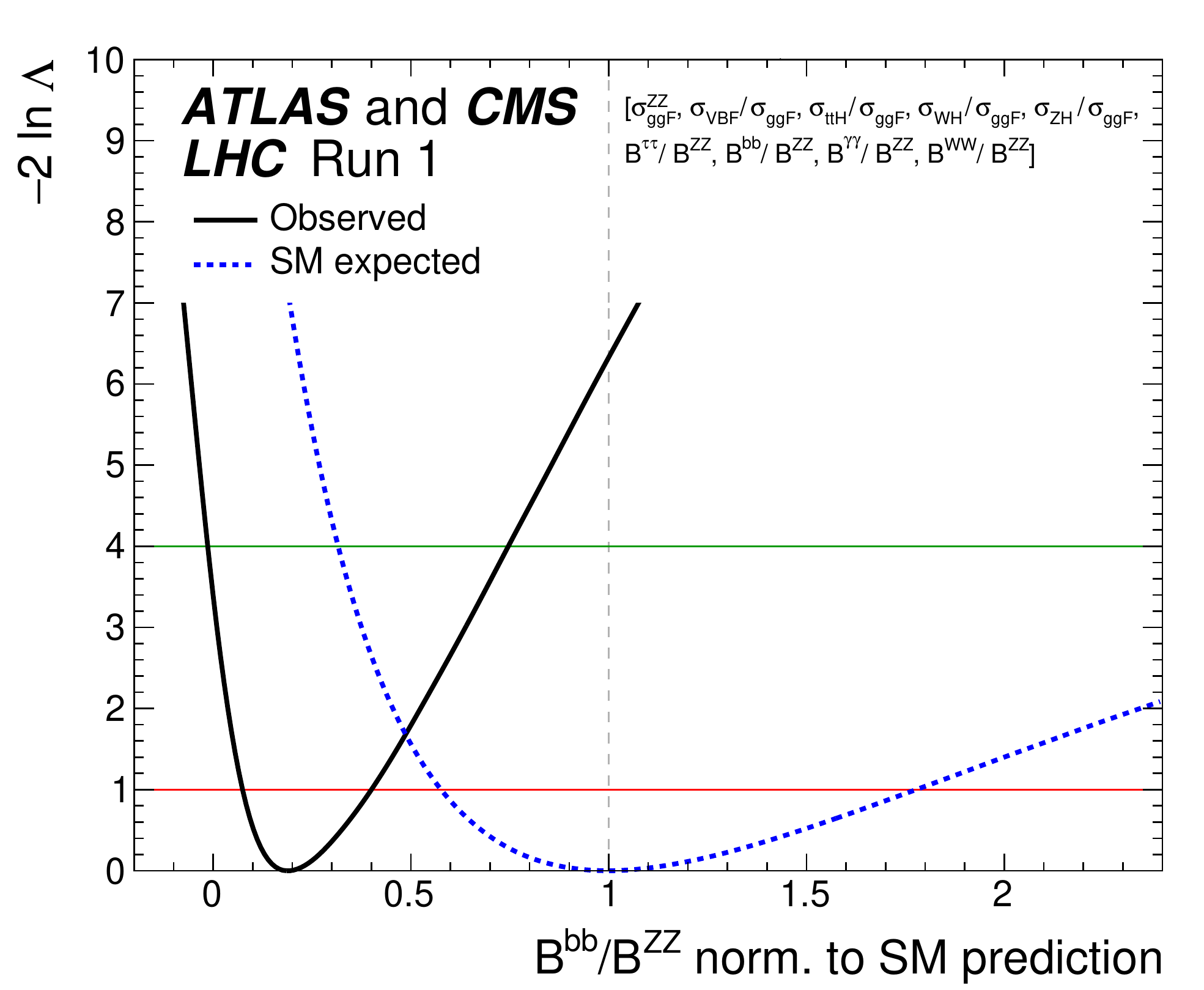}
\caption{Observed (solid line) and expected (dashed line) negative log-likelihood scan of the $\BR^{bb}/\BR^{ZZ}$ parameter normalised to the corresponding SM~prediction. All the other parameters of interest are also varied in the minimisation procedure. 
The red (green) horizontal line at the $-2\Delta\ln\Lambda$ value of 1 (4) indicates the value of the profile likelihood ratio corresponding to a $1\sigma$ ($2\sigma$) CL~interval for the parameter of interest, assuming the asymptotic $\chi^2$~distribution of the test statistic.
The vertical dashed line indicates the SM prediction.
}
\label{fig:rates_bb_likelihoodscan}
\end{figure}

\begin{table}[thbp]
  \centering
  \caption{Best fit values of $\Cc_{gZ}= \Cc_{g}\cdot\Cc_{Z} / \Cc_{H} $ and of the ratios of coupling modifiers, as defined in the parameterisation studied in the context of the $\Cc$-framework, from the combined analysis of the $\sqrt{s}=7$ and 8~TeV data. 
The results are shown for the combination of ATLAS and CMS and also separately for each experiment, together with their total uncertainties and their breakdown into statistical and systematic components.
The uncertainties in $\lambda_{tg}$ and $\lambda_{WZ}$, for which a negative solution is allowed, are calculated around the overall best fit value.
The combined $1\sigma$~CL~intervals are $\lambda_{tg} = [-2.00,-1.59] \cup [1.50,2.07]$ and $\lambda_{WZ} = [-0.96,-0.82] \cup [0.80,0.98]$.
The expected uncertainties in the measurements are displayed in parentheses. 
For those parameters with no sensitivity to the sign, only the absolute values are shown.
}
  \label{tab:genericcouplings}
  \begin{adjustbox}{width=\textwidth}%
    \setlength\extrarowheight{3pt}%
    \begin{tabular}{l|r@{\hskip 0.5ex}lcc|r@{\hskip 0.5ex}lcc|r@{\hskip 0.5ex}lcc}\hline\hline
       Parameter    & \multicolumn{2}{c}{Best fit} &  \multicolumn{2}{c|}{Uncertainty}  & \multicolumn{2}{c}{Best fit} &  \multicolumn{2}{c|}{Uncertainty} & \multicolumn{2}{c}{Best fit} &  \multicolumn{2}{c}{Uncertainty}     \\
                                             & \multicolumn{2}{c}{value}  & Stat                  & Syst                  &  \multicolumn{2}{c}{value} & Stat                  & Syst                  & \multicolumn{2}{c}{value} & Stat           & Syst                  \\
      \hline &  \multicolumn{4}{c|}{ATLAS+CMS} & \multicolumn{4}{c|}{ATLAS} &  \multicolumn{4}{c}{CMS}  \\
$\kappa_{gZ}$
& 1.09 & $       ^{+0.11}_{-0.11}     $ & $       ^{+0.09}_{-0.09}     $ & $       ^{+0.06}_{-0.06}     $          
& 1.20 & $       ^{+0.16}_{-0.15}     $ & $       ^{+0.14}_{-0.14}     $ & $       ^{+0.08}_{-0.07}     $          
& 0.99 & $       ^{+0.14}_{-0.13}     $ & $       ^{+0.12}_{-0.12}     $ & $       ^{+0.07}_{-0.06}     $ \\       
&      & $\Bgl({}^{+0.11}_{-0.11}\Big)$ & $\Big({}^{+0.09}_{-0.09}\Big)$ & $\Big({}^{+0.06}_{-0.05}\Big)$          
&      & $\Bgl({}^{+0.15}_{-0.15}\Big)$ & $\Big({}^{+0.14}_{-0.13}\Big)$ & $\Big({}^{+0.07}_{-0.06}\Big)$          
&      & $\Bgl({}^{+0.14}_{-0.14}\Big)$ & $\Big({}^{+0.13}_{-0.12}\Big)$ & $\Big({}^{+0.07}_{-0.06}\Big)$ \\[1mm]  
$\lambda_{Zg}$
& 1.27 & $       ^{+0.23}_{-0.20}     $ & $       ^{+0.18}_{-0.16}     $ & $       ^{+0.15}_{-0.12}     $          
& 1.07 & $       ^{+0.26}_{-0.22}     $ & $       ^{+0.21}_{-0.18}     $ & $       ^{+0.15}_{-0.11}     $          
& 1.47 & $       ^{+0.45}_{-0.34}     $ & $       ^{+0.35}_{-0.28}     $ & $       ^{+0.28}_{-0.20}     $ \\       
&      & $\Bgl({}^{+0.20}_{-0.17}\Big)$ & $\Big({}^{+0.15}_{-0.14}\Big)$ & $\Big({}^{+0.12}_{-0.10}\Big)$          
&      & $\Bgl({}^{+0.28}_{-0.23}\Big)$ & $\Big({}^{+0.23}_{-0.20}\Big)$ & $\Big({}^{+0.16}_{-0.11}\Big)$          
&      & $\Bgl({}^{+0.27}_{-0.23}\Big)$ & $\Big({}^{+0.21}_{-0.19}\Big)$ & $\Big({}^{+0.16}_{-0.13}\Big)$ \\[1mm]  
$\lambda_{tg}$
& 1.78 & $       ^{+0.30}_{-0.27}     $ & $       ^{+0.21}_{-0.20}     $ & $       ^{+0.21}_{-0.18}     $          
& 1.40 & $       ^{+0.34}_{-0.33}     $ & $       ^{+0.25}_{-0.24}     $ & $       ^{+0.23}_{-0.23}     $          
& -2.26 & $       ^{+0.50}_{-0.53}     $ & $       ^{+0.43}_{-0.39}     $ & $       ^{+0.26}_{-0.36}     $ \\       
&      & $\Bgl({}^{+0.28}_{-0.38}\Big)$ & $\Big({}^{+0.20}_{-0.30}\Big)$ & $\Big({}^{+0.20}_{-0.24}\Big)$          
&      & $\Bgl({}^{+0.38}_{-0.54}\Big)$ & $\Big({}^{+0.28}_{-0.39}\Big)$ & $\Big({}^{+0.25}_{-0.37}\Big)$          
&      & $\Bgl({}^{+0.42}_{-0.64}\Big)$ & $\Big({}^{+0.31}_{-0.42}\Big)$ & $\Big({}^{+0.28}_{-0.49}\Big)$ \\[1mm]  
$\lambda_{WZ}$
& 0.88 & $       ^{+0.10}_{-0.09}     $ & $       ^{+0.09}_{-0.08}     $ & $       ^{+0.04}_{-0.04}     $          
& 0.92 & $       ^{+0.14}_{-0.12}     $ & $       ^{+0.13}_{-0.11}     $ & $       ^{+0.05}_{-0.05}     $          
& -0.85 & $       ^{+0.13}_{-0.15}     $ & $       ^{+0.11}_{-0.13}     $ & $       ^{+0.06}_{-0.07}     $ \\       
&      & $\Bgl({}^{+0.12}_{-0.10}\Big)$ & $\Big({}^{+0.11}_{-0.09}\Big)$ & $\Big({}^{+0.05}_{-0.04}\Big)$          
&      & $\Bgl({}^{+0.18}_{-0.15}\Big)$ & $\Big({}^{+0.17}_{-0.13}\Big)$ & $\Big({}^{+0.06}_{-0.06}\Big)$          
&      & $\Bgl({}^{+0.17}_{-0.14}\Big)$ & $\Big({}^{+0.15}_{-0.13}\Big)$ & $\Big({}^{+0.07}_{-0.06}\Big)$ \\[1mm]  
$|\lambda_{\gamma Z}|$
& 0.89 & $       ^{+0.11}_{-0.10}     $ & $       ^{+0.10}_{-0.09}     $ & $       ^{+0.04}_{-0.03}     $          
& 0.87 & $       ^{+0.15}_{-0.13}     $ & $       ^{+0.15}_{-0.13}     $ & $       ^{+0.05}_{-0.04}     $          
& 0.91 & $       ^{+0.17}_{-0.14}     $ & $       ^{+0.16}_{-0.14}     $ & $       ^{+0.05}_{-0.04}     $ \\       
&      & $\Bgl({}^{+0.13}_{-0.12}\Big)$ & $\Big({}^{+0.13}_{-0.11}\Big)$ & $\Big({}^{+0.04}_{-0.03}\Big)$          
&      & $\Bgl({}^{+0.20}_{-0.17}\Big)$ & $\Big({}^{+0.20}_{-0.17}\Big)$ & $\Big({}^{+0.06}_{-0.04}\Big)$          
&      & $\Bgl({}^{+0.18}_{-0.16}\Big)$ & $\Big({}^{+0.18}_{-0.15}\Big)$ & $\Big({}^{+0.05}_{-0.04}\Big)$ \\[1mm]  
$|\lambda_{\tau Z}|$
& 0.85 & $       ^{+0.13}_{-0.12}     $ & $       ^{+0.12}_{-0.10}     $ & $       ^{+0.07}_{-0.06}     $          
& 0.96 & $       ^{+0.21}_{-0.18}     $ & $       ^{+0.18}_{-0.15}     $ & $       ^{+0.11}_{-0.09}     $          
& 0.78 & $       ^{+0.20}_{-0.17}     $ & $       ^{+0.17}_{-0.15}     $ & $       ^{+0.10}_{-0.09}     $ \\       
&      & $\Bgl({}^{+0.17}_{-0.15}\Big)$ & $\Big({}^{+0.14}_{-0.13}\Big)$ & $\Big({}^{+0.09}_{-0.08}\Big)$          
&      & $\Bgl({}^{+0.27}_{-0.23}\Big)$ & $\Big({}^{+0.23}_{-0.19}\Big)$ & $\Big({}^{+0.14}_{-0.12}\Big)$          
&      & $\Bgl({}^{+0.23}_{-0.20}\Big)$ & $\Big({}^{+0.19}_{-0.17}\Big)$ & $\Big({}^{+0.12}_{-0.11}\Big)$ \\[1mm]  
$|\lambda_{bZ}|$
& 0.58 & $       ^{+0.16}_{-0.20}     $ & $       ^{+0.12}_{-0.17}     $ & $       ^{+0.10}_{-0.10}     $          
& 0.61 & $       ^{+0.24}_{-0.24}     $ & $       ^{+0.20}_{-0.19}     $ & $       ^{+0.14}_{-0.16}     $          
& 0.47 & $       ^{+0.26}_{-0.17}     $ & $       ^{+0.23}_{-0.13}     $ & $       ^{+0.13}_{-0.12}     $ \\       
&      & $\Bgl({}^{+0.25}_{-0.22}\Big)$ & $\Big({}^{+0.21}_{-0.20}\Big)$ & $\Big({}^{+0.13}_{-0.10}\Big)$          
&      & $\Bgl({}^{+0.36}_{-0.29}\Big)$ & $\Big({}^{+0.31}_{-0.26}\Big)$ & $\Big({}^{+0.18}_{-0.13}\Big)$          
&      & $\Bgl({}^{+0.38}_{-0.37}\Big)$ & $\Big({}^{+0.32}_{-0.34}\Big)$ & $\Big({}^{+0.20}_{-0.16}\Big)$ \\[1mm]  

      \hline\hline
    \end{tabular}
  \end{adjustbox}
\end{table}

The $p$-value of the compatibility between the data and the SM predictions is~16\%. 
Most measurements are consistent with the SM predictions within less than~$2\sigma$;
however, the production cross section ratio $\sigma_{\attH}/\sigma_{\aggF}$ relative to the SM~ratio is measured to be $3.3^{+1.0}_{-0.9}$, corresponding to an excess of approximately~3.0$\sigma$ relative to the SM~prediction. 
This excess is mainly due to the multi-lepton categories. 
The ratio $\sigma_{\aZH}/\sigma_{\aggF}$ relative to the SM~ratio is measured to be $3.2^{+1.8}_{-1.4}$, with the observed excess mainly due to the~\aZH,~$\HWW$ measurements. 
The ratio of branching fractions $\BR^{bb}/\BR^{ZZ}$ is measured to be $0.19^{+0.21}_{-0.12}$ relative to the SM prediction. 
In this parameterisation, the high values found for the production cross section ratios for the $\aZH$ and $\attH$ processes induce a low value for the $\Hbb$ decay branching fraction because the $\Hbb$~decay mode does not contribute to the observed excesses. 
The likelihood scan of the $\BR^{bb}/\BR^{ZZ}$ parameter is very asymmetric, as shown in~Fig.~\ref{fig:rates_bb_likelihoodscan}, resulting in an overall deficit of approximately~2.5$\sigma$ relative to the SM~prediction.
This deviation is anticorrelated with the ones quoted above for the~$\sigma_{\attH}/\sigma_{\aggF}$ and $\sigma_{\aZH}/\sigma_{\aggF}$~production cross section ratios, as shown in Fig.~\ref{fig:correlationB1ZZ} of Appendix~\ref{sec:app_correlations}.

In the various fits, the combination of the 7 and 8 TeV data is performed assuming that
the ratios of the production cross sections at 7 and 8 TeV are the same as in the SM. One can introduce as free parameters in the fit the ratios of the production cross sections at~7 and 8~TeV of the five main production processes: $\sigma_i(7 \mathrm{TeV})/\sigma_i(8 \mathrm{TeV})$. Given the limited size of the data samples at~7~TeV, only the \aggF\ and \aVBF\ ratios can be extracted with a meaningful precision. The results are:
$\sigma_{\aggF}(7\TeV)/\sigma_{\aggF}(8\TeV) = 1.12^{+0.33}_{-0.29} $ and $\sigma_{\aVBF}(7\TeV)/\sigma_{\aVBF}(8\TeV) = 0.37^{+0.49}_{-0.43} $. Both values are consistent with the SM~predictions of $\sigma_{\aggF}(7\TeV)/\sigma_{\aggF}(8\TeV) = 0.78$ and $\sigma_{\aVBF}(7\TeV)/\sigma_{\aVBF}(8\TeV) = 0.77$.

\subsection{Parameterisation using ratios of coupling modifiers}
\label{sec:kappaParam}

\begin{figure}[h!]
\centering
\includegraphics[width=1.0\textwidth]{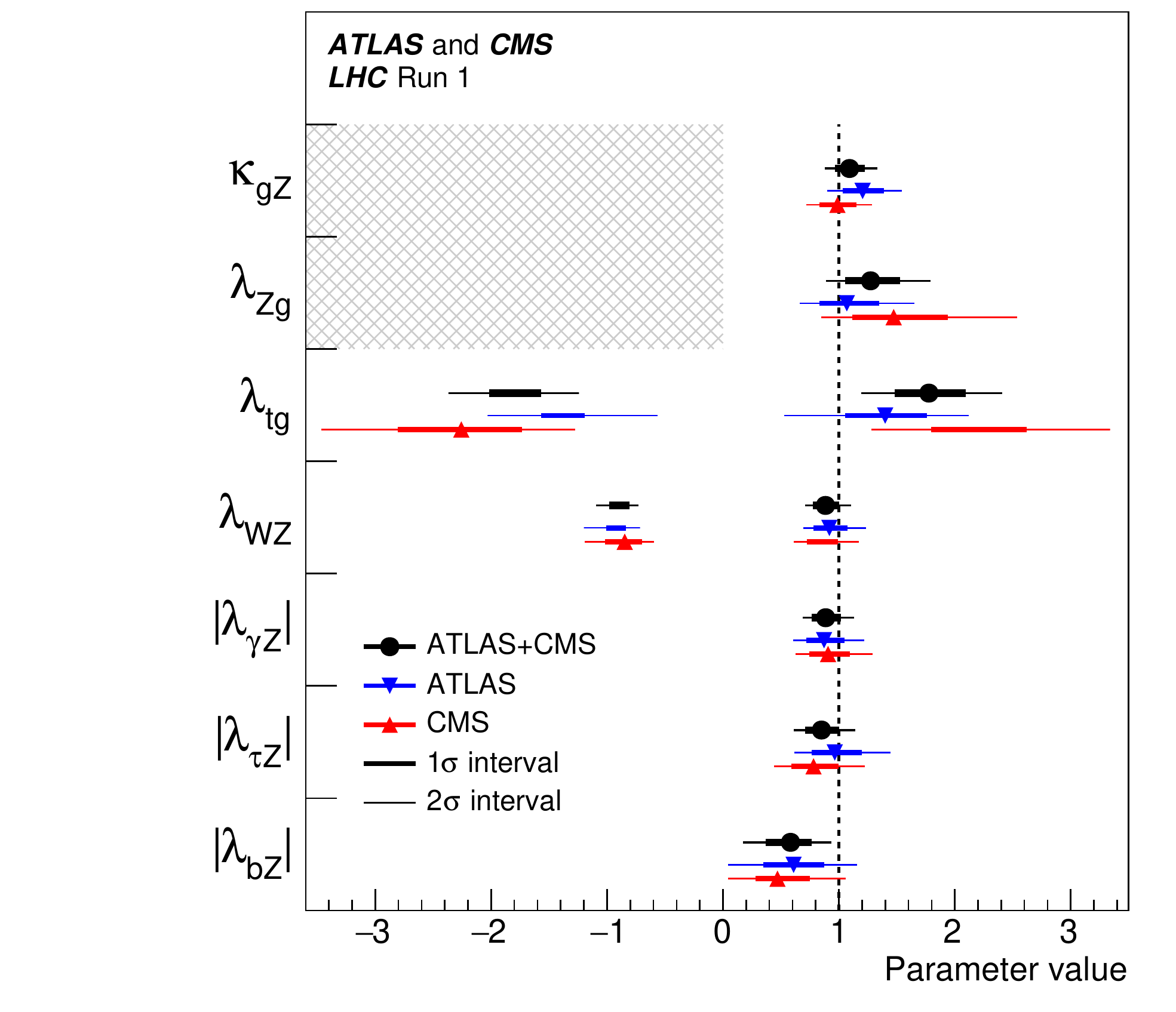}
\caption{Best fit values of ratios of Higgs boson coupling modifiers, as obtained from the generic parameterisation described in the text and as tabulated in Table~\ref{tab:genericcouplings} for the combination of the ATLAS and CMS measurements. 
Also shown are the results from each experiment. 
The error bars indicate the $1\sigma$~(thick lines) and $2\sigma$~(thin lines) intervals. 
The hatched areas indicate the non-allowed regions for the parameters that are assumed to be positive without loss of generality.
For those parameters with no sensitivity to the sign, only the absolute values are shown.
}
\label{fig:plot_L1}
\end{figure}

The parameterisation using the Higgs boson coupling modifiers is based on the $\Cc$-framework described in Section~\ref{sec:kappas} and the parameters of interest are listed in the right column of Table~\ref{tab:genericmodels}. 
The cross section times branching fraction for the $gg\to H\to ZZ$~channel is parameterised as a function of~$\Cc_{gZ} = \Cc_{g}\cdot\Cc_{Z} / \Cc_{H}$, where $\Cc_{g}$ is the effective coupling modifier of the Higgs boson to the gluon in \aggF\ production, which in the SM occurs mainly through loops involving top and bottom quarks. 
The $\Rr_{Zg} = \Cc_{Z} / \Cc_{g}$ parameter is probed by the measurements of \aVBF\ and \aZH\ production, while the measurements of the \attH\ production process are sensitive to~$\Rr_{tg} = \Cc_{t} / \Cc_{g}$. 
Three of the decay modes, namely $\HWW$, $\Htt$, and $\Hbb$, probe the three ratios $\Rr_{WZ} = \Cc_{W} / \Cc_{Z}$, $\Rr_{\tau Z} = \Cc_{\tau} / \Cc_{Z}$, and~$\Rr_{bZ} = \Cc_{b} / \Cc_{Z}$, through their respective ratios to the $\HZZ$~branching fraction. 
The remaining decay mode,~$\Hyy$, which in the~SM occurs through loops involving predominantly the top quark and the $W$ boson, is sensitive to the ratio~$\Rr_{\gamma Z} = \Cc_{\gamma} / \Cc_{Z}$.
In this parameterisation, $\Rr_{WZ} = \Cc_{W} / \Cc_{Z}$ is also probed by the \aVBF, \aWH, and \aZH~production processes. 
Without any loss of generality, the signs of $\Cc_{Z}$ and $\Cc_{g}$ can be assumed to be the same, constraining $\Rr_{Zg}$ and $\Cc_{gZ}$ to be positive.

Table~\ref{tab:genericcouplings} shows the results of the fit to the data with a breakdown of uncertainties into their statistical and systematic components, while the complete breakdown of the uncertainties into the four components is shown in~Table~\ref{tab:genericcouplings_full} of~Appendix~\ref{sec:app_B1WW}. The measured correlation matrix can be found in~Fig.~\ref{fig:correlationL1} of~Appendix~\ref{sec:app_correlations}. The coupling modifiers are assumed to be the same at the two centre-of-mass energies, as in the parameterisation based on the ratios of cross sections and branching fractions. 
Figure~\ref{fig:plot_L1} illustrates the complete ranges of allowed values with their total uncertainties, including the negative ranges allowed for~$\lambda_{WZ}$ and~$\lambda_{tg}$, the two parameters chosen to illustrate possible interference effects due to~\aggZH~or \atH~production. 
Figure~\ref{fig:lambdaWZ_lambda_tg} shows the likelihood scan results for these two parameters for the combination of ATLAS and CMS, both for the observed and expected results. As described in~Section~\ref{sec:kappas}, the interference terms are responsible for the small asymmetry between the likelihood curves for the positive and negative values of the parameters of interest. 
In both cases, the best fit values correspond to the positive sign, but the sensitivity to the interference terms remains small. 
Appendix~\ref{sec:app_negative}, with the specific example of~$\lambda_{bZ}$ (shown in~Fig.~\ref{fig:lambdaBZ-negatives}), describes how the four possible sign combinations between~$\lambda_{WZ}$ and~$\lambda_{tg}$ may impact the best fit value and the uncertainty in the other parameters of interest. 
The $p$-value of the compatibility between the data and the SM predictions is~13\%. 
All results are consistent with the SM~predictions within less than~2$\sigma$, except those for~$\Rr_{tg}$ and~$\Rr_{bZ}$, which exhibit deviations from the~SM similar to those reported and explained in~Section~\ref{sec:sigBR9} for the measurement of the ratios of the \attH\ and \aggF\ production cross sections, and of the ratios of the $bb$ and $ZZ$ decay branching fractions.

\begin{figure}[hbt!]
\vspace{4mm}
  \center
  \includegraphics[width=0.77\textwidth]{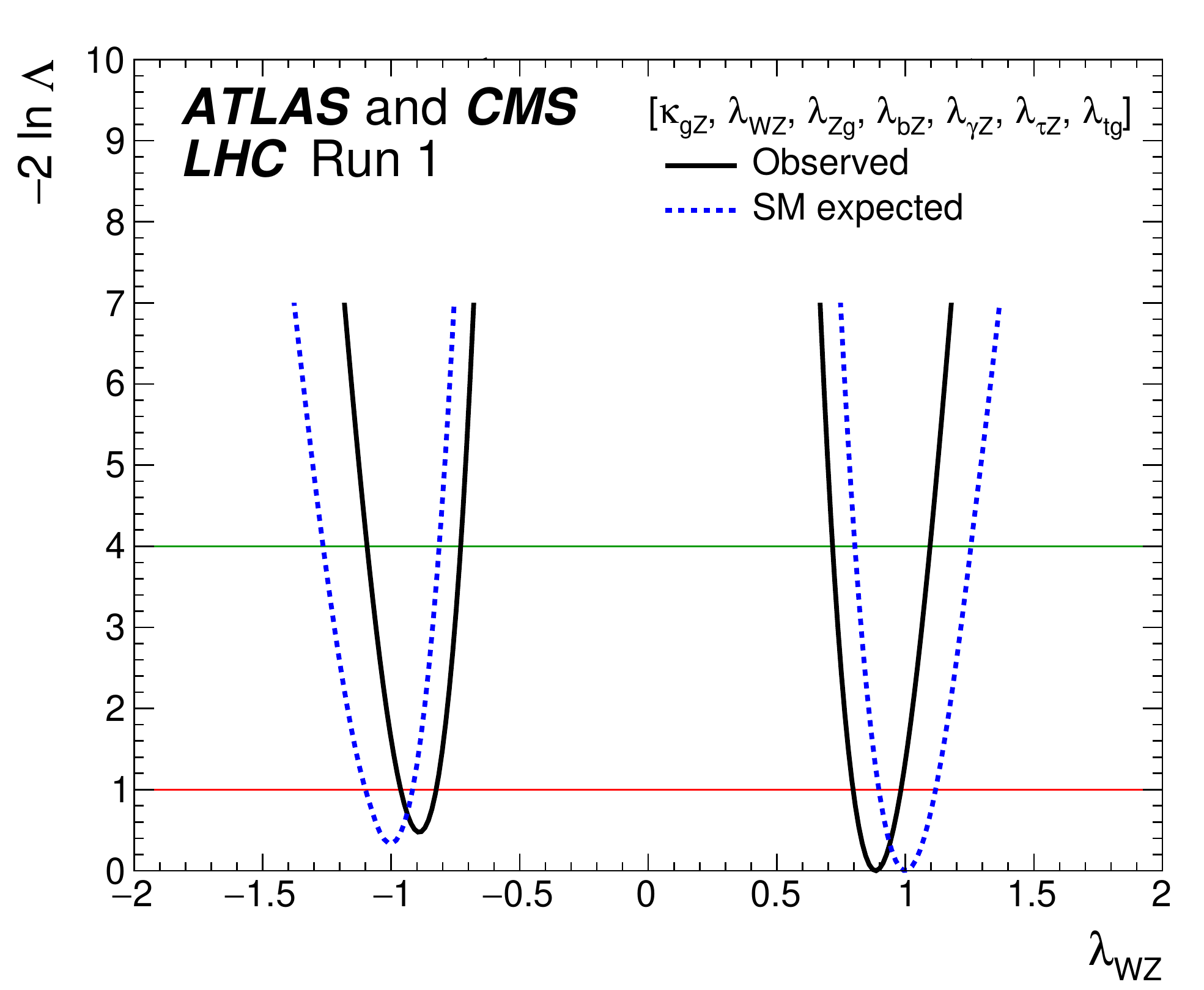}\\
  \includegraphics[width=0.77\textwidth]{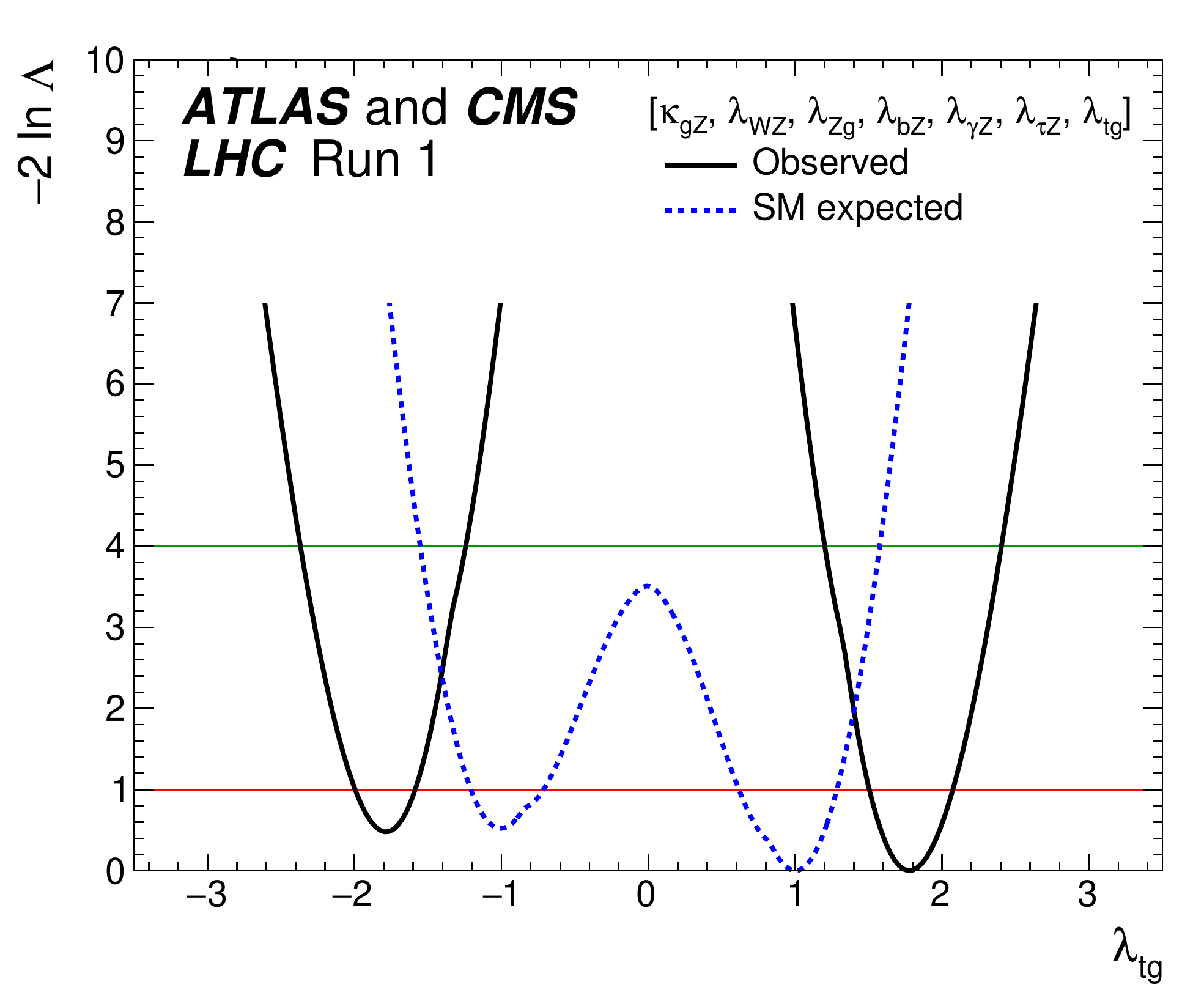}
  \vspace{-4mm}  
  \caption{Observed (solid line) and expected (dashed line) negative log-likelihood scans for $\Rr_{WZ}$ (top) and $\Rr_{tg}$ (bottom), the two parameters of~Fig.~\ref{fig:plot_L1} that are of interest in the negative range in the generic parameterisation of ratios of Higgs boson coupling modifiers described in the text. All the other parameters of interest from the list in the legend are also varied in the minimisation procedure. The red (green) horizontal lines at the $-2\Delta\ln\Lambda$ value of 1 (4) indicate the value of the profile likelihood ratio corresponding to a $1\sigma$ ($2\sigma$) CL~interval for the parameter of interest, assuming the asymptotic $\chi^2$~distribution of the test statistic.
}
\label{fig:lambdaWZ_lambda_tg}
\end{figure}

\clearpage

\begin{table}[htb]
\centering
\caption{Measured global signal strength $\mu$ and its total uncertainty, together with the breakdown of the uncertainty into its four components as defined in Section~\ref{sec:SystematicUncertainties}. The results are shown for the combination of ATLAS and CMS, and separately for each experiment. The expected uncertainty, with its breakdown, is also shown.
}
\setlength\extrarowheight{3pt}
\renewcommand{\arraystretch}{1.2}
\label{tab:muGlobal}
\begin{tabular}{lcc|cccc}\hline\hline
                       & Best fit $\mu$ & \multicolumn{5}{c}{Uncertainty}                                                                \\ \cline{3-7}
                       &        &  Total             & Stat               & Expt               & Thbgd              & Thsig              \\ \hline
ATLAS + CMS (measured) & $1.09$ & $^{+0.11}_{-0.10}$ & $^{+0.07}_{-0.07}$ & $^{+0.04}_{-0.04}$ & $^{+0.03}_{-0.03}$ & $^{+0.07}_{-0.06}$ \\
ATLAS + CMS (expected) &        & $^{+0.11}_{-0.10}$ & $^{+0.07}_{-0.07}$ & $^{+0.04}_{-0.04}$ & $^{+0.03}_{-0.03}$ & $^{+0.07}_{-0.06}$ \\[2pt] \hline
ATLAS (measured)       & $1.20$ & $^{+0.15}_{-0.14}$ & $^{+0.10}_{-0.10}$ & $^{+0.06}_{-0.06}$ & $^{+0.04}_{-0.04}$ & $^{+0.08}_{-0.07}$ \\
ATLAS (expected)       &        & $^{+0.14}_{-0.13}$ & $^{+0.10}_{-0.10}$ & $^{+0.06}_{-0.05}$ & $^{+0.04}_{-0.04}$ & $^{+0.07}_{-0.06}$ \\[2pt] \hline
CMS (measured)         & $0.97$ & $^{+0.14}_{-0.13}$ & $^{+0.09}_{-0.09}$ & $^{+0.05}_{-0.05}$ & $^{+0.04}_{-0.03}$ & $^{+0.07}_{-0.06}$ \\
CMS (expected)         &        & $^{+0.14}_{-0.13}$ & $^{+0.09}_{-0.09}$ & $^{+0.05}_{-0.05}$ & $^{+0.04}_{-0.03}$ & $^{+0.08}_{-0.06}$ \\[2pt]
\hline\hline
\end{tabular}
\end{table}


\section{Measurements of signal strengths}
\label{sec:SignalStrength}

Section~\ref{sec:sigBR} presents the results from generic parameterisations, expressed in terms of cross sections and branching fractions.
This section probes more specific parameterisations, with additional assumptions. 
Results for these parameterisations are presented, starting with the most restrictive one using a single parameter of interest, which was used to assess the sensitivity of the experimental analyses to the presence of a Higgs boson at the time of its discovery. Section~\ref{sec:DegenerateStates} describes the test of a hypothesis that two or more neutral Higgs bosons might be present with similar masses.

\subsection{Global signal strength}
\label{sec:GlobalMu}

The simplest and most restrictive signal strength parameterisation is to assume that the values of the signal strengths~$\mu_i^f$, as defined in~Eq.~(\ref{eq:muif}), are the same for all production processes~$i$ and decay channels~$f$. In this case, the SM~predictions of signal yields in all categories are scaled by a global signal strength~$\mu$. 
Such a parameterisation provides the simplest test of the compatibility of the experimental data with the SM~predictions. 
A fit to the ATLAS and CMS data at $\sqrt{s}=7$ and 8~TeV with $\mu$ as the parameter of interest results in the best fit value: 
$$\mu = 1.09^{+0.11}_{-0.10} = 1.09^{+0.07}_{-0.07}~\mathrm{(stat)}~^{+0.04}_{-0.04}~\mathrm{(expt)}~^{+0.03}_{-0.03}~\mathrm{(thbgd)} ^{+0.07}_{-0.06}~\mathrm{(thsig)},$$
where the breakdown of the uncertainties into their four components is performed as described in~Section~\ref{sec:SystematicUncertainties}. 
The overall systematic uncertainty of~${}^{+0.09}_{-0.08}$ is larger than the statistical uncertainty and its largest component is the theoretical uncertainty in the \aggF~cross section. 
This result is consistent with the SM prediction of~$\mu=1$ within less than~$1\sigma$ and the $p$-value of the compatibility between the data and the SM predictions is~40\%. 
This result is shown in~Table~\ref{tab:muGlobal}, together with that from each experiment, including the breakdown of the uncertainties into their four components. 
The expected uncertainties and their breakdown are also given.


\begin{table}[htb]
\centering
\caption{Measured signal strengths $\mu$ and their total uncertainties for different Higgs boson production processes. 
The results are shown for the combination of ATLAS and CMS, and separately for each experiment, for the combined $\sqrt{s}=7$~and~8~TeV data.
The expected uncertainties in the measurements are displayed in parentheses.
These results are obtained assuming that the Higgs boson branching fractions are the same as in the~SM.}
\label{tab:muProduction}
\setlength\extrarowheight{4pt}
\begin{tabular}{l|r@{\hskip 0.5ex}l|r@{\hskip 0.5ex}l|r@{\hskip 0.5ex}l}
\hline\hline
Production process & \multicolumn{2}{c|}{ATLAS+CMS} & \multicolumn{2}{c|}{ATLAS} & \multicolumn{2}{c}{CMS} \\
\hline
$\mu_{\aggF}$  & $1.03$ & $       ^{+0.16}_{-0.14}     $ & $1.26$ & $       ^{+0.23}_{-0.20}     $ & $0.84$ & $       ^{+0.18}_{-0.16}     $ \\
               &        & $\Bgl({}^{+0.16}_{-0.14}\Big)$ &        & $\Bgl({}^{+0.21}_{-0.18}\Big)$ &        & $\Bgl({}^{+0.20}_{-0.17}\Big)$ \\[1mm]
$\mu_{\aVBF}$  & $1.18$ & $       ^{+0.25}_{-0.23}     $ & $1.21$ & $       ^{+0.33}_{-0.30}     $ & $1.14$ & $       ^{+0.37}_{-0.34}     $ \\
               &        & $\Bgl({}^{+0.24}_{-0.23}\Big)$ &        & $\Bgl({}^{+0.32}_{-0.29}\Big)$ &        & $\Bgl({}^{+0.36}_{-0.34}\Big)$ \\[1mm]
$\mu_{\aWH}$   & $0.89$ & $       ^{+0.40}_{-0.38}     $ & $1.25$ & $       ^{+0.56}_{-0.52}     $ & $0.46$ & $       ^{+0.57}_{-0.53}     $ \\
               &        & $\Bgl({}^{+0.41}_{-0.39}\Big)$ &        & $\Bgl({}^{+0.56}_{-0.53}\Big)$ &        & $\Bgl({}^{+0.60}_{-0.57}\Big)$ \\[1mm]
$\mu_{\aZH}$   & $0.79$ & $       ^{+0.38}_{-0.36}     $ & $0.30$ & $       ^{+0.51}_{-0.45}     $ & $1.35$ & $       ^{+0.58}_{-0.54}     $ \\
               &        & $\Bgl({}^{+0.39}_{-0.36}\Big)$ &        & $\Bgl({}^{+0.55}_{-0.51}\Big)$ &        & $\Bgl({}^{+0.55}_{-0.51}\Big)$ \\[1mm]
$\mu_{\attH}$  & $2.3 $ & $       ^{+0.7} _{-0.6}      $ & $1.9 $ & $       ^{+0.8} _{-0.7}      $ & $2.9 $ & $       ^{+1.0} _{-0.9}      $ \\
               &        & $\Bgl({}^{+0.5} _{-0.5} \Big)$ &        & $\Bgl({}^{+0.7} _{-0.7} \Big)$ &        & $\Bgl({}^{+0.9} _{-0.8} \Big)$ \\[1mm]

\hline\hline
\end{tabular}
\end{table}

\subsection{Signal strengths of individual production processes and decay channels}
\label{sec:ProductionDecay}

The global signal strength is the most precisely measured Higgs boson coupling-related observable, but this simple parameterisation is very model dependent, since all Higgs boson production and decay measurements are combined assuming that all their ratios are the same as in the~SM. The compatibility of the measurements with the~SM can be tested in a less model-dependent way by relaxing these assumptions separately for the production cross sections and the decay branching fractions.

Assuming the SM values for the Higgs boson branching fractions, namely $\mu^f=1$ in~Eq.~(\ref{eq:nsig}), the five main Higgs boson production processes are explored with independent signal strengths: $\mu_{\aggF}$, $\mu_{\aVBF}$, $\mu_{\aWH}$, $\mu_{\aZH}$, and $\mu_{\attH}$.
A combined analysis of the ATLAS and CMS data is performed with these five signal strengths as parameters of interest. The results are shown in Table~\ref{tab:muProduction} for the combined $\sqrt{s}=7$ and 8~TeV data sets. 
The signal strengths at the two energies are assumed to be the same for each production process. 
Figure~\ref{fig:muProduction} illustrates these results with their total uncertainties. 
The $p$-value of the compatibility between the data and the SM predictions is~24\%.

\begin{figure}[htb]
\begin{center}
\includegraphics[width=0.95\textwidth]{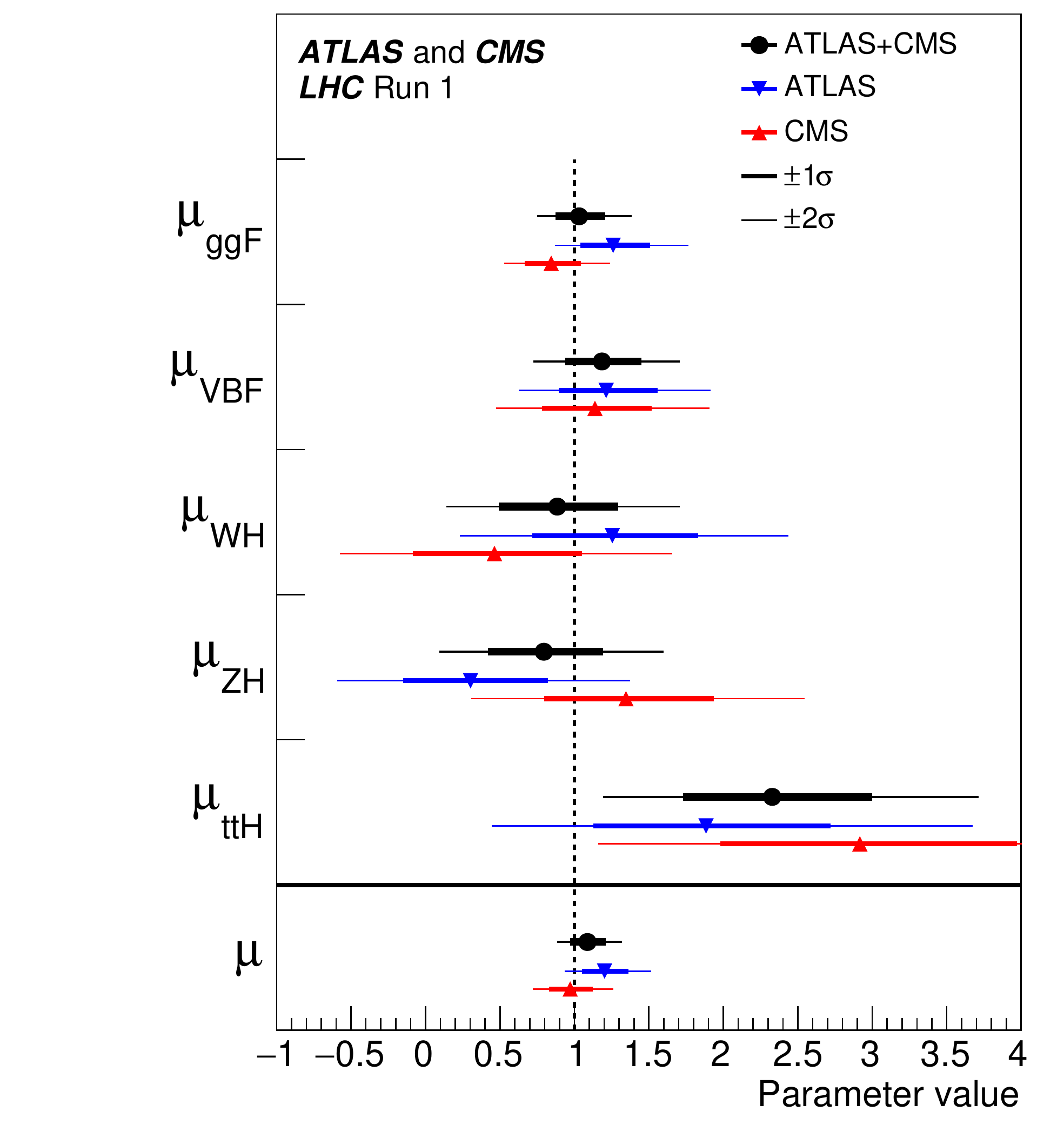}
\end{center}
\caption{Best fit results for the production signal strengths for the combination of ATLAS and CMS data. 
Also shown are the results from each experiment. 
The error bars indicate the $1\sigma$~(thick lines) and $2\sigma$~(thin lines) intervals. 
The measurements of the global signal strength~$\mu$ are also shown.}
\label{fig:muProduction}
\end{figure}

\clearpage

\begin{table}[htb]
\centering
\caption{Measured signal strengths $\mu$ and their total uncertainties for different Higgs boson decay channels. 
The results are shown for the combination of ATLAS and CMS, and separately for each experiment, for the combined $\sqrt{s}=7$~and~8~TeV data.
The expected uncertainties in the measurements are displayed in parentheses. 
These results are obtained assuming that the Higgs boson production process cross sections at $\sqrt{s} = 7$~and~8~TeV are the same as in the~SM.}
\label{tab:muDecay}
\setlength\extrarowheight{4pt}
\begin{tabular}{l|r@{\hskip 0.5ex}l|r@{\hskip 0.5ex}l|r@{\hskip 0.5ex}l}
\hline\hline
Decay channel & \multicolumn{2}{c|}{ATLAS+CMS} & \multicolumn{2}{c|}{ATLAS} & \multicolumn{2}{c}{CMS} \\
\hline
$\mu^{\gamma\gamma}$ & $1.14$ & $       ^{+0.19}_{-0.18}     $ & $1.14$ & $       ^{+0.27}_{-0.25}     $ & $1.11$ & $       ^{+0.25}_{-0.23}     $ \\
                     &        & $\Bgl({}^{+0.18}_{-0.17}\Big)$ &        & $\Bgl({}^{+0.26}_{-0.24}\Big)$ &        & $\Bgl({}^{+0.23}_{-0.21}\Big)$ \\[1mm]
$\mu^{ZZ}$           & $1.29$ & $       ^{+0.26}_{-0.23}     $ & $1.52$ & $       ^{+0.40}_{-0.34}     $ & $1.04$ & $       ^{+0.32}_{-0.26}     $ \\
                     &        & $\Bgl({}^{+0.23}_{-0.20}\Big)$ &        & $\Bgl({}^{+0.32}_{-0.27}\Big)$ &        & $\Bgl({}^{+0.30}_{-0.25}\Big)$ \\[1mm]
$\mu^{WW}$           & $1.09$ & $       ^{+0.18}_{-0.16}     $ & $1.22$ & $       ^{+0.23}_{-0.21}     $ & $0.90$ & $       ^{+0.23}_{-0.21}     $ \\
                     &        & $\Bgl({}^{+0.16}_{-0.15}\Big)$ &        & $\Bgl({}^{+0.21}_{-0.20}\Big)$ &        & $\Bgl({}^{+0.23}_{-0.20}\Big)$ \\[1mm]
$\mu^{\tau\tau}$     & $1.11$ & $       ^{+0.24}_{-0.22}     $ & $1.41$ & $       ^{+0.40}_{-0.36}     $ & $0.88$ & $       ^{+0.30}_{-0.28}     $ \\
                     &        & $\Bgl({}^{+0.24}_{-0.22}\Big)$ &        & $\Bgl({}^{+0.37}_{-0.33}\Big)$ &        & $\Bgl({}^{+0.31}_{-0.29}\Big)$ \\[1mm]
$\mu^{bb}$           & $0.70$ & $       ^{+0.29}_{-0.27}     $ & $0.62$ & $       ^{+0.37}_{-0.37}     $ & $0.81$ & $       ^{+0.45}_{-0.43}     $ \\
                     &        & $\Bgl({}^{+0.29}_{-0.28}\Big)$ &        & $\Bgl({}^{+0.39}_{-0.37}\Big)$ &        & $\Bgl({}^{+0.45}_{-0.43}\Big)$ \\[1mm]
$\mu^{\mu\mu}$       & $0.1 $ & $       ^{+2.5} _{-2.5}      $ & $-0.6$ & $       ^{+3.6} _{-3.6}      $ & $0.9 $ & $       ^{+3.6} _{-3.5}      $ \\
                     &        & $\Bgl({}^{+2.4} _{-2.3} \Big)$ &        & $\Bgl({}^{+3.6} _{-3.6} \Big)$ &        & $\Bgl({}^{+3.3} _{-3.2} \Big)$ \\[1mm]

\hline\hline
\end{tabular}
\end{table}

Higgs boson decays are also studied with six independent signal strengths, one for each decay channel included in the combination, assuming that the Higgs boson production cross sections are the same as in the~SM. 
Unlike the production signal strengths, these decay-based signal strengths are independent of the collision centre-of-mass energy and therefore the $\sqrt{s}=7$ and 8~TeV data sets can be combined without additional assumptions. 
Table~\ref{tab:muDecay} and Fig.~\ref{fig:muDecay} present the best fit results for the combination of ATLAS and CMS, and separately for each experiment (the results for $\mu^{\mu\mu}$ are only reported in~Table~\ref{tab:muDecay}).
The $p$-value of the compatibility between the data and the SM predictions is~75\%. 
 
From the combined likelihood scans it is possible to evaluate the significances for the observation of the different production processes and decay channels. 
The combination of the data from the two experiments
corresponds to summing their recorded integrated luminosities and consequently
increases the sensitivity by approximately a factor of~$\sqrt 2$, since the theoretical uncertainties in the Higgs boson signal  are only weakly relevant for this evaluation and all the other significant uncertainties are uncorrelated between the two experiments. 
The results are reported in Table~\ref{tab:muSignificance} for all production processes and decay channels, except for those that have already been clearly observed, namely the \aggF~production process and the $\HZZ$, $\HWW$, and $\Hyy$~decay channels. 
The combined significances for the observation of the \aVBF\ production process and of the $\Htt$~decay are each above~$5\sigma$, and the combined significance for the \aVH\ production process is above~$3\sigma$. 
The combined significance for the \attH\ process is $4.4\sigma$, whereas only~$2.0\sigma$ is expected, corresponding to a measured excess of~$2.3\sigma$ with respect to the SM~prediction.

\begin{figure}[htb]
\begin{center}
\includegraphics[width=0.95\textwidth]{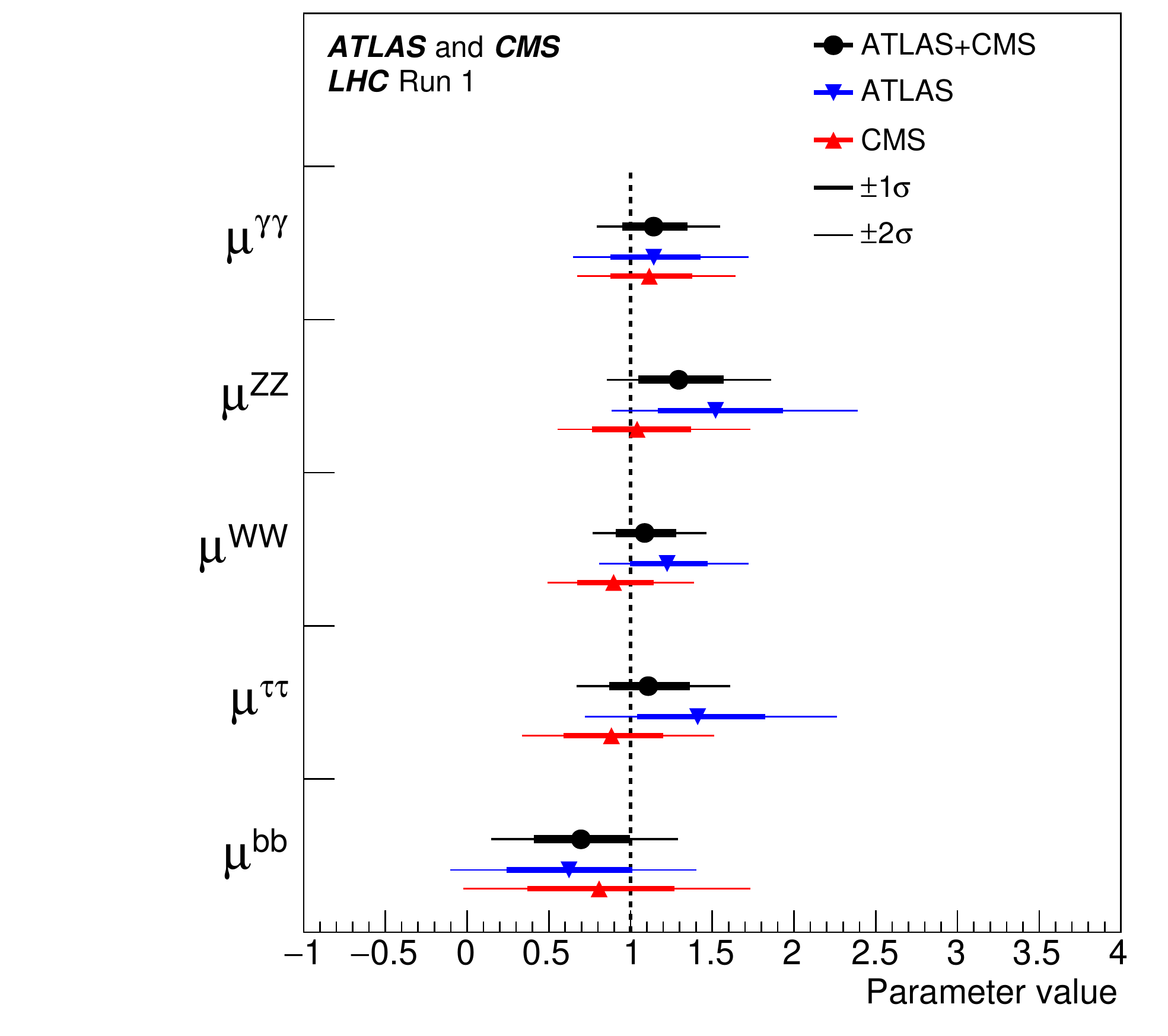}
\end{center}
\caption{Best fit results for the decay signal strengths for the combination of ATLAS and CMS data (the results for $\mu^{\mu\mu}$ are reported in~Table~\ref{tab:muDecay}). 
Also shown are the results from each experiment. 
The error bars indicate the $1\sigma$~(thick lines) and $2\sigma$~(thin lines) intervals.
}
\label{fig:muDecay}
\end{figure}

\clearpage

\begin{table}[htb]
\vspace{-4mm}
\caption{Measured and expected significances for the observation of Higgs boson production processes and decay channels for the combination of ATLAS and CMS. 
Not included are the \aggF~production process and the $\HZZ$, $\HWW$, and $\Hyy$~decay channels, which have already been clearly observed. 
All results are obtained constraining the decay branching fractions to their SM~values when considering the production processes, and constraining the production cross sections to their SM~values when studying the decays.}
\label{tab:muSignificance}
\begin{center}
\begin{tabular}{lcc}\hline\hline
  Production  process & Measured significance ($\sigma$) & Expected significance ($\sigma$) \\
  \aVBF               & 5.4                              & 4.6                              \\
  \aWH                & 2.4                              & 2.7                              \\
  \aZH                & 2.3                              & 2.9                              \\
  \aVH                & 3.5                              & 4.2                              \\
  \attH               & 4.4                              & 2.0                              \\ \hline
  Decay channel       &                                  &                                  \\
  $\Htt$              & 5.5                              & 5.0                              \\
  $\Hbb$              & 2.6                              & 3.7                              \\
\hline\hline
\end{tabular}
\end{center}
\end{table}

\begin{figure}[thb]
\begin{center}
  \includegraphics[width=0.95\textwidth]{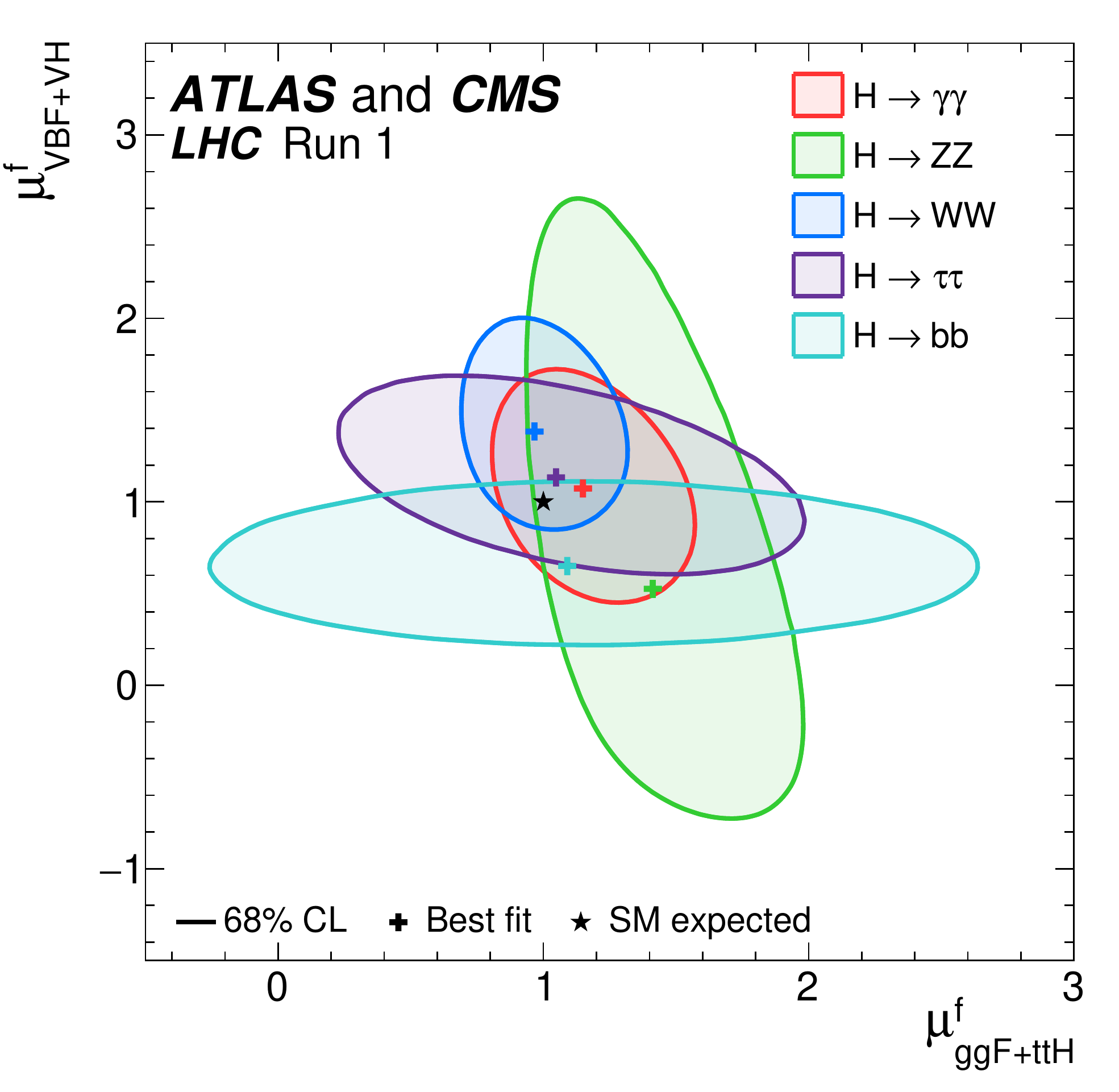}
\end{center}
\caption{Negative log-likelihood contours at~68\%~CL in the ($\mu_{\aggF+\attH}^f$, $\mu_{\aVBF+\aVH}^f$) plane for the combination of ATLAS and CMS, as obtained from the ten-parameter fit described in the text for each of the five decay channels $\Hzz$, $\Hww$, $\Hyy$, $\Htt$, and $\Hbb$. The best fit values obtained for each of the five decay channels are also shown, together with the SM~expectation. }
\label{fig:VFContour}
\end{figure}

\subsection{Boson- and fermion-mediated production processes}
\label{sec:BosonFermion}

The Higgs boson production processes can be associated with Higgs boson couplings to either fermions (\aggF\ and \attH) or vector bosons (\aVBF, \aWH, and \aZH). 
Potential deviations of these couplings from the SM predictions can be tested by using a parameterisation with two signal strengths for each decay channel~$f$: $\mu_F^f = \mu_{\aggF+\attH}^f$ for the fermion-mediated production processes and $\mu_V^f = \mu_{\aVBF+\aVH}^f$ for the vector-boson-mediated production processes. The branching fraction cancels in the ratio $\mu_V^f/\mu_F^f$ that can be formed for each Higgs boson decay channel.
Two fits are performed for the combination of ATLAS and CMS, and also separately for each experiment. 
The first is a ten-parameter fit of~$\mu_F^f$ and~$\mu_V^f$ for each of the five decay channels, while the second is a six-parameter fit of~$\mu_V/\mu_F$ and $\mu_F^f$ for each of the five decay channels. 

Figure~\ref{fig:VFContour} shows the 68\%~CL region for the ten-parameter fit of the five decay channels included in the combination of the ATLAS and CMS measurements. These results are obtained by combining the $\sqrt{s}=7$~and 8~TeV data, assuming that $\mu_F^f$ and $\mu_V^f$ are the same at the two energies. The SM predictions of $\mu_F^f = 1$ and~$\mu_V^f = 1$ lie within the~68\%~CL regions of all these measurements. 
Combinations of these regions would require assumptions about the branching fractions and are therefore not performed.
Table~\ref{tab:VFResults} reports the best fit values and the total uncertainties for all the parameters of the fits, together with the expected uncertainties for the combination of ATLAS and CMS. 
The $p$-values of the compatibility between the data and the SM predictions are~90\% and~75\% for the ten-parameter and six-parameter fits, respectively. The six-parameter fit, without any additional assumptions about the Higgs boson branching fractions, yields: $\mu_V/\mu_F = 1.09^{+0.36}_{-0.28}$, in agreement with the SM.

\subsection{Search for mass-degenerate states with different coupling structures}
\label{sec:DegenerateStates}

\begin{table}[htb]
\begin{center}
\caption{Results of the ten-parameter fit of $\mu_F^f = \mu_{\aggF+\attH}^f$ and $\mu_V^f = \mu_{\aVBF+\aVH}^f$ for each of the five decay channels, and of the six-parameter fit of the global ratio $\mu_V/\mu_F = \mu_{\aVBF+\aVH}/\mu_{\aggF+\attH}$ together with $\mu_F^f$ for each of the five decay channels. 
The results are shown for the combination of ATLAS and CMS, together with their measured and expected uncertainties. The measured results are also shown separately for each experiment.
 \label{tab:VFResults}
}
\setlength\extrarowheight{4pt}
\begin{tabular}{l|c|c|c|c}
\hline\hline
Parameter & ATLAS+CMS & ATLAS+CMS & ATLAS & CMS \\
          & Measured  & Expected uncertainty  & Measured  & Measured \\
\hline
\multicolumn{5}{c}{Ten-parameter fit of $\mu_F^f$ and $\mu_V^f$}\\
\hline
$\mu_{V}^{\gamma\gamma}$ & $1.05^{+0.44}_{-0.41}$ & $    ^{+0.41}_{-0.38}$ & $0.69^{+0.63}_{-0.58}$ & $1.37^{+0.62}_{-0.56}$ \\
$\mu_{V}^{ZZ}$           & $0.47^{+1.37}_{-0.92}$ & $    ^{+1.16}_{-0.84}$ & $0.24^{+1.60}_{-0.93}$ & $1.45^{+2.32}_{-2.29}$ \\
$\mu_{V}^{WW}$           & $1.38^{+0.41}_{-0.37}$ & $    ^{+0.38}_{-0.35}$ & $1.56^{+0.52}_{-0.46}$ & $1.08^{+0.65}_{-0.58}$ \\
$\mu_{V}^{\tau\tau}$     & $1.12^{+0.37}_{-0.35}$ & $    ^{+0.38}_{-0.35}$ & $1.29^{+0.58}_{-0.53}$ & $0.88^{+0.49}_{-0.45}$ \\
$\mu_{V}^{bb}$           & $0.65^{+0.31}_{-0.29}$ & $    ^{+0.32}_{-0.30}$ & $0.50^{+0.39}_{-0.37}$ & $0.85^{+0.47}_{-0.44}$ \\
$\mu_{F}^{\gamma\gamma}$ & $1.16^{+0.27}_{-0.24}$ & $    ^{+0.25}_{-0.23}$ & $1.30^{+0.37}_{-0.33}$ & $1.00^{+0.33}_{-0.30}$ \\
$\mu_{F}^{ZZ}$           & $1.42^{+0.37}_{-0.33}$ & $    ^{+0.29}_{-0.25}$ & $1.74^{+0.51}_{-0.44}$ & $0.96^{+0.53}_{-0.41}$ \\
$\mu_{F}^{WW}$           & $0.98^{+0.22}_{-0.20}$ & $    ^{+0.21}_{-0.19}$ & $1.10^{+0.29}_{-0.26}$ & $0.84^{+0.27}_{-0.24}$ \\
$\mu_{F}^{\tau\tau}$     & $1.06^{+0.60}_{-0.56}$ & $    ^{+0.56}_{-0.53}$ & $1.72^{+1.24}_{-1.12}$ & $0.89^{+0.67}_{-0.63}$ \\
$\mu_{F}^{bb}$           & $1.15^{+0.99}_{-0.94}$ & $    ^{+0.90}_{-0.86}$ & $1.52^{+1.16}_{-1.09}$ & $0.11^{+1.85}_{-1.90}$ \\[2pt]

\hline
\multicolumn{5}{c}{Six-parameter fit of global $\mu_V/\mu_F$ and of $\mu_F^f$} \\
\hline
$\mu_{V}/\mu_{F}$        & $1.09^{+0.36}_{-0.28}$ & $    ^{+0.34}_{-0.27}$ & $0.92^{+0.40}_{-0.30}$ & $1.31^{+0.68}_{-0.47}$ \\
$\mu_{F}^{\gamma\gamma}$ & $1.10^{+0.23}_{-0.21}$ & $    ^{+0.21}_{-0.19}$ & $1.17^{+0.32}_{-0.28}$ & $1.01^{+0.29}_{-0.25}$ \\
$\mu_{F}^{ZZ}$           & $1.27^{+0.28}_{-0.24}$ & $    ^{+0.24}_{-0.20}$ & $1.55^{+0.43}_{-0.35}$ & $0.98^{+0.32}_{-0.26}$ \\
$\mu_{F}^{WW}$           & $1.06^{+0.21}_{-0.18}$ & $    ^{+0.19}_{-0.17}$ & $1.25^{+0.28}_{-0.24}$ & $0.84^{+0.25}_{-0.21}$ \\
$\mu_{F}^{\tau\tau}$     & $1.05^{+0.33}_{-0.27}$ & $    ^{+0.33}_{-0.27}$ & $1.50^{+0.64}_{-0.48}$ & $0.74^{+0.38}_{-0.29}$ \\
$\mu_{F}^{bb}$           & $0.64^{+0.37}_{-0.28}$ & $    ^{+0.45}_{-0.34}$ & $0.67^{+0.58}_{-0.41}$ & $0.63^{+0.53}_{-0.35}$ \\[2pt]

\hline\hline
\end{tabular}
\end{center}
\end{table}

One important assumption underlying all the results reported elsewhere in this paper is that the observations are due to the presence of a single particle with well defined mass that has been precisely measured~\cite{ATLASCMSHmass}.
This section addresses the case in which the observed signal could be due to the presence of two or more particles with similar masses, such that they cannot be resolved within the current precision of the mass measurements in the different channels.
Several BSM models predict, for example, a superposition of states with indistinguishable mass values~\cite{Gunion:2012gc,Grzadkowski:2012ng,Drozd:2012vf,Ferreira:2012nv}, possibly with different coupling structures to the SM~particles.
With such an assumption, it may be possible to distinguish between single and multiple states by measuring the cross sections of individual production processes independently for each decay mode, as described in Section~\ref{sec:sigBR5x5}.
Several methods have been proposed to assess the compatibility of the data with a single state~\cite{GunionDegenerate,GrossmanDegenerate}.
A test for the possible presence of overlapping Higgs boson states is performed, based on a profile likelihood ratio suggested in~Ref.~\cite{MatrixRank}. This test accounts both for missing measurements, such as the $\Hbb$~decay mode in the \aggF\ and \aVBF\ production processes, and for uncertainties in the measurements, including their correlations.

The 25~possible combinations resulting from five production processes times five decay modes can be parameterised using a $5\times5$ matrix~\calM\ in two ways:
\begin{itemize}
 \item allowing full freedom for all yields except the two mentioned above, which are not addressed in the combined analyses, leading to 23~free parameters, similarly to the fit performed for all products of cross sections times branching fractions and presented in Section~\ref{sec:sigBR5x5};
\item assuming that all yields originate from five production processes and five decay modes, leading to nine~free parameters, as shown in~Table~\ref{tab:genericmodels}, similarly to the fit performed for one reference cross section times branching fraction and eight ratios of cross sections and branching fractions, described in Section~\ref{sec:sigBR9}.
\end{itemize}

\begin{table*}[htb]
\centering
\caption{The two signal parameterisations used to scale the expected yields of the $5\times5$ combinations of production processes and decay modes. 
The first parameterisation corresponds to the most general case with 25~independent parameters, while the second parameterisation corresponds to that expected for a single Higgs boson state. As explained in the text for the case of the general matrix parameterisation, the two parameters \muggfbb\ and \lvbfbb\ are set to unity in the fits, since the current analyses are not able to constrain them. 
}
\label{tab:degenerateSaturatedModel}
\renewcommand{\arraystretch}{1.1}
\begin{tabular}{c|ccccc}
\hline\hline
\multicolumn{6}{c}{General matrix parameterisation: \rankM = 5}\\
\hline
      &  $\Hyy$  &  $\Hzz$  &  $\Hww$  &  $\Htt$  &  $\Hbb$ \\
\hline
\aggF  & \muggfgg & \muggfzz & \muggfww & \muggftt & \muggfbb \\
\aVBF  & $\lvbfgg\:\muggfgg$ & $\lvbfzz\:\muggfzz$ & $\lvbfww\:\muggfww$ & $\lvbftt\:\muggftt$ & $\lvbfbb\:\muggfbb$ \\
\aWH   & $\lwhgg\:\muggfgg$ & $\lwhzz\:\muggfzz$ & $\lwhww\:\muggfww$ & $\lwhtt\:\muggftt$ & $\lwhbb\:\muggfbb$ \\
\aZH   & $\lzhgg\:\muggfgg$ & $\lzhzz\:\muggfzz$ & $\lzhww\:\muggfww$ & $\lzhtt\:\muggftt$ & $\lzhbb\:\muggfbb$ \\
\attH  & $\ltthgg\:\muggfgg$ & $\ltthzz\:\muggfzz$ & $\ltthww\:\muggfww$ & $\ltthtt\:\muggftt$ & $\ltthbb\:\muggfbb$ \\
\multicolumn{6}{c}{~~~}\\
\multicolumn{6}{c}{Single-state matrix parameterisation: \rankM = 1}\\
\hline
      &  $\Hyy$  &  $\Hzz$  &  $\Hww$  &  $\Htt$  &  $\Hbb$ \\
\hline
\aggF  & \muggfgg & \muggfzz & \muggfww & \muggftt & \muggfbb \\
\aVBF  & $\lvbf\:\muggfgg$ & $\lvbf\:\muggfzz$ & $\lvbf\:\muggfww$ & $\lvbf\:\muggftt$ & $\lvbf\:\muggfbb$ \\
\aWH   & $\lwh \:\muggfgg$ & $\lwh \:\muggfzz$ & $\lwh \:\muggfww$ & $\lwh \:\muggftt$ & $\lwh \:\muggfbb$ \\
\aZH   & $\lzh \:\muggfgg$ & $\lzh \:\muggfzz$ & $\lzh \:\muggfww$ & $\lzh \:\muggftt$ & $\lzh \:\muggfbb$ \\
\attH  & $\ltth\:\muggfgg$ & $\ltth\:\muggfzz$ & $\ltth\:\muggfww$ & $\ltth\:\muggftt$ & $\ltth\:\muggfbb$ \\
\hline\hline
\end{tabular}
\end{table*}

There is a direct relation between the rank of the $5\times5$ matrix \calM and the number of degenerate states. More specifically, if the observations are due to a single state, the matrix can be obtained from one of its rows by using one common multiplier per row and therefore $\rankM = 1$, in contrast to $\rankM = 5$ in the most general case. The two parameterisations of \calM used in this test are shown in~Table~\ref{tab:degenerateSaturatedModel}.  
They are both expressed in terms of $\mu_{\aggF}^{j}$, defined as in Eq.~(\ref{eq:muif}). Then, for the general case, the other parameters are $\lambda_i^j=\mu_{i}^j/\mu_{\aggF}^{j}$, whereas for the $\rankM = 1$ case the other parameters are $\lambda_i=\mu_{i}/\mu_{\aggF}$.
In this section, the index~$i$ runs over the \aVBF, \aWH, \aZH, and \attH\ production processes and the index~$j$ runs over the five decay modes. The two statistical parameterisations are nested since the second one can be obtained from the first by imposing $\lambda_i^j=\lambda_i$. The SM~prediction corresponds to the \rankM~=1~case, where $\mu_{\aggF}^{j}=\lambda_i=\lambda_i^j=1$.

In contrast to the fits described previously, all the parameters of interest are constrained to be positive for the fits performed in this section. This choice to restrict the parameter space to the physically meaningful region improves the convergence of the fits. 
The results of the fits to the data are consistent with those presented in~Section~\ref{sec:sigBR} for the two similar parameterisations and are not reported here.

In order to quantify the compatibility of the data with the single-state hypothesis, a profile likelihood ratio test statistic, $q_\lambda$, is built that compares the hypothesis of a single-state matrix with \rankM~=~1 to the general hypothesis with \rankM~=~5:
\begin{equation}
\label{eq:qlambda}
q_{\lambda}=-2 \ln\frac{
{L}(\mathrm{data}|\lambda^{j}_{i}=\hat{\lambda}_{i},{\hat{\mu}_{\aggF}^{j}})
}{
{L}(\mathrm{data}|\hat{\lambda}^{j}_{i},{\hat{\mu}_{\aggF}^{\prime j}})},
\end{equation}
where ${\hat{\mu}_{\aggF}^{j}}$ and ${\hat{\mu}_{\aggF}^{\prime j}}$ represent the best fit values of the parameters of interest, respectively for the single-state and general hypotheses. The observed value of~$q_{\lambda}$ in data is compared with the expected distribution, as obtained from pseudo-data samples randomly generated from the best fit values of the \rankM=1~hypothesis. The $p$-value of the data with the single-state hypothesis is~(29$\pm$2)\%, where the uncertainty reflects the finite number of pseudo-data samples generated, and does not show any significant departure from the single-state hypothesis. The $p$-values obtained for the individual experiments are~58\% and~33\% for ATLAS and CMS, respectively. These $p$-values can only be considered as the results of compatibility tests with the single-state hypothesis, represented by the \rankM~=~1~parameterisation described above.

\section{Constraints on Higgs boson couplings}
\label{sec:CouplingFits}

Section~\ref{sec:kappaParam} discusses the fit results from the most generic parameterisation in the context of the $\Cc$-framework.
This section probes more specific parameterisations with additional assumptions. 
In the following, results from a few selected parameterisations, with increasingly restrictive assumptions, are presented. 
The results are obtained from the combined fits to the $\sqrt{s}=7$ and 8~TeV data assuming that the coupling modifiers are the same at the two energies.

\subsection{Parameterisations allowing contributions from BSM particles in loops and in decays}
\label{sec:ModelK2}

As discussed in Sections~\ref{sec:TheoryFramework} and~\ref{sec:CombinationProcedure}, 
the rates of Higgs boson production in the various decay modes are inversely proportional to the Higgs boson width, which is sensitive to potential invisible or undetected decay modes predicted by BSM theories. 
To directly measure the individual coupling modifiers, an assumption about the Higgs boson width is necessary. 
Two possible scenarios are considered in this section: the first leaves $\BRbsm$~free, provided that $\BRbsm \ge 0$, but assumes that $|\Cc_{W}| \le 1$ and~$|\Cc_{Z}| \le 1$ and that the signs of $\Cc_{W}$ and $\Cc_{Z}$ are the same, assumptions denoted $|\Cc_{V}| \le 1$ in the following; the second assumes $\BRbsm = 0$. 
The constraints assumed in the first scenario are compatible with a wide range of BSM physics, which may become manifest in the loop-induced processes of $gg\to H$ production and $\Hyy$ decay. These processes are particularly sensitive to loop contributions from new heavy particles, carrying electric or colour charge, or both, and such new physics can be probed using the effective coupling modifiers $\Kg$~and~$\Ky$. 
Furthermore, potential deviations from the SM of the tree-level couplings to ordinary particles are parameterised with their respective coupling modifiers. The parameters of interest in the fits to data are thus the seven independent coupling modifiers, $\KZ,\, \KW,\, \Kt,\, \Ktau,\, \Kb,\, \Kg$, and $\Ky$, one for each SM~particle involved in the production processes and decay modes studied, plus $\BRbsm$ in the case of the first fit. Here and in Section~\ref{sec:ModelK1}, the coupling modifier $\Kt$ is assumed to be positive, without any loss of generality.

\begin{figure}[hbtp!]
  \center
\includegraphics[width=1.0\textwidth]{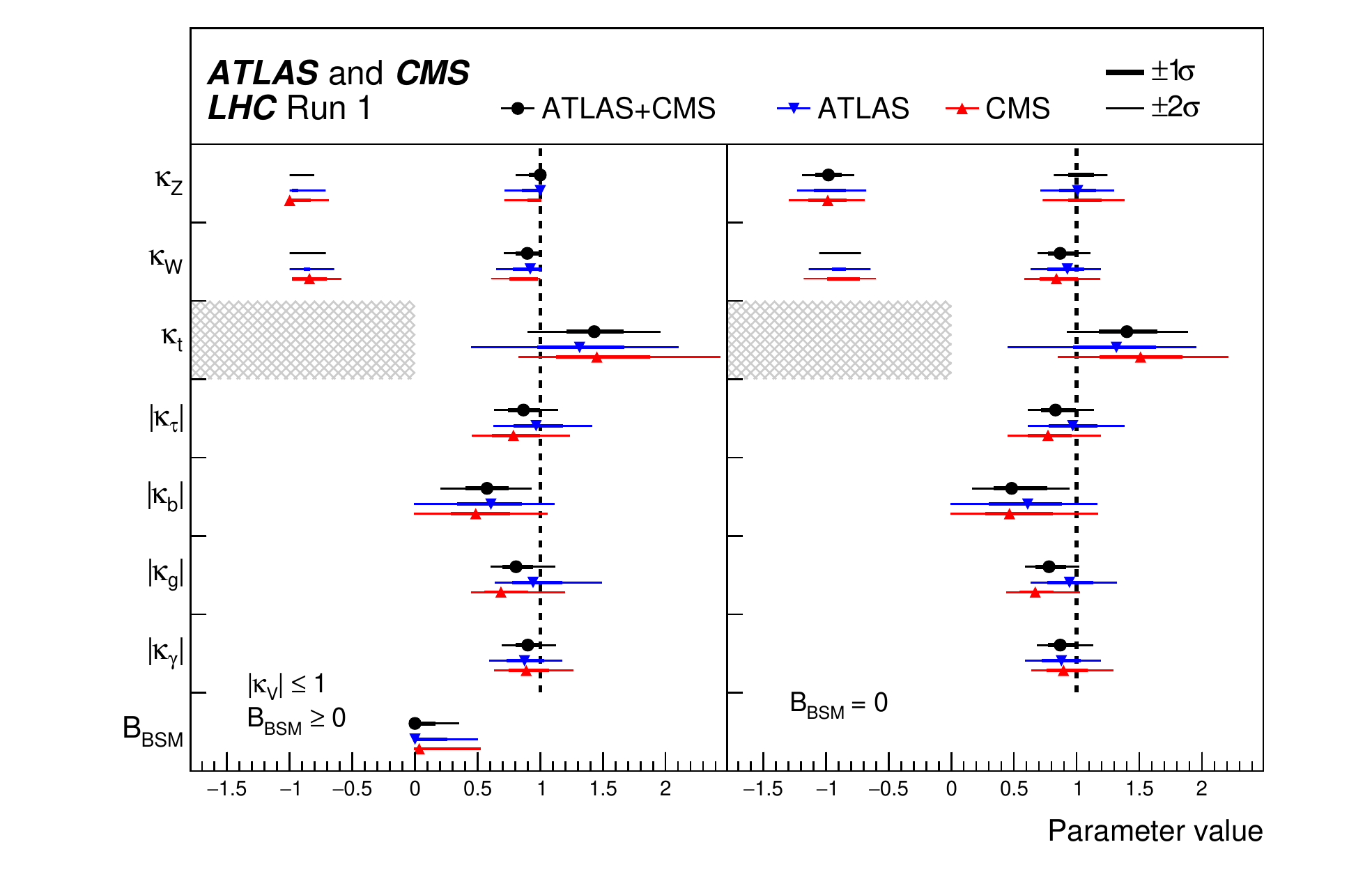}
\caption{Fit results for two parameterisations allowing BSM loop couplings discussed in the text: the first one assumes that $\BRbsm \ge 0$ and that $|\Cc_{V}| \le 1$, where $\Cc_{V}$ denotes $\Cc_{Z}$  or $\Cc_{W}$, and the second one assumes that there are no additional BSM~contributions to the Higgs boson width, i.e.~$\BRbsm=0$. The measured results for the combination of ATLAS and CMS are reported together with their uncertainties, as well as the individual results from each experiment. The hatched areas show the non-allowed regions for the \Kt\ parameter, which is assumed to be positive without loss of generality. 
The error bars indicate the $1\sigma$~(thick lines) and $2\sigma$~(thin lines) intervals. 
When a parameter is constrained and reaches a boundary, namely $|\Cc_{V}| = 1$ or $\BRbsm = 0$, the uncertainty is not defined beyond this boundary.
For those parameters with no sensitivity to the sign, only the absolute values are shown.
}
\label{fig:modelK2}
\end{figure}

Figure~\ref{fig:modelK2} and Table~\ref{tab:K1_K2_results} show the results of the two fits, assuming either~$|\Cc_{V}|\le1$ and~$\BRbsm \ge 0$ or~$\BRbsm=0$. In the former case, an upper limit of $\BRbsm=0.34$ at 95\%~CL is obtained, compared to an expected upper limit of~0.39.
The corresponding negative log-likelihood scan is shown in Fig.~\ref{fig:modelK2_BRinvscan}. 
Appendix~\ref{sec:app_negative} describes how the two possible sign combinations between~$\Cc_{W}$ and~$\Cc_{Z}$ impact the likelihood scan of~$\BRbsm$ for the observed and expected results, as illustrated in~Fig.~\ref{fig:BRBSM-signs}.   
The $p$-value of the compatibility between the data and the SM predictions is~11\% with the assumption that $\BRbsm=0$.

Another fit, motivated, for example, by BSM scenarios with new heavy particles that may contribute to loop processes in Higgs boson production or decay, assumes that all the couplings to SM~particles are the same as in the~SM, that there are no~BSM decays ($\BRbsm = 0$), and that only the gluon--gluon production and $\gamma\gamma$ decay loops may be affected by the presence of additional particles.
The results of this fit, which has only the effective coupling modifiers $\Ky$~and~$\Kg$ as free parameters, with all other coupling modifiers fixed to their SM~values of unity, are shown in Fig.~\ref{fig:kgkgamma}. 
The point $\Ky=1$ and $\Kg=1$ lies within the 68\% CL region and the $p$-value of the compatibility between the data and the SM predictions is~82\%.

\begin{table}[htb]
\centering
\caption{Fit results for two parameterisations allowing BSM loop couplings discussed in the text: the first one assumes that $|\Cc_{V}| \le 1$, where $\Cc_{V}$ denotes $\Cc_{Z}$  or $\Cc_{W}$, and that $\BRbsm \ge 0$, while the second one assumes that there are no additional BSM~contributions to the Higgs boson width, i.e.~$\BRbsm=0$. 
The results for the combination of ATLAS and CMS are reported with their measured and expected uncertainties.
Also shown are the results from each experiment.
For the parameters with both signs allowed, the $1\sigma$~intervals are shown on a second line.
When a parameter is constrained and reaches a boundary, namely~$\BRbsm = 0$, the uncertainty is not indicated.
For those parameters with no sensitivity to the sign, only the absolute values are shown.
\label{tab:K1_K2_results}
}
\setlength\extrarowheight{4pt}
\begin{tabular}{l|c|c|c|c}
\hline\hline
Parameter & ATLAS+CMS & ATLAS+CMS & ATLAS & CMS \\
          & Measured  & Expected uncertainty  & Measured  & Measured \\
\hline
\multicolumn{5}{c}{Parameterisation assuming $|\kappa_{V}|\leq 1$ and $\BRbsm\geq 0$}\\
\hline
$\kappa_{Z}$        & $ 1.00                $ &                         & $ 1.00                $ & $-1.00                $ \\[-2pt]
                    & $[0.92,1.00]          $ & $[-1.00,-0.89] \cup   $ & $[-0.97,-0.94] \cup   $ & $[-1.00,-0.84] \cup   $ \\[-2pt]
                    &                         & $[0.89,1.00]          $ & $[0.86,1.00]          $ & $[0.90,1.00]          $ \\[5pt]
$\kappa_{W}$        & $ 0.90                $ &                         & $ 0.92                $ & $-0.84                $ \\[-2pt]
                    & $[0.81,0.99]          $ & $[-1.00,-0.90] \cup   $ & $[-0.88,-0.84] \cup   $ & $[-1.00,-0.71] \cup   $ \\[-2pt]
                    &                         & $[0.89,1.00]          $ & $[0.79,1.00]          $ & $[0.76,0.98]          $ \\[5pt]
$\kappa_{t}$        & $ 1.43^{+0.23}_{-0.22}$ & $     ^{+0.27}_{-0.32}$ & $ 1.31^{+0.35}_{-0.33}$ & $ 1.45^{+0.42}_{-0.32}$ \\[5pt]
$|\kappa_{\tau}|$   & $ 0.87^{+0.12}_{-0.11}$ & $     ^{+0.14}_{-0.15}$ & $ 0.97^{+0.21}_{-0.17}$ & $ 0.79^{+0.20}_{-0.16}$ \\[5pt]
$|\kappa_{b}|$      & $ 0.57^{+0.16}_{-0.16}$ & $     ^{+0.19}_{-0.23}$ & $ 0.61^{+0.24}_{-0.26}$ & $ 0.49^{+0.26}_{-0.19}$ \\[5pt]
$|\kappa_{g}|$      & $ 0.81^{+0.13}_{-0.10}$ & $     ^{+0.17}_{-0.14}$ & $ 0.94^{+0.23}_{-0.16}$ & $ 0.69^{+0.21}_{-0.13}$ \\[5pt]
$|\kappa_{\gamma}|$ & $ 0.90^{+0.10}_{-0.09}$ & $     ^{+0.10}_{-0.12}$ & $ 0.87^{+0.15}_{-0.14}$ & $ 0.89^{+0.17}_{-0.13}$ \\[5pt]
$\BRbsm$            & $ 0.00^{+0.16}        $ & $     ^{+0.19}        $ & $ 0.00^{+0.25}        $ & $ 0.03^{+0.26}        $ \\[2pt]

\hline
\multicolumn{5}{c}{Parameterisation assuming $\BRbsm=0$}\\
\hline
$\kappa_{Z}$        & $-0.98                $ &                         & $ 1.01                $ & $-0.99                $ \\[-2pt]
                    & $[-1.08,-0.88] \cup   $ & $[-1.01,-0.87] \cup   $ & $[-1.09,-0.85] \cup   $ & $[-1.14,-0.84] \cup   $ \\[-2pt]
                    & $[0.94,1.13]          $ & $[0.89,1.11]          $ & $[0.87,1.15]          $ & $[0.94,1.19]          $ \\[5pt]
$\kappa_{W}$        & $ 0.87                $ &                         & $ 0.92                $ & $ 0.84                $ \\[-2pt]
                    & $[0.78,1.00]          $ & $[-1.08,-0.90] \cup   $ & $[-0.94,-0.85] \cup   $ & $[-0.99,-0.74] \cup   $ \\[-2pt]
                    &                         & $[0.88,1.11]          $ & $[0.78,1.05]          $ & $[0.71,1.01]          $ \\[5pt]
$\kappa_{t}$        & $ 1.40^{+0.24}_{-0.21}$ & $     ^{+0.26}_{-0.39}$ & $ 1.32^{+0.31}_{-0.33}$ & $ 1.51^{+0.33}_{-0.32}$ \\[5pt]
$|\kappa_{\tau}|$   & $ 0.84^{+0.15}_{-0.11}$ & $     ^{+0.16}_{-0.15}$ & $ 0.97^{+0.19}_{-0.19}$ & $ 0.77^{+0.18}_{-0.15}$ \\[5pt]
$|\kappa_{b}|$      & $ 0.49^{+0.27}_{-0.15}$ & $     ^{+0.25}_{-0.28}$ & $ 0.61^{+0.26}_{-0.31}$ & $ 0.47^{+0.34}_{-0.19}$ \\[5pt]
$|\kappa_{g}|$      & $ 0.78^{+0.13}_{-0.10}$ & $     ^{+0.17}_{-0.14}$ & $ 0.94^{+0.18}_{-0.17}$ & $ 0.67^{+0.14}_{-0.12}$ \\[5pt]
$|\kappa_{\gamma}|$ & $ 0.87^{+0.14}_{-0.09}$ & $     ^{+0.12}_{-0.13}$ & $ 0.88^{+0.15}_{-0.15}$ & $ 0.89^{+0.19}_{-0.13}$ \\[2pt]

\hline\hline
\end{tabular}
\end{table}

\clearpage

\begin{figure}[hbtp!]
  \center
\includegraphics[width=0.8\textwidth]{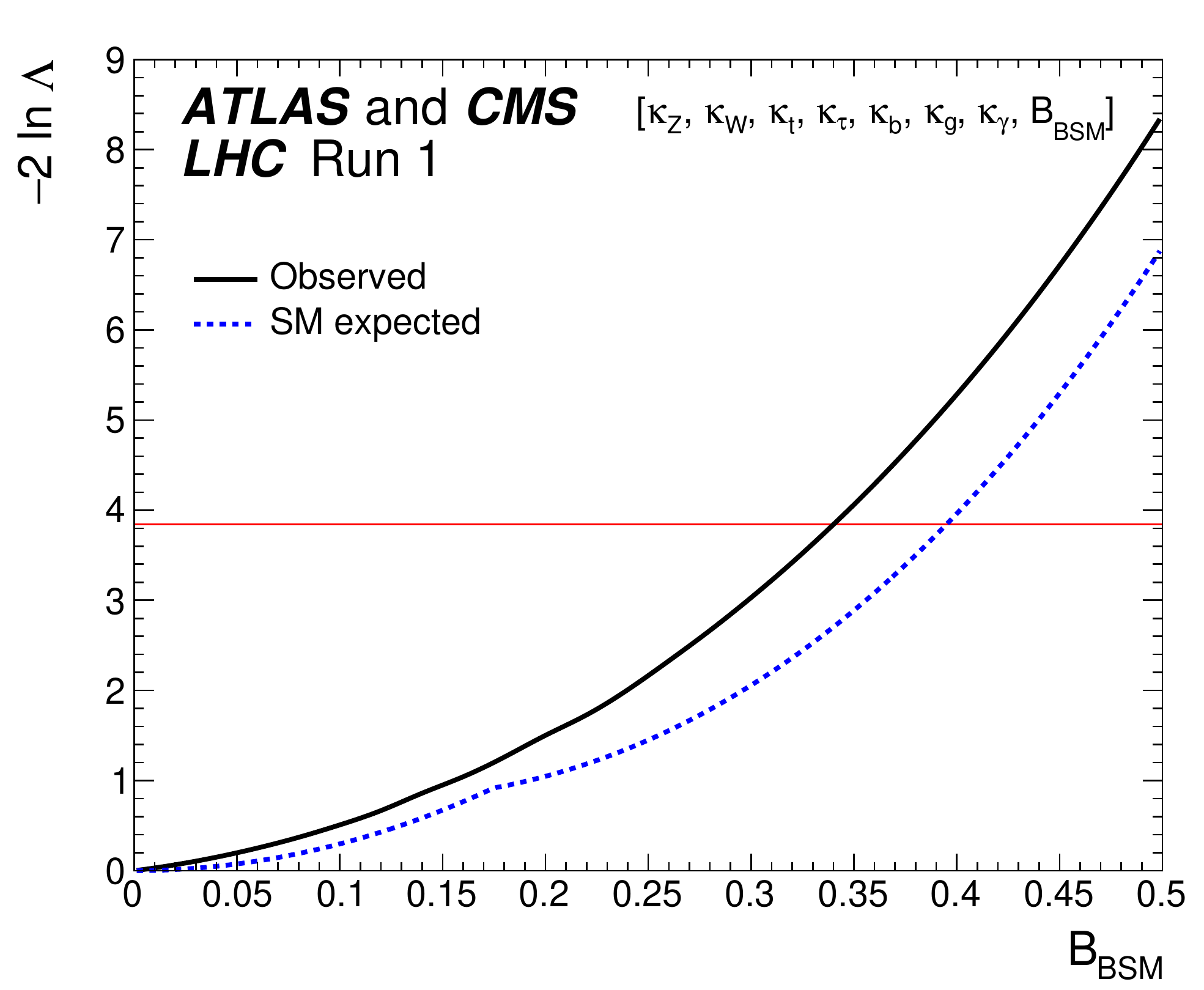}
  \caption{Observed (solid line) and expected (dashed line) negative log-likelihood scan of $\BRbsm$, shown for the combination of ATLAS and CMS when allowing additional BSM~contributions to the Higgs boson width. The results are shown for the parameterisation with the assumptions that $|\Cc_{V}|\le 1$ and $\BRbsm~\ge~0$ in~Fig.~\ref{fig:modelK2}. All the other parameters of interest from the list in the legend are also varied in the minimisation procedure. The red horizontal line at 3.84 indicates the log-likelihood variation corresponding to the 95\% CL upper limit, 
as discussed in Section~\ref{sec:CombinationStatistics}.
} 
\label{fig:modelK2_BRinvscan}
\end{figure}

\begin{figure}[hbtp!]
  \includegraphics[width=0.8\textwidth]{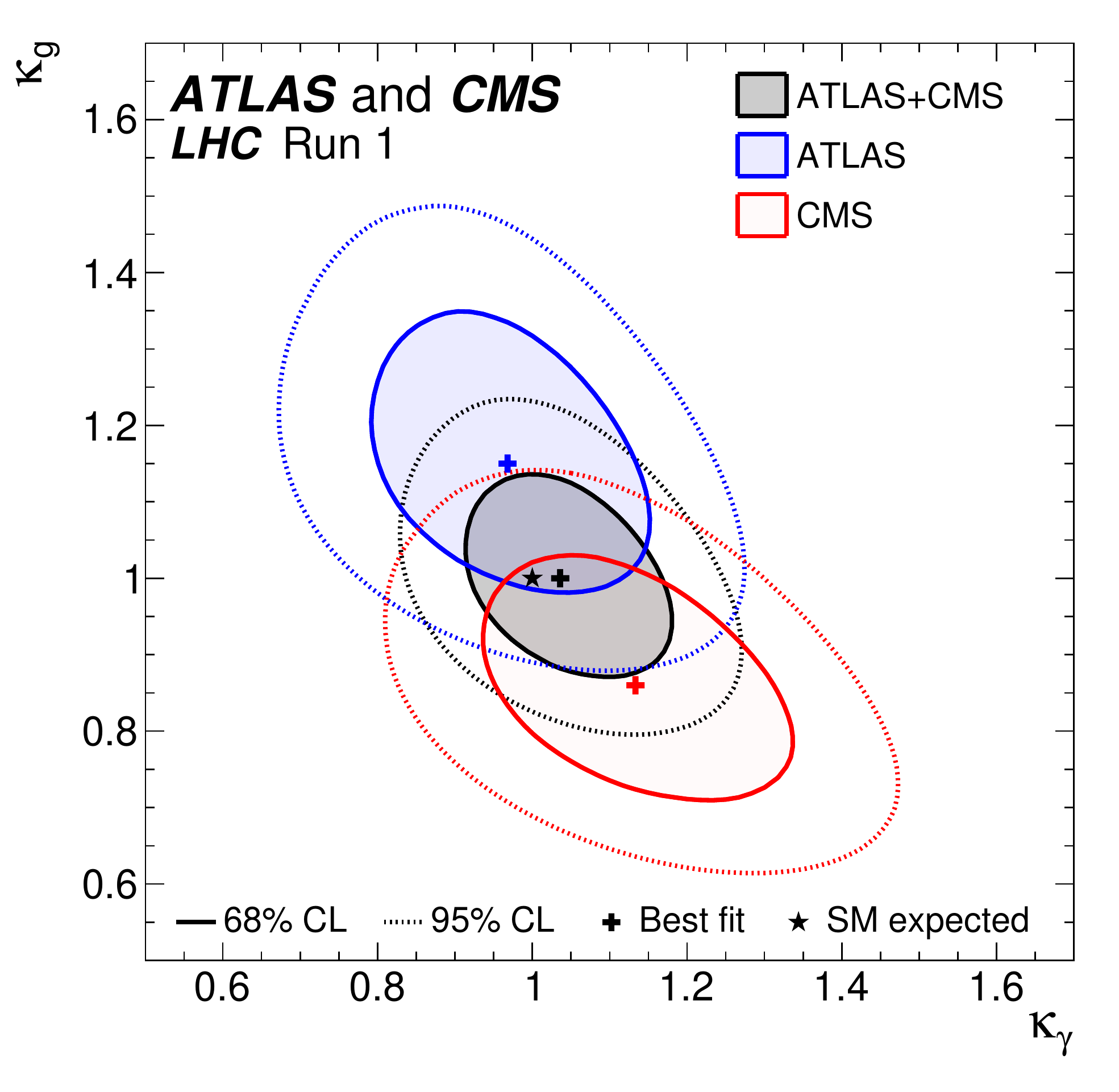}
\caption{Negative log-likelihood contours at~68\% and 95\%~CL in the ($\Ky$,~$\Kg$) plane for the combination of ATLAS and CMS and for each experiment separately, as obtained from the fit to the parameterisation constraining all the other coupling modifiers to their SM~values and assuming~$\BRbsm = 0$. }
\label{fig:kgkgamma}
\end{figure}

\subsection{Parameterisation assuming SM structure of the loops and no BSM decays}
\label{sec:ModelK1}

In this section it is assumed that there are no new particles in the loops entering \aggF\ production and $\Hyy$ decay. This assumption is supported by the measurements of the effective coupling modifiers $\Kg$ and $\Ky$, which are consistent with the SM~predictions.
The cross section for \aggF\ production and the branching fraction for the $\Hyy$ decay are expressed in terms of the coupling modifiers of the SM particles in the loops, as indicated in~Table~\ref{tab:kexpr}. 
This leads to a parameterisation with six free coupling modifiers: $\KW,\, \KZ,\, \Kt,\, \Ktau$, $\Kb$, and $\Kmu$; the results of the $\Hmm$~analysis are included for this specific case. 
In this more constrained fit, it is also assumed that~$\BRbsm=0$.

\begin{figure}[hbt!]
  \center
    \includegraphics[width=0.7\textwidth]{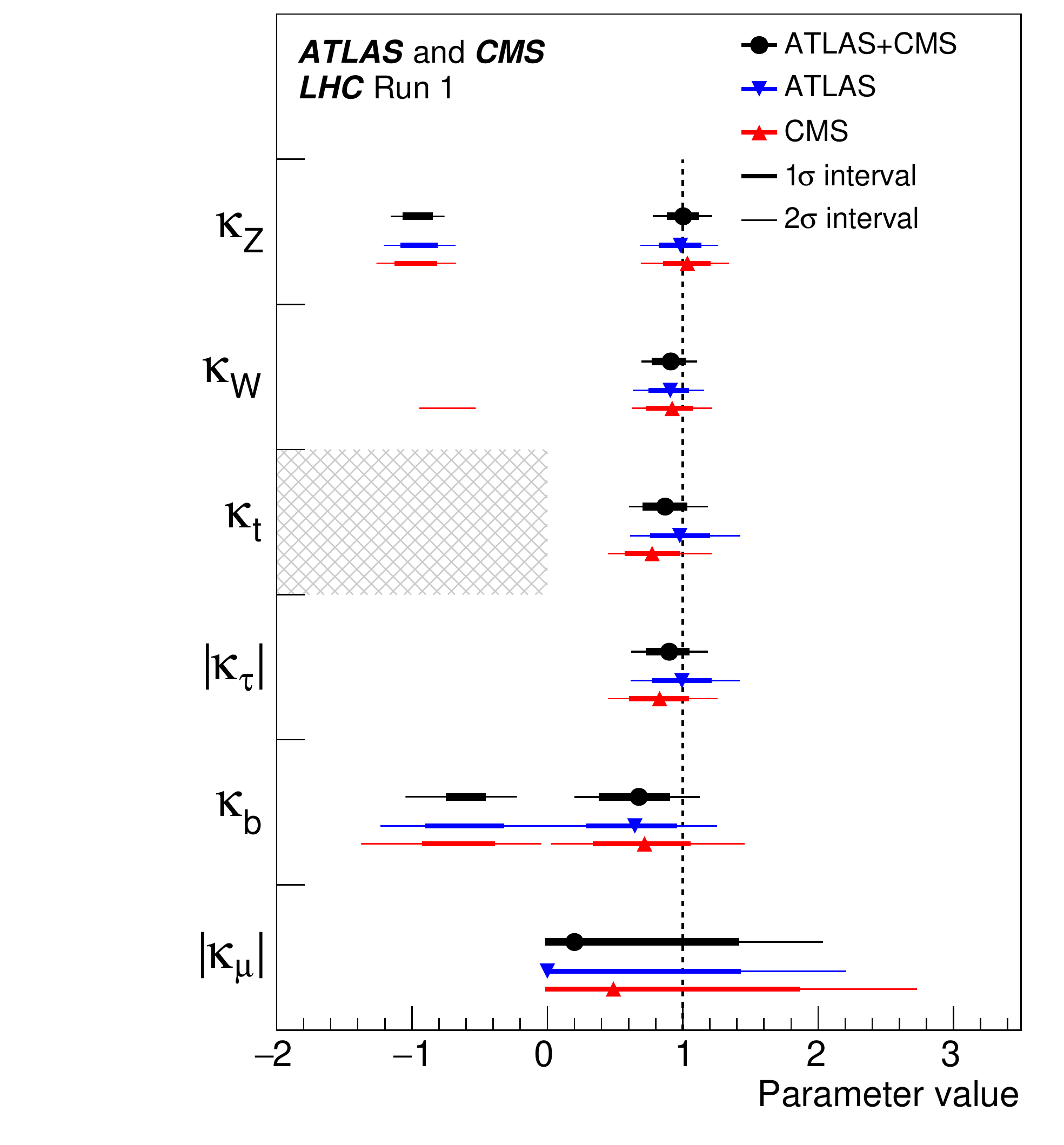}
  \caption{Best fit values of parameters for the combination of ATLAS and CMS data, and separately for each experiment, for the parameterisation assuming the absence of BSM particles in the loops, $\BRbsm=0$. 
The hatched area indicates the non-allowed region for the parameter that is assumed to be positive without loss of generality. The error bars indicate the $1\sigma$~(thick lines) and $2\sigma$~(thin lines) intervals. When a parameter is constrained and reaches a boundary, namely $|\kappa_\mu| = 0$, the uncertainty is not defined beyond this boundary. 
For those parameters with no sensitivity to the sign, only the absolute values are shown.
}
  \label{fig:bm:GenMod1:par}
\end{figure}

Figure~\ref{fig:bm:GenMod1:par} and Table~\ref{tab:K1_results} show the results of the fit for the combination of ATLAS and CMS, and separately for each experiment. 
Compared to the results from the fitted decay signal strengths (Table~\ref{tab:muDecay}) or the global signal strength $\mu = 1.09 \pm 0.11$ (Section~\ref{sec:GlobalMu}), this fit yields values of the coupling modifiers lower than those predicted by the~SM. 
This is a consequence of the low value of~$\Cc_b$, as measured by the combination of ATLAS and CMS and by each experiment. 
A low value of~$\Cc_b$ decreases the total Higgs boson width through the dominant $\Gamma^{bb}$ partial decay width, and, as a consequence, the measured values of all the coupling modifiers decrease, such that the values of~$\sigma_{\mathit{i}}(\vec\Cc)\cdot \BR^{\mathit{f}}$ remain consistent with the observed signal yields. 
The $p$-value of the compatibility between the data and the SM predictions is~74\%.

\begin{table}[htb]
\centering
\vspace{3mm}
\caption{Fit results for the parameterisation assuming the absence of BSM particles in the loops ($\BRbsm=0$). The results with their measured and expected uncertainties are reported for the combination of ATLAS and CMS, together with the individual results from each experiment. 
For the parameters with both signs allowed, the $1\sigma$~CL~intervals are shown on a second line.
When a parameter is constrained and reaches a boundary, namely $|\kappa_\mu| = 0$, the uncertainty is not indicated.
For those parameters with no sensitivity to the sign, only the absolute values are shown.
}
\label{tab:K1_results}
\setlength\extrarowheight{4pt}
\begin{tabular}{l|c|c|c|c}
\hline\hline
Parameter & ATLAS+CMS & ATLAS+CMS & ATLAS & CMS \\
          & Measured  & Expected uncertainty  & Measured  & Measured \\
\hline
$\kappa_{Z}$      & $1.00                $ &                        & $0.98                $ & $1.03                $ \\[-2pt]
                  & $[-1.05,-0.86] \cup  $ & $[-1.00,-0.88] \cup  $ & $[-1.07,-0.83] \cup  $ & $[-1.11,-0.83] \cup  $ \\[-2pt]
                  & $[0.90,1.11]         $ & $[0.90,1.10]         $ & $[0.84,1.12]         $ & $[0.87,1.19]         $ \\[5pt]
$\kappa_{W}$      & $0.91^{+0.10}_{-0.12}$ & $    ^{+0.10}_{-0.11}$ & $0.91^{+0.12}_{-0.15}$ & $0.92^{+0.14}_{-0.17}$ \\[5pt]
$\kappa_{t}$      & $0.87^{+0.15}_{-0.15}$ & $    ^{+0.15}_{-0.18}$ & $0.98^{+0.21}_{-0.20}$ & $0.77^{+0.20}_{-0.18}$ \\[5pt]
$|\kappa_{\tau}|$ & $0.90^{+0.14}_{-0.16}$ & $    ^{+0.15}_{-0.14}$ & $0.99^{+0.20}_{-0.20}$ & $0.83^{+0.20}_{-0.21}$ \\[5pt]
$\kappa_{b}$      & $0.67                $ &                        & $0.64                $ & $0.71                $ \\[-2pt]
                  & $[-0.73,-0.47] \cup  $ & $[-1.24,-0.76] \cup  $ & $[-0.89,-0.33] \cup  $ & $[-0.91,-0.40] \cup  $ \\[-2pt]
                  & $[0.40,0.89]         $ & $[0.74,1.24]         $ & $[0.30,0.94]         $ & $[0.35,1.04]         $ \\[5pt]
$|\kappa_{\mu}|$  & $0.2 ^{+1.2}         $ & $    ^{+0.9}         $ & $0.0 ^{+1.4}         $ & $0.5 ^{+1.4}         $ \\[2pt]

\hline\hline
\end{tabular}
\end{table}

A different view of the relation between the fitted coupling modifiers and the SM~predictions is presented in Fig.~\ref{fig:bm:GenMod1:prplot}.
New parameters are derived from the coupling modifiers, to make explicit the dependence on the particle masses: linear for the Yukawa couplings to the fermions and quadratic for the gauge couplings of the Higgs boson to the weak vector bosons.
These new parameters are all assumed in this case to be positive. For fermions with mass $m_{F,i}$, the parameters are $\Cc_{F,i} \cdot~y_{F,i}/\sqrt{2} = \Cc_{F,i} \cdot~m_{F,i}/v$, where $y_{F,i}$ is the Yukawa coupling strength, assuming a SM~Higgs boson with a mass of 125.09~GeV, and $v = 246$~GeV is the vacuum expectation value of the Higgs field. For the weak vector bosons with mass~$m_{V,i}$, the new parameters are $\sqrt{ \Cc_{V,i} \cdot~g_{V,i}/2v} = \sqrt{\Cc_{V,i}} \cdot~m_{V,i}/v$, where $g_{V,i}$ is the absolute Higgs boson gauge coupling strength. The linear scaling of these new parameters as a function of the particle masses observed in~Fig.~\ref{fig:bm:GenMod1:prplot} indicates qualitatively the compatibility of the measurements with the~SM. For the $b$~quark, the running mass evaluated at a scale equal to~$\mH$, $m_b(\mH)=2.76$~GeV, is used.

Following the phenomenological model suggested in Ref.~\cite{MepsilonTheory}, the coupling modifiers can also be expressed as a function of a mass scaling parameter~$\epsilon$, with a value $\epsilon = 0$ in the~SM, and a free parameter~$M$, equal to $v$ in the~SM: $\Cc_{F,i} = v\cdot~m_{F,i}^{\epsilon}/M^{1+\epsilon}$ and $\Cc_{V,i}  = v \cdot~m_{V,i}^{2\epsilon}/M^{1+2\epsilon}$. A fit is then performed with the same assumptions as those of Table~\ref{tab:K1_results} with $\epsilon$ and $M$ as parameters of interest. The results for the combination of ATLAS and CMS are $\epsilon = 0.023^{+0.029}_{-0.027}$ and $M = 233^{+13}_{-12}$~GeV, and are compatible with the SM~predictions. Figure~\ref{fig:bm:GenMod1:prplot} shows the results of this fit with its corresponding 68\%~and 95\%~CL~bands.

\begin{figure}[hbt!]
  \center
 \includegraphics[width=0.7\textwidth]{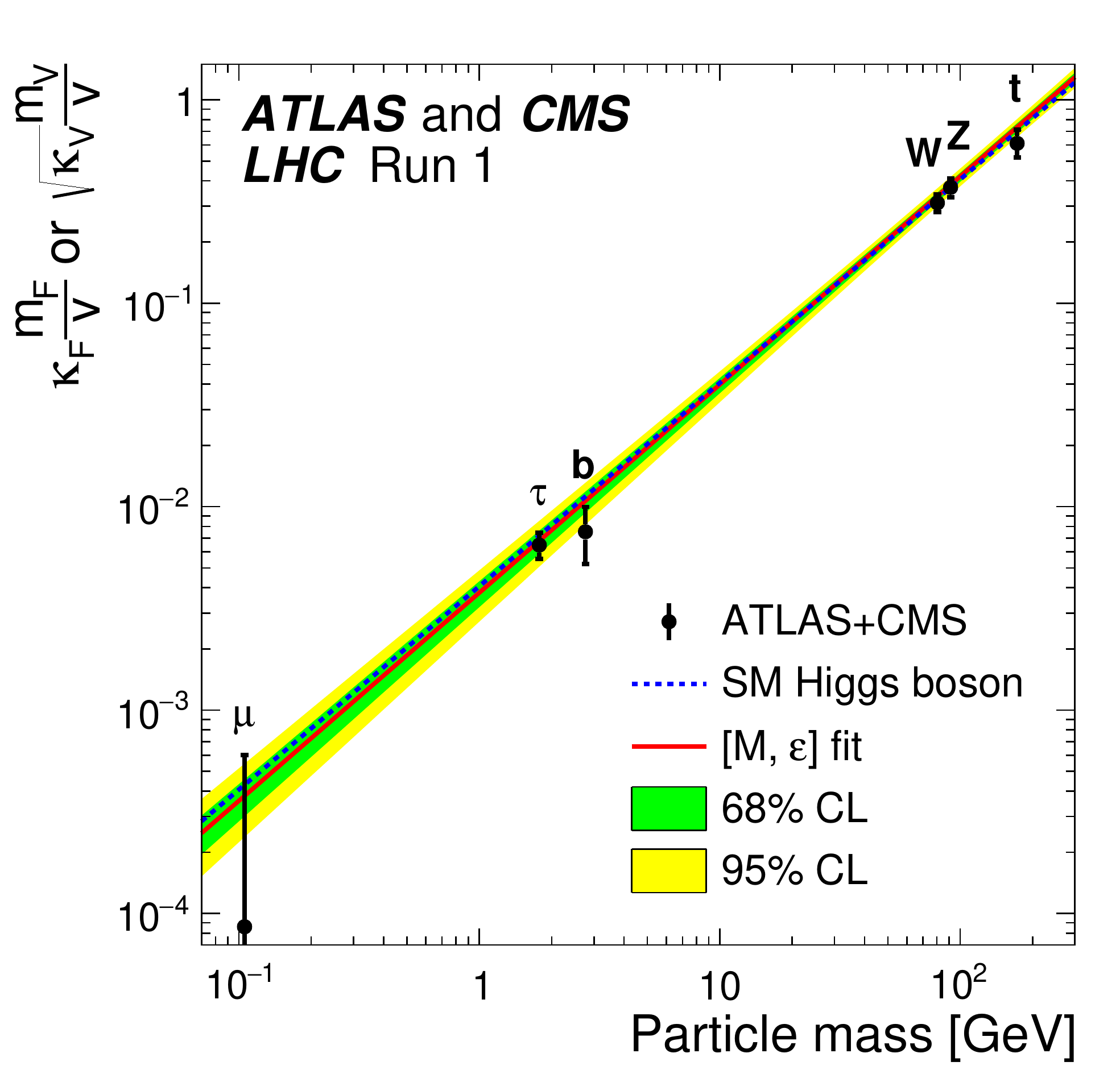}
  \caption{Best fit values as a function of particle mass for the combination of ATLAS and CMS data in the case of the parameterisation described in the text, with parameters defined as $\Cc_{F} \cdot\ m_{F}/v$ for the fermions, and as $\sqrt{\Cc_{V}} \cdot\ m_{V}/v$ for the weak vector bosons, where $v = 246$~GeV is the vacuum expectation value of the Higgs field. The dashed (blue) line indicates the predicted dependence on the particle mass in the case of the SM~Higgs boson. The solid (red) line indicates the best fit result to the $[M,\epsilon]$~phenomenological model of~Ref.~\cite{MepsilonTheory} with the corresponding 68\%~and 95\%~CL~bands.
}
\label{fig:bm:GenMod1:prplot}
\end{figure}

\begin{table}[htb]
\centering
\caption{\normalsize Summary of fit results for the two parameterisations probing the ratios of coupling modifiers for up-type versus down-type fermions and for leptons versus quarks. 
The results for the combination of ATLAS and CMS are reported together with their measured and expected uncertainties.
Also shown are the results from each experiment. The parameters~$\kappa_{uu}$ and~$\kappa_{qq}$ are both positive definite since $\KH$~is always assumed to be positive. For the parameter~$\lambda_{du}$, for which both signs are allowed, the $1\sigma$~CL~intervals are shown on a second line. For the parameter $\lambda_{lq}$, for which there is no sensitivity to the sign, only the absolute values are shown. Negative values for the parameters~$\lambda_{Vu}$ and~$\lambda_{Vq}$ are excluded by more than~$4\sigma$.  
\label{tab:ud_lq_results}
}
\setlength\extrarowheight{4pt}
\begin{tabular}{l|c|c|c|c}
\hline\hline
Parameter & ATLAS+CMS & ATLAS+CMS & ATLAS & CMS \\
          & Measured  & Expected uncertainty  & Measured  & Measured \\
\hline
$\lambda_{du}$ & $0.92                $ &                        & $0.86                $ & $1.01                $ \\[-2pt]
               & $[0.80,1.04]         $ & $[-1.21,-0.92] \cup  $ & $[-1.03,-0.78] \cup  $ & $[-1.20,-0.94] \cup  $ \\[-2pt]
               &                        & $[0.87,1.14]         $ & $[0.73,1.01]         $ & $[0.83,1.21]         $ \\[5pt]
$\lambda_{Vu}$ & $1.00^{+0.13}_{-0.12}$ & $    ^{+0.20}_{-0.12}$ & $0.88^{+0.18}_{-0.14}$ & $1.16^{+0.23}_{-0.19}$ \\[5pt]
$\kappa_{uu}$  & $1.07^{+0.22}_{-0.18}$ & $    ^{+0.20}_{-0.27}$ & $1.33^{+0.35}_{-0.34}$ & $0.82^{+0.24}_{-0.21}$ \\[2pt]

\hline
$|\lambda_{lq}|$ & $1.06^{+0.15}_{-0.14}$ & $    ^{+0.16}_{-0.14}$ & $1.10^{+0.20}_{-0.18}$ & $1.05^{+0.24}_{-0.22}$ \\[5pt]
$\lambda_{Vq}$   & $1.09^{+0.14}_{-0.13}$ & $    ^{+0.13}_{-0.12}$ & $1.01^{+0.17}_{-0.15}$ & $1.18^{+0.22}_{-0.19}$ \\[5pt]
$\kappa_{qq}$    & $0.93^{+0.17}_{-0.15}$ & $    ^{+0.18}_{-0.16}$ & $1.07^{+0.24}_{-0.21}$ & $0.80^{+0.22}_{-0.18}$ \\[2pt]

\hline\hline
\end{tabular}
\end{table}

\subsection{Parameterisations related to the fermion sector}
\label{sec:ModelL2}

Common coupling modifications for up-type fermions versus down-type fermions or for leptons versus quarks are predicted by many extensions of the SM. One such class of theoretically well motivated models is the 2HDM~\cite{Lee:1973iz}. 

The ratios of the coupling modifiers are tested in the most generic parameterisation proposed in Ref.~\cite{Heinemeyer:2013tqa}, in which the total Higgs boson width is also allowed to vary.
The main parameters of interest for these tests are $\lambda_{du}=\Kd/\Ku$ for the up- and down-type fermion symmetry, and $\lambda_{lq}=\Kl/\Kq$ for the lepton and quark symmetry, where both are allowed to be positive or negative. 
In this parameterisation, the loops are resolved in terms of their expected SM contributions.

\subsubsection{Probing the up- and down-type fermion symmetry}
\label{sec:lambdadu}

The free parameters for this test are: $\lambda_{du}=\Kd/\Ku$, $\lambda_{Vu}=\KV/\Ku$, and $\kappa_{uu} = \Ku\cdot\Ku/\KH$, where this latter term is positive definite since $\KH$~is always assumed to be positive. 
The up-type fermion couplings are mainly probed by the \aggF\ production process, the $\Hyy$ decay channel, and to a certain extent the \attH\ production process. 
The down-type fermion couplings are mainly probed by the $\Hbb$ and $\Htt$ decays. A small sensitivity to the relative sign arises from the interference between top and bottom quarks in the gluon fusion loop.

The results of the fit are reported in~Table~\ref{tab:ud_lq_results} and Fig.~\ref{fig:lambda_du}. 
The $p$-value of the compatibility between the data and the SM predictions is~72\%.
The likelihood scan for the $\lambda_{du}$~parameter is shown in Fig.~\ref{fig:likelihood_lambda_du} for the combination of ATLAS and CMS. Negative values for the parameter~$\lambda_{Vu}$ are excluded by more than~$4\sigma$. 

\begin{figure}[h!]
\begin{center}
\vspace{-3mm}
\includegraphics[width=0.6\textwidth]{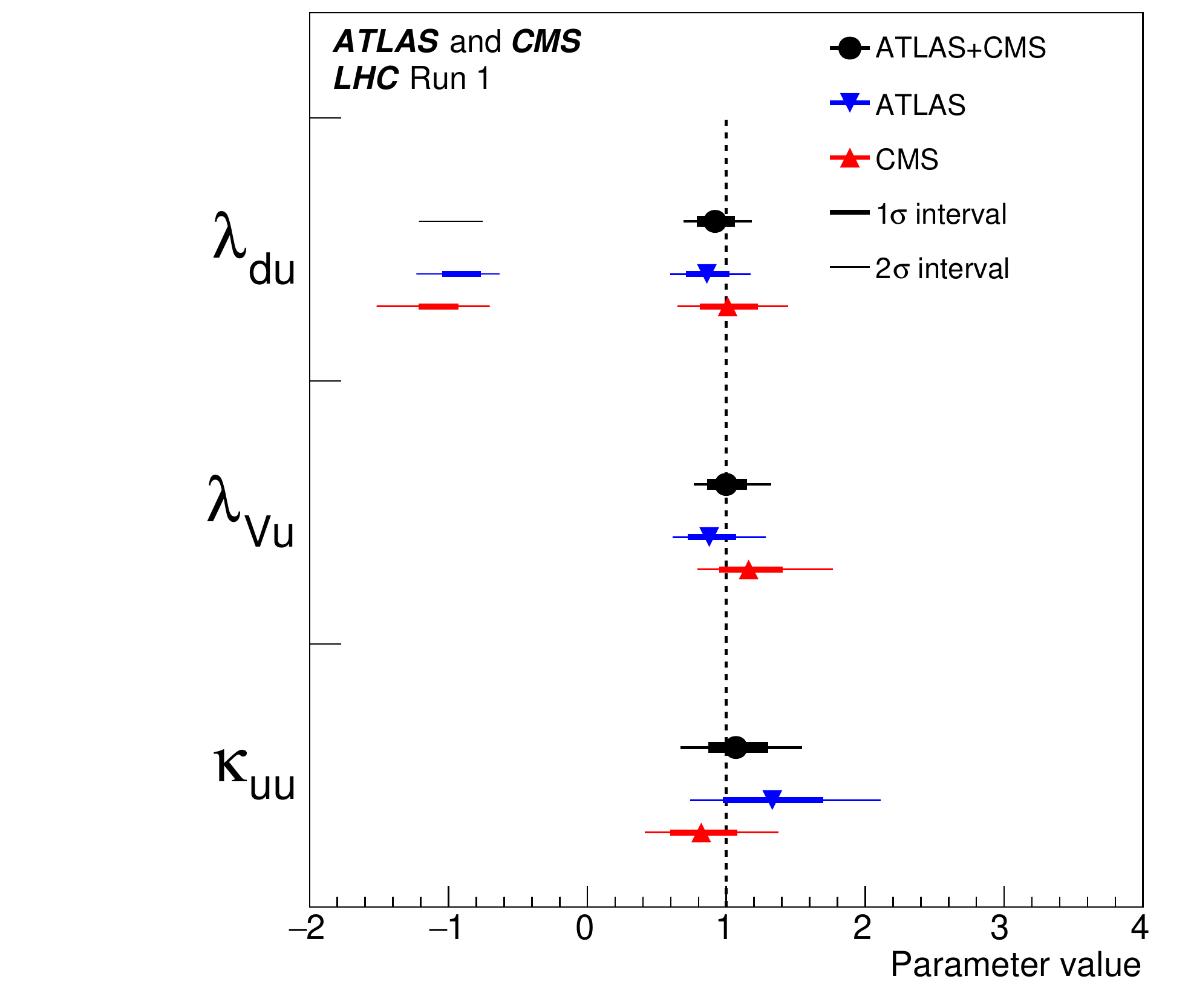} 
\end{center}
\vspace{-5mm}
\caption{Best fit values of parameters for the combination of ATLAS and CMS data, and separately for each experiment, for the parameterisation testing the up- and down-type fermion coupling ratios. The error bars indicate the $1\sigma$~(thick lines) and $2\sigma$~(thin lines) intervals. The parameter~$\kappa_{uu}$ is positive definite since $\KH$~is always assumed to be positive. Negative values for the parameter~$\lambda_{Vu}$ are excluded by more than~$4\sigma$. 
}
\label{fig:lambda_du}
\end{figure}

\begin{figure}[h!]
\begin{center}
\vspace{-3mm}
\includegraphics[width=0.6\textwidth]{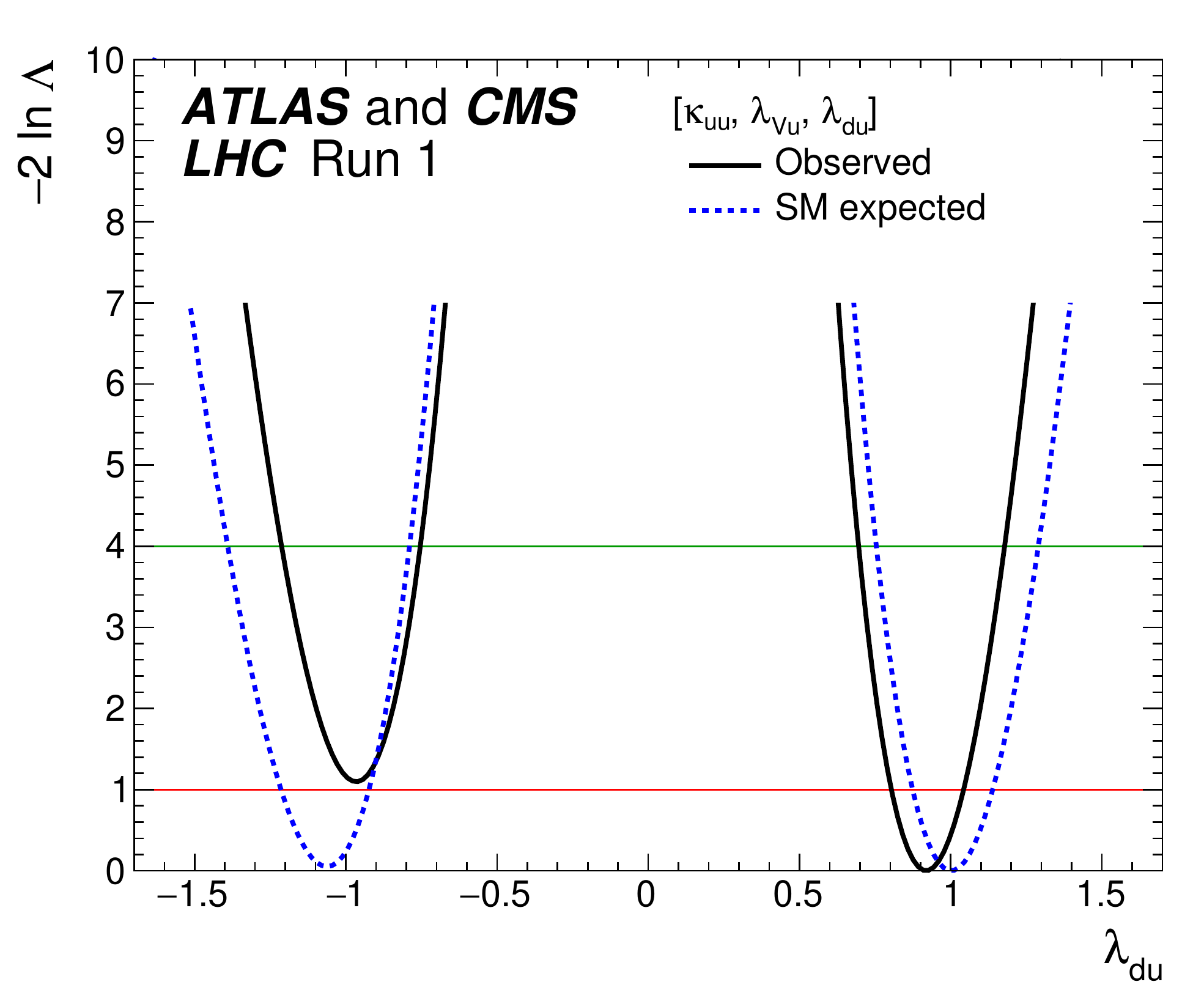} 
\end{center}
\vspace{2mm}
\caption{Observed (solid line) and expected (dashed line) negative log-likelihood scan of the $\lambda_{du}$ parameter, probing the ratios of coupling modifiers for up-type versus down-type fermions for the combination of ATLAS and CMS. The other parameters of interest from the list in the legend are also varied in the minimisation procedure. The red (green) horizontal line at the $-2\Delta\ln\Lambda$ value of 1 (4) indicates the value of the profile likelihood ratio corresponding to a $1\sigma$ ($2\sigma$) CL~interval for the parameter of interest, assuming the asymptotic $\chi^2$~distribution of the test statistic.
}
\label{fig:likelihood_lambda_du}
\end{figure}

\subsubsection{Probing the lepton and quark symmetry}

The parameterisation for this test is very similar to that of Section~\ref{sec:lambdadu}, which probes the up- and down-type fermion symmetry.
In this case, the free parameters are $\lambda_{lq}=\Kl/\Kq$, $\lambda_{Vq}=\KV/\Kq$, and $\kappa_{qq} = \Kq\cdot\Kq/\KH$, where the latter term is positive definite, like $\kappa_{uu}$. 
The quark couplings are mainly probed by the \aggF\ process, the $\Hyy$ and $\Hbb$ decays, and to a lesser extent by the \attH\ process.
The lepton couplings are probed by the $\Htt$ decays. The results are expected, however, to be insensitive to the relative sign of the couplings, because there is no sizeable lepton--quark interference in any of the relevant Higgs boson production processes and decay modes. Only the absolute value of the $\lambda_{lq}$~parameter is therefore considered in the fit.

The results of the fit are reported in~Table~\ref{tab:ud_lq_results} and Fig.~\ref{fig:lambda_lq}. 
The $p$-value of the compatibility between the data and the SM predictions is~79\%.
The likelihood scan for the $\lambda_{lq}$~parameter is shown in Fig.~\ref{fig:likelihood_lambda_lq} for the combination of ATLAS and CMS. Negative values for the parameter~$\lambda_{Vq}$ are excluded by more than~$4\sigma$. 

\begin{figure}
\begin{center}
\includegraphics[width=0.6\textwidth]{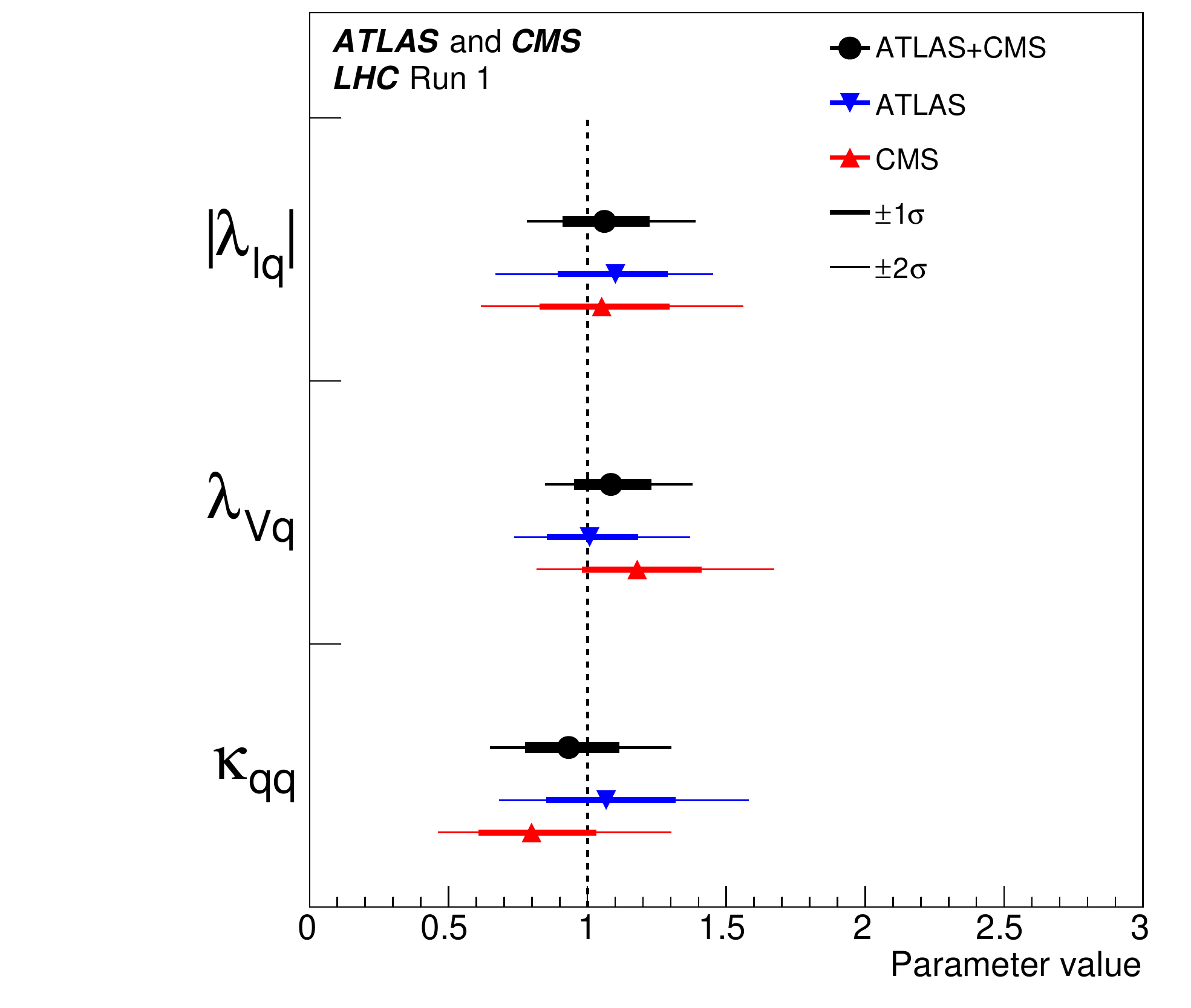} 
\end{center}
\caption{Best fit values of parameters for the combination of ATLAS and CMS data, and separately for each experiment, for the parameterisation testing the lepton and quark coupling ratios. The error bars indicate the $1\sigma$~(thick lines) and $2\sigma$~(thin lines) intervals. For the parameter $\lambda_{lq}$, for which there is no sensitivity to the sign, only the absolute values are shown. The parameter~$\kappa_{qq}$ is positive definite since $\KH$~is always assumed to be positive. Negative values for the parameter~$\lambda_{Vq}$ are excluded by more than~$4\sigma$. 
}
\label{fig:lambda_lq}
\end{figure}

\begin{figure}
\begin{center}
\includegraphics[width=0.6\textwidth]{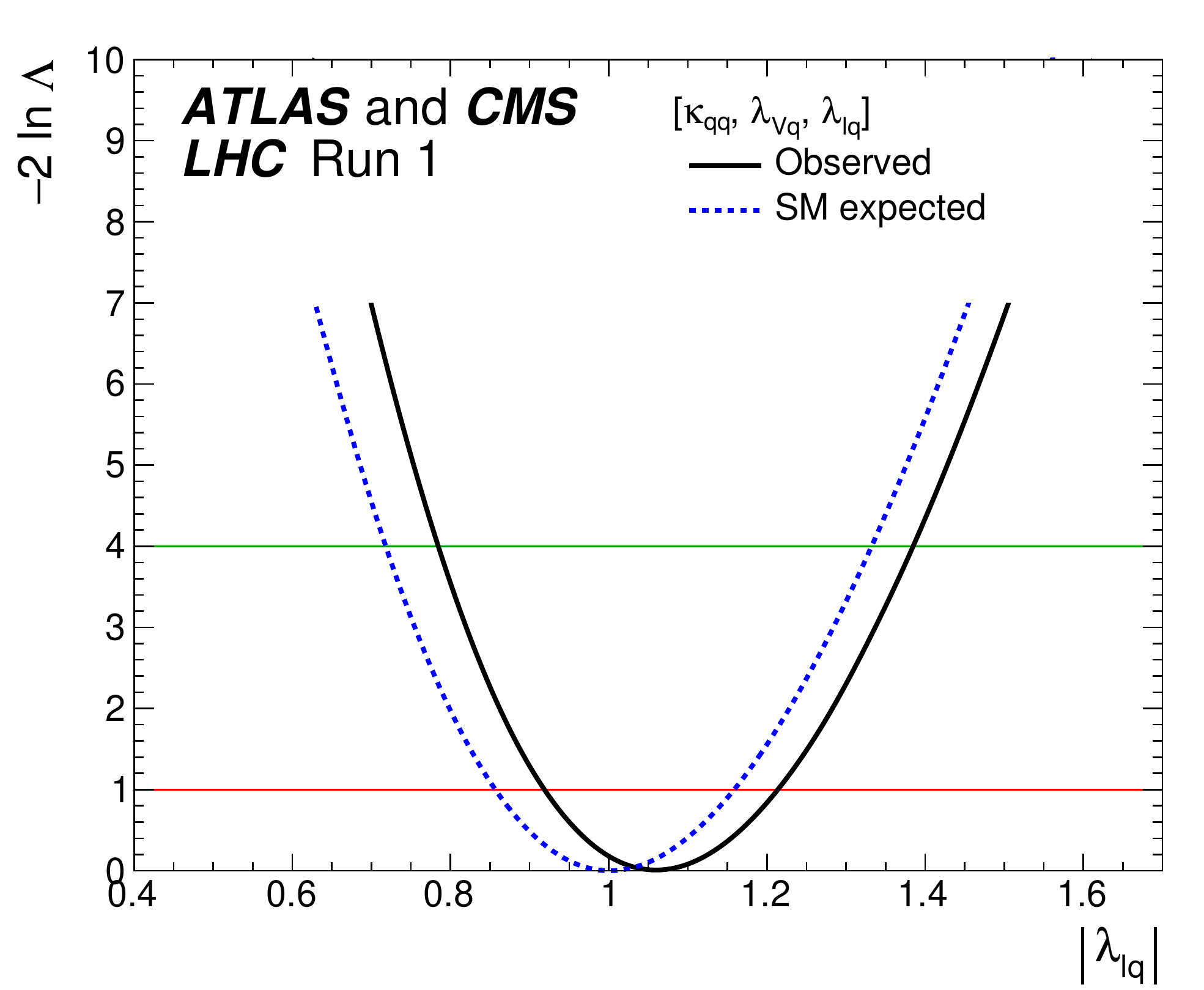} 
\end{center}
\caption{Observed (solid line) and expected (dashed line) negative log-likelihood scan of the $\lambda_{lq}$ parameter, probing the ratios of coupling modifiers for leptons versus quarks for the combination of ATLAS and CMS. The other parameters of interest from the list in the legend are also varied in the minimisation procedure. The red (green) horizontal line at the $-2\Delta\ln\Lambda$ value of 1 (4) indicates the value of the profile likelihood ratio corresponding to a $1\sigma$ ($2\sigma$) CL~interval for the parameter of interest, assuming the asymptotic $\chi^2$~distribution of the test statistic.
}
\label{fig:likelihood_lambda_lq}
\end{figure}

\subsection{Fermion and vector boson couplings}
\label{sec:ModelK3}

The last and most constrained parameterisation studied in this section is motivated by the intrinsic difference between the Higgs boson couplings to weak vector bosons, which originate from the breaking of the EW symmetry, and the Yukawa couplings to the fermions. 
Similarly to~Section~\ref{sec:ModelK1}, it is assumed in this section that there are no new particles in the loops (\aggF~production process and $\Hyy$~decay mode) and that there are no BSM decays, i.e.~$\BRbsm=0$. 
Vector and fermion coupling modifiers, $\kappa_V$~and~$\kappa_F$, are defined such that $\kappa_Z = \kappa_W = \kappa_V$ and~$\kappa_t = \kappa_\tau = \kappa_b = \kappa_F$. 
These definitions can be applied either globally, yielding two parameters, or separately for each of the five decay channels, yielding ten parameters $\kappa_V^f$~and~$\kappa_F^f$ (following the notation related to Higgs boson decays used for the signal strength parameterisation). 
Two fits are performed: a two-parameter fit as a function of~$\kappa_V$~and~$\kappa_F$, and a ten-parameter fit as a function of~$\kappa_V^f$~and~$\kappa_F^f $ for each decay channel.  

As explained in~Section~\ref{sec:kappas} and shown explicitly in~Table~\ref{tab:kexpr}, the Higgs boson production cross sections and partial decay widths are only sensitive to products of coupling modifiers and not to their absolute sign. 
Any sensitivity to the relative sign between~$\kappa_V$ and~$\kappa_F$ can only occur through interference terms, either in the $\Hyy$ decays, through the $t$--$W$ interference in the $\gamma\gamma$ decay loop, or in~\aggZH\ or \atH~production. Without any loss of generality, this parameterisation assumes that one of the two coupling modifiers, namely~$\kappa_V$ (or $\kappa_V^f$), is positive.  

\begin{figure}[htb]
\begin{center}
\includegraphics[height=0.7\textheight,width=\textwidth]{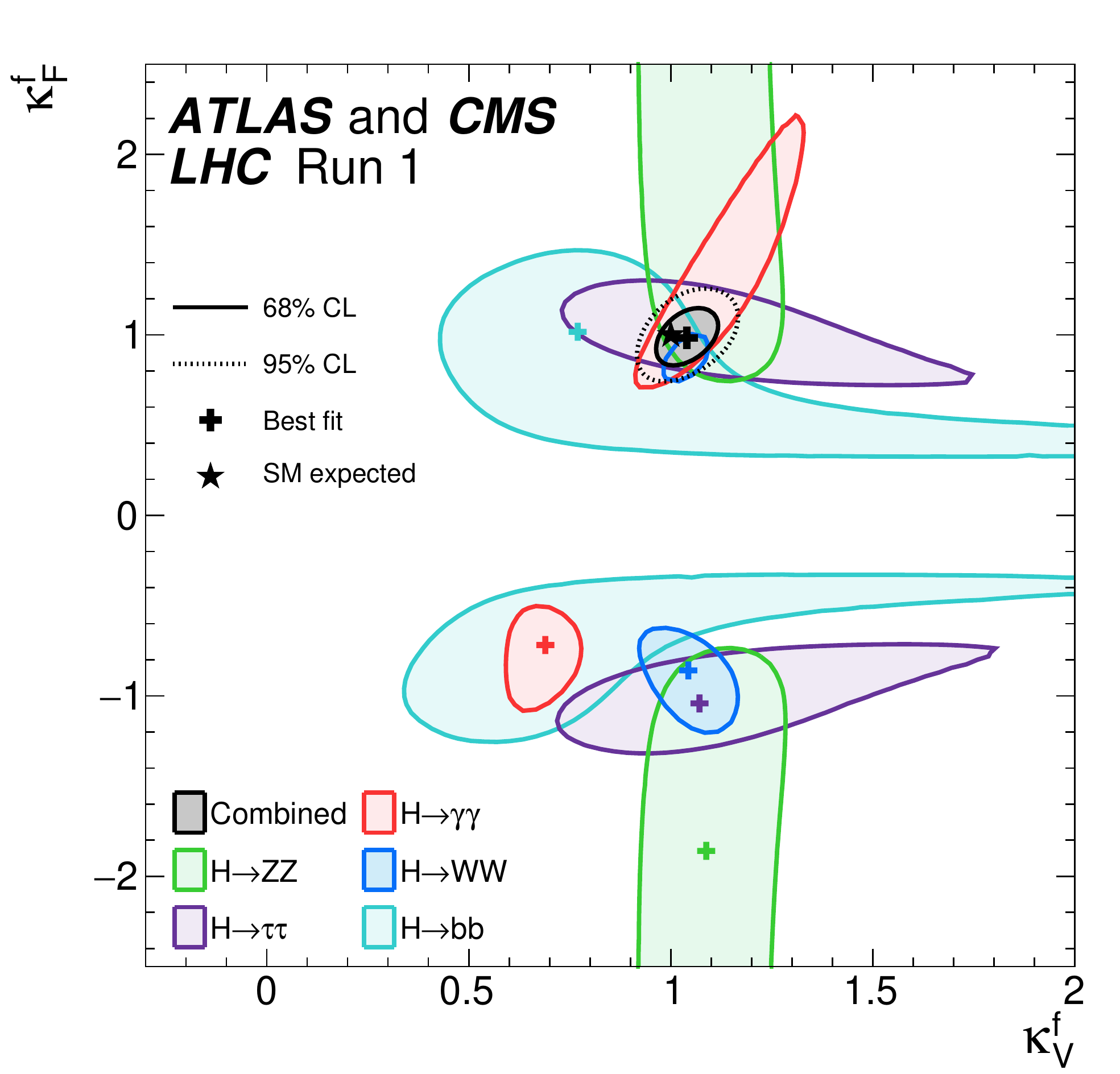}
\end{center}
\caption{Negative log-likelihood contours at~68\% and~95\%~CL in the ($\kappa_F^f$,~$\kappa_V^f$) plane for the combination of ATLAS and CMS and for the individual decay channels, as well as for their combination ($\kappa_F$ versus $\kappa_V$ shown in black), without any assumption about the sign of the coupling modifiers. 
The other two quadrants (not shown) are symmetric with respect to the point (0,0).}
\label{fig:kVkFall}
\end{figure}

The combined ATLAS and CMS results are shown in~Fig.~\ref{fig:kVkFall} for the individual channels and their combination. The individual decay channels are seen to be compatible with each other only for positive values of~$\kappa_F^f$. The incompatibility between the channels for negative values of~$\kappa_F^f$ arises mostly from the $\Hyy$,~$\HWW$, and~$\HZZ$ channels. Nonetheless, the best fit values for most of the individual channels correspond to negative values of~$\kappa_F^f$. However, the best fit value from the global fit yields~$\kappa_F \ge 0$, a result that is driven by the large asymmetry between the positive and negative coupling ratios in the case of $\Hyy$~decays. 

\begin{figure}[ht!]
\centering
\begin{tabular}{cc}
\includegraphics[width=0.46\textwidth,trim=20pt 20pt 20pt 0pt]{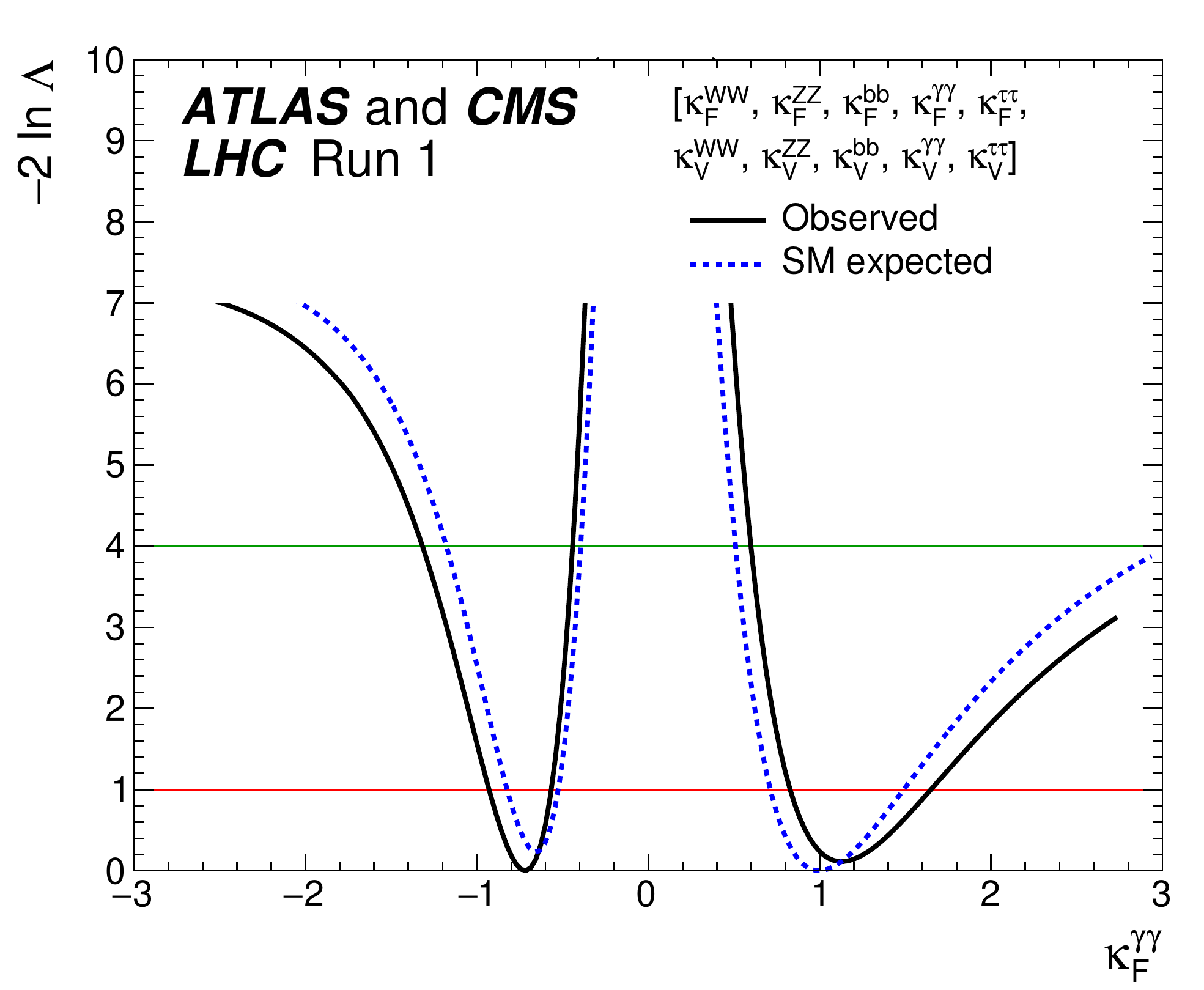}&
\includegraphics[width=0.46\textwidth,trim=20pt 20pt 20pt 0pt]{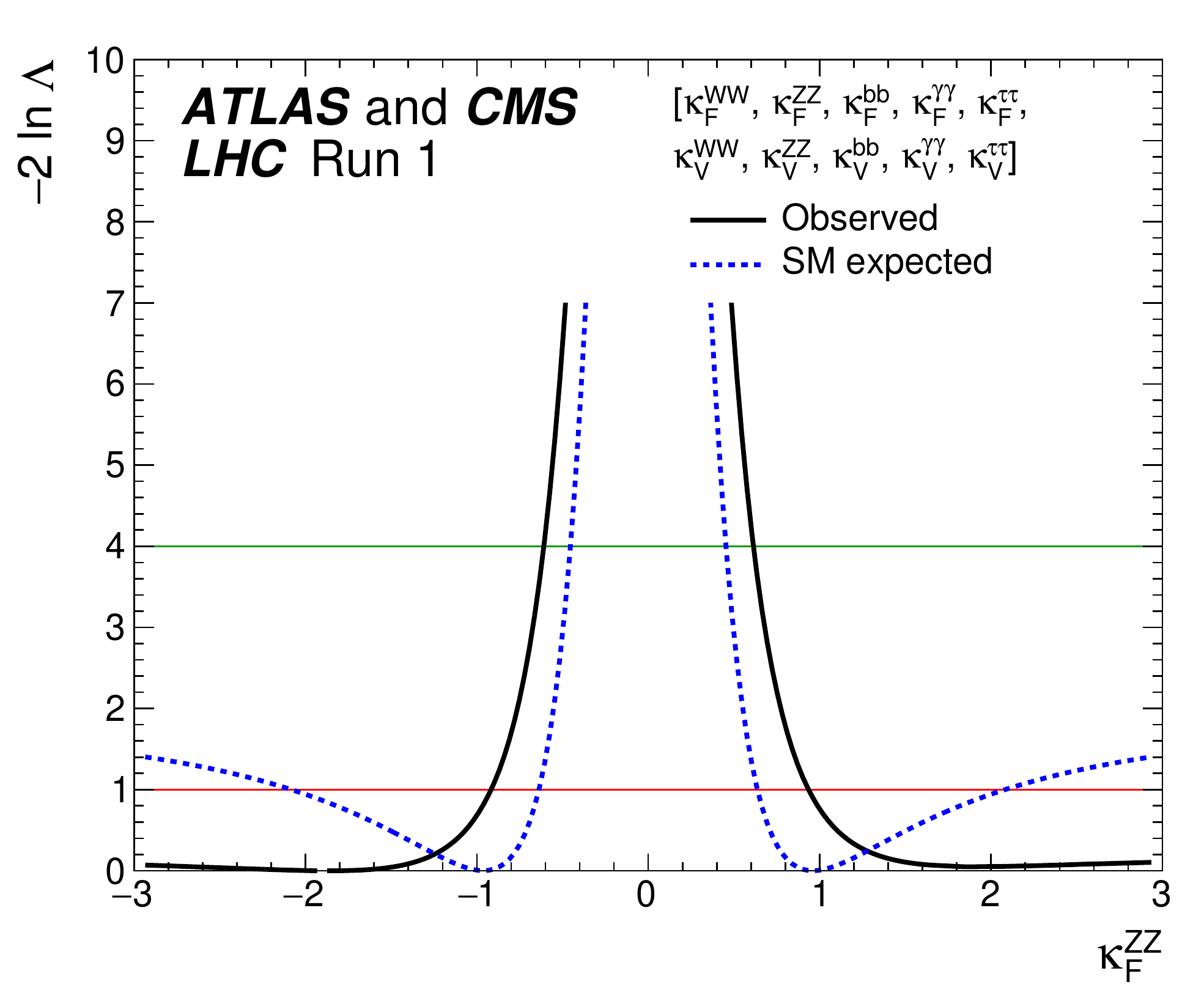}\\
(a)&(b)\\
\includegraphics[width=0.46\textwidth,trim=20pt 20pt 20pt 0pt]{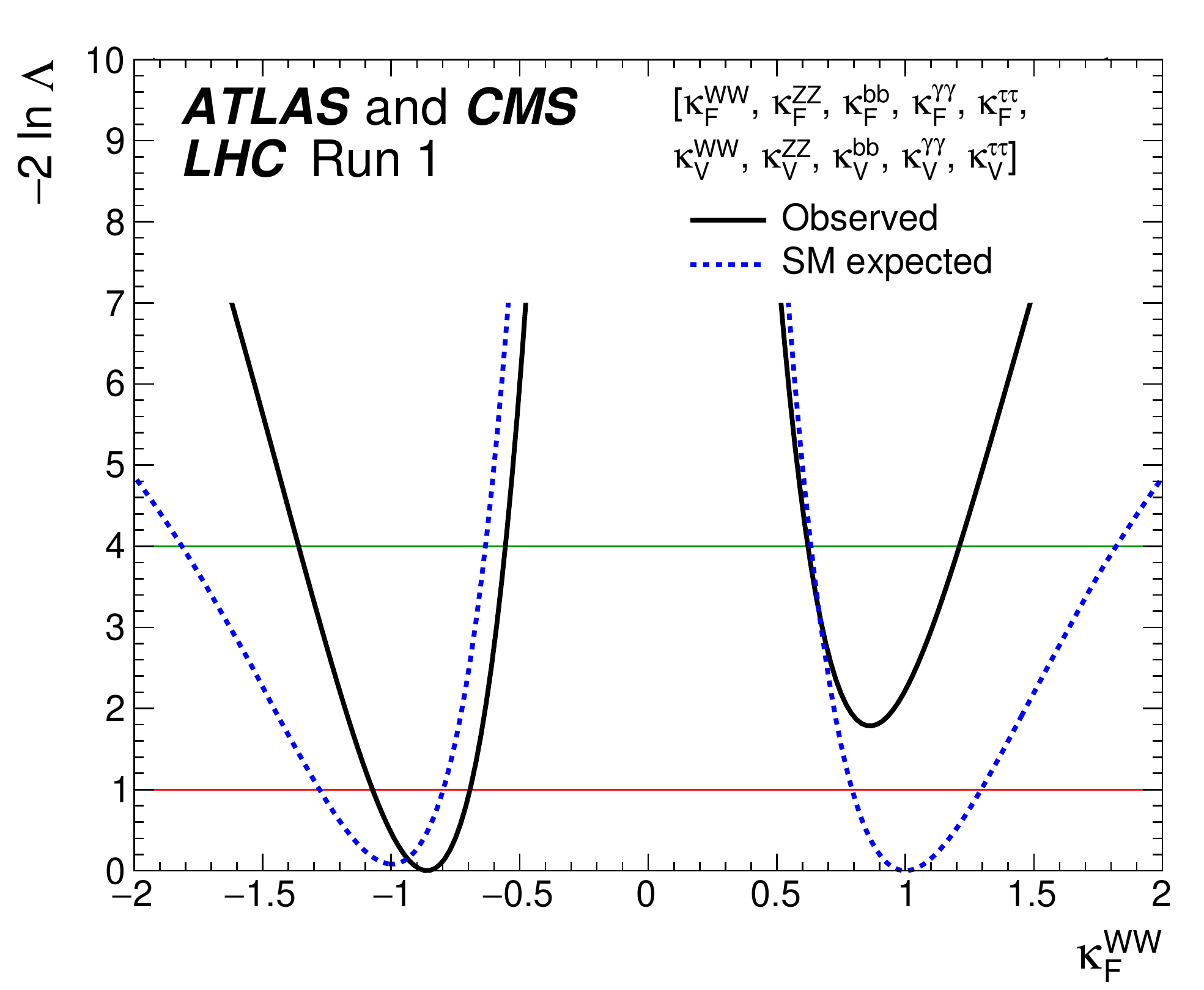}&
\includegraphics[width=0.46\textwidth,trim=20pt 20pt 20pt 0pt]{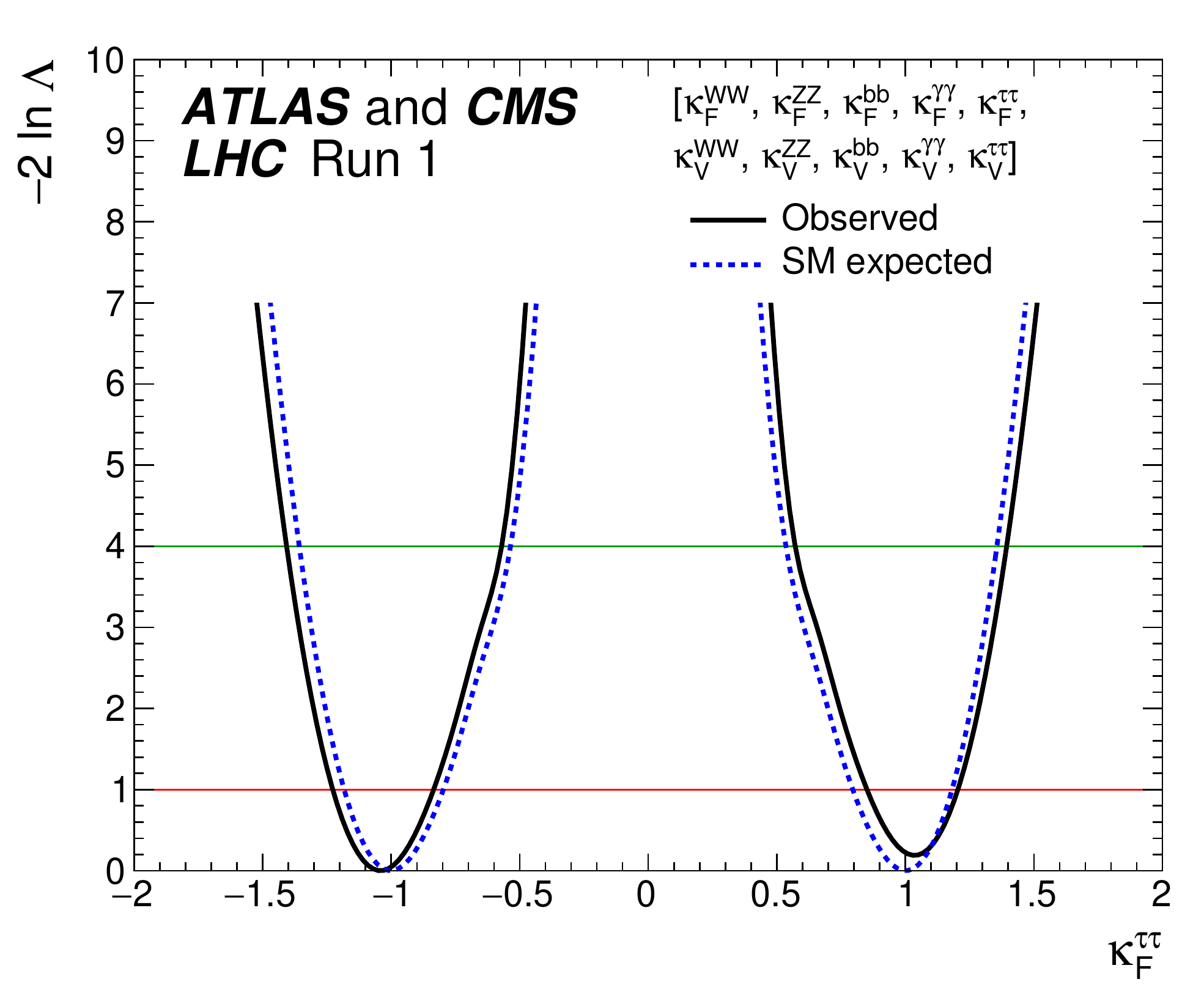}\\
(c)&(d)\\
\includegraphics[width=0.46\textwidth,trim=20pt 20pt 20pt 0pt]{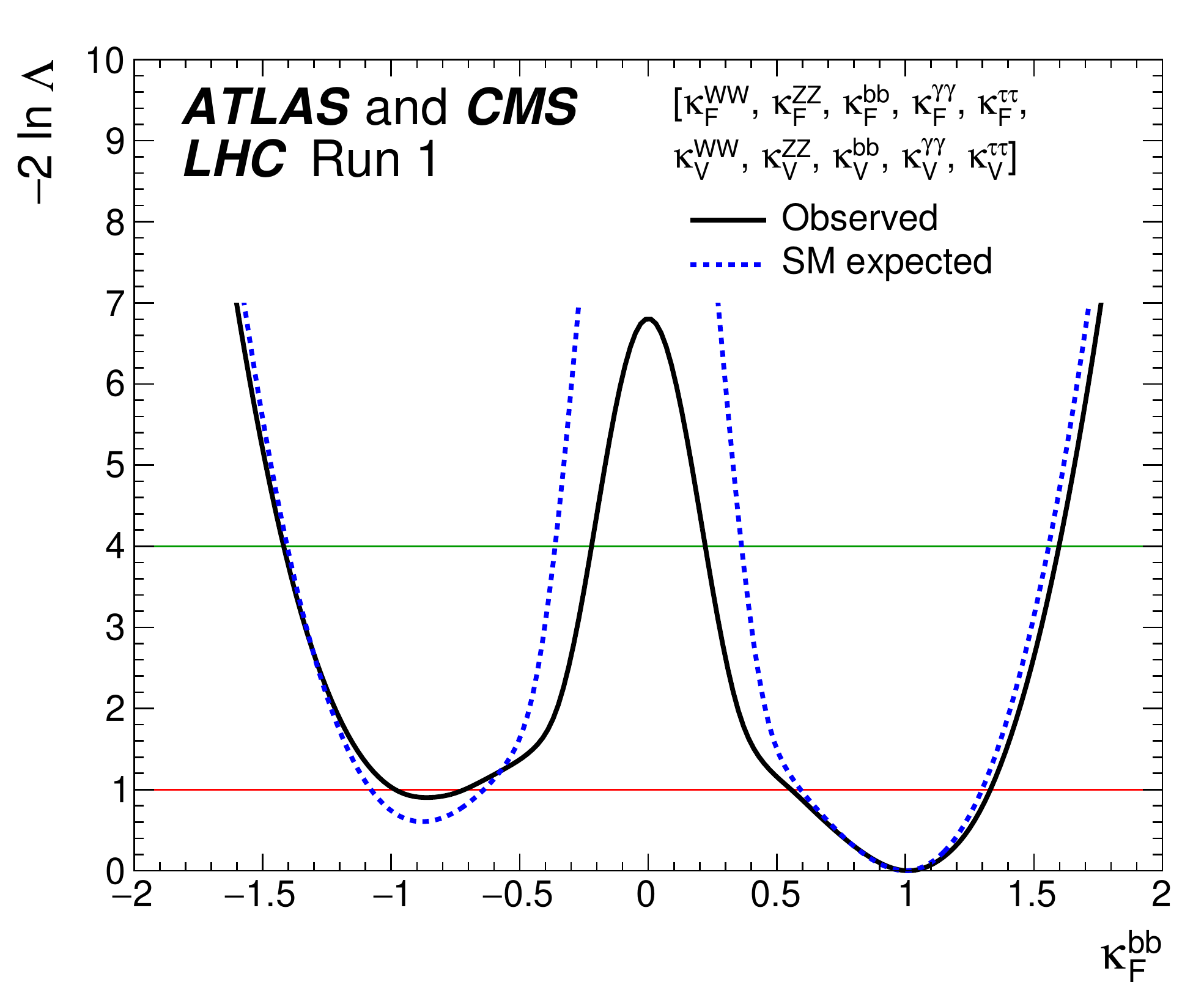}&
\includegraphics[width=0.46\textwidth,trim=20pt 20pt 20pt 0pt]{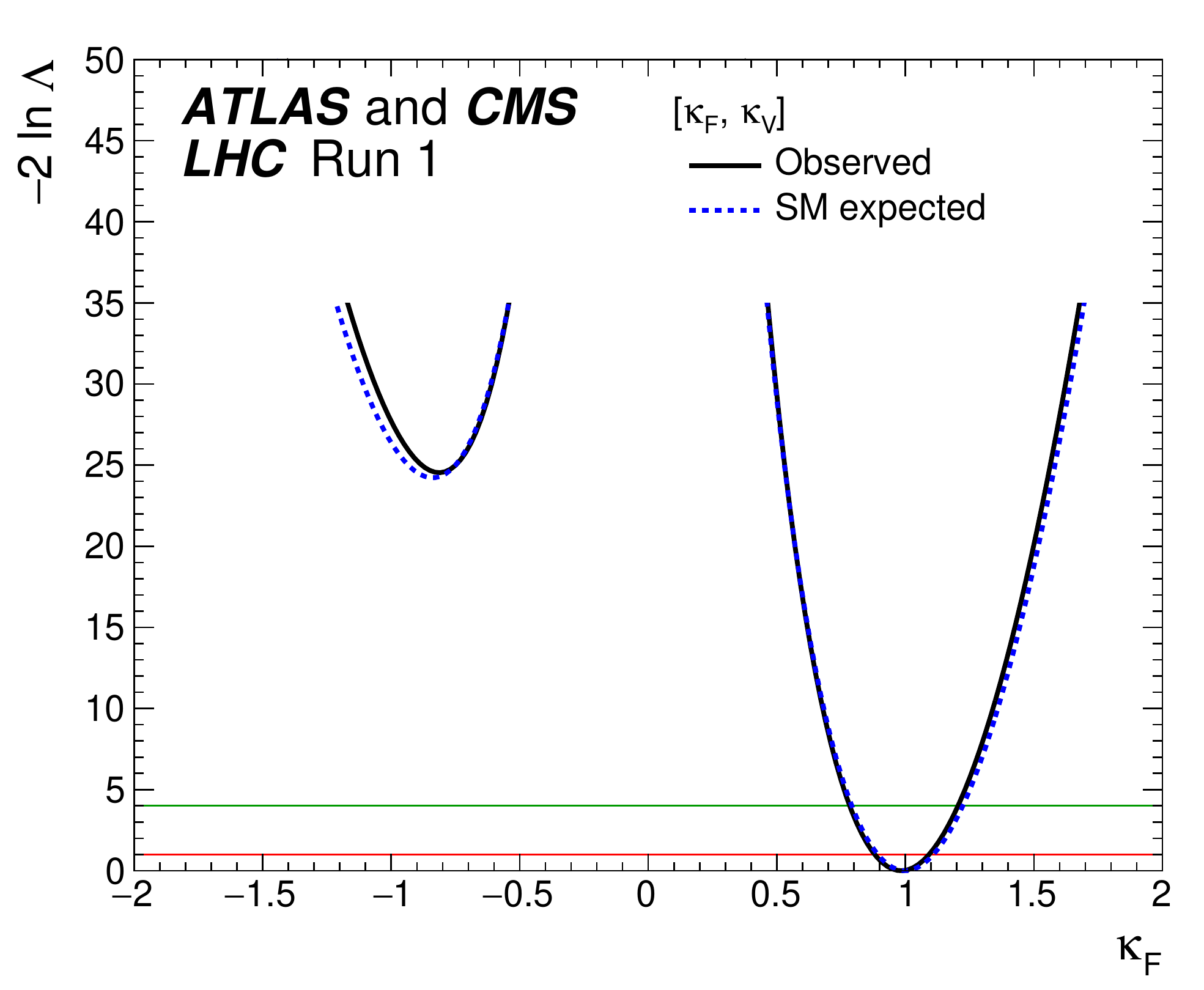}\\
(e)&(f)\\
\end{tabular}
\caption{Observed (solid line) and expected (dashed line) negative log-likelihood scans for the five $\kappa_F^f$ parameters, corresponding to each individual decay channel, and for the global $\kappa_F$ parameter, corresponding to the combination of all decay channels: 
(a) $\kappa_{F}^{\gamma\gamma}$,
(b) $\kappa_{F}^{ZZ}$,
(c) $\kappa_{F}^{WW}$,
(d) $\kappa_{F}^{\tau\tau}$,
(e) $\kappa_{F}^{bb}$, and
(f) $\kappa_F$.
All the other parameters of interest from the list in the legends are also varied in the minimisation procedure.
The red (green) horizontal lines at the $-2\Delta\ln\Lambda$ value of 1 (4) indicate the value of the profile likelihood ratio corresponding to a $1\sigma$ ($2\sigma$) CL~interval for the parameter of interest, assuming the asymptotic $\chi^2$~distribution of the test statistic.
}
\label{fig:scan_K3}
\end{figure}

\begin{figure}[htb]
\begin{center}
\vspace{-3mm}
\includegraphics[width=0.7\textwidth]{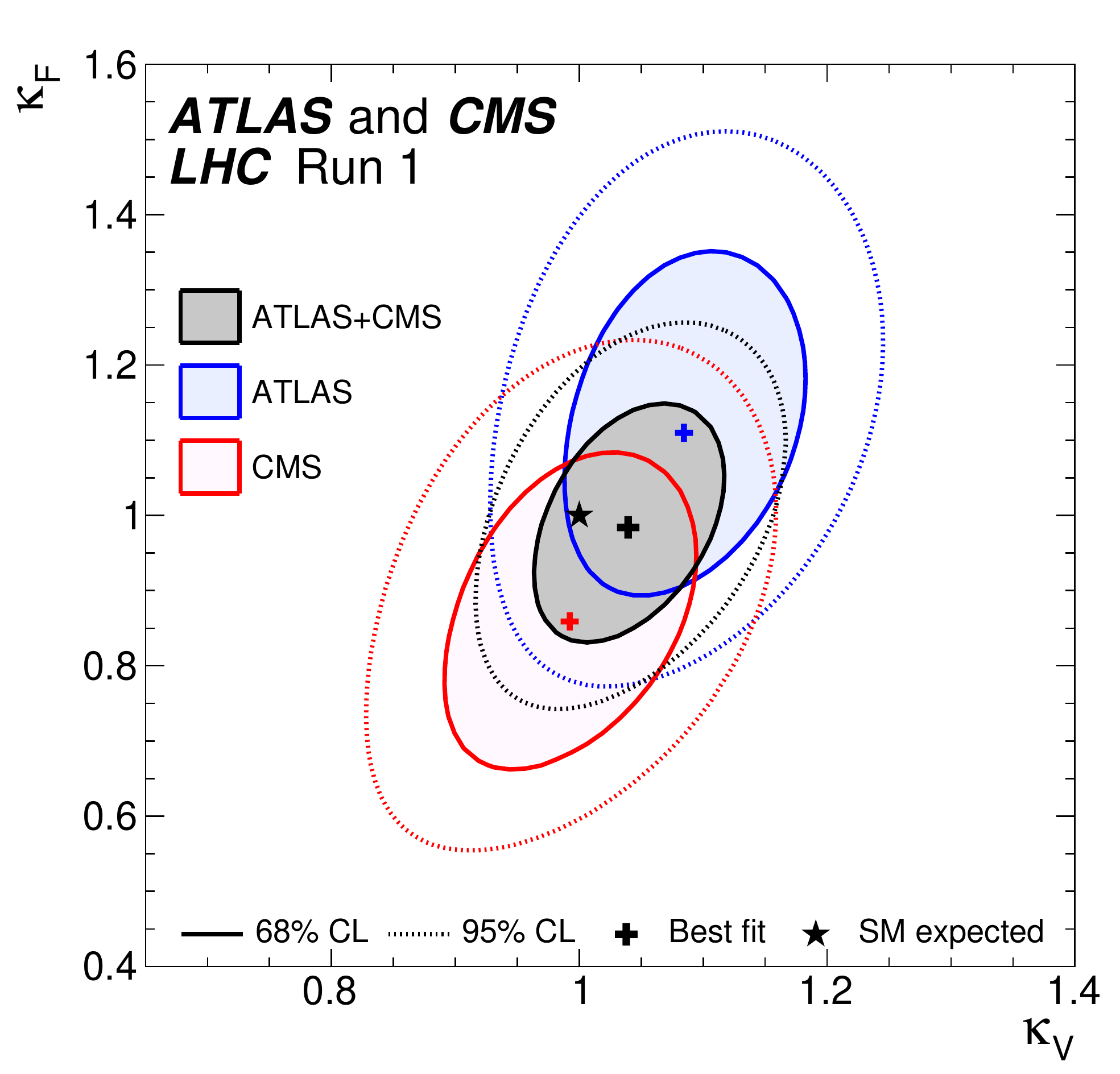}
\includegraphics[width=0.7\textwidth]{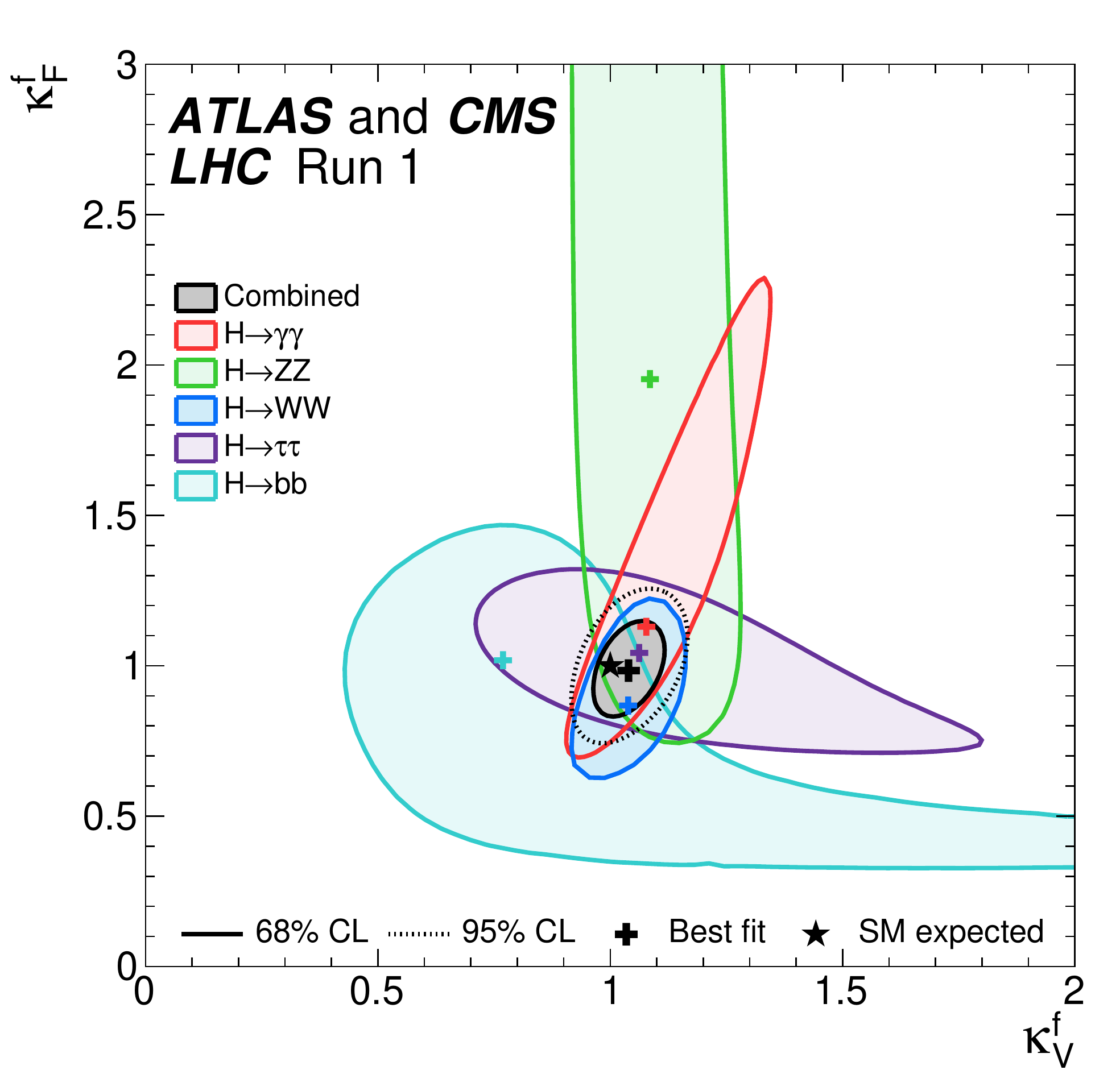}\\
\end{center}
\vspace{-7mm}
\caption{Top: negative log-likelihood contours at~68\% and~95\%~CL in the ($\kappa_F$,~$\kappa_V$)~plane on an enlarged scale for the combination of ATLAS and CMS and for the global fit of all channels. Also shown are the contours obtained for each experiment separately. Bottom: negative log-likelihood contours at~68\%~CL in the ($\kappa_F^f$,~$\kappa_V^f$) plane for the combination of ATLAS and CMS and for the individual decay channels as well as for their global combination ($\kappa_F$ versus $\kappa_V$), assuming that all coupling modifiers are positive. }
\label{fig:kVkFpos}
\end{figure}

The fact that, for four of the five individual channels, the best fit values correspond to~$\kappa_F^f \le 0$ is not significant, as shown by the likelihood curves in~Figs.~\ref{fig:scan_K3}~(a-e). The $\Hbb$~decay channel displays the largest expected sensitivity, mostly arising from the contribution of the \aggZH\ process, and the best fit value of $\kappa_{F}^{bb}$ is positive. 
For all other decay modes, a small sensitivity arises because of the \atH~process. The excess observed in the combination of the two experiments for the \attH~production process induces a preference for a relative negative sign between the two coupling modifiers, which increases significantly the \atH~cross section and thereby provides a better fit to the data. The only visible difference between the two minima at positive and negative values of~$\kappa_F^f$ is observed for the $\Hww$~channel. 

As stated above, the channel most affected by the relative sign of the couplings is the $\Hyy$~decay channel: because of the negative $t$--$W$~interference in the $\gamma\gamma$~loop, the $\Hyy$~partial width would be much larger if the sign of~$\kappa_F^{\gamma\gamma}$ were opposite to that of~$\kappa_V^{\gamma\gamma}$. When combining the $\Hyy$~decay channel with all the other channels, the opposite sign case is excluded by almost~$5\sigma$, as can be inferred from Fig.~\ref{fig:scan_K3}~(f).

Figure~\ref{fig:kVkFpos}~(top) presents, on an enlarged scale, the results of the scan for the global coupling modifiers as well as those obtained separately for each experiment. For completeness, additional likelihood scans are performed for the two global coupling modifiers and for those of each decay channel, assuming in all cases that $\kappa_F$ and $\kappa_V$ are both positive. The results of these scans are shown in~Fig.~\ref{fig:kVkFpos}~(bottom). The most precise determination of~$\kappa_F^f$ and~$\kappa_V^f$ is obtained from the $\HWW$~decay channel because it is the only one that provides significant constraints on both parameters, through the measurements of the~\aggF\ and \aVBF~production processes. The difference in size between the $\Hww$~confidence regions obtained for~$\kappa_F^f \ge 0$ in~Fig.~\ref{fig:kVkFall} and~Fig.~\ref{fig:kVkFpos}~(bottom), where it is explicitly assumed that $\kappa_F^f \ge 0$, is due to the fact that the negative log-likelihood contours are evaluated using as a reference the minima obtained from different likelihood fits. The combination of all decay modes provides significant additional constraints. All results are in agreement with the SM~prediction, $\kappa_F^f = 1$ and~$\kappa_V^f = 1$, and the~$p$-value of the compatibility between the data and the SM~predictions is~59\%.


\section{Summary}
\label{sec:Conclusion}

An extensive set of combined ATLAS and CMS measurements of the Higgs boson production and decay rates is presented, and a number of constraints on its couplings to vector bosons and fermions are derived based on various sets of assumptions. The combination is based on the analysis of approximately 600~categories of selected events, concerning five production processes, \aggF, \aVBF, \aWH, \aZH, and \attH, where \aggF\ and \aVBF\ refer, respectively, to production through the gluon fusion and vector boson fusion processes; and six decay channels, $H \to ZZ, WW, \gamma\gamma, \tau\tau, bb$, and $\mu\mu$. All results are reported assuming a value of~$125.09$~GeV for the Higgs boson mass, the result of the combined Higgs boson mass measurement by the two experiments~\cite{ATLASCMSHmass}. The analysis uses the LHC proton-proton collision data sets recorded by the ATLAS and CMS detectors in 2011 and 2012, corresponding to integrated luminosities per experiment of approximately~5~\ifb\ at $\sqrt{s}=7$ TeV and 20~\ifb\ at $\sqrt{s} = 8$~TeV. This paper presents the final Higgs boson coupling combined results from ATLAS and CMS based on the LHC \runone\ data.  

The combined analysis is sensitive to the couplings of the Higgs boson to the weak vector bosons and to the heavier fermions (top quarks,~$b$ quarks,~$\tau$ leptons, and~-- marginally~-- muons). The analysis is also sensitive to the effective couplings of the Higgs boson to the photon and the gluon. At the LHC, only products of cross sections and branching fractions are measured, so the width of the Higgs boson cannot be probed without assumptions beyond the main one used for all measurements presented here, namely that the Higgs boson production and decay kinematics are close to those predicted by the Standard Model~(SM). 
In general, the combined analysis presented in this paper provides a significant improvement with respect to the individual combinations published by each experiment separately. The precision of the results improves in most cases by a factor of approximately~$1/\sqrt 2$, as one would expect for the combination of two largely uncorrelated measurements based on similar-size data samples. A few illustrative results are summarised below.

For the first time, results are shown for the most generic parameterisation of the observed event yields in terms of products of Higgs boson production cross sections times branching fractions, separately for each of 20~measurable $(\sigma_i$,~$\BR^f$)~pairs of production processes and decay modes. These measurements do not rely on theoretical predictions for the inclusive cross sections and the uncertainties are mostly dominated by their statistical component. In the context of this parameterisation, one can test whether the observed yields arise from more than one Higgs boson, all with experimentally indistinguishable masses, but possibly with different coupling structures to the SM~particles. The data are compatible with the hypothesis of a single Higgs boson, yielding a $p$-value of~29\%.

Fits to the observed event yields are also performed without any assumption about the Higgs boson width in the context of two other generic parameterisations. The first parameterisation is in terms of ratios of production cross sections and branching fractions, together with the reference cross section of the process~$gg \to H \to ZZ$. All results are compatible with the~SM. The best relative precision, of about~30\%, is achieved for the ratio of cross sections $\sigma_{\aVBF}/\sigma_{\aggF}$ and for the ratios of branching fractions $\BR^{WW}/\BR^{ZZ}$ and~$\BR^{\gamma\gamma}/\BR^{ZZ}$. A relative precision of around~40\% is achieved for the ratio of branching fractions $\BR^{\tau\tau}/\BR^{ZZ}$. The second parameterisation is in terms of ratios of coupling modifiers, together with one parameter expressing the $gg \to H \to ZZ$ reference process in terms of these modifiers. The ratios of coupling modifiers are measured with precisions of approximately~$10$--$20\%$, where the improvement in precision in this second parameterisation arises because the signal yields are expressed as squares or products of these coupling modifiers.

All measurements based on the generic parameterisations are compatible between the two experiments and with the predictions of the~SM. The potential presence of physics beyond the~SM (BSM) is also probed using specific parameterisations. With minimal additional assumptions, the overall branching fraction of the Higgs boson into BSM~decays is determined to be less than~34\% at~95\%~CL. This constraint applies to invisible decays into BSM particles, decays into BSM particles that are not detected as such, and modifications of the decays into SM~particles that are not directly measured by the experiments. 

The combined signal yield relative to the SM~expectation is measured to be $1.09 \pm 0.07$~(stat)~$\pm 0.08$~(syst), where the systematic uncertainty is dominated by the theoretical uncertainty in the inclusive cross sections. The measured (expected) significance for the direct observation of the \aVBF\ production process is at the level of~$5.4\sigma~(4.6\sigma)$, while that for the $\Htt$~decay channel is at the level of~$5.5\sigma~(5.0\sigma)$.

\clearpage

\section*{Acknowledgments}
We thank CERN for the very successful operation of the LHC, as well as the support staff from our institutions without whom ATLAS and CMS
could not be operated efficiently.

We acknowledge the support of ANPCyT (Argentina); YerPhI (Armenia); ARC (Australia); BMWFW and FWF (Austria); ANAS (Azerbaijan); SSTC (Belarus); FNRS and FWO (Belgium); CNPq, CAPES, FAPERJ, and FAPESP (Brazil); MES (Bulgaria); NSERC, NRC, and CFI (Canada); CERN; CONICYT (Chile); CAS, MoST, and NSFC (China); COLCIENCIAS (Colombia); MSES and CSF (Croatia); RPF (Cyprus); MSMT CR, MPO CR, and VSC CR (Czech Republic); DNRF and DNSRC (Denmark); MoER, ERC IUT, and ERDF (Estonia); Academy of Finland, MEC, and HIP (Finland); CEA and CNRS/IN2P3 (France); GNSF (Georgia); BMBF , DFG, HGF, and MPG (Germany); GSRT (Greece); RGC (Hong Kong SAR, China); OTKA and NIH (Hungary); DAE and DST (India); IPM (Iran); SFI (Ireland); ISF, I-CORE, and Benoziyo Center (Israel); INFN (Italy); MEXT and JSPS (Japan); JINR; MSIP, and NRF (Republic of Korea); LAS (Lithuania); MOE and UM (Malaysia); BUAP, CINVESTAV, CONACYT, LNS, SEP, and UASLP-FAI (Mexico); CNRST (Morocco); FOM and NWO (Netherlands); MBIE (New Zealand); RCN (Norway); PAEC (Pakistan); MNiSW, MSHE, NCN, and NSC (Poland); FCT (Portugal); MNE/IFA (Romania); MES of Russia, MON, NRC KI, RosAtom, RAS, and RFBR  (Russian Federation); MESTD (Serbia); MSSR (Slovakia); ARRS and MIZ\v{S} (Slovenia); DST/NRF (South Africa); MINECO, SEIDI, and CPAN (Spain); SRC and Wallenberg Foundation (Sweden); ETH Board, ETH Zurich, PSI, SERI, SNSF, UniZH, and Cantons of Bern, Geneva and Zurich (Switzerland); MOST  (Taipei); ThEPCenter, IPST, STAR, and NSTDA (Thailand); TUBITAK and TAEK (Turkey); NASU and SFFR (Ukraine); STFC (United Kingdom); DOE and NSF (United States of America). 

In addition, individual groups and members have received support from BELSPO, FRIA, and IWT (Belgium); BCKDF, the Canada Council, CANARIE, CRC, Compute Canada, FQRNT, and the Ontario Innovation Trust (Canada); the Leventis Foundation (Cyprus); MEYS (Czech Republic); EPLANET, ERC, FP7, Horizon 2020, and Marie Sk{\l}odowska-Curie Actions (European Union); Investissements d'Avenir Labex and Idex, ANR, R{\'e}gion Auvergne and Fondation Partager le Savoir (France); AvH Foundation (Germany); the Herakleitos, Thales, and Aristeia programmes co-financed by EU-ESF and the Greek NSRF (Greece); CSIR (India); BSF, GIF, and Minerva (Israel); BRF (Norway); the HOMING PLUS programme of the FPS, co-financed from the EU Regional Development Fund, the Mobility Plus programme of the MSHE, and the OPUS programme of the NSC (Poland); the NPRP by Qatar NRF (Qatar); Generalitat de Catalunya, Generalitat Valenciana, and the Programa Clar\'in-COFUND del Principado de Asturias (Spain); the Rachadapisek Sompot Fund for Postdoctoral Fellowship, Chulalongkorn University, and the Chulalongkorn Academic into Its 2nd Century Project Advancement Project (Thailand); the Royal Society and Leverhulme Trust (United Kingdom); the A.\ P.\ Sloan Foundation and the Welch Foundation (United States of America).

The crucial computing support from all WLCG partners is acknowledged gratefully, in particular from CERN and the Tier-1 facilities at TRIUMF (Canada), NDGF (Denmark, Norway, Sweden), CC-IN2P3 (France), KIT/GridKA (Germany), INFN-CNAF (Italy), NL-T1 (Netherlands), RRC-KI and JINR (Russian Federation), PIC (Spain), ASGC (Taipei), RAL (UK), and BNL and FNAL (USA), and from the Tier-2 facilities worldwide.


   




\clearpage
\appendix
\part*{Appendix}
\addcontentsline{toc}{part}{Appendix}


\section{Correlation matrices}
\label{sec:app_correlations}

Figures~\ref{fig:correlation5x5},~\ref{fig:correlationB1ZZ} and~\ref{fig:correlationL1} show the correlation matrices obtained from the fits to the generic parameterisations described respectively in Sections~\ref{sec:sigBR5x5},~\ref{sec:sigBR9}, and~\ref{sec:kappaParam}. The correlation coefficients are evaluated around the best fit values, using the second derivatives of the negative log-likelihood ratio. 

In the case of the parameterisation using 23~products of cross sections times branching fractions, most of the parameters are uncorrelated, as shown in~Fig.~\ref{fig:correlation5x5}. Some significant anticorrelations are present however, because of cross-contamination between different channels. These can be seen in the \aggF\ versus \aVBF\ production processes for all decay modes, in the \aWH\ versus \aZH\ production processes for the $\Hyy$~decay mode, and in the~$\HWW$ versus $\Htt$~decay modes for the \attH\ production process.

In contrast, for the two parameterisations based on ratios shown in Figs.~\ref{fig:correlationB1ZZ} and~\ref{fig:correlationL1}, correlations are present for all pairs of parameters. For example, in each of these parameterisations, the first parameter is anticorrelated to most of the others, which are all expressed as ratios of cross sections, branching fractions, or coupling modifiers, because it is directly correlated to the denominators of these ratios. 

These correlation matrices are constructed as symmetric at the observed best fit values of the parameters of interest, and therefore are not fully representative of the asymmetric uncertainties observed in certain parameterisations, as shown for example in~Fig.~\ref{fig:rates_bb_likelihoodscan}. The derivation of the results for a specific parameterisation, with additional assumptions compared to a more generic one, from the fit results and the covariance matrix of this more generic parameterisation, is therefore not straightforward. This is one of the reasons for quoting the best fit results in Sections~\ref{sec:SignalStrength} and~\ref{sec:CouplingFits} for a wide range of parameterisations, beyond the more generic ones discussed in~Section~\ref{sec:GenericParameterisation}.

\section{Breakdown of systematic uncertainties}
\label{sec:app_B1WW}

The results of the generic parameterisation of~Section~\ref{sec:sigBR9}, in terms of ratios of cross sections and branching fractions, with $gg \to H \to ZZ$ as the reference channel, are shown with the full breakdown of the uncertainties in~Table~\ref{tab:B1ZZ_results_full}. The corresponding results for a similar parameterisation, with $gg \to H \to WW$ as reference, are shown in~Table~\ref{tab:B1WW_results} and illustrated in~Fig.~\ref{fig:plot_B1WW}. The parameters corresponding to ratios of cross sections are identical in each of these parameterisations, and they are included in both tables for convenience, as are the two ratios,~$\BR^{WW}/\BR^{ZZ}$ and~$\BR^{ZZ}/\BR^{WW}$. Finally, the results of the generic parameterisation of~Section~\ref{sec:kappaParam}, in terms of ratios of coupling modifiers, are shown with the full breakdown of the uncertainties in~Table~\ref{tab:genericcouplings_full}.

\begin{figure}[hbt!]
  \center
   \includegraphics[width=\textwidth]{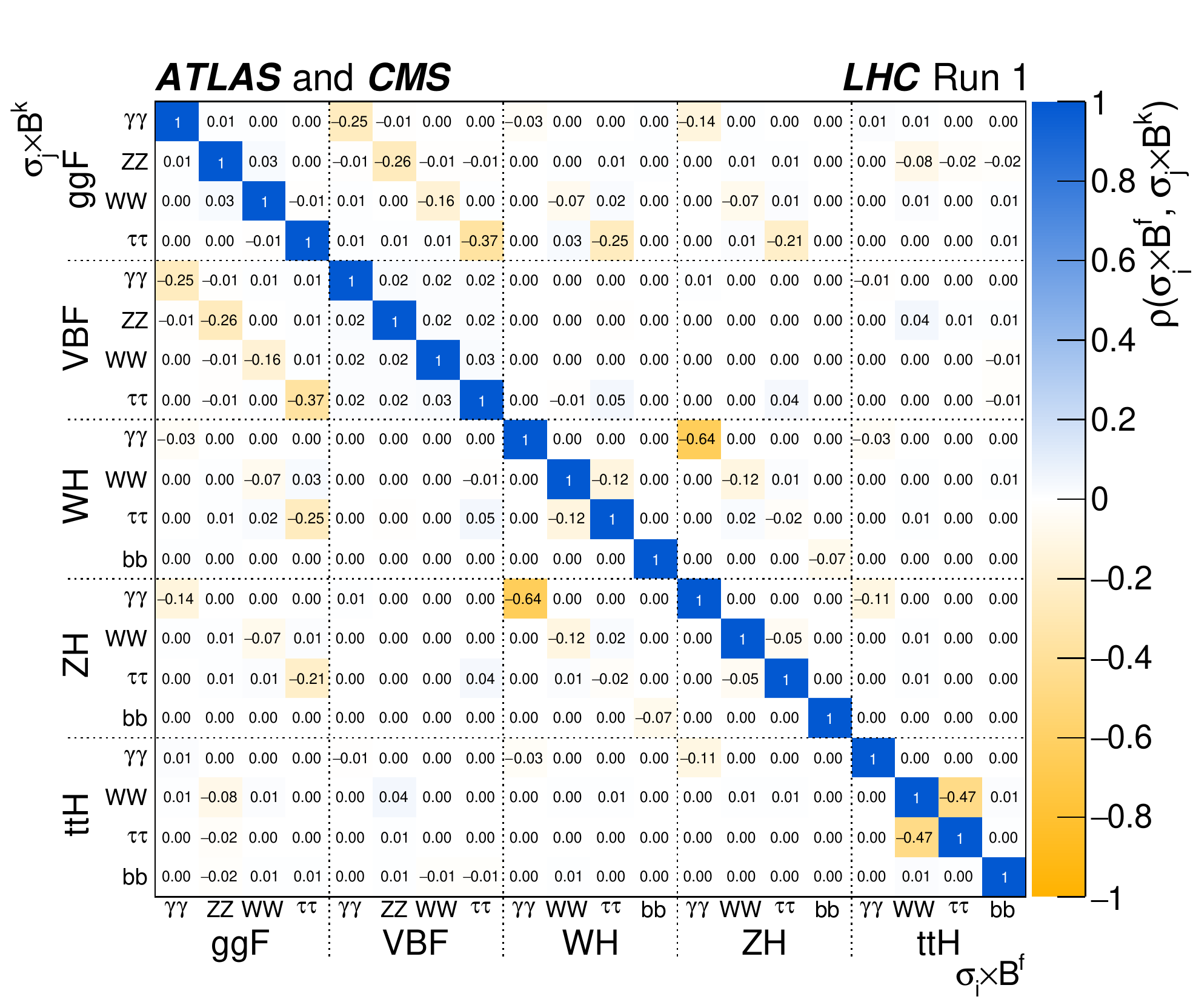}
  \caption{Correlation matrix obtained from the fit combining the ATLAS and CMS data using the generic parameterisation with 23 parameters described in~Section~\ref{sec:sigBR5x5}.
    Only 20~parameters are shown because the other three, corresponding to the $\Hzz$ decay channel for the \aWH,~\aZH, and \attH~production processes, are not measured with a meaningful precision.}
\label{fig:correlation5x5}
\end{figure}

\begin{figure}[hbt!]
  \center
  \includegraphics[width=0.67\textwidth]{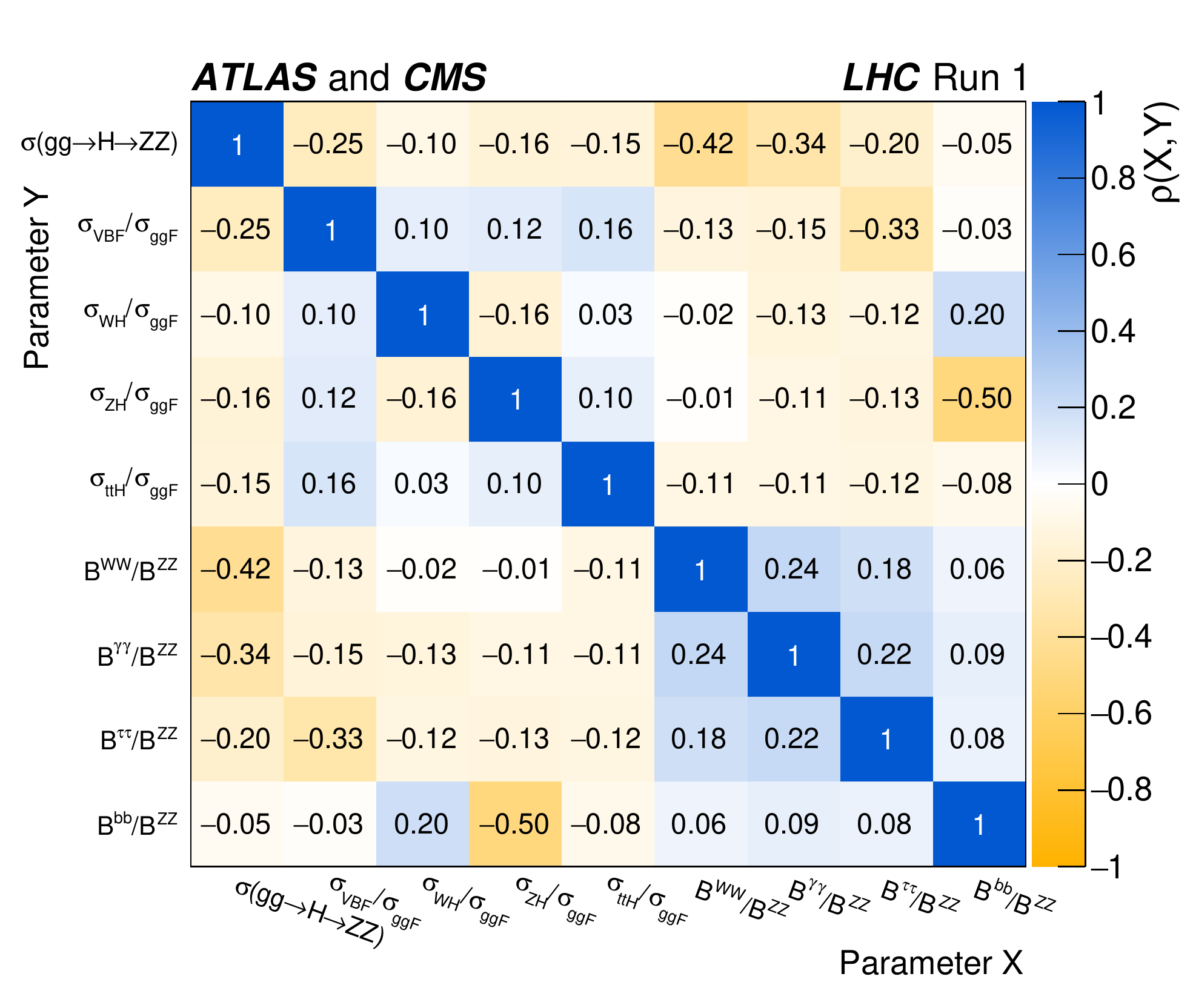}
  \caption{Correlation matrix obtained from the fit combining the ATLAS and CMS data using the generic parameterisation with nine parameters described in~Section~\ref{sec:sigBR9}.}
\label{fig:correlationB1ZZ}
\end{figure}

\begin{figure}[hbt!]
  \center
  \includegraphics[width=0.67\textwidth]{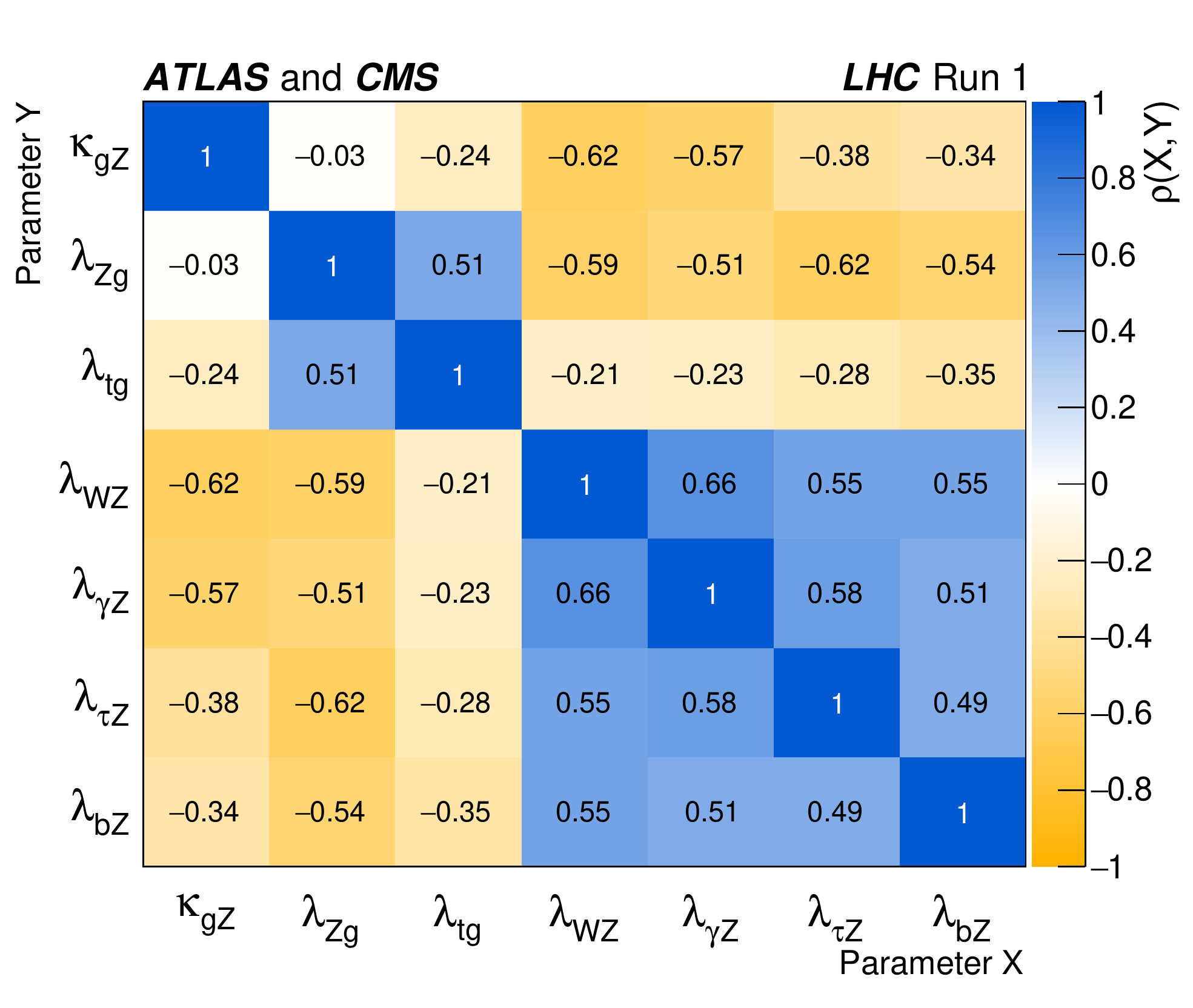}
  \caption{Correlation matrix obtained from the fit combining the ATLAS and CMS data using the generic parameterisation with seven parameters described in~Section~\ref{sec:kappaParam}. 
}
\label{fig:correlationL1}
\end{figure}

\clearpage

\begin{sidewaystable}[hbtp]
  \centering
  \caption{Best fit values of $\sigma(gg\to H\to ZZ)$, $\sigma_i/\sigma_{\aggF}$, and $\BR^f/\BR^{ZZ}$ from the combined analysis of the $\sqrt{s}=7$ and 8~TeV data. 
The values involving cross sections are given for $\sqrt{s}=8$~TeV, assuming the SM values for $\sigma_i(7~\TeV)/\sigma_i(8~\TeV)$. 
The results are shown for the combination of ATLAS and CMS, and also separately for each experiment, together with their total uncertainties and their breakdown into the four components described in the text. 
The expected total uncertainties in the measurements are also shown in parentheses. 
The SM predictions~\protect\cite{Heinemeyer:2013tqa} are shown with their total uncertainties. 
  }
  \label{tab:B1ZZ_results_full}
  \begin{adjustbox}{width=\textwidth}%
    \setlength\extrarowheight{3pt}%
    \setlength\tabcolsep{3pt}%
    \begin{tabular}{l|r@{\hskip 0.5ex}l|r@{\hskip 0.5ex}lcccc|r@{\hskip 0.5ex}lcccc|r@{\hskip 0.5ex}lcccc}\hline\hline
       Parameter  & \multicolumn{2}{c|}{SM prediction}   & \multicolumn{2}{c}{Best fit} &  \multicolumn{4}{c|}{Uncertainty}  & \multicolumn{2}{c}{Best fit} &  \multicolumn{4}{c|}{Uncertainty} & \multicolumn{2}{c}{Best fit} &  \multicolumn{4}{c}{Uncertainty}     \\[-3pt]
                                        &  &       & \multicolumn{2}{c}{value}  & Stat                  & Expt                  &  Thbgd   & Thsig  &  \multicolumn{2}{c}{value} & Stat                  & Expt                  &  Thbgd   & Thsig&  \multicolumn{2}{c}{value} & Stat           & Expt                  &  Thbgd   & Thsig   \\
      \hline & & & \multicolumn{6}{c|}{ATLAS+CMS} & \multicolumn{6}{c|}{ATLAS} &  \multicolumn{6}{c}{CMS}  \\
\multirow{2}{*}{\begin{tabular}{@{}l}$\sigma(gg\to$\\$H\to ZZ)$ [pb]\end{tabular}} & $0.51$   & $\pm 0.06$                                                                                                 
& 0.59   & $       ^{+0.11}  _{-0.10}       $ & $       ^{+0.11}  _{-0.10}       $ & $       ^{+0.02}  _{-0.01}       $ & $       ^{+0.01}  _{-0.01}       $ & $       ^{+0.01}  _{-0.01}       $          
& 0.77   & $       ^{+0.19}  _{-0.17}       $ & $       ^{+0.19}  _{-0.16}       $ & $       ^{+0.04}  _{-0.03}       $ & $       ^{+0.02}  _{-0.02}       $ & $       ^{+0.01}  _{-0.01}       $          
& 0.44   & $       ^{+0.14}  _{-0.12}       $ & $       ^{+0.13}  _{-0.11}       $ & $       ^{+0.04}  _{-0.03}       $ & $       ^{+0.01}  _{-0.01}       $ & $       ^{+0.02}  _{-0.01}       $ \\       
& &&     & $\Bgl({}^{+0.11}  _{-0.10}  \Big)$ & $\Big({}^{+0.11}  _{-0.09}  \Big)$ & $\Big({}^{+0.02}  _{-0.02}  \Big)$ & $\Big({}^{+0.01}  _{-0.01}  \Big)$ & $\Big({}^{+0.01}  _{-0.01}  \Big)$          
&        & $\Bgl({}^{+0.16}  _{-0.14}  \Big)$ & $\Big({}^{+0.16}  _{-0.13}  \Big)$ & $\Big({}^{+0.03}  _{-0.02}  \Big)$ & $\Big({}^{+0.01}  _{-0.01}  \Big)$ & $\Big({}^{+0.01}  _{-0.01}  \Big)$          
&        & $\Bgl({}^{+0.15}  _{-0.13}  \Big)$ & $\Big({}^{+0.15}  _{-0.13}  \Big)$ & $\Big({}^{+0.03}  _{-0.02}  \Big)$ & $\Big({}^{+0.01}  _{-0.01}  \Big)$ & $\Big({}^{+0.02}  _{-0.01}  \Big)$ \\[1mm]  
$\sigma_{\aVBF}/\sigma_{\aggF}$                                    & $0.082$  & $\pm 0.009$                                                                                                                
& 0.109  & $       ^{+0.034} _{-0.027}      $ & $       ^{+0.029} _{-0.024}      $ & $       ^{+0.013} _{-0.009}      $ & $       ^{+0.006} _{-0.004}      $ & $       ^{+0.010} _{-0.008}      $          
& 0.079  & $       ^{+0.035} _{-0.026}      $ & $       ^{+0.030} _{-0.023}      $ & $       ^{+0.014} _{-0.009}      $ & $       ^{+0.008} _{-0.005}      $ & $       ^{+0.009} _{-0.006}      $          
& 0.138  & $       ^{+0.073} _{-0.051}      $ & $       ^{+0.061} _{-0.046}      $ & $       ^{+0.033} _{-0.019}      $ & $       ^{+0.014} _{-0.006}      $ & $       ^{+0.015} _{-0.010}      $ \\       
& &&     & $\Bgl({}^{+0.029} _{-0.024} \Big)$ & $\Big({}^{+0.024} _{-0.020} \Big)$ & $\Big({}^{+0.012} _{-0.009} \Big)$ & $\Big({}^{+0.005} _{-0.003} \Big)$ & $\Big({}^{+0.009} _{-0.007} \Big)$          
&        & $\Bgl({}^{+0.042} _{-0.031} \Big)$ & $\Big({}^{+0.036} _{-0.028} \Big)$ & $\Big({}^{+0.016} _{-0.010} \Big)$ & $\Big({}^{+0.010} _{-0.006} \Big)$ & $\Big({}^{+0.011} _{-0.008} \Big)$          
&        & $\Bgl({}^{+0.043} _{-0.033} \Big)$ & $\Big({}^{+0.037} _{-0.029} \Big)$ & $\Big({}^{+0.020} _{-0.012} \Big)$ & $\Big({}^{+0.006} _{-0.003} \Big)$ & $\Big({}^{+0.010} _{-0.008} \Big)$ \\[1mm]  
$\sigma_{\aWH}/\sigma_{\aggF}$                                     & $0.037$  & $\pm 0.004$                                                                                                                
& 0.031  & $       ^{+0.028} _{-0.026}      $ & $       ^{+0.024} _{-0.022}      $ & $       ^{+0.012} _{-0.012}      $ & $       ^{+0.008} _{-0.008}      $ & $       ^{+0.003} _{-0.002}      $          
& 0.054  & $       ^{+0.036} _{-0.026}      $ & $       ^{+0.031} _{-0.023}      $ & $       ^{+0.012} _{-0.008}      $ & $       ^{+0.014} _{-0.009}      $ & $       ^{+0.007} _{-0.004}      $          
& 0.005  & $       ^{+0.044} _{-0.037}      $ & $       ^{+0.037} _{-0.028}      $ & $       ^{+0.021} _{-0.023}      $ & $       ^{+0.010} _{-0.008}      $ & $       ^{+0.003} _{-0.001}      $ \\       
& &&     & $\Bgl({}^{+0.021} _{-0.017} \Big)$ & $\Big({}^{+0.019} _{-0.015} \Big)$ & $\Big({}^{+0.008} _{-0.005} \Big)$ & $\Big({}^{+0.007} _{-0.005} \Big)$ & $\Big({}^{+0.002} _{-0.002} \Big)$          
&        & $\Bgl({}^{+0.033} _{-0.022} \Big)$ & $\Big({}^{+0.029} _{-0.020} \Big)$ & $\Big({}^{+0.010} _{-0.006} \Big)$ & $\Big({}^{+0.011} _{-0.006} \Big)$ & $\Big({}^{+0.005} _{-0.002} \Big)$          
&        & $\Bgl({}^{+0.032} _{-0.022} \Big)$ & $\Big({}^{+0.027} _{-0.020} \Big)$ & $\Big({}^{+0.014} _{-0.008} \Big)$ & $\Big({}^{+0.009} _{-0.006} \Big)$ & $\Big({}^{+0.003} _{-0.001} \Big)$ \\[1mm]  
$\sigma_{\aZH}/\sigma_{\aggF}$                                     & $0.0216$ & $\pm 0.0024$                                                                                                               
& 0.066  & $       ^{+0.039} _{-0.031}      $ & $       ^{+0.032} _{-0.025}      $ & $       ^{+0.018} _{-0.012}      $ & $       ^{+0.014} _{-0.012}      $ & $       ^{+0.005} _{-0.003}      $          
& 0.013  & $       ^{+0.028} _{-0.014}      $ & $       ^{+0.021} _{-0.012}      $ & $       ^{+0.013} _{-0.005}      $ & $       ^{+0.013} _{-0.005}      $ & $       ^{+0.003} _{-0.002}      $          
& 0.123  & $       ^{+0.076} _{-0.053}      $ & $       ^{+0.063} _{-0.046}      $ & $       ^{+0.038} _{-0.022}      $ & $       ^{+0.019} _{-0.013}      $ & $       ^{+0.009} _{-0.005}      $ \\       
& &&     & $\Bgl({}^{+0.016} _{-0.011} \Big)$ & $\Big({}^{+0.014} _{-0.010} \Big)$ & $\Big({}^{+0.006} _{-0.003} \Big)$ & $\Big({}^{+0.006} _{-0.003} \Big)$ & $\Big({}^{+0.002} _{-0.001} \Big)$          
&        & $\Bgl({}^{+0.027} _{-0.014} \Big)$ & $\Big({}^{+0.023} _{-0.013} \Big)$ & $\Big({}^{+0.009} _{-0.003} \Big)$ & $\Big({}^{+0.011} _{-0.004} \Big)$ & $\Big({}^{+0.003} _{-0.001} \Big)$          
&        & $\Bgl({}^{+0.024} _{-0.013} \Big)$ & $\Big({}^{+0.020} _{-0.012} \Big)$ & $\Big({}^{+0.010} _{-0.004} \Big)$ & $\Big({}^{+0.009} _{-0.004} \Big)$ & $\Big({}^{+0.002} _{-0.001} \Big)$ \\[1mm]  
$\sigma_{\attH}/\sigma_{\aggF}$                                    & $0.0067$ & $\pm 0.0010$                                                                                                               
& 0.0220 & $       ^{+0.0068}_{-0.0057}     $ & $       ^{+0.0055}_{-0.0048}     $ & $       ^{+0.0031}_{-0.0023}     $ & $       ^{+0.0023}_{-0.0020}     $ & $       ^{+0.0014}_{-0.0010}     $          
& 0.0126 & $       ^{+0.0066}_{-0.0053}     $ & $       ^{+0.0052}_{-0.0042}     $ & $       ^{+0.0031}_{-0.0023}     $ & $       ^{+0.0024}_{-0.0020}     $ & $       ^{+0.0013}_{-0.0007}     $          
& 0.0340 & $       ^{+0.0158}_{-0.0116}     $ & $       ^{+0.0121}_{-0.0097}     $ & $       ^{+0.0085}_{-0.0051}     $ & $       ^{+0.0048}_{-0.0036}     $ & $       ^{+0.0026}_{-0.0015}     $ \\       
& &&     & $\Bgl({}^{+0.0042}_{-0.0035}\Big)$ & $\Big({}^{+0.0033}_{-0.0027}\Big)$ & $\Big({}^{+0.0018}_{-0.0013}\Big)$ & $\Big({}^{+0.0020}_{-0.0019}\Big)$ & $\Big({}^{+0.0005}_{-0.0003}\Big)$          
&        & $\Bgl({}^{+0.0061}_{-0.0045}\Big)$ & $\Big({}^{+0.0047}_{-0.0035}\Big)$ & $\Big({}^{+0.0026}_{-0.0017}\Big)$ & $\Big({}^{+0.0027}_{-0.0022}\Big)$ & $\Big({}^{+0.0008}_{-0.0004}\Big)$          
&        & $\Bgl({}^{+0.0066}_{-0.0054}\Big)$ & $\Big({}^{+0.0051}_{-0.0038}\Big)$ & $\Big({}^{+0.0027}_{-0.0016}\Big)$ & $\Big({}^{+0.0032}_{-0.0034}\Big)$ & $\Big({}^{+0.0006}_{-0.0002}\Big)$ \\[1mm]  
$\BR^{WW}/\BR^{ZZ}$                                                & $8.09$   & $\pm <0.01$                                                                                                                
& 6.7    & $       ^{+1.6}   _{-1.3}        $ & $       ^{+1.5}   _{-1.2}        $ & $       ^{+0.4}   _{-0.3}        $ & $       ^{+0.4}   _{-0.3}        $ & $       ^{+0.3}   _{-0.2}        $          
& 6.5    & $       ^{+2.1}   _{-1.6}        $ & $       ^{+2.0}   _{-1.4}        $ & $       ^{+0.6}   _{-0.4}        $ & $       ^{+0.5}   _{-0.4}        $ & $       ^{+0.3}   _{-0.2}        $          
& 7.1    & $       ^{+2.9}   _{-2.1}        $ & $       ^{+2.6}   _{-1.8}        $ & $       ^{+1.0}   _{-0.7}        $ & $       ^{+0.7}   _{-0.5}        $ & $       ^{+0.4}   _{-0.3}        $ \\       
& &&     & $\Bgl({}^{+2.2}   _{-1.7}   \Big)$ & $\Big({}^{+2.0}   _{-1.6}   \Big)$ & $\Big({}^{+0.7}   _{-0.5}   \Big)$ & $\Big({}^{+0.5}   _{-0.4}   \Big)$ & $\Big({}^{+0.3}   _{-0.2}   \Big)$          
&        & $\Bgl({}^{+3.5}   _{-2.4}   \Big)$ & $\Big({}^{+3.3}   _{-2.2}   \Big)$ & $\Big({}^{+0.9}   _{-0.6}   \Big)$ & $\Big({}^{+0.8}   _{-0.6}   \Big)$ & $\Big({}^{+0.4}   _{-0.3}   \Big)$          
&        & $\Bgl({}^{+3.2}   _{-2.2}   \Big)$ & $\Big({}^{+2.9}   _{-2.0}   \Big)$ & $\Big({}^{+1.1}   _{-0.8}   \Big)$ & $\Big({}^{+0.7}   _{-0.5}   \Big)$ & $\Big({}^{+0.5}   _{-0.4}   \Big)$ \\[1mm]  
$\BR^{\gamma\gamma}/\BR^{ZZ}$                                      & $0.0854$ & $\pm 0.0010$                                                                                                               
& 0.069  & $       ^{+0.018} _{-0.014}      $ & $       ^{+0.018} _{-0.014}      $ & $       ^{+0.003} _{-0.002}      $ & $       ^{+0.002} _{-0.001}      $ & $       ^{+0.002} _{-0.002}      $          
& 0.062  & $       ^{+0.024} _{-0.018}      $ & $       ^{+0.023} _{-0.017}      $ & $       ^{+0.007} _{-0.004}      $ & $       ^{+0.002} _{-0.001}      $ & $       ^{+0.003} _{-0.002}      $          
& 0.079  & $       ^{+0.034} _{-0.023}      $ & $       ^{+0.032} _{-0.023}      $ & $       ^{+0.009} _{-0.005}      $ & $       ^{+0.003} _{-0.002}      $ & $       ^{+0.004} _{-0.003}      $ \\       
& &&     & $\Bgl({}^{+0.025} _{-0.019} \Big)$ & $\Big({}^{+0.024} _{-0.019} \Big)$ & $\Big({}^{+0.005} _{-0.003} \Big)$ & $\Big({}^{+0.002} _{-0.001} \Big)$ & $\Big({}^{+0.003} _{-0.002} \Big)$          
&        & $\Bgl({}^{+0.040} _{-0.027} \Big)$ & $\Big({}^{+0.039} _{-0.027} \Big)$ & $\Big({}^{+0.008} _{-0.005} \Big)$ & $\Big({}^{+0.003} _{-0.002} \Big)$ & $\Big({}^{+0.004} _{-0.003} \Big)$          
&        & $\Bgl({}^{+0.035} _{-0.025} \Big)$ & $\Big({}^{+0.034} _{-0.024} \Big)$ & $\Big({}^{+0.007} _{-0.004} \Big)$ & $\Big({}^{+0.002} _{-0.001} \Big)$ & $\Big({}^{+0.004} _{-0.003} \Big)$ \\[1mm]  
$\BR^{\tau\tau}/\BR^{ZZ}$                                          & $2.36$   & $\pm 0.05$                                                                                                                 
& 1.77   & $       ^{+0.59}  _{-0.46}       $ & $       ^{+0.52}  _{-0.41}       $ & $       ^{+0.27}  _{-0.20}       $ & $       ^{+0.05}  _{-0.04}       $ & $       ^{+0.06}  _{-0.04}       $          
& 2.17   & $       ^{+1.07}  _{-0.74}       $ & $       ^{+0.89}  _{-0.64}       $ & $       ^{+0.53}  _{-0.35}       $ & $       ^{+0.16}  _{-0.10}       $ & $       ^{+0.17}  _{-0.09}       $          
& 1.56   & $       ^{+0.90}  _{-0.61}       $ & $       ^{+0.78}  _{-0.54}       $ & $       ^{+0.45}  _{-0.26}       $ & $       ^{+0.07}  _{-0.05}       $ & $       ^{+0.07}  _{-0.04}       $ \\       
& &&     & $\Bgl({}^{+0.90}  _{-0.68}  \Big)$ & $\Big({}^{+0.75}  _{-0.58}  \Big)$ & $\Big({}^{+0.47}  _{-0.33}  \Big)$ & $\Big({}^{+0.08}  _{-0.06}  \Big)$ & $\Big({}^{+0.10}  _{-0.06}  \Big)$          
&        & $\Bgl({}^{+1.54}  _{-0.98}  \Big)$ & $\Big({}^{+1.30}  _{-0.86}  \Big)$ & $\Big({}^{+0.76}  _{-0.44}  \Big)$ & $\Big({}^{+0.22}  _{-0.12}  \Big)$ & $\Big({}^{+0.22}  _{-0.10}  \Big)$          
&        & $\Bgl({}^{+1.23}  _{-0.86}  \Big)$ & $\Big({}^{+1.03}  _{-0.73}  \Big)$ & $\Big({}^{+0.66}  _{-0.44}  \Big)$ & $\Big({}^{+0.04}  _{-0.03}  \Big)$ & $\Big({}^{+0.12}  _{-0.07}  \Big)$ \\[1mm]  
$\BR^{bb}/\BR^{ZZ}$                                                & $21.5$   & $\pm 1.0$                                                                                                                  
& 4.2    & $       ^{+4.4}   _{-2.6}        $ & $       ^{+2.8}   _{-2.0}        $ & $       ^{+2.3}   _{-1.1}        $ & $       ^{+2.5}   _{-1.2}        $ & $       ^{+0.4}   _{-0.2}        $          
& 9.6    & $       ^{+10.1}  _{-5.7}        $ & $       ^{+7.4}   _{-4.4}        $ & $       ^{+4.5}   _{-2.4}        $ & $       ^{+5.1}   _{-2.7}        $ & $       ^{+1.3}   _{-0.5}        $          
& 3.7    & $       ^{+4.1}   _{-2.4}        $ & $       ^{+3.1}   _{-2.0}        $ & $       ^{+1.8}   _{-0.9}        $ & $       ^{+1.9}   _{-1.1}        $ & $       ^{+0.4}   _{-0.2}        $ \\       
& &&     & $\Bgl({}^{+16.8}  _{-9.0}   \Big)$ & $\Big({}^{+13.9}  _{-7.9}   \Big)$ & $\Big({}^{+6.3}   _{-2.8}   \Big)$ & $\Big({}^{+6.7}   _{-3.3}   \Big)$ & $\Big({}^{+2.1}   _{-0.9}   \Big)$          
&        & $\Bgl({}^{+29.3}  _{-11.8}  \Big)$ & $\Big({}^{+24.2}  _{-10.5}  \Big)$ & $\Big({}^{+10.9}  _{-3.3}   \Big)$ & $\Big({}^{+11.8}  _{-4.0}   \Big)$ & $\Big({}^{+4.0}   _{-1.2}   \Big)$          
&        & $\Bgl({}^{+29.4}  _{-11.9}  \Big)$ & $\Big({}^{+23.4}  _{-10.4}  \Big)$ & $\Big({}^{+12.7}  _{-3.8}   \Big)$ & $\Big({}^{+12.2}  _{-4.4}   \Big)$ & $\Big({}^{+2.5}   _{-0.9}   \Big)$ \\[1mm]  

      \hline\hline
    \end{tabular}
  \end{adjustbox}
\end{sidewaystable}

\begin{sidewaystable}[hbtp]
  \centering
  \caption{Best fit values of $\sigma(gg\to H\to WW)$, $\sigma_i/\sigma_{\aggF}$, and $\BR^f/\BR^{WW}$ from the combined analysis of the $\sqrt{s}=7$ and 8~TeV data. 
The values involving cross sections are given for $\sqrt{s}=8$~TeV, assuming the SM values for $\sigma_i(7~\TeV)/\sigma_i(8~\TeV)$.
The results are shown for the combination of ATLAS and CMS, and also separately for each experiment, together with their total uncertainties and their breakdown into the four components described in the text. 
The expected total uncertainties in the measurements are also shown in parentheses. 
The SM predictions~\protect\cite{Heinemeyer:2013tqa} are shown with their total uncertainties. 
  }
  \label{tab:B1WW_results}
  \begin{adjustbox}{width=\textwidth}%
    \setlength\extrarowheight{3pt}%
    \setlength\tabcolsep{3pt}%
    \begin{tabular}{l|r@{\hskip 0.5ex}l|r@{\hskip 0.5ex}lcccc|r@{\hskip 0.5ex}lcccc|r@{\hskip 0.5ex}lcccc}\hline\hline
       Parameter  & \multicolumn{2}{c|}{SM prediction}   & \multicolumn{2}{c}{Best fit} &  \multicolumn{4}{c|}{Uncertainty}  & \multicolumn{2}{c}{Best fit} &  \multicolumn{4}{c|}{Uncertainty} & \multicolumn{2}{c}{Best fit} &  \multicolumn{4}{c}{Uncertainty}     \\[-3pt]
                                        &  &       & \multicolumn{2}{c}{value}  & Stat                  & Expt                  &  Thbgd   & Thsig  &  \multicolumn{2}{c}{value} & Stat                  & Expt                  &  Thbgd   & Thsig&  \multicolumn{2}{c}{value} & Stat           & Expt                  &  Thbgd   & Thsig   \\
      \hline & & & \multicolumn{6}{c|}{ATLAS+CMS} & \multicolumn{6}{c|}{ATLAS} &  \multicolumn{6}{c}{CMS}  \\
\multirow{2}{*}{\begin{tabular}{@{}l}$\sigma(gg\to$\\$H\to WW)$ [pb]\end{tabular}} & $4.1$    & $\pm 0.5$                                                                                                  
& 4.0    & $       ^{+0.6}   _{-0.6}        $ & $       ^{+0.5}   _{-0.5}        $ & $       ^{+0.3}   _{-0.3}        $ & $       ^{+0.2}   _{-0.2}        $ & $       ^{+0.2}   _{-0.1}        $          
& 5.0    & $       ^{+1.0}   _{-0.9}        $ & $       ^{+0.7}   _{-0.7}        $ & $       ^{+0.4}   _{-0.4}        $ & $       ^{+0.4}   _{-0.4}        $ & $       ^{+0.2}   _{-0.2}        $          
& 3.1    & $       ^{+0.8}   _{-0.8}        $ & $       ^{+0.6}   _{-0.6}        $ & $       ^{+0.5}   _{-0.4}        $ & $       ^{+0.3}   _{-0.3}        $ & $       ^{+0.2}   _{-0.1}        $ \\       
& &&     & $\Bgl({}^{+0.7}   _{-0.6}   \Big)$ & $\Big({}^{+0.5}   _{-0.5}   \Big)$ & $\Big({}^{+0.3}   _{-0.3}   \Big)$ & $\Big({}^{+0.3}   _{-0.3}   \Big)$ & $\Big({}^{+0.2}   _{-0.1}   \Big)$          
&        & $\Bgl({}^{+0.9}   _{-0.9}   \Big)$ & $\Big({}^{+0.7}   _{-0.7}   \Big)$ & $\Big({}^{+0.4}   _{-0.4}   \Big)$ & $\Big({}^{+0.4}   _{-0.4}   \Big)$ & $\Big({}^{+0.2}   _{-0.1}   \Big)$          
&        & $\Bgl({}^{+0.9}   _{-0.9}   \Big)$ & $\Big({}^{+0.6}   _{-0.6}   \Big)$ & $\Big({}^{+0.5}   _{-0.5}   \Big)$ & $\Big({}^{+0.4}   _{-0.3}   \Big)$ & $\Big({}^{+0.2}   _{-0.2}   \Big)$ \\[1mm]  
$\sigma_{\aVBF}/\sigma_{\aggF}$                                    & $0.082$  & $\pm 0.009$                                                                                                                
& 0.109  & $       ^{+0.034} _{-0.027}      $ & $       ^{+0.028} _{-0.024}      $ & $       ^{+0.013} _{-0.009}      $ & $       ^{+0.006} _{-0.004}      $ & $       ^{+0.011} _{-0.008}      $          
& 0.079  & $       ^{+0.035} _{-0.026}      $ & $       ^{+0.030} _{-0.023}      $ & $       ^{+0.015} _{-0.009}      $ & $       ^{+0.008} _{-0.005}      $ & $       ^{+0.009} _{-0.006}      $          
& 0.137  & $       ^{+0.072} _{-0.051}      $ & $       ^{+0.061} _{-0.046}      $ & $       ^{+0.032} _{-0.018}      $ & $       ^{+0.014} _{-0.006}      $ & $       ^{+0.016} _{-0.010}      $ \\       
& &&     & $\Bgl({}^{+0.029} _{-0.024} \Big)$ & $\Big({}^{+0.024} _{-0.020} \Big)$ & $\Big({}^{+0.012} _{-0.009} \Big)$ & $\Big({}^{+0.005} _{-0.003} \Big)$ & $\Big({}^{+0.009} _{-0.007} \Big)$          
&        & $\Bgl({}^{+0.042} _{-0.031} \Big)$ & $\Big({}^{+0.036} _{-0.028} \Big)$ & $\Big({}^{+0.016} _{-0.010} \Big)$ & $\Big({}^{+0.010} _{-0.006} \Big)$ & $\Big({}^{+0.011} _{-0.008} \Big)$          
&        & $\Bgl({}^{+0.043} _{-0.033} \Big)$ & $\Big({}^{+0.037} _{-0.029} \Big)$ & $\Big({}^{+0.020} _{-0.012} \Big)$ & $\Big({}^{+0.006} _{-0.003} \Big)$ & $\Big({}^{+0.010} _{-0.008} \Big)$ \\[1mm]  
$\sigma_{\aWH}/\sigma_{\aggF}$                                     & $0.037$  & $\pm 0.004$                                                                                                                
& 0.030  & $       ^{+0.028} _{-0.026}      $ & $       ^{+0.024} _{-0.022}      $ & $       ^{+0.012} _{-0.012}      $ & $       ^{+0.008} _{-0.008}      $ & $       ^{+0.003} _{-0.002}      $          
& 0.054  & $       ^{+0.037} _{-0.026}      $ & $       ^{+0.031} _{-0.023}      $ & $       ^{+0.012} _{-0.008}      $ & $       ^{+0.014} _{-0.009}      $ & $       ^{+0.007} _{-0.003}      $          
& 0.005  & $       ^{+0.043} _{-0.037}      $ & $       ^{+0.037} _{-0.028}      $ & $       ^{+0.021} _{-0.023}      $ & $       ^{+0.010} _{-0.008}      $ & $       ^{+0.003} _{-0.001}      $ \\       
& &&     & $\Bgl({}^{+0.021} _{-0.017} \Big)$ & $\Big({}^{+0.019} _{-0.015} \Big)$ & $\Big({}^{+0.008} _{-0.005} \Big)$ & $\Big({}^{+0.007} _{-0.005} \Big)$ & $\Big({}^{+0.003} _{-0.002} \Big)$          
&        & $\Bgl({}^{+0.032} _{-0.022} \Big)$ & $\Big({}^{+0.029} _{-0.020} \Big)$ & $\Big({}^{+0.010} _{-0.006} \Big)$ & $\Big({}^{+0.011} _{-0.006} \Big)$ & $\Big({}^{+0.004} _{-0.002} \Big)$          
&        & $\Bgl({}^{+0.032} _{-0.022} \Big)$ & $\Big({}^{+0.027} _{-0.020} \Big)$ & $\Big({}^{+0.014} _{-0.008} \Big)$ & $\Big({}^{+0.009} _{-0.006} \Big)$ & $\Big({}^{+0.003} _{-0.001} \Big)$ \\[1mm]  
$\sigma_{\aZH}/\sigma_{\aggF}$                                     & $0.0216$ & $\pm 0.0024$                                                                                                               
& 0.066  & $       ^{+0.039} _{-0.031}      $ & $       ^{+0.032} _{-0.025}      $ & $       ^{+0.018} _{-0.013}      $ & $       ^{+0.014} _{-0.012}      $ & $       ^{+0.005} _{-0.003}      $          
& 0.013  & $       ^{+0.028} _{-0.013}      $ & $       ^{+0.021} _{-0.008}      $ & $       ^{+0.013} _{-0.011}      $ & $       ^{+0.013} _{-0.006}      $ & $       ^{+0.003} _{-0.002}      $          
& 0.123  & $       ^{+0.075} _{-0.052}      $ & $       ^{+0.062} _{-0.046}      $ & $       ^{+0.037} _{-0.021}      $ & $       ^{+0.018} _{-0.013}      $ & $       ^{+0.009} _{-0.005}      $ \\       
& &&     & $\Bgl({}^{+0.016} _{-0.011} \Big)$ & $\Big({}^{+0.014} _{-0.010} \Big)$ & $\Big({}^{+0.006} _{-0.003} \Big)$ & $\Big({}^{+0.006} _{-0.003} \Big)$ & $\Big({}^{+0.002} _{-0.001} \Big)$          
&        & $\Bgl({}^{+0.027} _{-0.014} \Big)$ & $\Big({}^{+0.023} _{-0.013} \Big)$ & $\Big({}^{+0.009} _{-0.003} \Big)$ & $\Big({}^{+0.011} _{-0.004} \Big)$ & $\Big({}^{+0.003} _{-0.001} \Big)$          
&        & $\Bgl({}^{+0.024} _{-0.013} \Big)$ & $\Big({}^{+0.020} _{-0.012} \Big)$ & $\Big({}^{+0.010} _{-0.004} \Big)$ & $\Big({}^{+0.009} _{-0.004} \Big)$ & $\Big({}^{+0.002} _{-0.001} \Big)$ \\[1mm]  
$\sigma_{\attH}/\sigma_{\aggF}$                                    & $0.0067$ & $\pm 0.0010$                                                                                                               
& 0.0220 & $       ^{+0.0068}_{-0.0057}     $ & $       ^{+0.0055}_{-0.0048}     $ & $       ^{+0.0031}_{-0.0023}     $ & $       ^{+0.0023}_{-0.0020}     $ & $       ^{+0.0014}_{-0.0010}     $          
& 0.0126 & $       ^{+0.0066}_{-0.0053}     $ & $       ^{+0.0052}_{-0.0042}     $ & $       ^{+0.0031}_{-0.0023}     $ & $       ^{+0.0024}_{-0.0020}     $ & $       ^{+0.0013}_{-0.0007}     $          
& 0.0340 & $       ^{+0.0157}_{-0.0116}     $ & $       ^{+0.0121}_{-0.0096}     $ & $       ^{+0.0085}_{-0.0050}     $ & $       ^{+0.0048}_{-0.0036}     $ & $       ^{+0.0026}_{-0.0015}     $ \\       
& &&     & $\Bgl({}^{+0.0042}_{-0.0035}\Big)$ & $\Big({}^{+0.0033}_{-0.0027}\Big)$ & $\Big({}^{+0.0018}_{-0.0013}\Big)$ & $\Big({}^{+0.0020}_{-0.0019}\Big)$ & $\Big({}^{+0.0005}_{-0.0003}\Big)$          
&        & $\Bgl({}^{+0.0060}_{-0.0045}\Big)$ & $\Big({}^{+0.0047}_{-0.0035}\Big)$ & $\Big({}^{+0.0026}_{-0.0017}\Big)$ & $\Big({}^{+0.0027}_{-0.0022}\Big)$ & $\Big({}^{+0.0008}_{-0.0004}\Big)$          
&        & $\Bgl({}^{+0.0067}_{-0.0054}\Big)$ & $\Big({}^{+0.0051}_{-0.0038}\Big)$ & $\Big({}^{+0.0027}_{-0.0016}\Big)$ & $\Big({}^{+0.0032}_{-0.0034}\Big)$ & $\Big({}^{+0.0006}_{-0.0002}\Big)$ \\[1mm]  
$\BR^{ZZ}/\BR^{WW}$                                                & $0.124$  & $\pm <0.001$                                                                                                               
& 0.148  & $       ^{+0.035} _{-0.029}      $ & $       ^{+0.032} _{-0.027}      $ & $       ^{+0.010} _{-0.007}      $ & $       ^{+0.009} _{-0.006}      $ & $       ^{+0.006} _{-0.004}      $          
& 0.155  & $       ^{+0.050} _{-0.039}      $ & $       ^{+0.045} _{-0.036}      $ & $       ^{+0.016} _{-0.009}      $ & $       ^{+0.014} _{-0.008}      $ & $       ^{+0.007} _{-0.005}      $          
& 0.140  & $       ^{+0.057} _{-0.041}      $ & $       ^{+0.049} _{-0.038}      $ & $       ^{+0.023} _{-0.012}      $ & $       ^{+0.016} _{-0.008}      $ & $       ^{+0.009} _{-0.005}      $ \\       
& &&     & $\Bgl({}^{+0.033} _{-0.027} \Big)$ & $\Big({}^{+0.029} _{-0.025} \Big)$ & $\Big({}^{+0.011} _{-0.007} \Big)$ & $\Big({}^{+0.009} _{-0.006} \Big)$ & $\Big({}^{+0.005} _{-0.003} \Big)$          
&        & $\Bgl({}^{+0.050} _{-0.037} \Big)$ & $\Big({}^{+0.045} _{-0.035} \Big)$ & $\Big({}^{+0.014} _{-0.008} \Big)$ & $\Big({}^{+0.014} _{-0.007} \Big)$ & $\Big({}^{+0.006} _{-0.004} \Big)$          
&        & $\Bgl({}^{+0.048} _{-0.035} \Big)$ & $\Big({}^{+0.041} _{-0.033} \Big)$ & $\Big({}^{+0.019} _{-0.010} \Big)$ & $\Big({}^{+0.013} _{-0.007} \Big)$ & $\Big({}^{+0.009} _{-0.005} \Big)$ \\[1mm]  
$\BR^{\gamma\gamma}/\BR^{WW}$                                      & $0.01056$ & $\pm 0.00010$                                                                                                             
& 0.0102 & $       ^{+0.0022}_{-0.0019}     $ & $       ^{+0.0020}_{-0.0017}     $ & $       ^{+0.0008}_{-0.0005}     $ & $       ^{+0.0006}_{-0.0004}     $ & $       ^{+0.0004}_{-0.0003}     $          
& 0.0097 & $       ^{+0.0031}_{-0.0025}     $ & $       ^{+0.0026}_{-0.0023}     $ & $       ^{+0.0013}_{-0.0008}     $ & $       ^{+0.0008}_{-0.0005}     $ & $       ^{+0.0006}_{-0.0004}     $          
& 0.0111 & $       ^{+0.0039}_{-0.0029}     $ & $       ^{+0.0033}_{-0.0027}     $ & $       ^{+0.0018}_{-0.0010}     $ & $       ^{+0.0012}_{-0.0006}     $ & $       ^{+0.0006}_{-0.0004}     $ \\       
& &&     & $\Bgl({}^{+0.0025}_{-0.0020}\Big)$ & $\Big({}^{+0.0021}_{-0.0019}\Big)$ & $\Big({}^{+0.0009}_{-0.0006}\Big)$ & $\Big({}^{+0.0007}_{-0.0005}\Big)$ & $\Big({}^{+0.0005}_{-0.0003}\Big)$          
&        & $\Bgl({}^{+0.0037}_{-0.0029}\Big)$ & $\Big({}^{+0.0032}_{-0.0027}\Big)$ & $\Big({}^{+0.0014}_{-0.0008}\Big)$ & $\Big({}^{+0.0011}_{-0.0006}\Big)$ & $\Big({}^{+0.0007}_{-0.0005}\Big)$          
&        & $\Bgl({}^{+0.0036}_{-0.0027}\Big)$ & $\Big({}^{+0.0030}_{-0.0025}\Big)$ & $\Big({}^{+0.0015}_{-0.0008}\Big)$ & $\Big({}^{+0.0011}_{-0.0005}\Big)$ & $\Big({}^{+0.0007}_{-0.0004}\Big)$ \\[1mm]  
$\BR^{\tau\tau}/\BR^{WW}$                                          & $0.292$  & $\pm 0.006$                                                                                                                
& 0.26   & $       ^{+0.08}  _{-0.06}       $ & $       ^{+0.06}  _{-0.05}       $ & $       ^{+0.04}  _{-0.03}       $ & $       ^{+0.02}  _{-0.01}       $ & $       ^{+0.01}  _{-0.01}       $          
& 0.34   & $       ^{+0.14}  _{-0.10}       $ & $       ^{+0.11}  _{-0.09}       $ & $       ^{+0.08}  _{-0.06}       $ & $       ^{+0.03}  _{-0.02}       $ & $       ^{+0.03}  _{-0.02}       $          
& 0.22   & $       ^{+0.11}  _{-0.08}       $ & $       ^{+0.09}  _{-0.07}       $ & $       ^{+0.06}  _{-0.04}       $ & $       ^{+0.02}  _{-0.01}       $ & $       ^{+0.01}  _{-0.01}       $ \\       
& &&     & $\Bgl({}^{+0.09}  _{-0.08}  \Big)$ & $\Big({}^{+0.07}  _{-0.06}  \Big)$ & $\Big({}^{+0.06}  _{-0.04}  \Big)$ & $\Big({}^{+0.02}  _{-0.01}  \Big)$ & $\Big({}^{+0.01}  _{-0.01}  \Big)$          
&        & $\Bgl({}^{+0.15}  _{-0.11}  \Big)$ & $\Big({}^{+0.12}  _{-0.09}  \Big)$ & $\Big({}^{+0.09}  _{-0.06}  \Big)$ & $\Big({}^{+0.03}  _{-0.02}  \Big)$ & $\Big({}^{+0.03}  _{-0.01}  \Big)$          
&        & $\Bgl({}^{+0.13}  _{-0.10}  \Big)$ & $\Big({}^{+0.10}  _{-0.08}  \Big)$ & $\Big({}^{+0.08}  _{-0.06}  \Big)$ & $\Big({}^{+0.03}  _{-0.01}  \Big)$ & $\Big({}^{+0.02}  _{-0.01}  \Big)$ \\[1mm]  
$\BR^{bb}/\BR^{WW}$                                                & $2.66$   & $\pm 0.12$                                                                                                                 
& 0.61   & $       ^{+0.64}  _{-0.38}       $ & $       ^{+0.39}  _{-0.29}       $ & $       ^{+0.34}  _{-0.16}       $ & $       ^{+0.37}  _{-0.18}       $ & $       ^{+0.05}  _{-0.02}       $          
& 1.50   & $       ^{+1.54}  _{-0.90}       $ & $       ^{+1.06}  _{-0.66}       $ & $       ^{+0.73}  _{-0.42}       $ & $       ^{+0.82}  _{-0.43}       $ & $       ^{+0.20}  _{-0.09}       $          
& 0.52   & $       ^{+0.54}  _{-0.34}       $ & $       ^{+0.39}  _{-0.27}       $ & $       ^{+0.25}  _{-0.13}       $ & $       ^{+0.28}  _{-0.14}       $ & $       ^{+0.05}  _{-0.02}       $ \\       
& &&     & $\Bgl({}^{+2.01}  _{-1.09}  \Big)$ & $\Big({}^{+1.63}  _{-0.93}  \Big)$ & $\Big({}^{+0.78}  _{-0.35}  \Big)$ & $\Big({}^{+0.84}  _{-0.42}  \Big)$ & $\Big({}^{+0.23}  _{-0.10}  \Big)$          
&        & $\Bgl({}^{+3.49}  _{-1.38}  \Big)$ & $\Big({}^{+2.83}  _{-1.20}  \Big)$ & $\Big({}^{+1.37}  _{-0.42}  \Big)$ & $\Big({}^{+1.45}  _{-0.52}  \Big)$ & $\Big({}^{+0.45}  _{-0.14}  \Big)$          
&        & $\Bgl({}^{+3.57}  _{-1.46}  \Big)$ & $\Big({}^{+2.78}  _{-1.25}  \Big)$ & $\Big({}^{+1.58}  _{-0.49}  \Big)$ & $\Big({}^{+1.56}  _{-0.58}  \Big)$ & $\Big({}^{+0.26}  _{-0.10}  \Big)$ \\[1mm]  

      \hline\hline
    \end{tabular}
  \end{adjustbox}
\end{sidewaystable}

\clearpage

\begin{figure}[h!]
\centering
\includegraphics[width=1.0\textwidth]{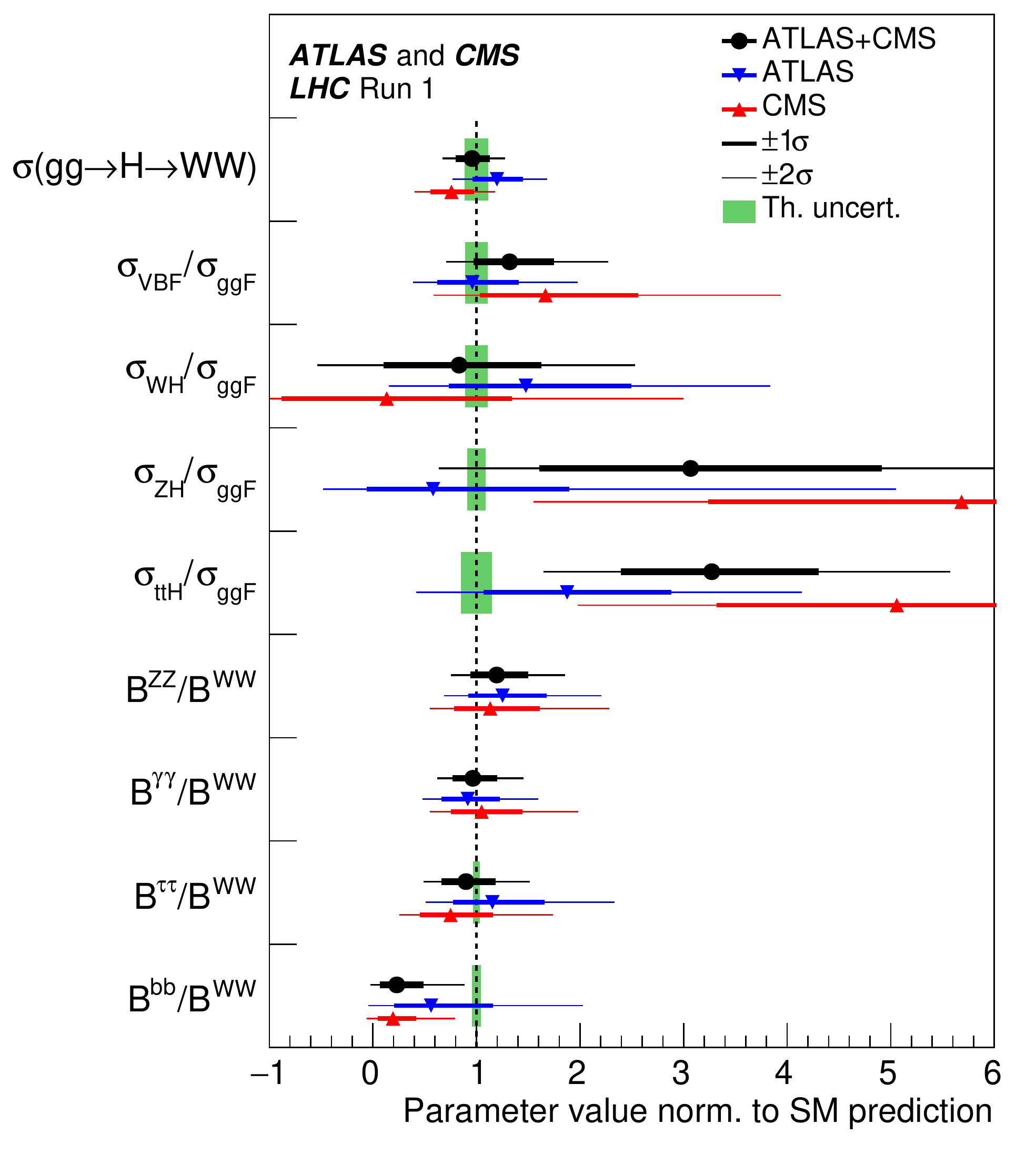}
\caption{Best fit values of the $gg\to H\to WW$ cross section and of ratios of cross sections and branching fractions, as obtained from the generic parameterisation described in~Section~\ref{sec:sigBR9} and as tabulated in~Table~\ref{tab:B1WW_results} for the combination of the ATLAS and CMS measurements. Also shown are the results from each experiment. 
The values involving cross sections are given for $\sqrt{s}=8$~TeV, assuming the SM values for $\sigma_i(7~\TeV)/\sigma_i(8~\TeV)$.
The error bars indicate the $1\sigma$~(thick lines) and $2\sigma$~(thin lines) intervals. 
In this figure, the fit results are normalised to the SM predictions for the various parameters and the shaded bands indicate the theoretical uncertainties in these predictions.
}
\label{fig:plot_B1WW}
\end{figure}

\clearpage

\begin{sidewaystable}[hbtp]
  \centering
  \caption{Best fit values of $\Cc_{gZ}= \Cc_{g}\cdot\Cc_{Z} / \Cc_{H} $ and of the ratios of coupling modifiers, as defined in the most generic parameterisation described in the context of the $\Cc$~framework, from the combined analysis of the $\sqrt{s}=7$ and 8~TeV data.
The results are shown for the combination of ATLAS and CMS and also separately for each experiment, together with their total uncertainties and their breakdown into the four components described in the text.
The uncertainties in $\lambda_{tg}$ and $\lambda_{WZ}$, for which a negative solution is allowed, are calculated around the overall best fit value. The combined $1\sigma$~CL intervals are $\lambda_{tg} = [-2.00,-1.59] \cup [1.50,2.07]$ and $\lambda_{WZ} = [-0.96,-0.82] \cup [0.80,0.98]$.
The expected total uncertainties in the measurements are also shown in parentheses. 
For those parameters with no sensitivity to the sign, only the absolute values are shown.
}
  \label{tab:genericcouplings_full}
  \begin{adjustbox}{width=\textwidth}%
    \setlength\extrarowheight{3pt}%
    \begin{tabular}{l|r@{\hskip 0.5ex}lcccc|r@{\hskip 0.5ex}lcccc|r@{\hskip 0.5ex}lcccc}\hline\hline
       Parameter    & \multicolumn{2}{c}{Best fit} &  \multicolumn{4}{c|}{Uncertainty}  & \multicolumn{2}{c}{Best fit} &  \multicolumn{4}{c|}{Uncertainty} & \multicolumn{2}{c}{Best fit} &  \multicolumn{4}{c}{Uncertainty}     \\[-3pt]
                                             & \multicolumn{2}{c}{value}  & Stat                  & Expt                  &  Thbgd   & Thsig  &  \multicolumn{2}{c}{value} & Stat                  & Expt                  &  Thbgd   & Thsig&  \multicolumn{2}{c}{value} & Stat           & Expt                  &  Thbgd   & Thsig   \\
      \hline &  \multicolumn{6}{c|}{ATLAS+CMS} & \multicolumn{6}{c|}{ATLAS} &  \multicolumn{6}{c}{CMS}  \\
$\kappa_{gZ}$
& 1.09 & $       ^{+0.11}_{-0.11}     $ & $       ^{+0.09}_{-0.09}     $ & $       ^{+0.02}_{-0.02}     $ & $       ^{+0.00}_{-0.01}     $ & $       ^{+0.06}_{-0.05}     $          
& 1.20 & $       ^{+0.16}_{-0.15}     $ & $       ^{+0.14}_{-0.14}     $ & $       ^{+0.03}_{-0.03}     $ & $       ^{+0.02}_{-0.02}     $ & $       ^{+0.07}_{-0.06}     $          
& 0.99 & $       ^{+0.14}_{-0.13}     $ & $       ^{+0.12}_{-0.12}     $ & $       ^{+0.03}_{-0.04}     $ & $       ^{+0.01}_{-0.01}     $ & $       ^{+0.06}_{-0.04}     $ \\       
&      & $\Bgl({}^{+0.11}_{-0.11}\Big)$ & $\Big({}^{+0.09}_{-0.09}\Big)$ & $\Big({}^{+0.02}_{-0.02}\Big)$ & $\Big({}^{+0.01}_{-0.01}\Big)$ & $\Big({}^{+0.06}_{-0.05}\Big)$          
&      & $\Bgl({}^{+0.15}_{-0.15}\Big)$ & $\Big({}^{+0.14}_{-0.13}\Big)$ & $\Big({}^{+0.03}_{-0.03}\Big)$ & $\Big({}^{+0.01}_{-0.02}\Big)$ & $\Big({}^{+0.06}_{-0.05}\Big)$          
&      & $\Bgl({}^{+0.14}_{-0.14}\Big)$ & $\Big({}^{+0.13}_{-0.12}\Big)$ & $\Big({}^{+0.03}_{-0.03}\Big)$ & $\Big({}^{+0.01}_{-0.01}\Big)$ & $\Big({}^{+0.06}_{-0.05}\Big)$ \\[1mm]  
$\lambda_{Zg}$
& 1.27 & $       ^{+0.23}_{-0.20}     $ & $       ^{+0.18}_{-0.16}     $ & $       ^{+0.10}_{-0.07}     $ & $       ^{+0.06}_{-0.05}     $ & $       ^{+0.10}_{-0.08}     $          
& 1.07 & $       ^{+0.26}_{-0.22}     $ & $       ^{+0.21}_{-0.18}     $ & $       ^{+0.10}_{-0.06}     $ & $       ^{+0.07}_{-0.06}     $ & $       ^{+0.09}_{-0.07}     $          
& 1.47 & $       ^{+0.45}_{-0.34}     $ & $       ^{+0.35}_{-0.28}     $ & $       ^{+0.22}_{-0.14}     $ & $       ^{+0.11}_{-0.10}     $ & $       ^{+0.13}_{-0.09}     $ \\       
&      & $\Bgl({}^{+0.20}_{-0.17}\Big)$ & $\Big({}^{+0.15}_{-0.14}\Big)$ & $\Big({}^{+0.08}_{-0.06}\Big)$ & $\Big({}^{+0.05}_{-0.04}\Big)$ & $\Big({}^{+0.08}_{-0.07}\Big)$          
&      & $\Bgl({}^{+0.28}_{-0.23}\Big)$ & $\Big({}^{+0.23}_{-0.20}\Big)$ & $\Big({}^{+0.10}_{-0.07}\Big)$ & $\Big({}^{+0.09}_{-0.05}\Big)$ & $\Big({}^{+0.09}_{-0.07}\Big)$          
&      & $\Bgl({}^{+0.27}_{-0.23}\Big)$ & $\Big({}^{+0.21}_{-0.19}\Big)$ & $\Big({}^{+0.12}_{-0.09}\Big)$ & $\Big({}^{+0.07}_{-0.05}\Big)$ & $\Big({}^{+0.09}_{-0.07}\Big)$ \\[1mm]  
$\lambda_{tg}$
& 1.78 & $       ^{+0.30}_{-0.27}     $ & $       ^{+0.21}_{-0.20}     $ & $       ^{+0.13}_{-0.11}     $ & $       ^{+0.09}_{-0.09}     $ & $       ^{+0.14}_{-0.11}     $          
& 1.40 & $       ^{+0.34}_{-0.33}     $ & $       ^{+0.25}_{-0.24}     $ & $       ^{+0.14}_{-0.15}     $ & $       ^{+0.12}_{-0.14}     $ & $       ^{+0.14}_{-0.09}     $          
& -2.26 & $       ^{+0.50}_{-0.53}     $ & $       ^{+0.43}_{-0.39}     $ & $       ^{+0.22}_{-0.23}     $ & $       ^{+0.04}_{-0.18}     $ & $       ^{+0.14}_{-0.21}     $ \\       
&      & $\Bgl({}^{+0.28}_{-0.38}\Big)$ & $\Big({}^{+0.20}_{-0.30}\Big)$ & $\Big({}^{+0.11}_{-0.13}\Big)$ & $\Big({}^{+0.14}_{-0.20}\Big)$ & $\Big({}^{+0.09}_{-0.05}\Big)$          
&      & $\Bgl({}^{+0.38}_{-0.54}\Big)$ & $\Big({}^{+0.28}_{-0.39}\Big)$ & $\Big({}^{+0.14}_{-0.22}\Big)$ & $\Big({}^{+0.18}_{-0.29}\Big)$ & $\Big({}^{+0.11}_{-0.06}\Big)$          
&      & $\Bgl({}^{+0.42}_{-0.64}\Big)$ & $\Big({}^{+0.31}_{-0.42}\Big)$ & $\Big({}^{+0.16}_{-0.22}\Big)$ & $\Big({}^{+0.21}_{-0.43}\Big)$ & $\Big({}^{+0.11}_{-0.06}\Big)$ \\[1mm]  
$\lambda_{WZ}$
& 0.88 & $       ^{+0.10}_{-0.09}     $ & $       ^{+0.09}_{-0.08}     $ & $       ^{+0.03}_{-0.03}     $ & $       ^{+0.03}_{-0.02}     $ & $       ^{+0.02}_{-0.01}     $          
& 0.92 & $       ^{+0.14}_{-0.12}     $ & $       ^{+0.13}_{-0.11}     $ & $       ^{+0.04}_{-0.03}     $ & $       ^{+0.03}_{-0.03}     $ & $       ^{+0.02}_{-0.02}     $          
& -0.85 & $       ^{+0.13}_{-0.15}     $ & $       ^{+0.11}_{-0.13}     $ & $       ^{+0.05}_{-0.06}     $ & $       ^{+0.04}_{-0.04}     $ & $       ^{+0.01}_{-0.03}     $ \\       
&      & $\Bgl({}^{+0.12}_{-0.10}\Big)$ & $\Big({}^{+0.11}_{-0.09}\Big)$ & $\Big({}^{+0.04}_{-0.03}\Big)$ & $\Big({}^{+0.03}_{-0.03}\Big)$ & $\Big({}^{+0.02}_{-0.01}\Big)$          
&      & $\Bgl({}^{+0.18}_{-0.15}\Big)$ & $\Big({}^{+0.17}_{-0.13}\Big)$ & $\Big({}^{+0.04}_{-0.04}\Big)$ & $\Big({}^{+0.04}_{-0.04}\Big)$ & $\Big({}^{+0.02}_{-0.02}\Big)$          
&      & $\Bgl({}^{+0.17}_{-0.14}\Big)$ & $\Big({}^{+0.15}_{-0.13}\Big)$ & $\Big({}^{+0.06}_{-0.05}\Big)$ & $\Big({}^{+0.03}_{-0.03}\Big)$ & $\Big({}^{+0.03}_{-0.02}\Big)$ \\[1mm]  
$|\lambda_{\gamma Z}|$
& 0.89 & $       ^{+0.11}_{-0.10}     $ & $       ^{+0.10}_{-0.09}     $ & $       ^{+0.03}_{-0.02}     $ & $       ^{+0.01}_{-0.02}     $ & $       ^{+0.02}_{-0.01}     $          
& 0.87 & $       ^{+0.15}_{-0.13}     $ & $       ^{+0.15}_{-0.13}     $ & $       ^{+0.04}_{-0.04}     $ & $       ^{+0.02}_{-0.01}     $ & $       ^{+0.02}_{-0.02}     $          
& 0.91 & $       ^{+0.17}_{-0.14}     $ & $       ^{+0.16}_{-0.14}     $ & $       ^{+0.04}_{-0.03}     $ & $       ^{+0.02}_{-0.02}     $ & $       ^{+0.02}_{-0.02}     $ \\       
&      & $\Bgl({}^{+0.13}_{-0.12}\Big)$ & $\Big({}^{+0.13}_{-0.11}\Big)$ & $\Big({}^{+0.03}_{-0.02}\Big)$ & $\Big({}^{+0.01}_{-0.01}\Big)$ & $\Big({}^{+0.02}_{-0.01}\Big)$          
&      & $\Bgl({}^{+0.20}_{-0.17}\Big)$ & $\Big({}^{+0.20}_{-0.17}\Big)$ & $\Big({}^{+0.05}_{-0.03}\Big)$ & $\Big({}^{+0.03}_{-0.01}\Big)$ & $\Big({}^{+0.02}_{-0.02}\Big)$          
&      & $\Bgl({}^{+0.18}_{-0.16}\Big)$ & $\Big({}^{+0.18}_{-0.15}\Big)$ & $\Big({}^{+0.04}_{-0.03}\Big)$ & $\Big({}^{+0.01}_{-0.01}\Big)$ & $\Big({}^{+0.02}_{-0.02}\Big)$ \\[1mm]  
$|\lambda_{\tau Z}|$
& 0.85 & $       ^{+0.13}_{-0.12}     $ & $       ^{+0.12}_{-0.10}     $ & $       ^{+0.07}_{-0.06}     $ & $       ^{+0.01}_{-0.02}     $ & $       ^{+0.02}_{-0.01}     $          
& 0.96 & $       ^{+0.21}_{-0.18}     $ & $       ^{+0.18}_{-0.15}     $ & $       ^{+0.10}_{-0.09}     $ & $       ^{+0.04}_{-0.03}     $ & $       ^{+0.03}_{-0.02}     $          
& 0.78 & $       ^{+0.20}_{-0.17}     $ & $       ^{+0.17}_{-0.15}     $ & $       ^{+0.10}_{-0.08}     $ & $       ^{+0.01}_{-0.02}     $ & $       ^{+0.02}_{-0.01}     $ \\       
&      & $\Bgl({}^{+0.17}_{-0.15}\Big)$ & $\Big({}^{+0.14}_{-0.13}\Big)$ & $\Big({}^{+0.09}_{-0.08}\Big)$ & $\Big({}^{+0.02}_{-0.02}\Big)$ & $\Big({}^{+0.02}_{-0.02}\Big)$          
&      & $\Bgl({}^{+0.27}_{-0.23}\Big)$ & $\Big({}^{+0.23}_{-0.19}\Big)$ & $\Big({}^{+0.13}_{-0.11}\Big)$ & $\Big({}^{+0.04}_{-0.03}\Big)$ & $\Big({}^{+0.04}_{-0.02}\Big)$          
&      & $\Bgl({}^{+0.23}_{-0.20}\Big)$ & $\Big({}^{+0.19}_{-0.17}\Big)$ & $\Big({}^{+0.12}_{-0.11}\Big)$ & $\Big({}^{+0.02}_{-0.01}\Big)$ & $\Big({}^{+0.02}_{-0.02}\Big)$ \\[1mm]  
$|\lambda_{bZ}|$
& 0.58 & $       ^{+0.16}_{-0.20}     $ & $       ^{+0.12}_{-0.17}     $ & $       ^{+0.07}_{-0.06}     $ & $       ^{+0.07}_{-0.07}     $ & $       ^{+0.03}_{-0.04}     $          
& 0.61 & $       ^{+0.24}_{-0.24}     $ & $       ^{+0.20}_{-0.19}     $ & $       ^{+0.09}_{-0.12}     $ & $       ^{+0.10}_{-0.10}     $ & $       ^{+0.04}_{-0.03}     $          
& 0.47 & $       ^{+0.26}_{-0.17}     $ & $       ^{+0.23}_{-0.13}     $ & $       ^{+0.06}_{-0.07}     $ & $       ^{+0.11}_{-0.08}     $ & $       ^{+0.00}_{-0.04}     $ \\       
&      & $\Bgl({}^{+0.25}_{-0.22}\Big)$ & $\Big({}^{+0.21}_{-0.20}\Big)$ & $\Big({}^{+0.09}_{-0.07}\Big)$ & $\Big({}^{+0.08}_{-0.05}\Big)$ & $\Big({}^{+0.06}_{-0.05}\Big)$          
&      & $\Bgl({}^{+0.36}_{-0.29}\Big)$ & $\Big({}^{+0.31}_{-0.26}\Big)$ & $\Big({}^{+0.12}_{-0.09}\Big)$ & $\Big({}^{+0.11}_{-0.08}\Big)$ & $\Big({}^{+0.08}_{-0.05}\Big)$          
&      & $\Bgl({}^{+0.38}_{-0.37}\Big)$ & $\Big({}^{+0.32}_{-0.34}\Big)$ & $\Big({}^{+0.15}_{-0.11}\Big)$ & $\Big({}^{+0.11}_{-0.09}\Big)$ & $\Big({}^{+0.07}_{-0.07}\Big)$ \\[1mm]  

      \hline\hline
    \end{tabular}
  \end{adjustbox}
\end{sidewaystable}

\clearpage

\section{Likelihood scans for coupling modifier parameterisations}
\label{sec:app_negative}
For the results based on certain coupling modifier parameterisations described in Sections \ref{sec:kappaParam} and \ref{sec:CouplingFits}, it is necessary to account for the relative signs between parameters that modify the rates of certain signal production processes and decay modes through interference effects. For example, in the generic parameterisation in terms of ratios of coupling modifiers, the signs of $\lambda_{Zg}$, $\lambda_{WZ}$, and $\lambda_{tg}$ affect the rates of $\atH$ production through $t$--$W$~interference, and of $\aggZH$ production through $t$--$Z$~interference. The parameters~$\lambda_{Zg}$ and, as a consequence,~$\kappa_{gZ}$ are assumed to be positive without loss of generality. From this follows that there are four relevant sign combinations of~$\lambda_{WZ}$ and~$\lambda_{tg}$, which must be evaluated when performing all likelihood scans. An example is given in~Fig.~\ref{fig:lambdaBZ-negatives} for~$|\lambda_{bZ}|$, with a separate curve shown for each sign combination. Each sign hypothesis gives rise to a distinct local minimum. As the negative log-likelihood for each case is determined relative to a common reference point, it is possible to identify the global minimum. For this parameterisation, the SM~sign hypothesis~($\lambda_{WZ}>0,\lambda_{tg}>0$) corresponds to this global minimum. 
A new negative log-likelihood curve is defined as the envelope of the ones obtained for the different sign hypotheses, by taking the smallest value of~$-2\ln \Lambda$ from the different sign hypotheses as a function of the parameter being considered in the scan. This curve, indicated by the solid line in~Fig.~\ref{fig:lambdaBZ-negatives}, is used to determine the uncertainties and the confidence intervals. In the case of the example chosen here, but also more generally, this procedure results in larger confidence intervals than would be found from the ($\lambda_{WZ} > 0, \lambda_{tg} > 0$)~hypothesis alone.

Another example of the effect of the different sign combinations is given in~Fig.~\ref{fig:BRBSM-signs}, which shows the observed and expected negative log-likelihood scans for~$\BRbsm$ using the parameterisation described in~Section~\ref{sec:ModelK2}. Given that in this parameterisation, opposite signs of $\kappa_{W}$ and $\kappa_{Z}$ are not considered, only two sign combination hypotheses, $\kappa_{W} > 0, \kappa_{Z} >0$ and $\kappa_{W} < 0, \kappa_{Z} < 0$, are evaluated. In the expected negative log-likelihood curve, a transition between the two hypotheses occurs at $\BRbsm\approx 0.175$. This has the effect of increasing the expected 95\%~CL~upper limit on~$\BRbsm$ from~0.35, when considering only the $\kappa_{W} > 0, \kappa_{Z} >0$ case, to~0.39, once both sign combinations are considered.

\begin{figure}[hbt!]
  \center
   \includegraphics[width=0.8\textwidth]{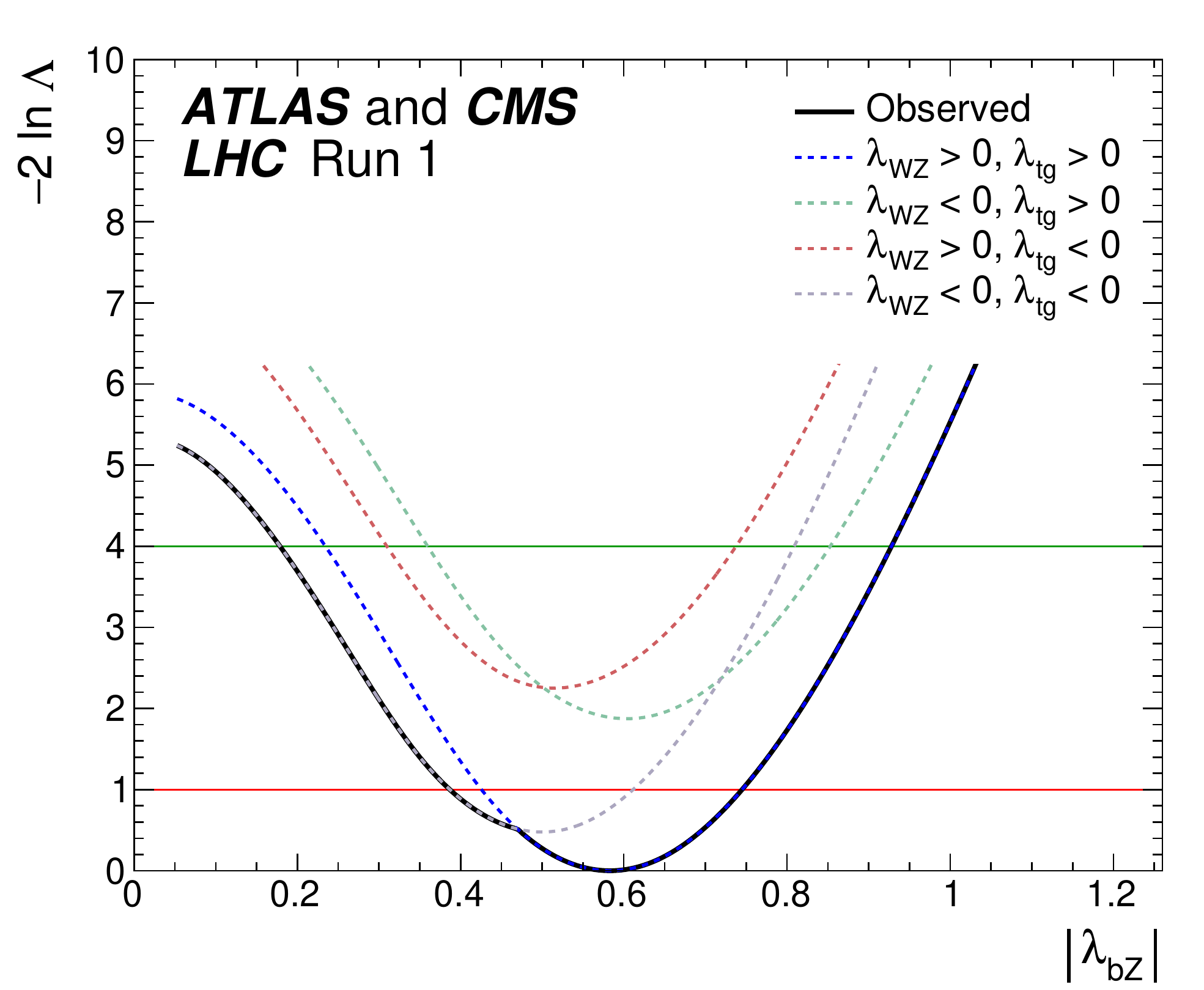}
  \caption{Negative log-likelihood scan for $\lambda_{bZ}$ showing the minima obtained when considering all sign combinations (solid line) and each specific one separately (dashed lines).}
\label{fig:lambdaBZ-negatives}
\end{figure}

\begin{figure}[hbt!]
  \center
   \includegraphics[width=0.48\textwidth]{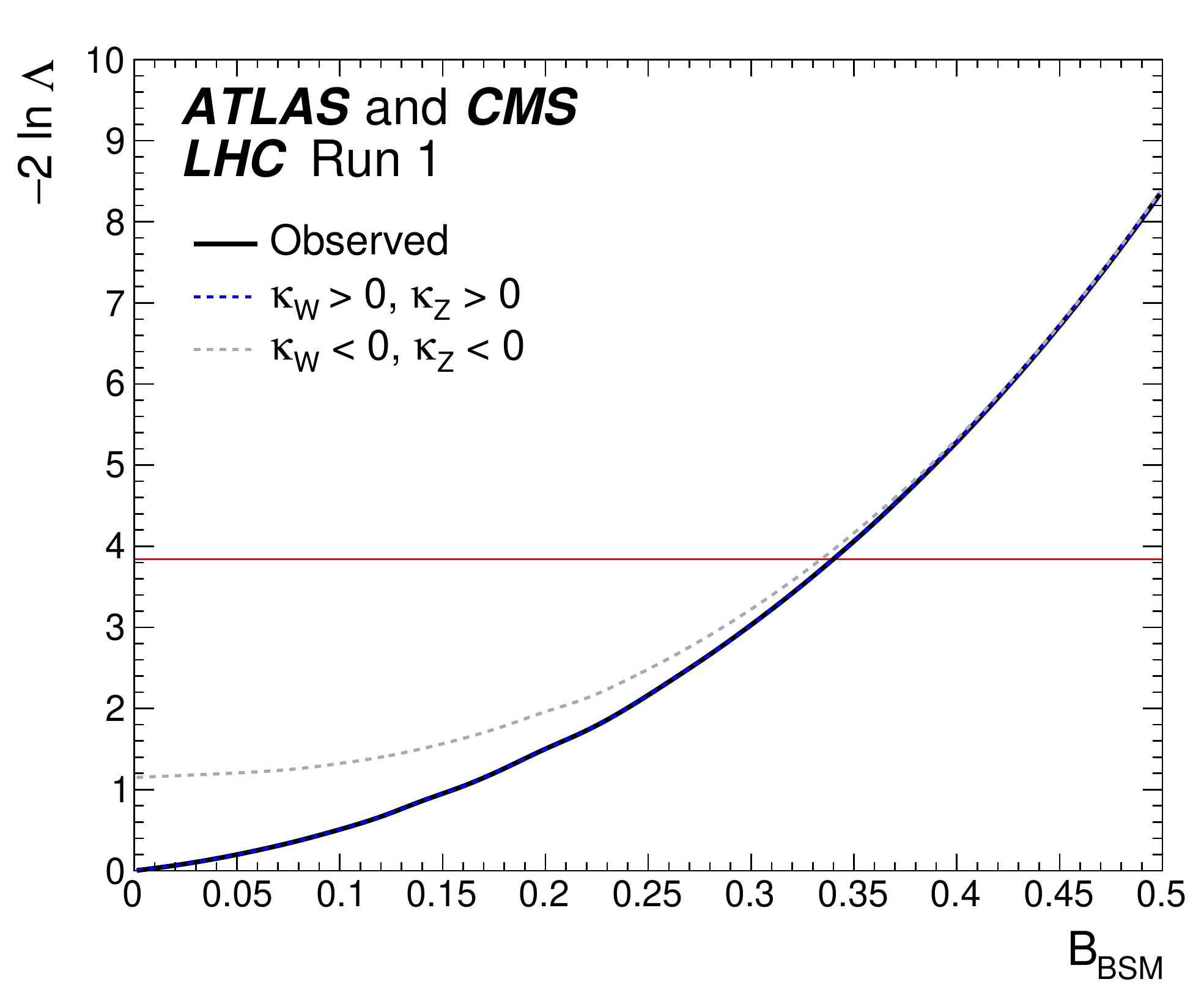}
   \includegraphics[width=0.48\textwidth]{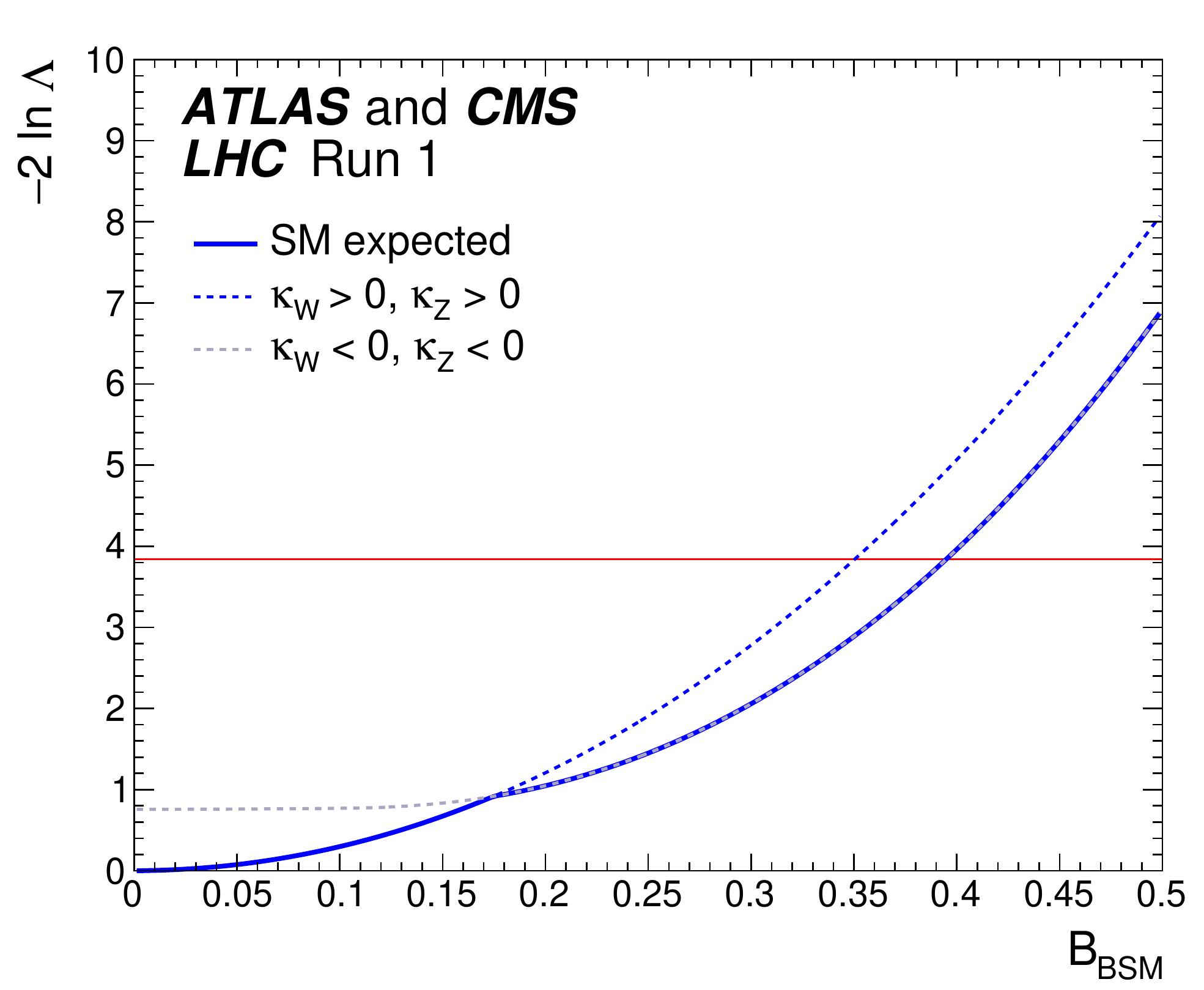}
  \caption{Observed (left) and expected (right) negative log-likelihood scan for $\BRbsm$, the minima obtained when considering both sign combinations (solid line) and each specific one separately (dashed lines).}
\label{fig:BRBSM-signs}
\end{figure}

\clearpage


\bibliographystyle{bibtex/bst/atlasBibStyleWithTitle}
\bibliography{Coupling2015}


\newpage 
\begin{flushleft}
{\Large The ATLAS Collaboration}

\bigskip

G.~Aad$^\textrm{\scriptsize 87}$,
B.~Abbott$^\textrm{\scriptsize 114}$,
J.~Abdallah$^\textrm{\scriptsize 65}$,
O.~Abdinov$^\textrm{\scriptsize 12}$,
B.~Abeloos$^\textrm{\scriptsize 118}$,
R.~Aben$^\textrm{\scriptsize 108}$,
O.S.~AbouZeid$^\textrm{\scriptsize 138}$,
N.L.~Abraham$^\textrm{\scriptsize 150}$,
H.~Abramowicz$^\textrm{\scriptsize 154}$,
H.~Abreu$^\textrm{\scriptsize 153}$
% ...
\bigskip
\\
% ...
$^{12}$ Institute of Physics, Azerbaijan Academy of Sciences, Baku, Azerbaijan\\
% ...
$^{65}$ University of Iowa, Iowa City IA, United States of America\\
% ...
$^{87}$ CPPM, Aix-Marseille Universit{\'e} and CNRS/IN2P3, Marseille, France\\
% ...
$^{108}$ Nikhef National Institute for Subatomic Physics and University of Amsterdam, Amsterdam, Netherlands\\
% ...
$^{114}$ Homer L. Dodge Department of Physics and Astronomy, University of Oklahoma, Norman OK, United States of America\\
% ...
$^{118}$ LAL, Univ. Paris-Sud, CNRS/IN2P3, Universit{\'e} Paris-Saclay, Orsay, France\\
% ...
$^{138}$ Santa Cruz Institute for Particle Physics, University of California Santa Cruz, Santa Cruz CA, United States of America\\
% ...
$^{150}$ Department of Physics and Astronomy, University of Sussex, Brighton, United Kingdom\\
% ...
$^{153}$ Department of Physics, Technion: Israel Institute of Technology, Haifa, Israel\\
$^{154}$ Raymond and Beverly Sackler School of Physics and Astronomy, Tel Aviv University, Tel Aviv, Israel\\
% ...

\end{flushleft}

\newpage \newskip{\cmsinstskip} \cmsinstskip=0pt plus 4pt
\newskip{\cmsauthskip} \cmsauthskip=16pt

\begin{flushleft}
{\Large The CMS Collaboration}

\bigskip

\textbf{Yerevan~Physics~Institute,~Yerevan,~Armenia}\\*[0pt]
V.~Khachatryan,~A.M.~Sirunyan,~A.~Tumasyan
\vskip\cmsinstskip
\textbf{Institut~f\"{u}r~Hochenergiephysik~der~OeAW,~Wien,~Austria}\\*[0pt]
W.~Adam,~E.~Asilar,~T.~Bergauer,~J.~Brandstetter,~E.~Brondolin,~M.~Dragicevic,~J.~Er\"{o},~M.~Flechl,~M.~Friedl,~R.~Fr\"{u}hwirth\cmsAuthorMark{1},~V.M.~Ghete,~C.~Hartl,~N.~H\"{o}rmann,~J.~Hrubec,~M.~Jeitler\cmsAuthorMark{1},~A.~K\"{o}nig,~I.~Kr\"{a}tschmer,~D.~Liko,~T.~Matsushita,~I.~Mikulec,~D.~Rabady,~N.~Rad,~B.~Rahbaran,~H.~Rohringer,~J.~Schieck\cmsAuthorMark{1},~J.~Strauss,~W.~Treberer-Treberspurg,~W.~Waltenberger,~C.-E.~Wulz\cmsAuthorMark{1}
% ...
\vskip\cmsinstskip
% ...
1:~~Also at~Vienna~University~of~Technology,~Vienna,~Austria\\
% ...
12:~Also at~Fayoum~University,~El-Fayoum,~Egypt\\
% ...
65:~Also at~Utah~Valley~University,~Orem,~USA\\
% ...

\end{flushleft}



\end{document}